\documentclass[useAMS,usenatbib]{mn2e} 

\usepackage{times,array,graphicx,amssymb}

\newcommand{\ion}[2]{{#1~\small#2}}
\newcommand{\lya}{Ly$\alpha$}

\newcommand{\kms}{$\rm km~s^{-1}$}
\newcommand{\nhi}{$N_{\rm HI}$}

\title[Imaging DLAs at $z>2$ (II)]{Directly imaging damped Ly$\alpha$ 
galaxies at $z>2$. II: Imaging and spectroscopic observations of 32
quasar fields.}

\author[Fumagalli et al.]{Michele Fumagalli$^{1,2}$\thanks{E-mail: michele.fumagalli@durham.ac.uk}, 
 John M. O'Meara$^{3}$, J. Xavier Prochaska$^{4,5}$, Nissim
 Kanekar$^{6}$, \and Arthur M. Wolfe$^7$  \\
$^{1}$Institute for Computational Cosmology, Department of Physics, Durham University, 
South Road, Durham, DH1 3LE, UK \\
$^{2}$Carnegie Observatories, 813 Santa Barbara Street, Pasadena, CA 91101, USA \\
$^{3}$Department of Chemistry and Physics, Saint Michael's College, One Winooski Park, 
     Colchester, VT 05439, USA \\
$^{4}$Department of Astronomy and Astrophysics, University of California, 1156 High Street, 
     Santa Cruz, CA 95064 USA \\
$^{5}$University of California Observatories, Lick Observatory 1156 High Street, Santa Cruz, CA 95064 USA \\
$^{6}$National Centre for Radio Astrophysics, TIFR, Post Bag 3, Ganeshkhind, Pune 411 007, India \\ 
$^{7}$Department of Physics, and Center for Astrophysics and Space Sciences, University of California, San Diego, 9500 Gilman Dr., La Jolla, CA 92093-0424, USA \\}

\begin{document}

\date{Accepted xxxx. Received xxxx; in original form xxxx}

\pagerange{\pageref{firstpage}--\pageref{lastpage}} \pubyear{xxxx}

\maketitle

\label{firstpage}

\begin{abstract}
Damped Ly$\alpha$ absorbers (DLAs) are a well-studied class of absorption line systems,
and yet the properties of their host galaxies remain largely unknown. To investigate the 
origin of these systems, we have conducted an imaging survey of 32 quasar fields with intervening 
DLAs between $z\sim 1.9-3.8$, leveraging a technique that allows us to image galaxies at any small
angular separation from the background quasars. In this paper, we present the properties of the 
targeted DLA sample, new imaging observations of the quasar fields, and the analysis of new and 
archival spectra of the background quasars. 
\end{abstract}


\section{Introduction}

Damped Ly$\alpha$ absorbers (DLAs), the strongest \ion{H}{I} absorption line systems 
detected in the foreground of UV-bright sources, are easily identifiable in the spectra of 
high redshift quasars, to the point that recent surveys have uncovered more than 6000 DLAs 
\citep[e.g.][]{pro09,not09,not12} above $z \gtrsim 2$. Thanks to this large parent sample and dedicated 
follow-up observations, the statistical properties of DLAs, including their hydrogen 
distribution, their metallicity, and kinematics, are currently well measured 
between $z \sim 2-4$ \citep[e.g.][]{pro03,pro07,pro09,not09,not12,raf12,jor13,raf14,nee13,mol13,zaf13}. 
However, despite this detailed knowledge of the DLA properties, a fundamental question still remains 
open: what is the typical galaxy population that gives rise to DLAs?

\begin{figure*}
\centering
\includegraphics[scale=0.6]{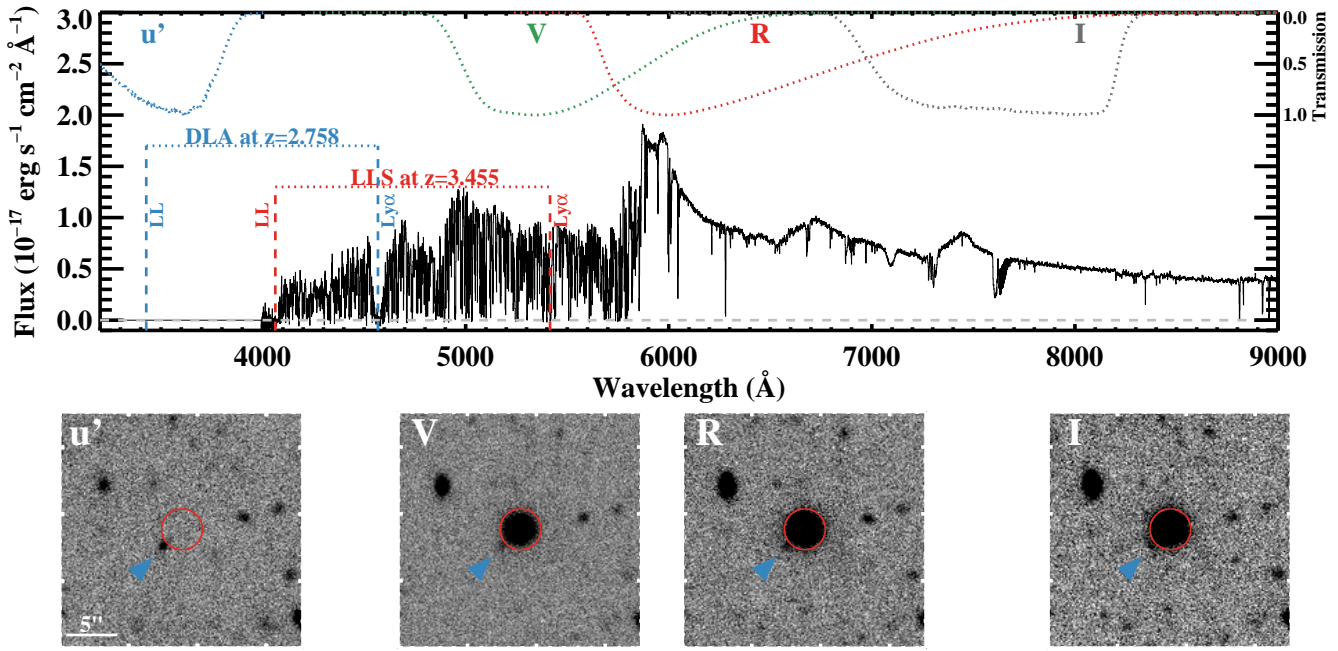}
\caption{{\it (Top:)} ESI/Keck spectrum of the quasar J094927+111518 with, superimposed, the 
  mirrored transmission curves of the $u'$, $V$, $R$, and $I$ filters of LRIS at Keck.
  The corresponding y-axis of the transmission curves is shown on the top right. The 
  \lya\ line and the Lyman limit of the targeted DLA and of the LLS that acts as a 
  blocking filter are also marked. {\it (Bottom:)} LRIS/Keck images of
  a $20'' \times 20''$ region centered at the quasar position in these four filters. Because of the 
  intervening LLS, the quasar light which dominates the inner $\sim 2''$ (red circles) in the redder filters is 
  instead completely absorbed in the $u'$ band. This enables the detection of faint 
  galaxies at all impact parameters, including objects that are aligned or in close 
  proximity to the quasar location. The blue arrow highlights one such galaxy.}\label{design}
\end{figure*}

The quest to find DLA host galaxies\footnote{We will often refer 
to the system seen in absorption as ``DLA gas'' or simply DLA, while we will 
refer to the (candidate) host galaxy as ``DLA galaxy''.} 
is a long-standing one \citep[for a review, see][]{wol05}. More than 40 years after the discovery of DLAs 
\citep{wol86}, many attempts to identify DLA galaxies have been pursued 
\citep[see e.g. Appendix B in][]{fum10b}, most of which have been unsuccessful. 
At least two reasons can justify the current lack of large samples of DLA galaxies. First, in order to 
detect $z\gtrsim 2$ galaxies in proximity or superimposed to bright ($m \lesssim 20$) quasars, 
observers face the challenging task of detecting faint sources against background fluxes that are 
at least a few orders of magnitude brighter \citep[e.g.][]{mol98}. 
This is especially true for galaxies lacking strong \lya\ emission. 
Second, if in fact a substantial fraction of DLA galaxies are fainter than $\sim 25$ mag,
as suggested by most theoretical studies \citep{nag07,cen12} and
some observations \citep[e.g.][]{fyn99,rau08}, then very deep imaging
surveys are needed. And while current $8-10$ m telescopes can reach
sensitive detection limits for imaging ($m \gtrsim 26-27$ mag), at
these magnitudes, spectroscopic redshifts for the candidate DLA
galaxies are extremely difficult (if not impossible) to obtain. Furthermore,
the number of candidates detected in proximity to the quasars rapidly
increases as one probes the fainter end of the galaxy luminosity
function, increasing the number of low- and high-redshift interlopers.  

Despite these challenges, previous efforts and especially 
more recent searches that have employed efficient spectroscopic
techniques \citep[e.g.][]{fyn10,per11,not12b,jor13b}, have resulted in
a dozen confirmed DLA galaxies. While useful for some investigations \citep{kro12}, 
this sample is obviously small compared to the known DLAs or compared to the galaxy populations 
that are selected with imaging techniques (e.g. the Lyman break galaxies or LBGs). 
Furthermore, the sample of confirmed host galaxies includes both serendipitous discoveries
and targeted observations of DLAs, which have been pre-selected according to their
absorption properties. It is therefore difficult to establish a rigorous census of 
non detections, critical to empirically constrain the luminosity function of DLA hosts.

To overcome some of these limitations, 
we have undertaken a new imaging survey that targets 32 quasar fields with 
intervening DLAs between $z\sim 1.9-3.8$. As we discuss in the following sections, 
this sample represents an unbiased selection with respect to DLA hydrogen 
column densities and metallicities. Moreover, we have targeted fields blindly, that is without prior knowledge 
of the presence of DLA galaxy candidates near the quasars. Therefore, 
the census of candidate DLA galaxies in these fields is representative of the generic 
population of DLAs, simply defined as absorbers with 
$\log N_{\rm HI} \ge 20.3~\rm cm^{-2}$.
Our survey resembles some of the previous HST imaging searches that have targeted 
quasar fields known to host absorbers which are representative of the general DLA population. 
Among those, we recall the survey conducted by \citet{war01}, who imaged 16 quasar fields 
with $z\sim 1.8-4.0$ DLAs across a wide range of column densities, extending the pioneering search 
for intermediate-redshift absorbers by \citet{leb97}. A novel key element in our survey, however, 
is the use of the technique discussed in Section \ref{sample}, which takes advantage of the 
presence of high-redshift absorption line systems along the line of sight to ``block'' the quasar 
glare \citep{ste92,ome06,chr09,fum10b}. This technique allows us to achieve the same sensitivity at any 
distance from the quasars, including angular separations as small as $\sim 0.2-1''$. 
However, as it will become clear from the following analysis, our deep imaging observations uncover galaxies 
that are fainter than $m \sim 25$ mag, the limit beyond which spectroscopic follow-up is currently too 
expensive even with the largest ground-based telescopes. Therefore, our study will be limited to a statistical 
analysis, which nevertheless will offer unique constraints on the properties of 
DLA galaxies.

This paper presents the results of our imaging campaign based on the mentioned technique 
which avoids the contamination from background quasars. 
The design of the survey has been presented in the first paper of the series \citep{fum10b}.
Here, after a brief review of the adopted technique and of the sample selection
(Section \ref{sample}), we present new ground-based and space-based imaging observations
for the 32 quasar fields, together with new and archival spectra of the studied quasars
(Section \ref{obserimg} and Section \ref{obserspe}). 
In the third paper of the series (Fumagalli et al. in prep.), we will 
use these data to study the {\it in-situ} SFRs of DLAs and the connection 
between DLAs and star-forming galaxies, also in comparison to previous work.

In this work, unless otherwise noted, distances are in proper units and magnitudes are in the 
AB system, and we adopt the following cosmological parameters: $H_0=70.4~\rm km~s^{-1}~Mpc^{-1}$, 
$\Omega_{\rm m}=0.27$ and $\Omega_\Lambda=0.73$ \citep{kom11}.

\begin{table*}
\caption{Summary of the sample properties.}\label{tabprop}
\centering
\begin{tabular}{l c c c c c c c c c c c}
\hline
Field$^{a}$     & R.A.$^{b}$       & Dec.$^{b}$      &$z_{\rm qso}^{c}$&$z_{\rm dla}^d$&$z_{\rm lls}^e$&$\lambda_{\rm LL,dla}^f$&$\lambda_{\rm LL,lls}^g$&$N_{\rm HI,dla}^h$       &$N_{\rm HI,lls}^i$       &$\rm [X/H]_{dla}^l$& Element$^m$\\
          & (J2000)    & (J2000)   &             &             &             &(\AA)             &(\AA)	           &$(\log~\rm cm^{-2})$  &$(\log~\rm cm^{-2})$ &    &  \\
\hline										  
    1:G1  & 21:14:43.95&-00:55:32.7&3.424 &2.9181 & 3.4420&3572 & 4050 &$20.25\pm 0.10$ & $ > 20.0^*        $ &$-0.63\pm0.11$   & S     \\ 
    2:G2  & 07:31:49.50&+28:54:48.6&3.676 &2.6878 & 3.6080&3362 & 4201 &$20.60\pm 0.10$ & $ > 17.6          $ &$-1.45\pm0.17$   & Si	 \\
    3:G3  & 09:56:04.43&+34:44:15.5&3.427 &2.3887 & 3.3958&3090 & 4008 &$21.10\pm 0.15$ & $ > 17.5          $ &$-1.00\pm0.17$   & Zn	 \\
    4:G4  & 23:43:49.41&-10:47:42.0&3.616 &2.6878 & 3.3652&3362 & 3980 &$20.60\pm 0.10$ & $ > 17.5          $ &$-1.27\pm0.20$   & Si,Zn \\
    5:G5  & 03:43:00.88&-06:22:29.9&3.623 &2.5713 & 3.5071&3256 & 4109 &$20.75\pm 0.20$ & $   19.95\pm0.15  $ &$-2.02\pm0.26$   & Fe    \\
    6:G6  & 23:51:52.80&+16:00:48.9&4.694 &3.7861 & 4.5835&4364 & 5091 &$20.85\pm 0.10$ & $ > 17.7     	    $ &$-2.03\pm0.20$   & Fe    \\
    7:G7  & 00:42:19.74&-10:20:09.4&3.880 &2.7544 & 3.6287&3423 & 4220 &$20.20\pm 0.10$ & $ > 17.7          $ &$-0.96\pm0.16$   & Fe    \\
    8:G9  & 09:49:27.88&+11:15:18.2&3.824 &2.7584 & 3.4559&3427 & 4063 &$20.85\pm 0.10$ & $ > 17.6     	    $ &$-0.95\pm0.10$   & Si    \\
    9:G10 & 10:18:06.28&+31:06:27.2&3.629 &2.4592 & 3.4812&3154 & 4086 &$20.35\pm 0.10$ & $   20.10\pm0.10  $ &$-1.19\pm0.28$   & Si,Zn \\
    10:G11& 08:51:43.72&+23:32:08.9&4.499 &3.5297 & 4.4671&4130 & 4985 &$21.10\pm 0.10$ & $ > 17.8          $ &$-1.05\pm0.15$   & Zn    \\
    11:G12& 09:56:05.09&+14:48:54.7&3.435 &2.6606 & 3.4759&3338 & 4081 &$20.85\pm 0.10$ & $   20.70\pm0.10^*$ &$-1.46\pm0.12$   & Si    \\
    12:G13& 11:51:30.48&+35:36:25.0&3.581 &2.5978 & 3.4193&3280 & 4029 &$20.90\pm 0.10$ & $ > 17.5          $ &$-1.28\pm0.11$   & Si    \\
    13:H1 & 21:23:57.56&-00:53:50.1&3.583 &2.7803 & 3.6251&3447 & 4217 &$20.70\pm 0.10$ & $ > 20.6^*        $ &$-1.59\pm0.15$   & Si,Zn \\
    14:H2 & 04:07:18.06&-44:10:14.0&3.000 &1.9127 & 2.6215&2656 & 3302 &$20.55\pm 0.10$ & $   20.45 \pm0.10 $ &$-0.77\pm0.11$   & Si    \\
    15:H3 & 02:55:18.58&+00:48:47.6&3.996 &3.2530 & 3.9147&3878 & 4481 &$20.60\pm 0.10$ & $ > 21.0^*        $ &$-0.80\pm0.11$   & Si    \\
    16:H4 & 08:16:18.99&+48:23:28.4&3.582 &2.7067 & 3.4366&3380 & 4045 &$20.70\pm 0.15$ & $   20.70\pm0.15  $ &$-2.36\pm0.15$   & Si    \\
    17:H5 & 09:30:51.93&+60:23:01.1&3.719 &3.0010 & 3.6373&3648 & 4228 &$21.05\pm 0.15$ & $   20.40\pm0.20  $ &-  	        & -     \\
    18:H6 & 09:08:10.36&+02:38:18.7&3.710 &2.9586 & 3.4071&3609 & 4018 &$21.10\pm 0.10$ & $   20.80\pm0.20  $ &$-0.93\pm0.12$   & Si    \\
    19:H7 & 12:20:21.39&+09:21:35.7&4.133 &3.3069 & 4.1215&3927 & 4670 &$20.40\pm 0.20$ & $ > 17.5	    $ &$-2.48\pm0.22$   & -     \\	   
    20:H8 & 14:42:33.01&+49:52:42.6&3.175 &2.6320 & 3.1124&3312 & 3750 &$20.35\pm 0.15$ & $   20.25\pm0.20  $ &-  	        & -     \\
    21:H9 & 08:44:24.24&+12:45:46.7&2.482 &1.8639 & 2.4762&2611 & 3169 &$21.00\pm 0.10$ & $   20.80\pm0.10  $ &$-1.54\pm0.12$   & Si    \\
    22:H10& 07:51:55.10&+45:16:19.6&3.341 &2.6826 & 3.2554&3358 & 3880 &$20.50\pm 0.10$ & $ > 17.5	    $ &$-1.16\pm0.13$   & Si    \\
    23:H11& 08:18:13.14&+07:20:54.9&4.177 &3.2332 & 3.8399&3860 & 4413 &$21.15\pm 0.10$ & $ > 17.5	    $ &$-1.41\pm0.25$   & Si,Zn \\
    24:H12& 08:18:13.05&+26:31:36.9&4.179 &3.5629 & 4.1629&4160 & 4707 &$20.65\pm 0.10$ & $   20.90\pm0.15^*$ &$-0.93\pm0.24$   & Si,Zn \\
    25:H13& 08:11:14.32&+39:36:33.2&3.073 &2.6500 & 3.0427&3328 & 3686 &$20.70\pm 0.15$ & $ > 20.0^*  	    $ &$-1.44\pm0.15$   & Si,Zn \\
    26:H14& 15:08:51.94&+51:56:27.7&3.804 &2.7333 & 3.5865&3404 & 4182 &$20.30\pm 0.20$ & $   20.80\pm0.20  $ &-  	        & -     \\			    
    27:H15& 10:54:30.07&+49:19:47.1&3.998 &2.9236 & 3.7016&3577 & 4287 &$20.45\pm 0.15$ & $ > 17.4          $ &-  	        & -     \\			   
    28:H16& 09:56:25.16&+47:34:42.5&4.478 &3.4035 & 4.2441&4015 & 4781 &$21.05\pm 0.10$ & $   20.80\pm0.15  $ &($-2.09,-1.50$)$^1$  & Si,Ni \\				   
    29:H17& 14:41:47.52&+54:15:38.1&3.467 &2.6289 & 3.3302&3309 & 3948 &$20.70\pm 0.15$ & $   20.30\pm0.15  $ &-  	        & -     \\			   
    30:H18& 11:55:38.60&+05:30:50.5&3.475 &2.6079 & 3.3260&3290 & 3944 &$20.35\pm 0.15$ & $   21.00\pm0.10  $ &$-1.60 \pm 0.16$ & Si    \\			   
    31:H19& 15:24:13.35&+43:05:37.4&3.920 &2.8721 & 3.8802&3530 & 4450 &$20.40\pm 0.15$ & $   20.65\pm0.15  $ &-  	        & -     \\			   
    32:H20& 13:20:05.97&+13:10:15.3&3.352 &2.6722 & 3.3411&3348 & 3958 &$20.30\pm 0.10$ & $   19.50\pm0.15  $ &$-2.30\pm0.10$   & Si    \\  
\hline 
\end{tabular}
\\\flushleft{$^a$ ID of the field. $^b$ Right ascension and declination of the quasar. $^c$ Quasar redshift. $^d$ Redshift of the targeted DLA.
$^e$  Redshift of the LLS acting as blocking filter. $^f$ Wavelength corresponding to the Lyman limit of the targeted DLA. 
$^g$ Wavelength corresponding to the Lyman limit of the higher redshift LLS. $^h$ \ion{H}{I} column density of the DLA. 
$^i$ \ion{H}{I} column density of the higher redshift LLS.  Asterisks mark DLAs that are proximite to the quasars and 
for which measurements are more uncertain. $^l$ DLA metallicity. $^m$ Tracer element used for the metallicity measurement.
$^1$ The listed values bracket the range of allowed metallicity given upper/lower limits.}   
\end{table*}

\section{Technique and Sample Selection}\label{sample}

To overcome the glare of the background quasar that would preclude the detection of 
faint galaxies at small projected separations, we select quasar fields that host both 
a DLA and a second optically-thick absorber along the line of sight. This
``Lyman limit technique'' has been already discussed elsewhere \citep{ste92,ome06,chr09,fum10b} and it is 
only briefly summarized in this section. 

We select quasar fields in which there are two optically-thick absorbers, the targeted 
DLA at redshift $z_{\rm dla}$ and a second optically-thick absorber at redshift $z_{\rm lls}$ 
that acts as a ``blocking filter'' for the quasar light. For this survey, we primarily select 
sightlines from the Sloan Digital Sky Survey (SDSS) DR5 quasar catalogue \citep{sch07}, with 
the addition of few quasars that lie outside of the SDSS footprint, but for which archival 
spectra are available.  To be considered for imaging follow-up, a quasar has to exhibit a 
DLA in its optical spectrum, which results in a redshift lower limit of $z_{\rm dla} \gtrsim 2.3$ 
for SDSS spectra, and a lower limit of $z_{\rm dla} \gtrsim 1.8$  for archival spectra with 
extended blue coverage down to the atmospheric cutoff. Further, each sightline has to have 
a second, optically-thick absorber with redshift $z_{\rm lls}$ such that the Lyman limit of 
this system ($\lambda_{\rm LL,lls}$) falls redward to the transmission curve of the blue filters available 
at ground-based observatories, or in the UVIS channel of the  Wide Field Camera 3 
(WFC3) camera on board of the \emph{Hubble Space Telescope} (HST). 
The filter choice is further dictated by the DLA redshift in order to prevent the Lyman limit 
of the DLA ($\lambda_{\rm LL,dla}$) from entering the transmission curve of the selected filter, or at least to 
minimize the impact of the intrinsic Lyman limit absorption on the final images.

\begin{figure*}
\begin{tabular}{cc}
\includegraphics[scale=0.32]{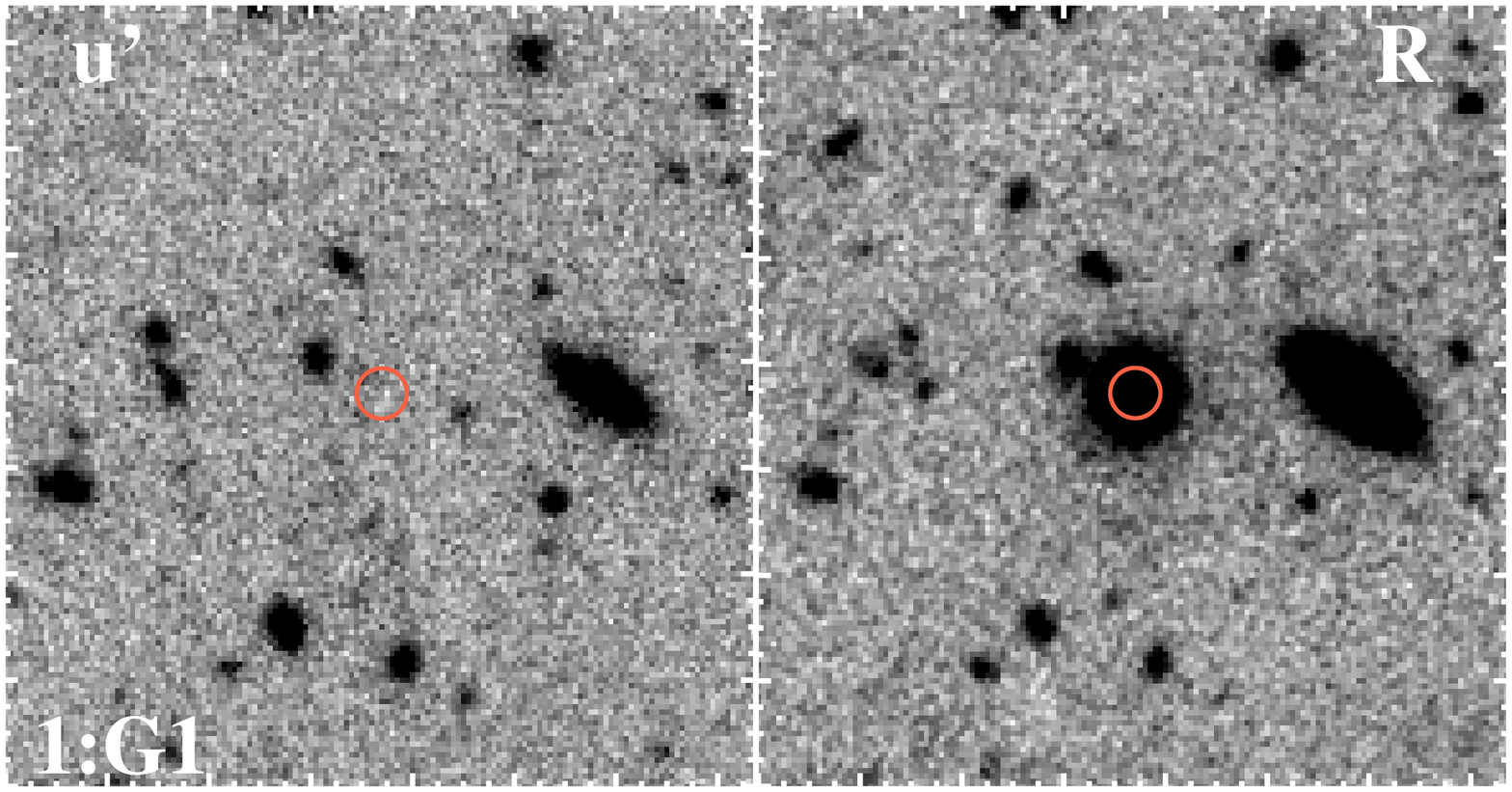}&
\includegraphics[scale=0.32]{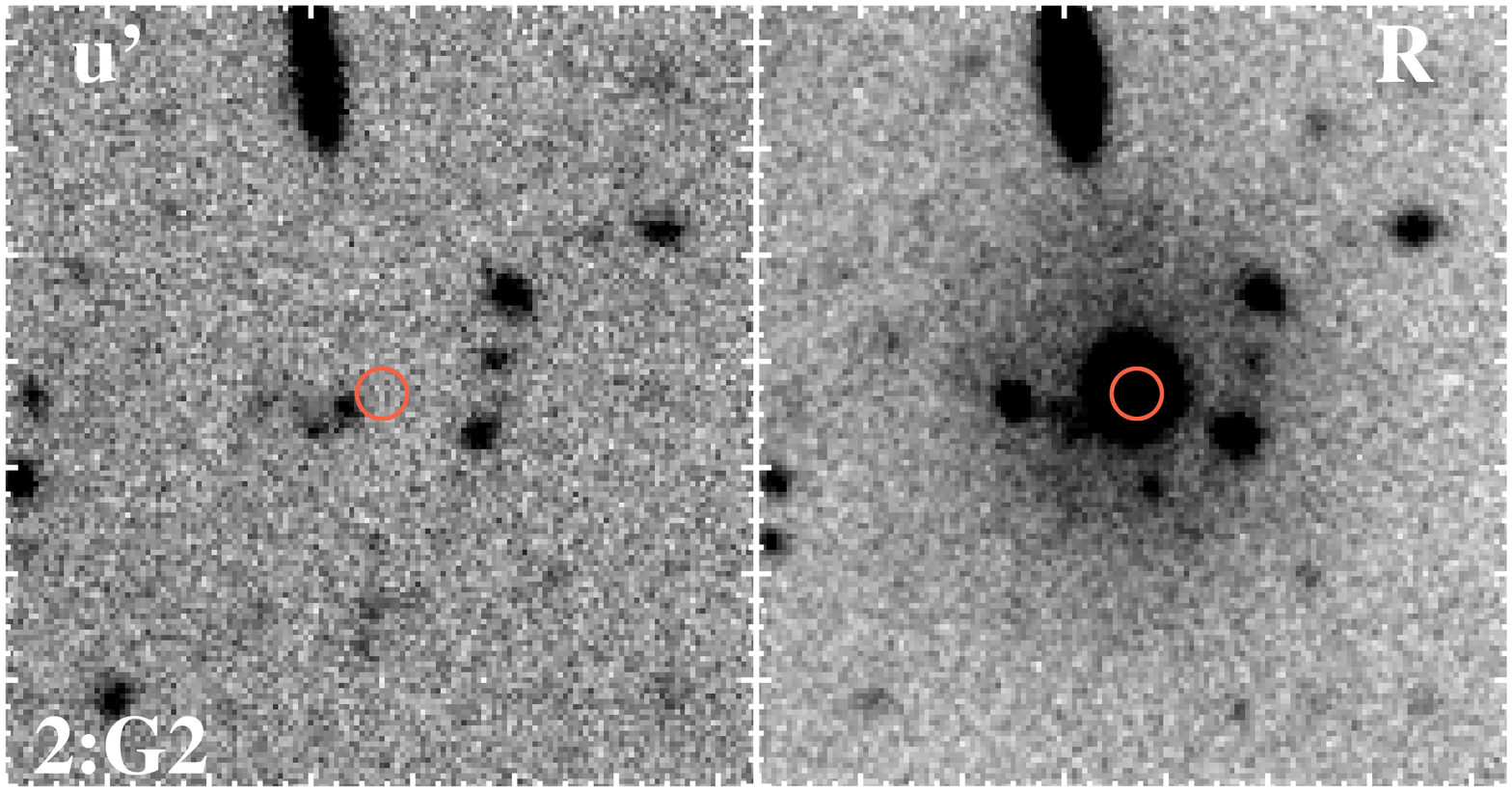}\\
\includegraphics[scale=0.32]{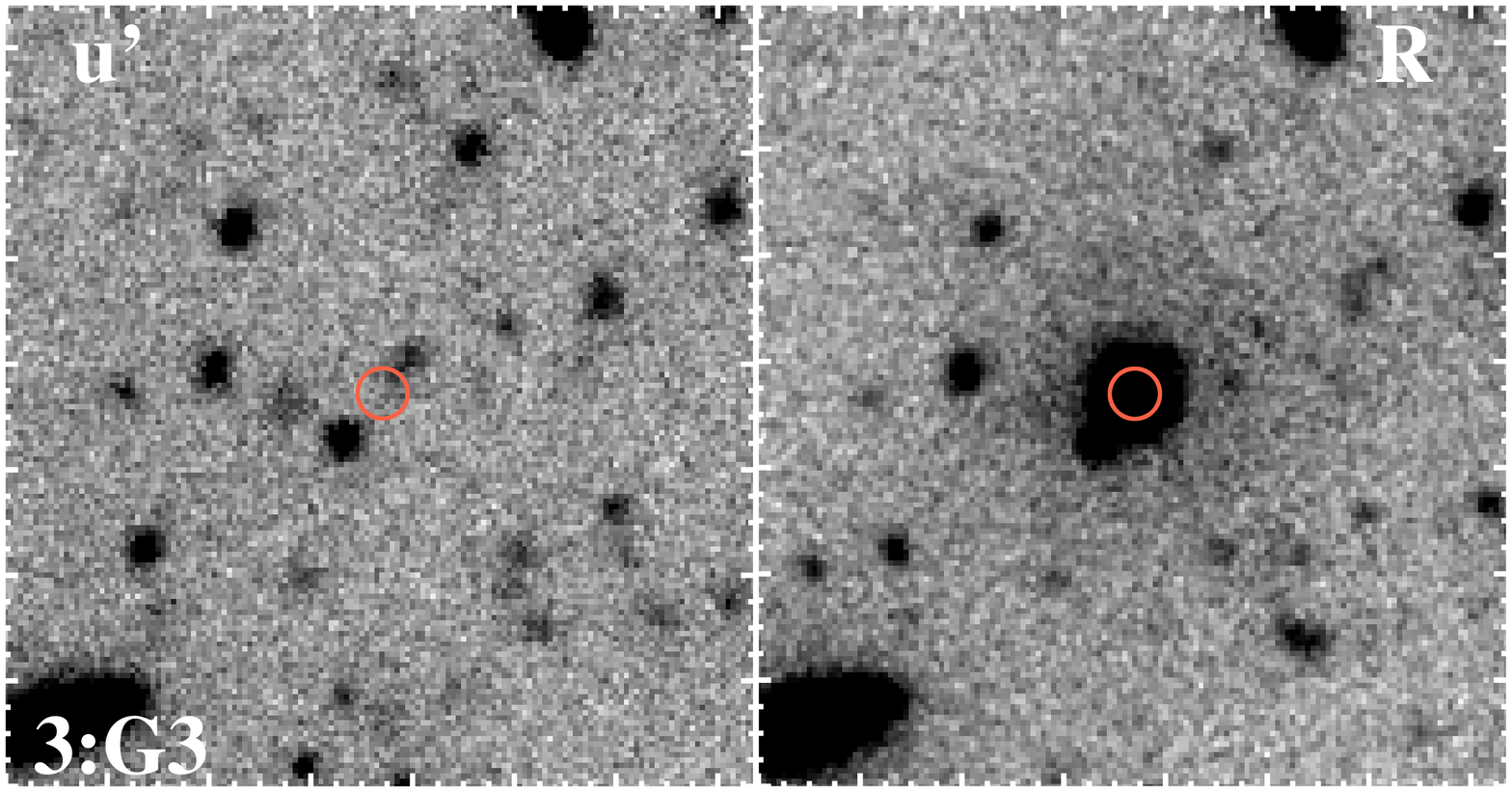}&
\includegraphics[scale=0.32]{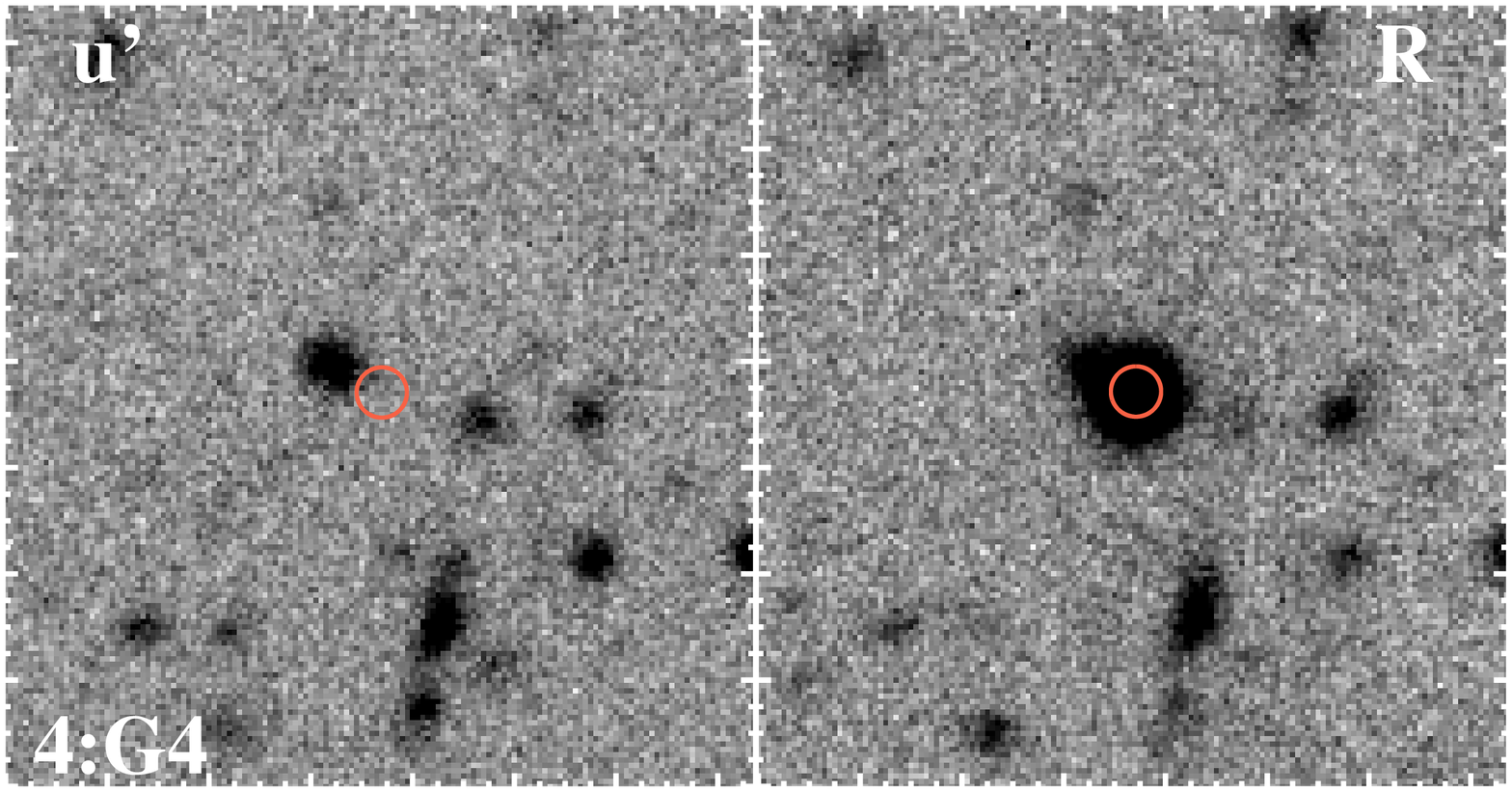}\\
\includegraphics[scale=0.32]{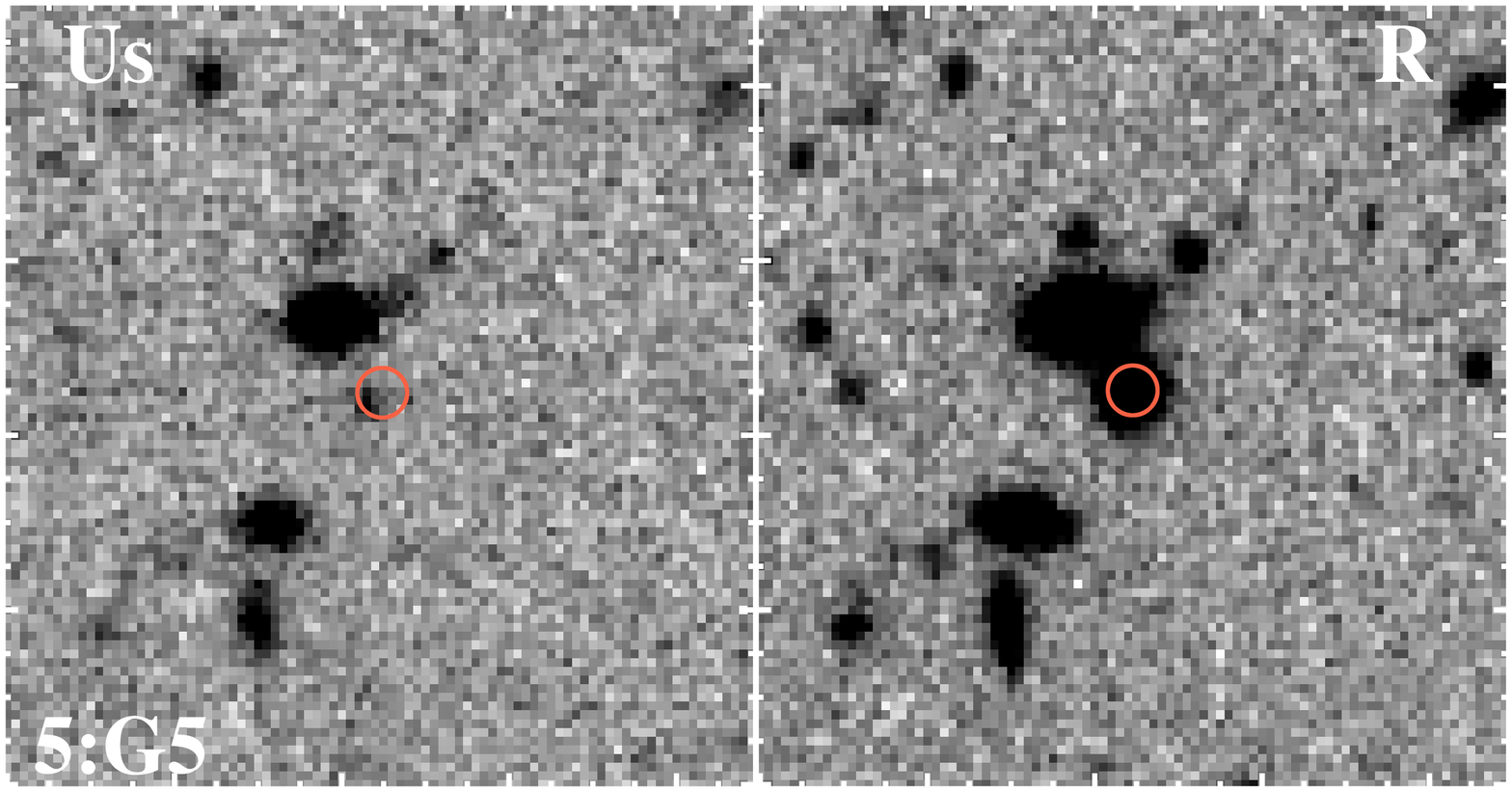}&
\includegraphics[scale=0.32]{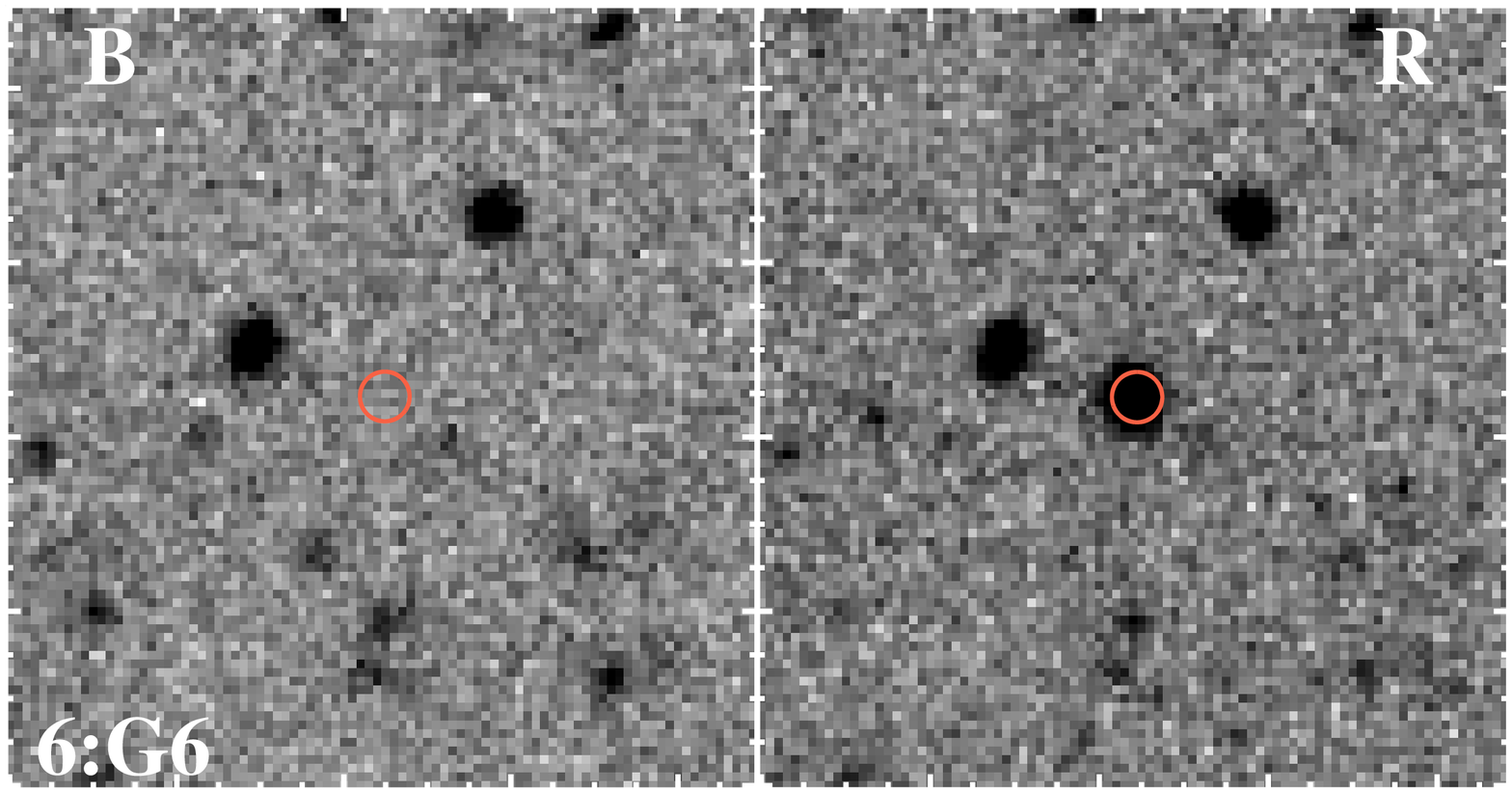}\\
\includegraphics[scale=0.32]{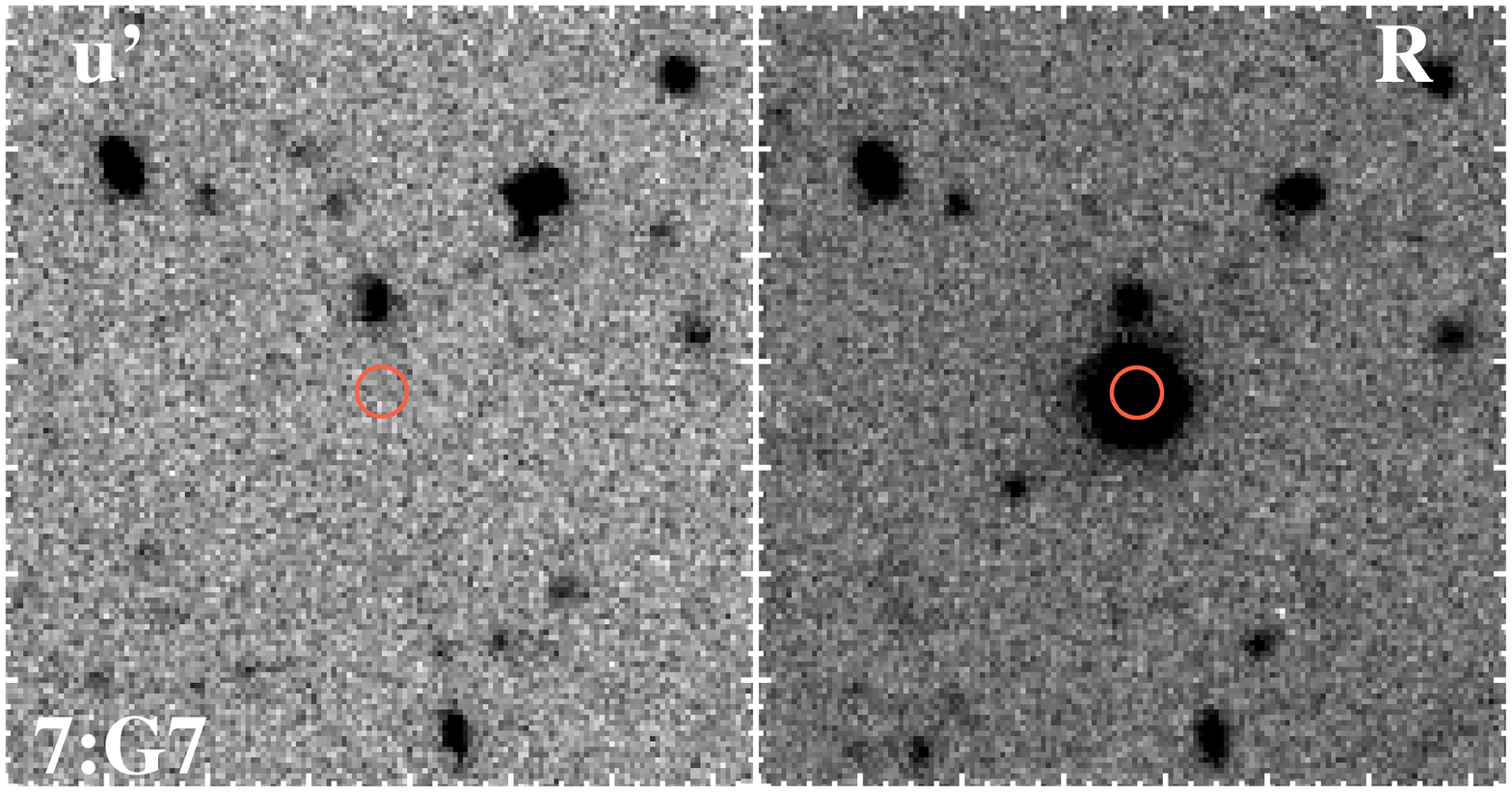}&
\includegraphics[scale=0.32]{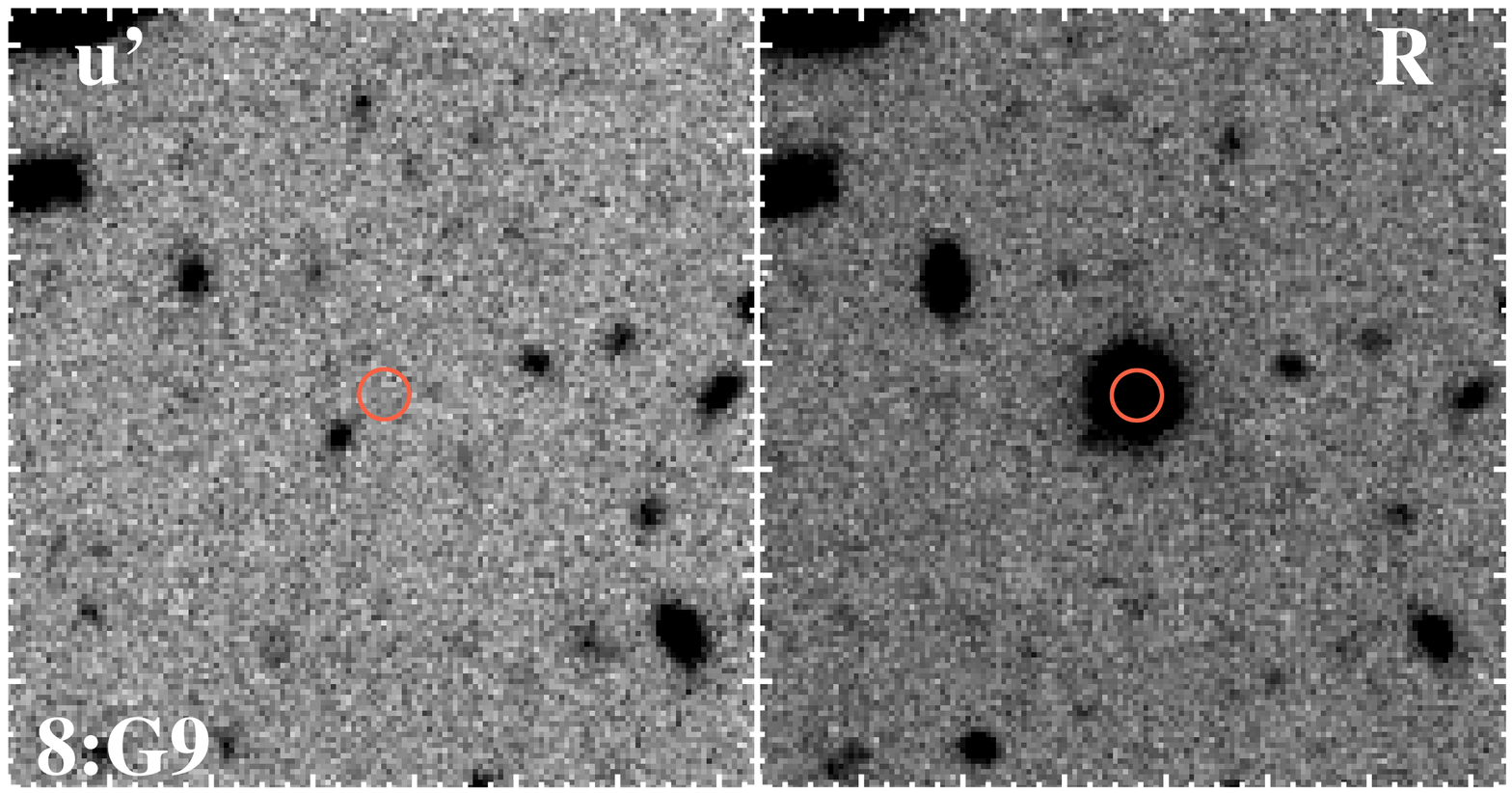}\\
\includegraphics[scale=0.32]{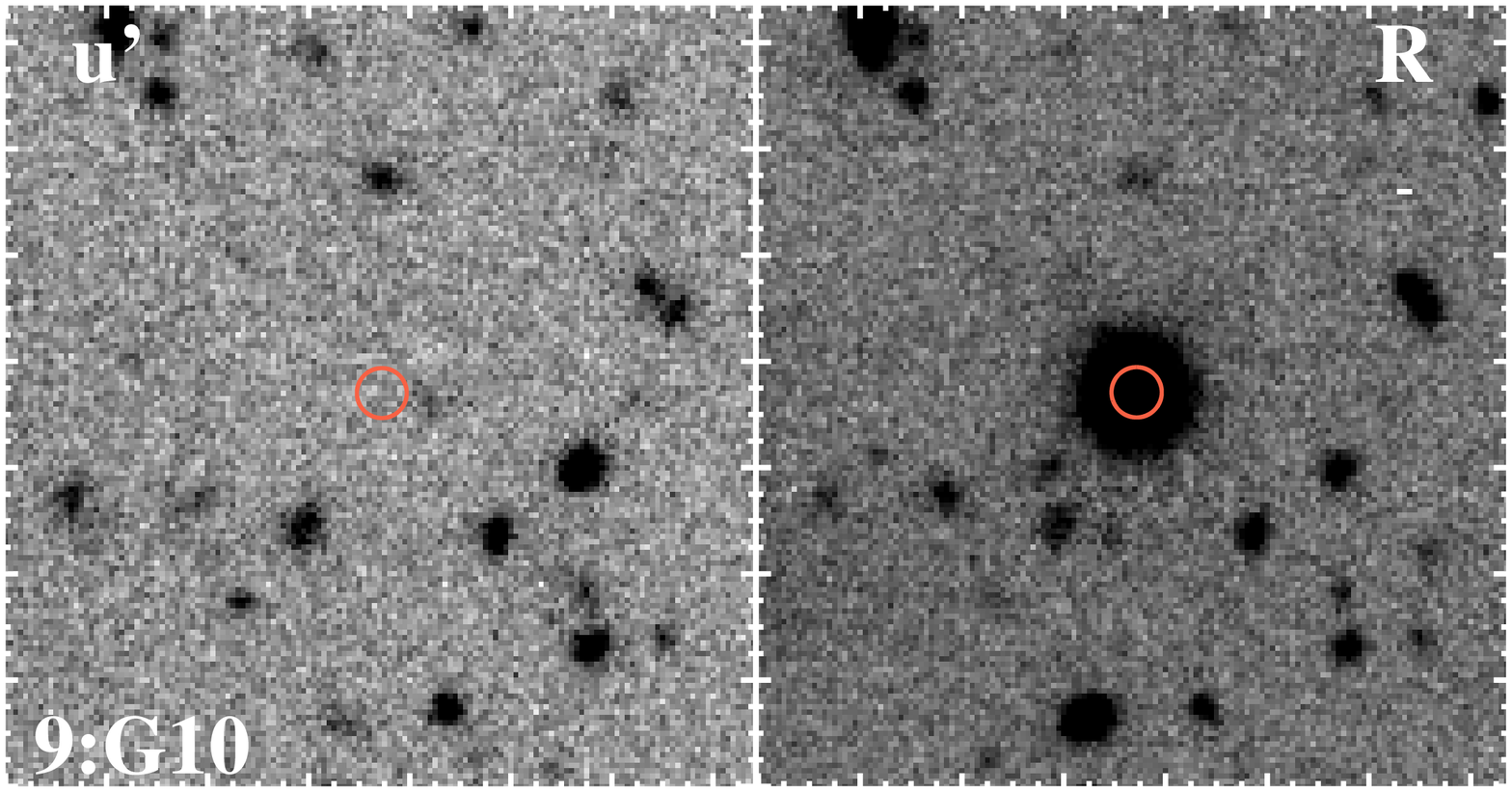}&
\includegraphics[scale=0.32]{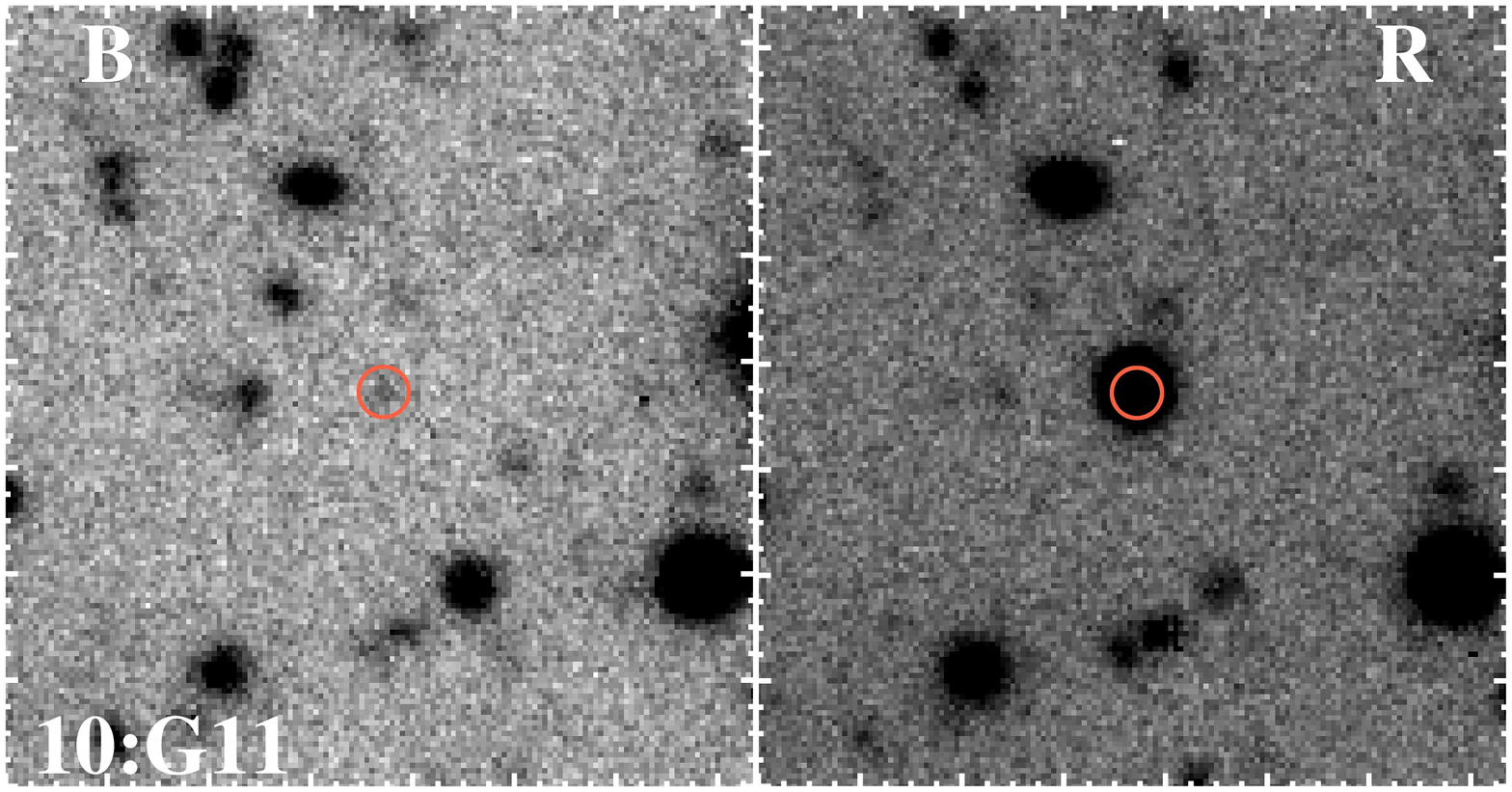}\\
\includegraphics[scale=0.32]{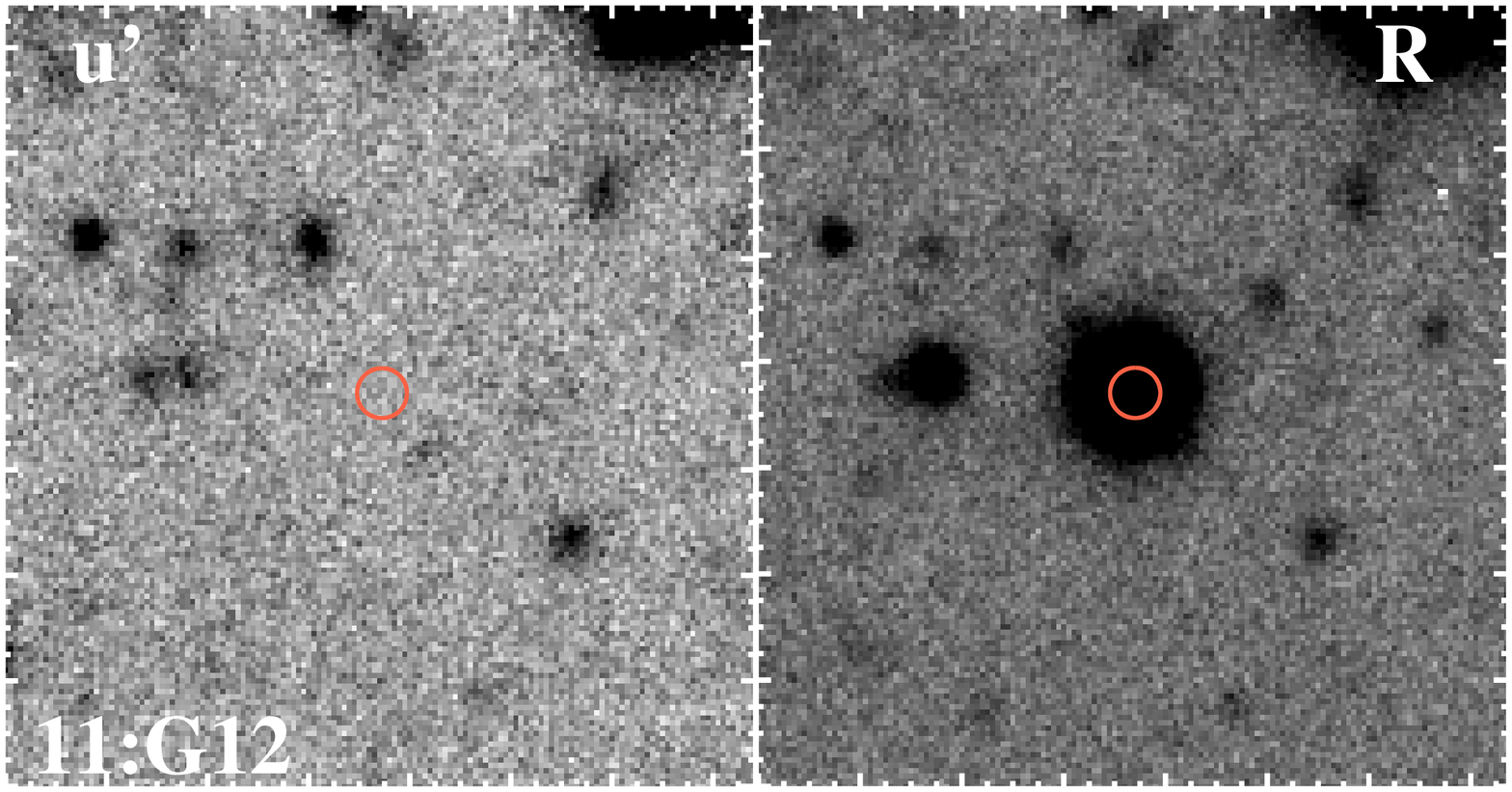}&
\includegraphics[scale=0.32]{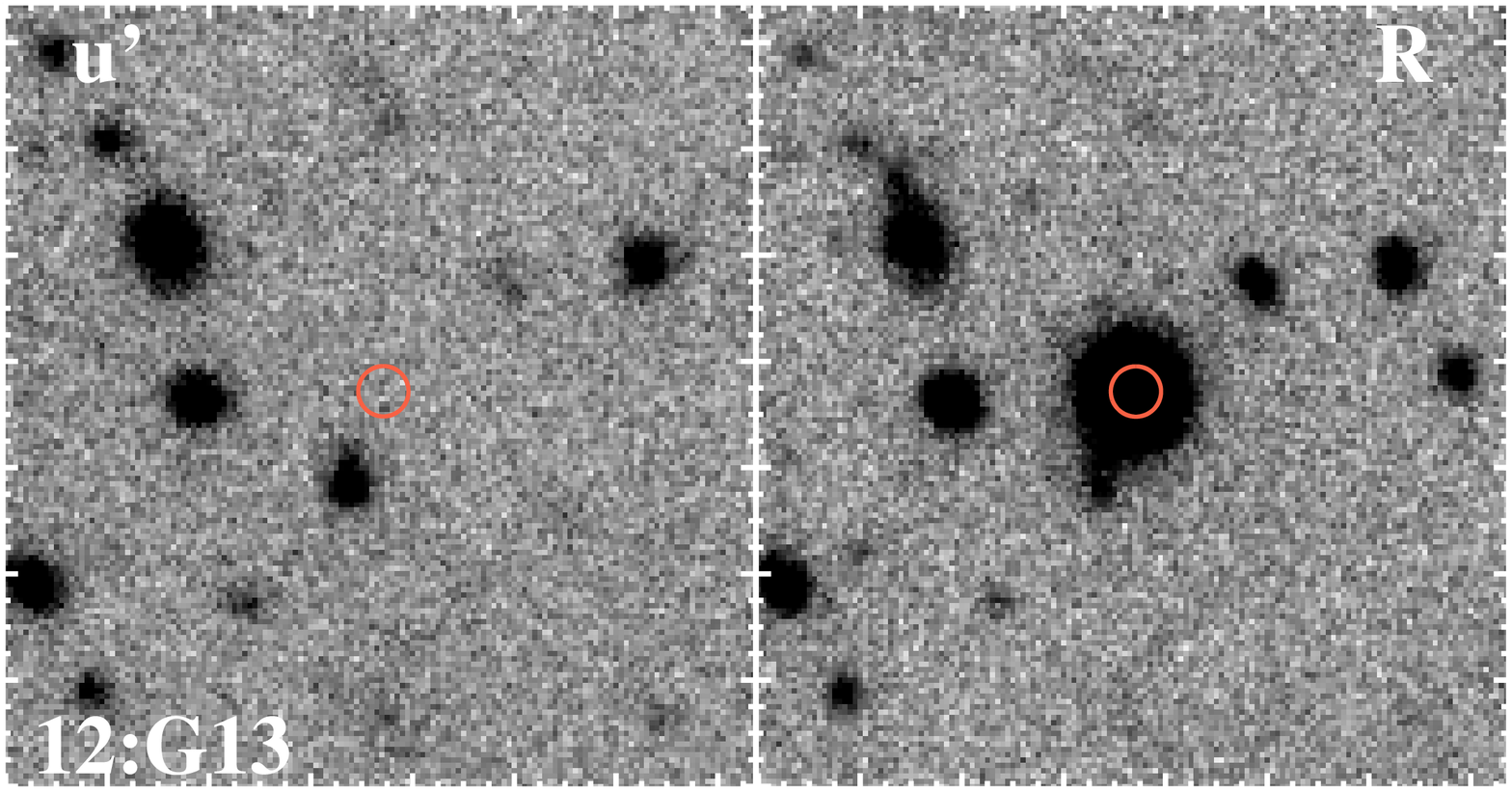}\\
\end{tabular}
\caption{Gallery of the HST and ground-based imaging. For each quasar field, we show on the left imaging in the bluest avilable filter and on the right imaging in the $R-$band filter. Each panel is $30''$ on a side, with North up and East to the left. The quasar position is marked by a red circle of $1''$ in radius.\label{fig:imggallery}}
\end{figure*}
\begin{figure*}
\begin{tabular}{cc}
\includegraphics[scale=0.32]{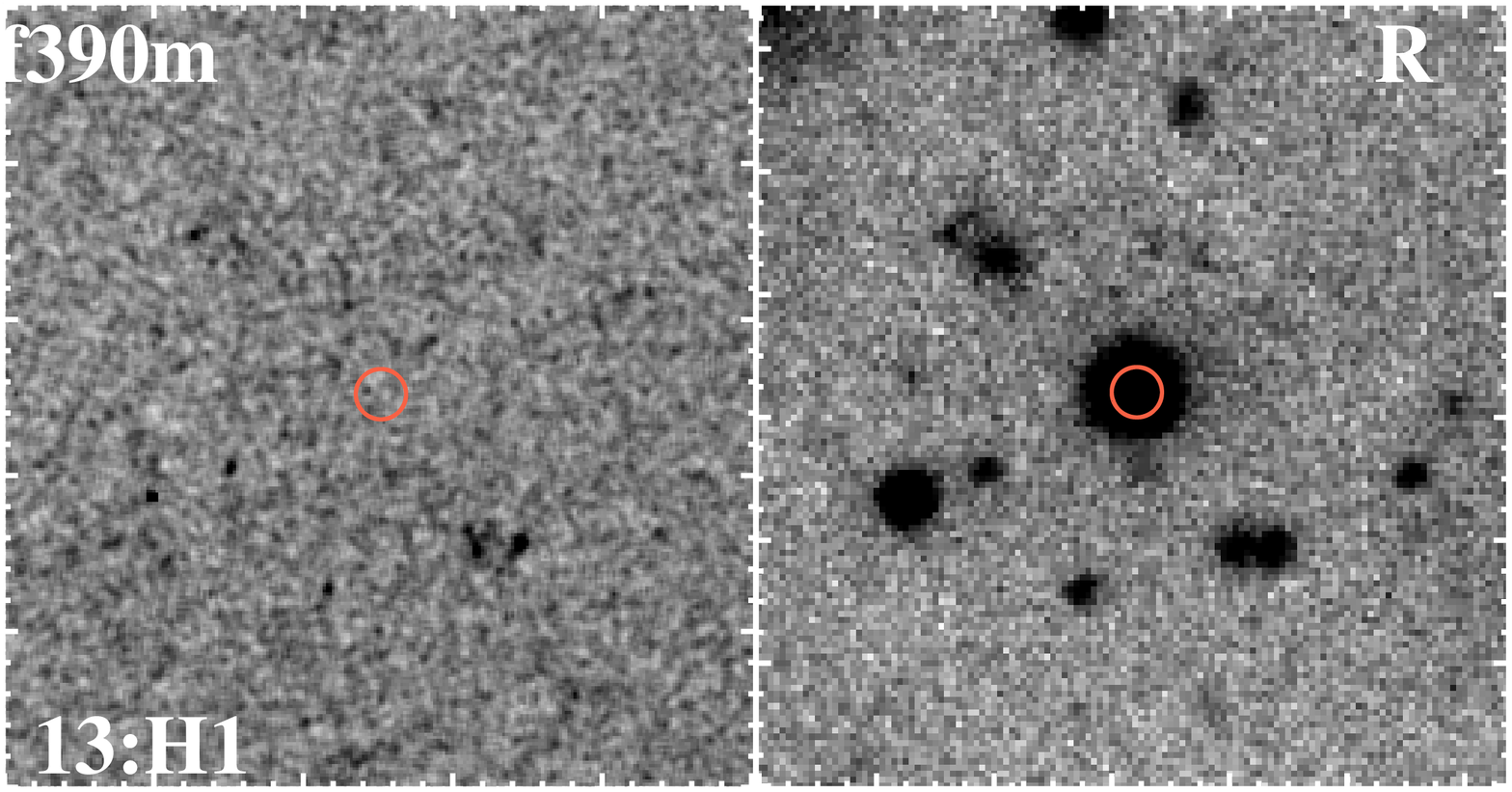}&
\includegraphics[scale=0.32]{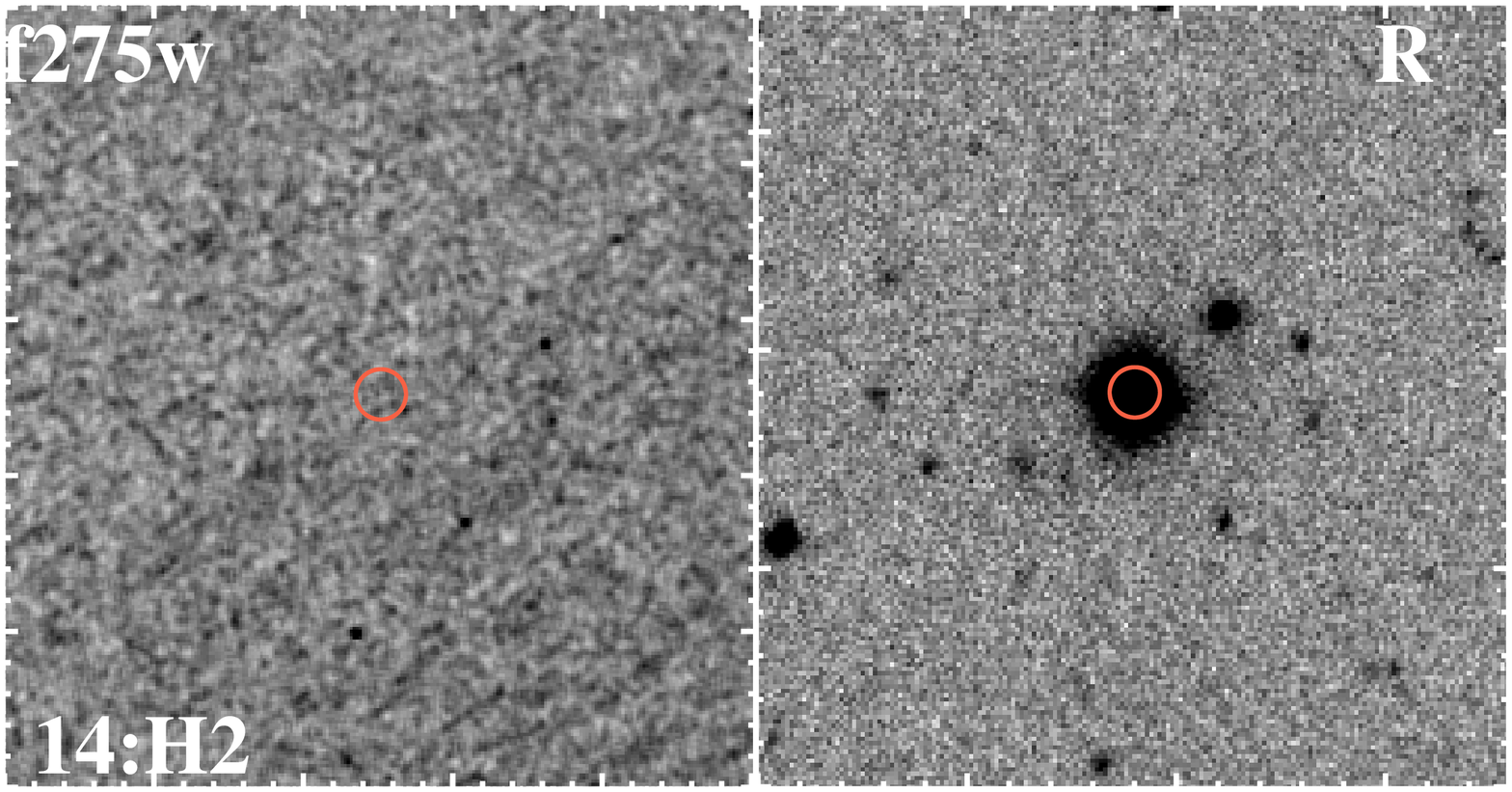}\\
\includegraphics[scale=0.32]{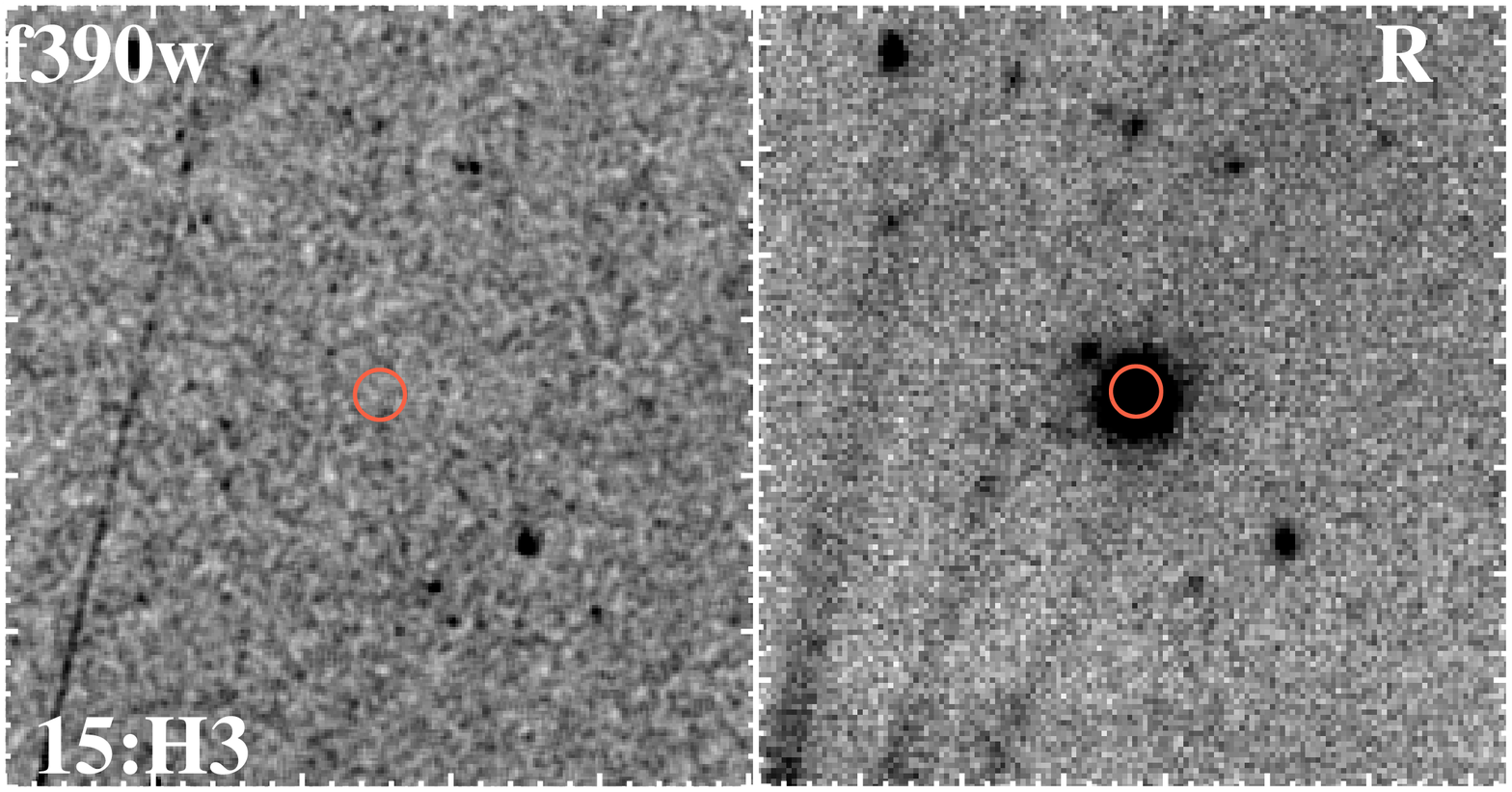}&
\includegraphics[scale=0.32]{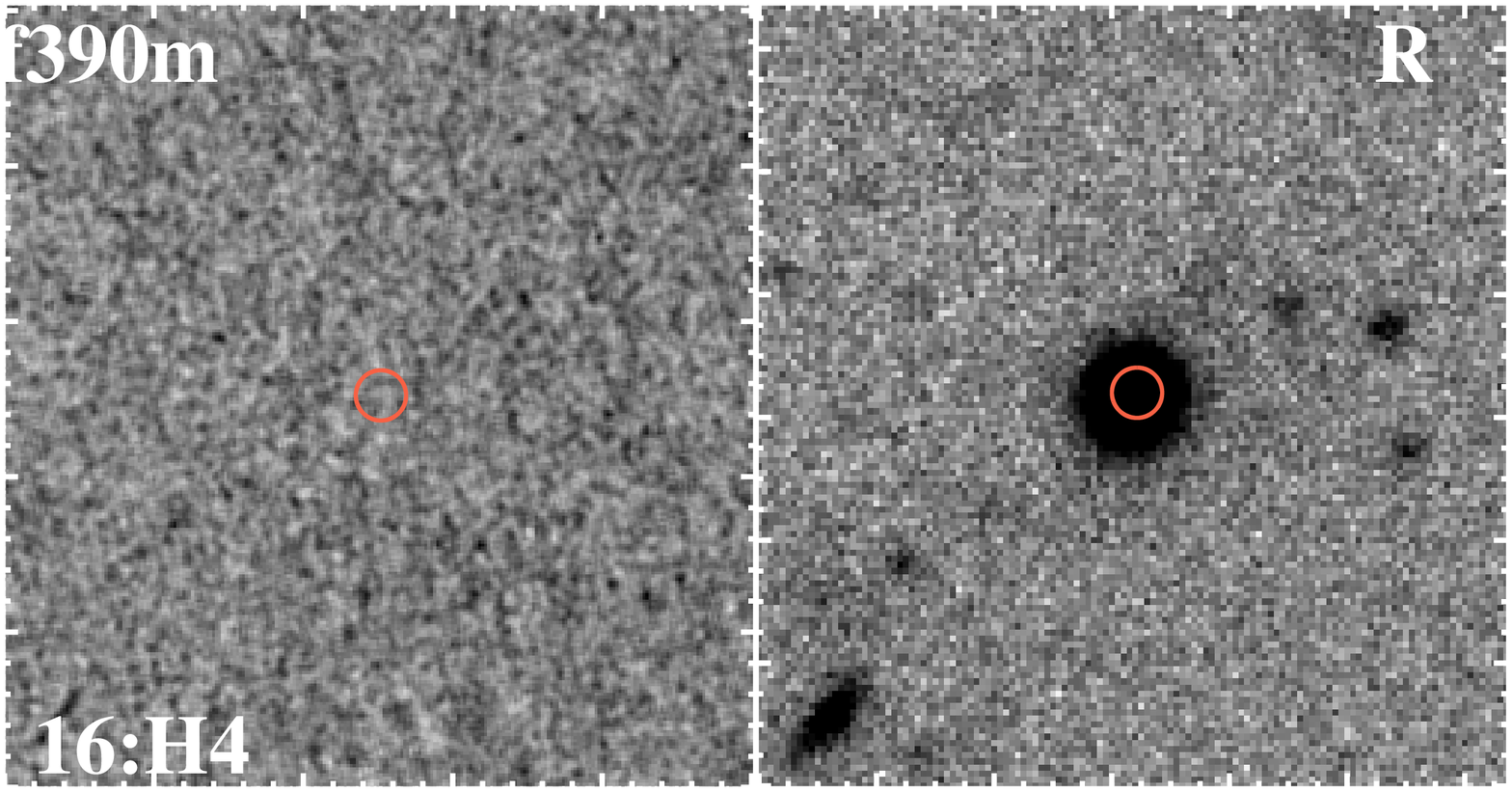}\\
\includegraphics[scale=0.32]{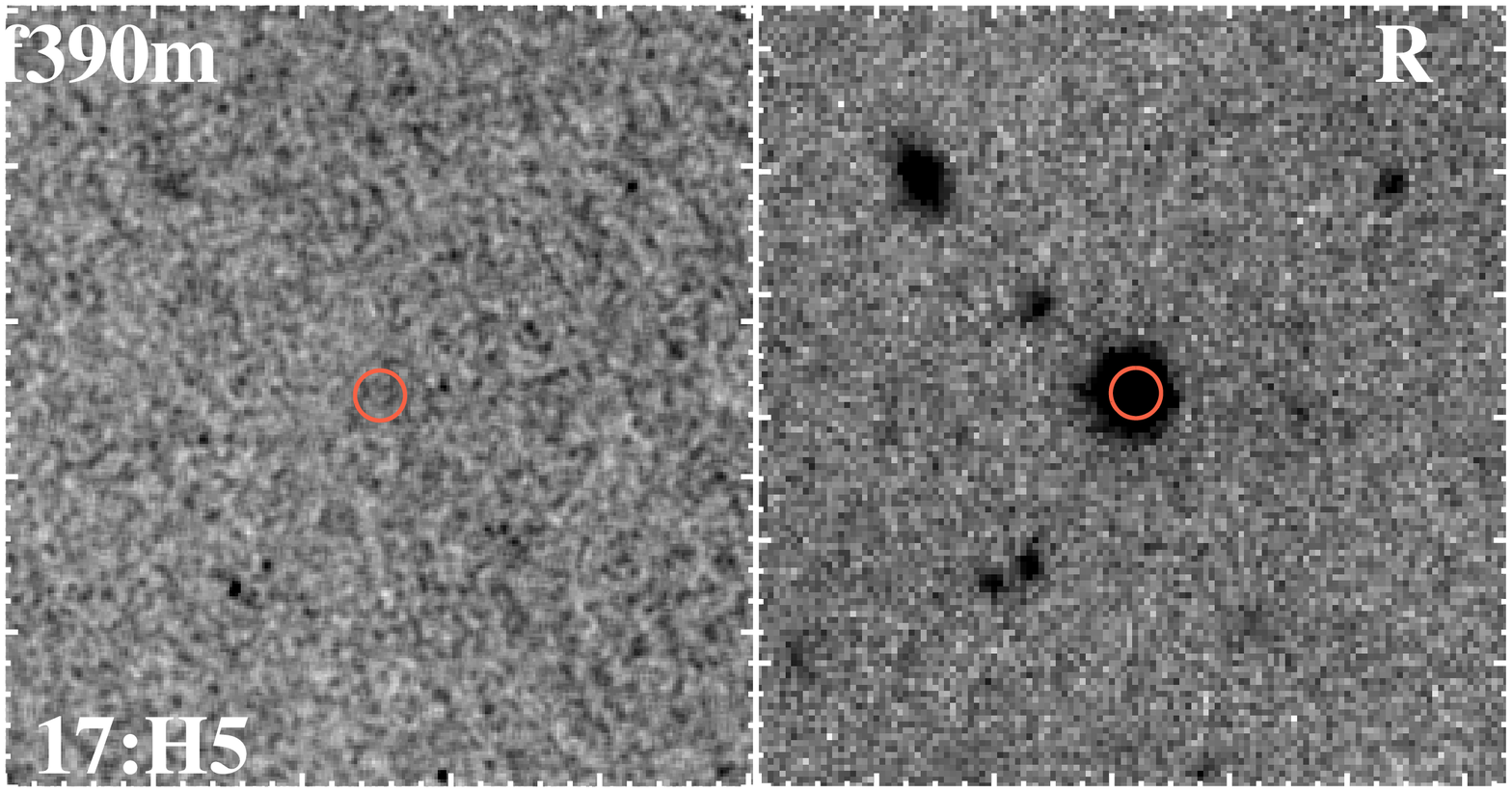}&
\includegraphics[scale=0.32]{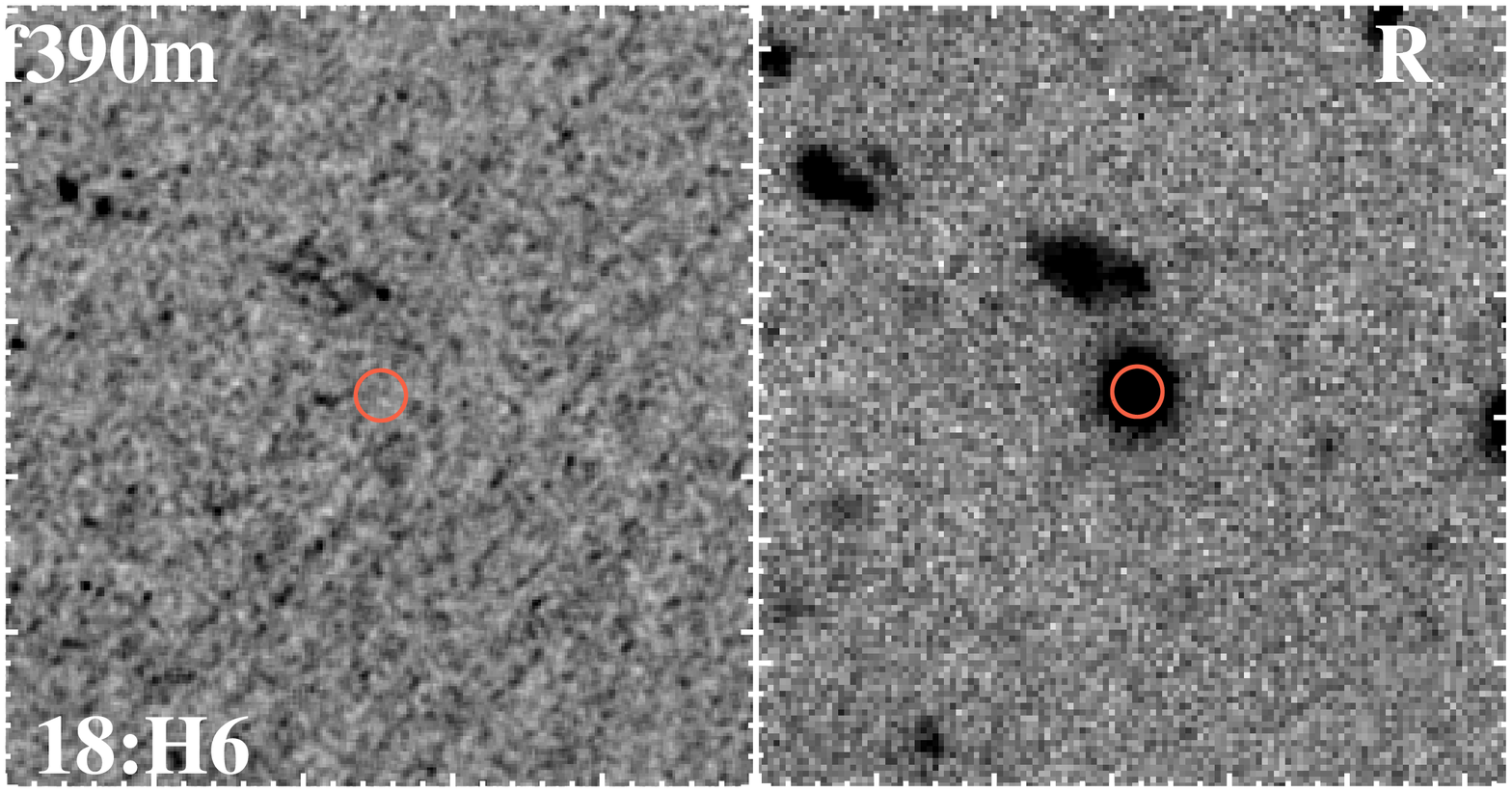}\\
\includegraphics[scale=0.32]{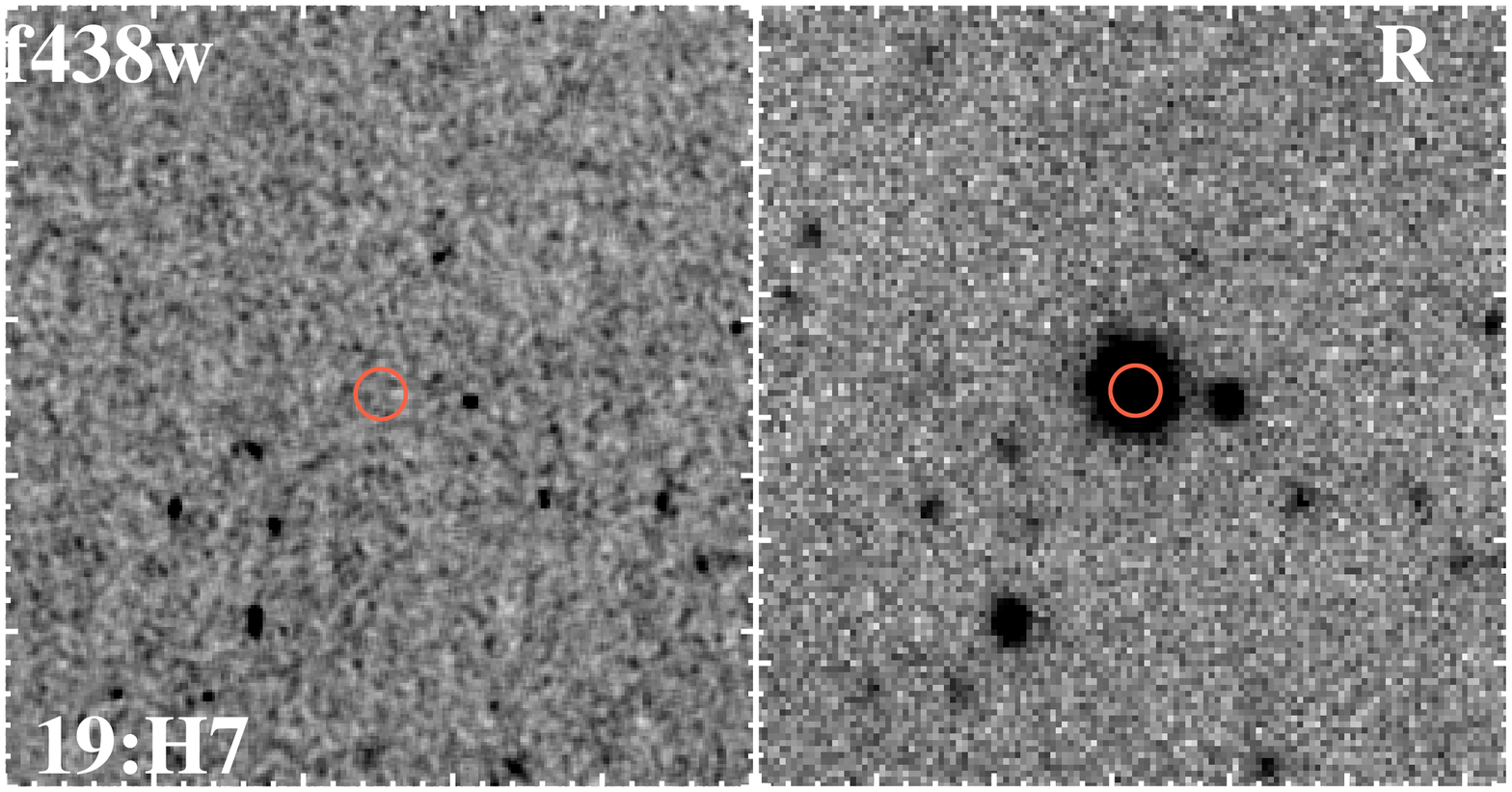}&
\includegraphics[scale=0.32]{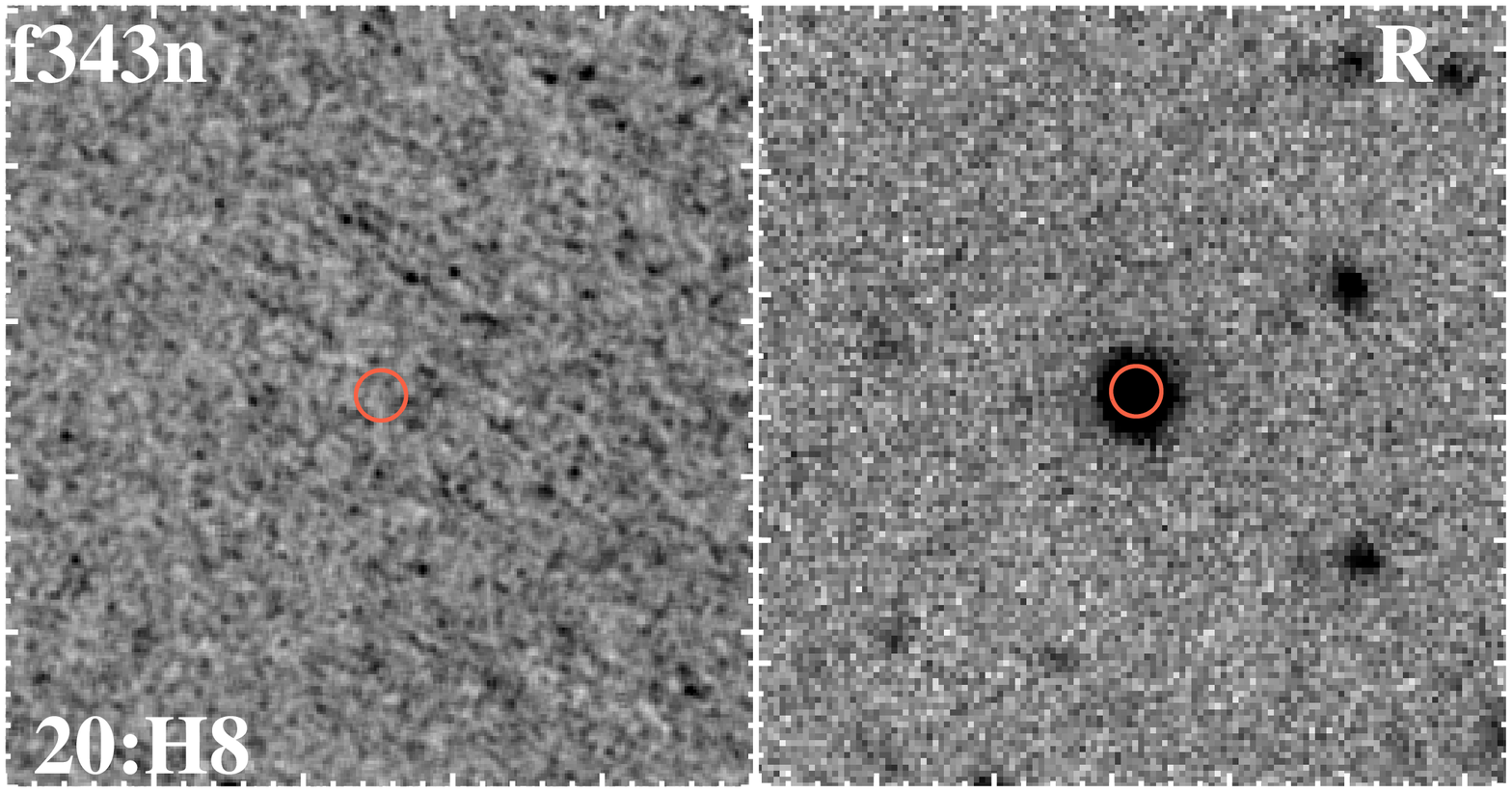}\\
\includegraphics[scale=0.32]{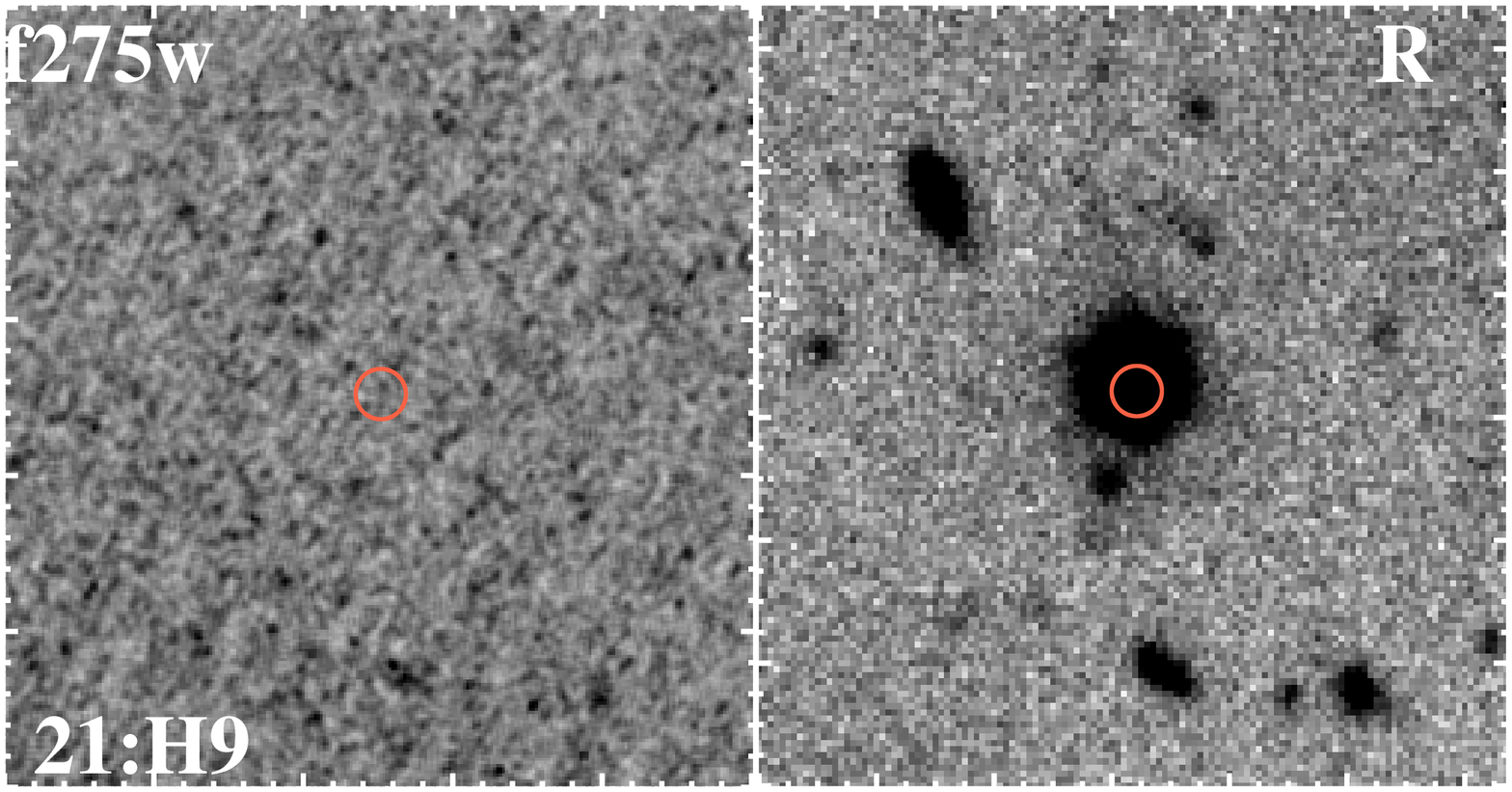}&
\includegraphics[scale=0.32]{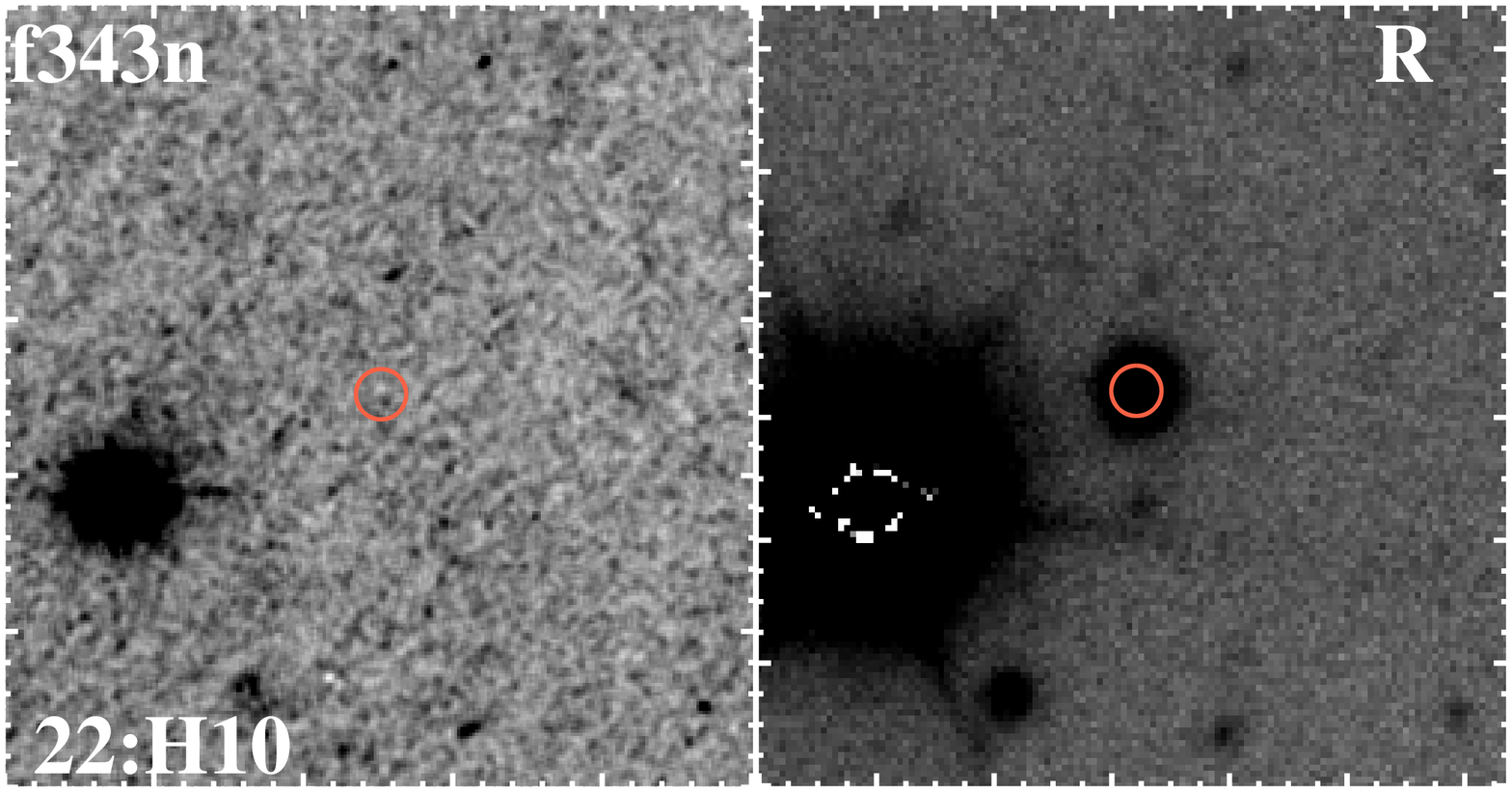}\\
\includegraphics[scale=0.32]{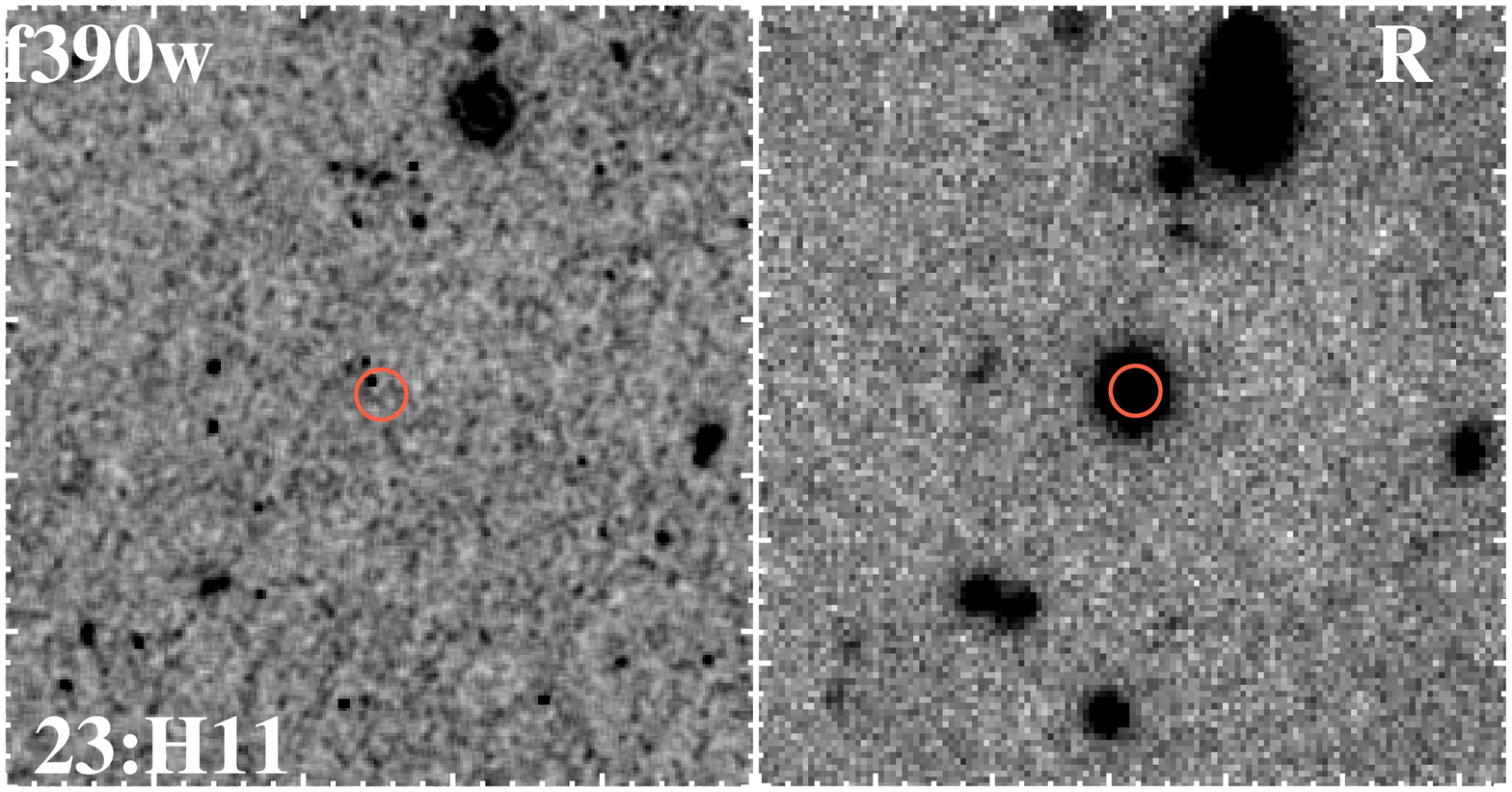}&
\includegraphics[scale=0.32]{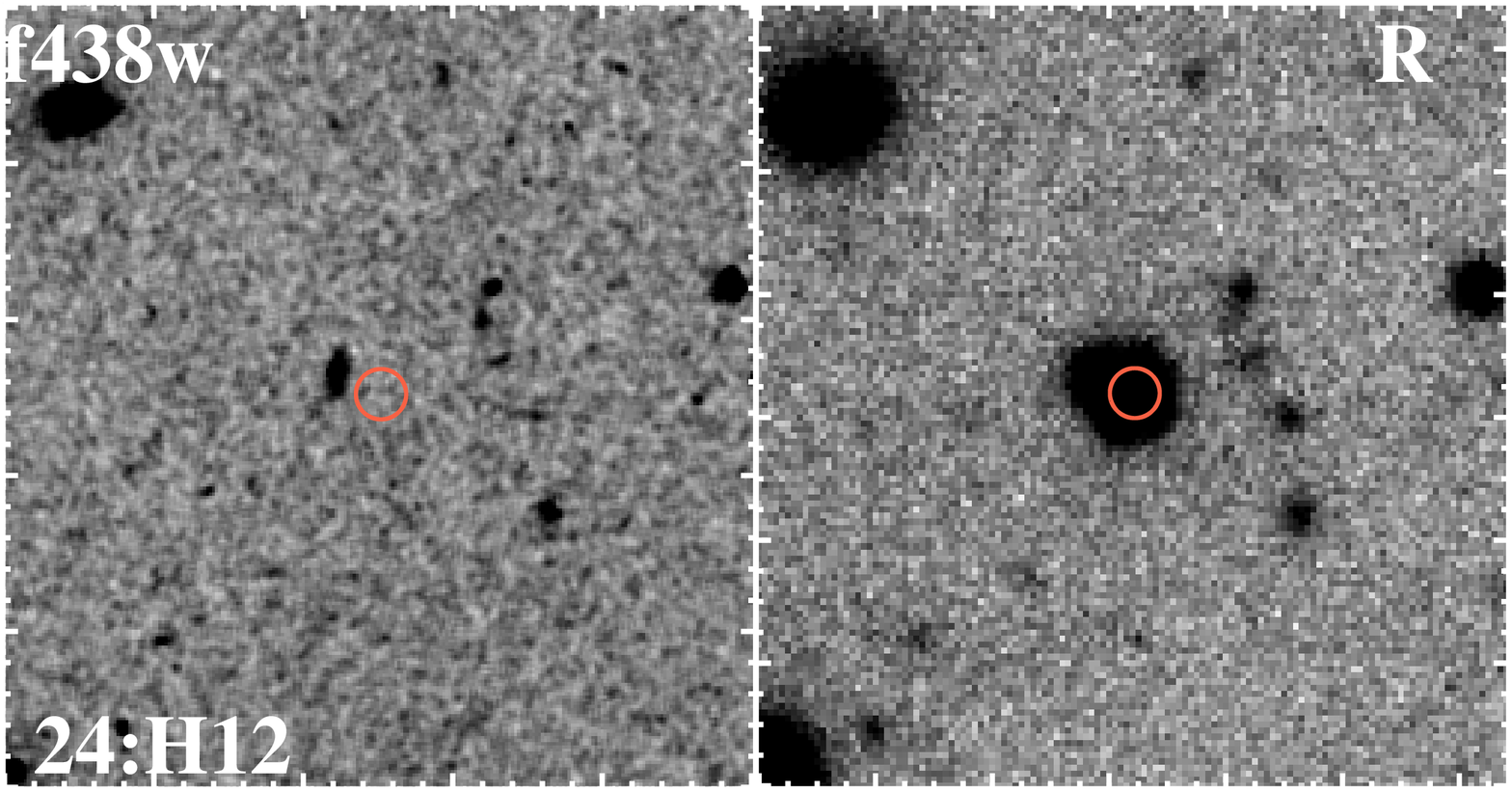}\\
\end{tabular}
\contcaption{Gallery of the HST and ground-based imaging. For each quasar field, we show on the left imaging in the bluest avilable filter and on the right imaging in the $R-$band filter. Each panel is $30''$ on a side, with North up and East to the left. The quasar position is marked by a red circle of $1''$ in radius.}
\end{figure*}
\begin{figure*}
\begin{tabular}{cc}
\includegraphics[scale=0.32]{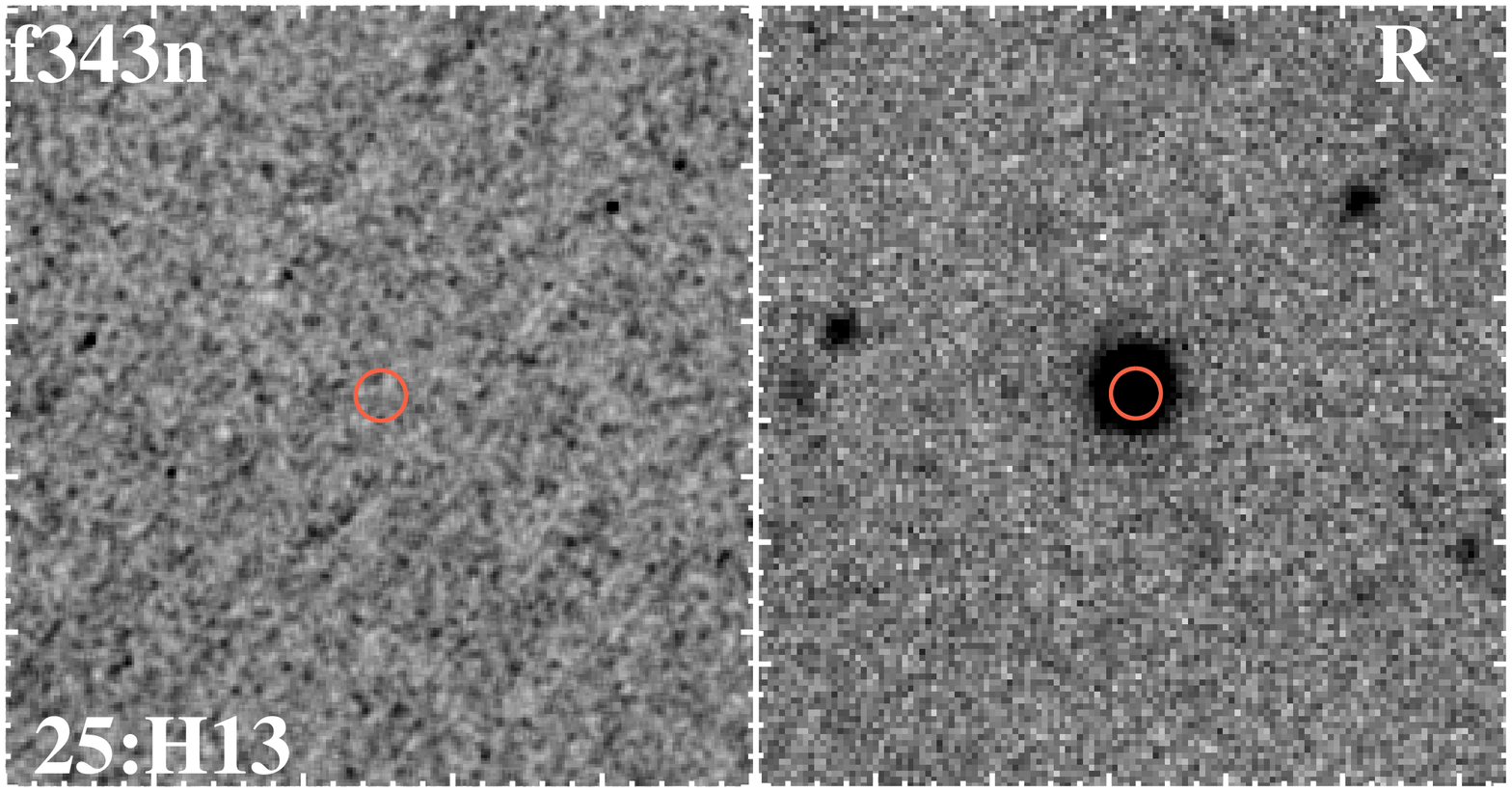}&
\includegraphics[scale=0.32]{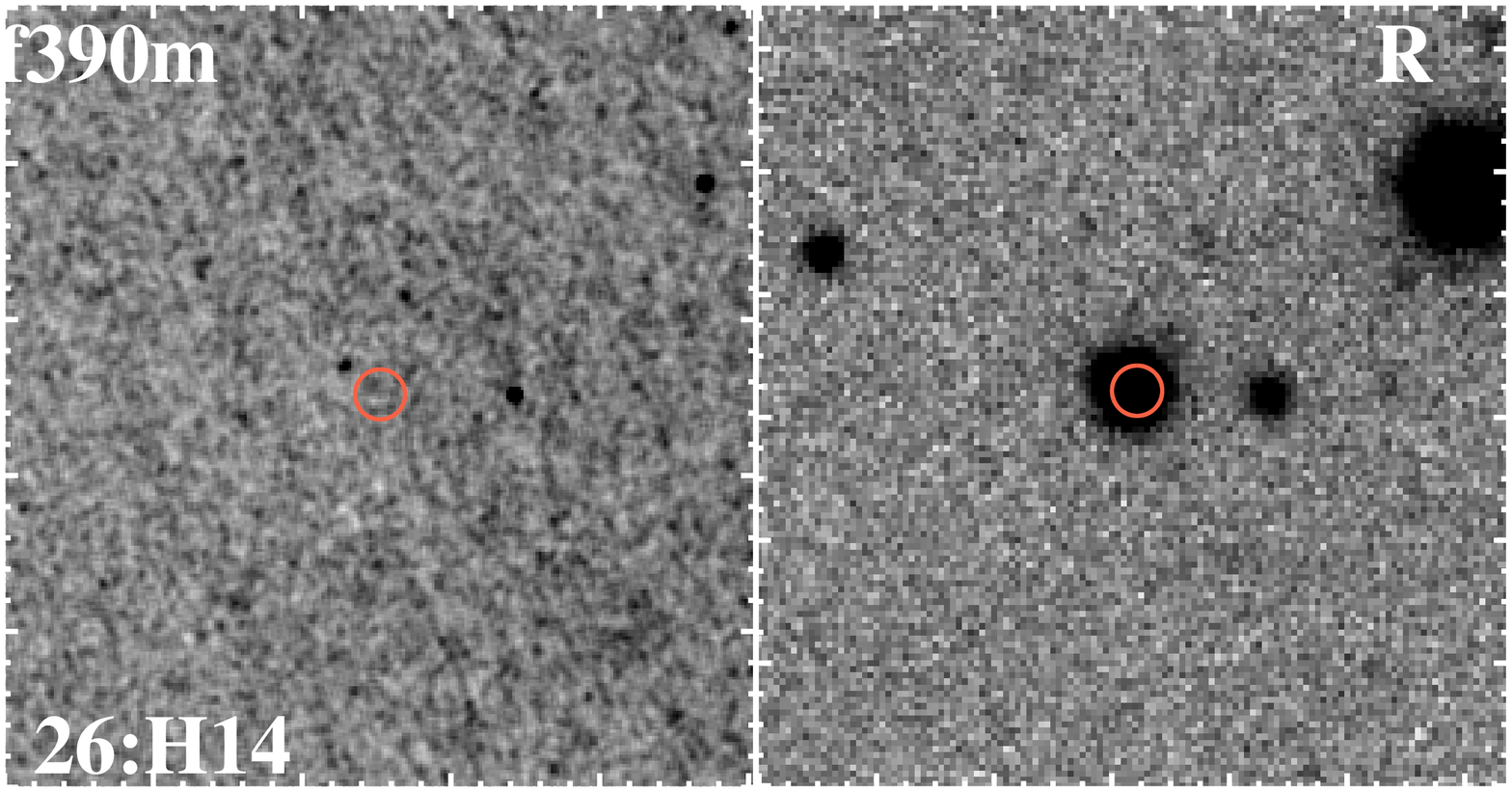}\\
\includegraphics[scale=0.32]{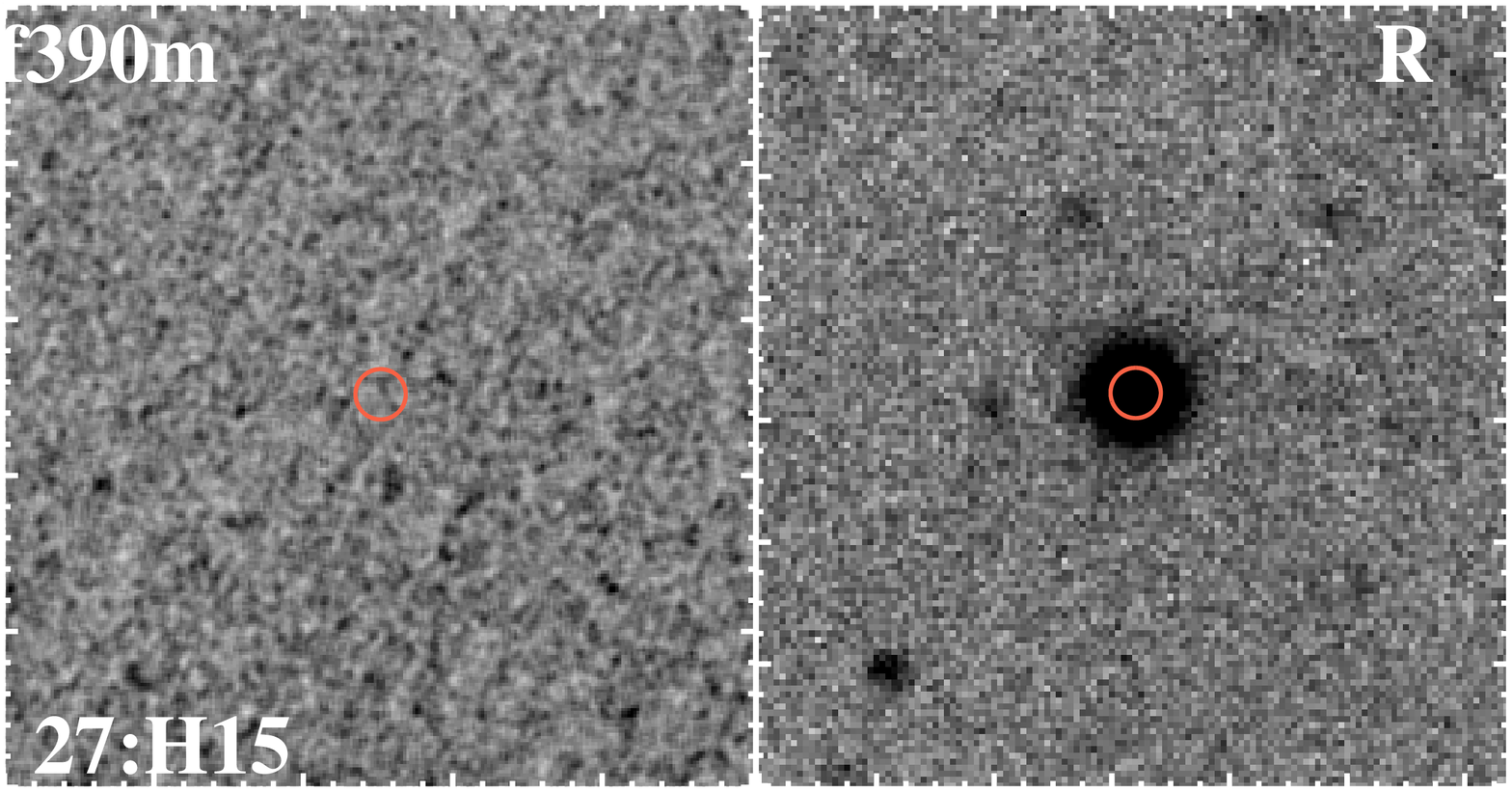}&
\includegraphics[scale=0.32]{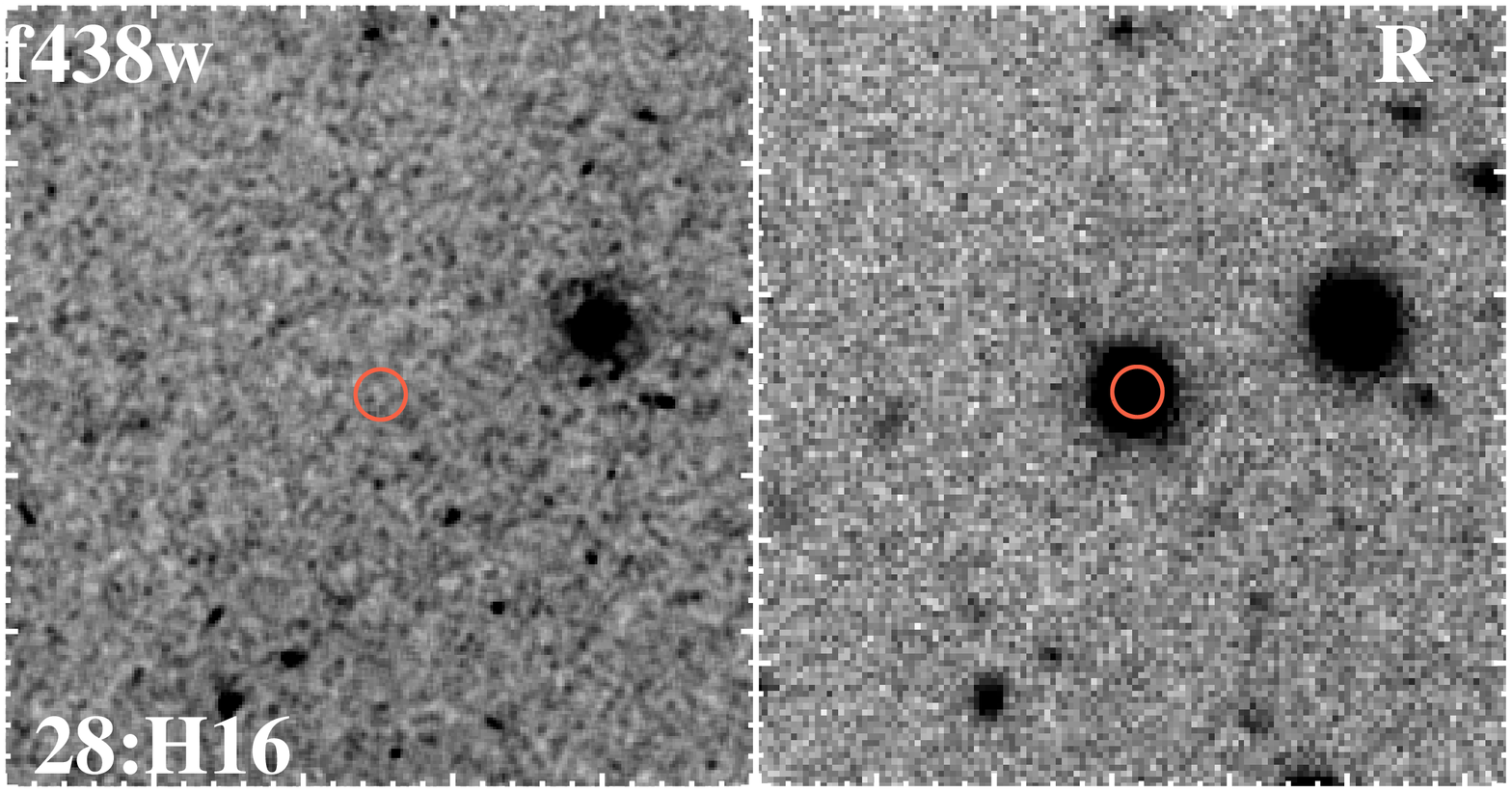}\\
\includegraphics[scale=0.32]{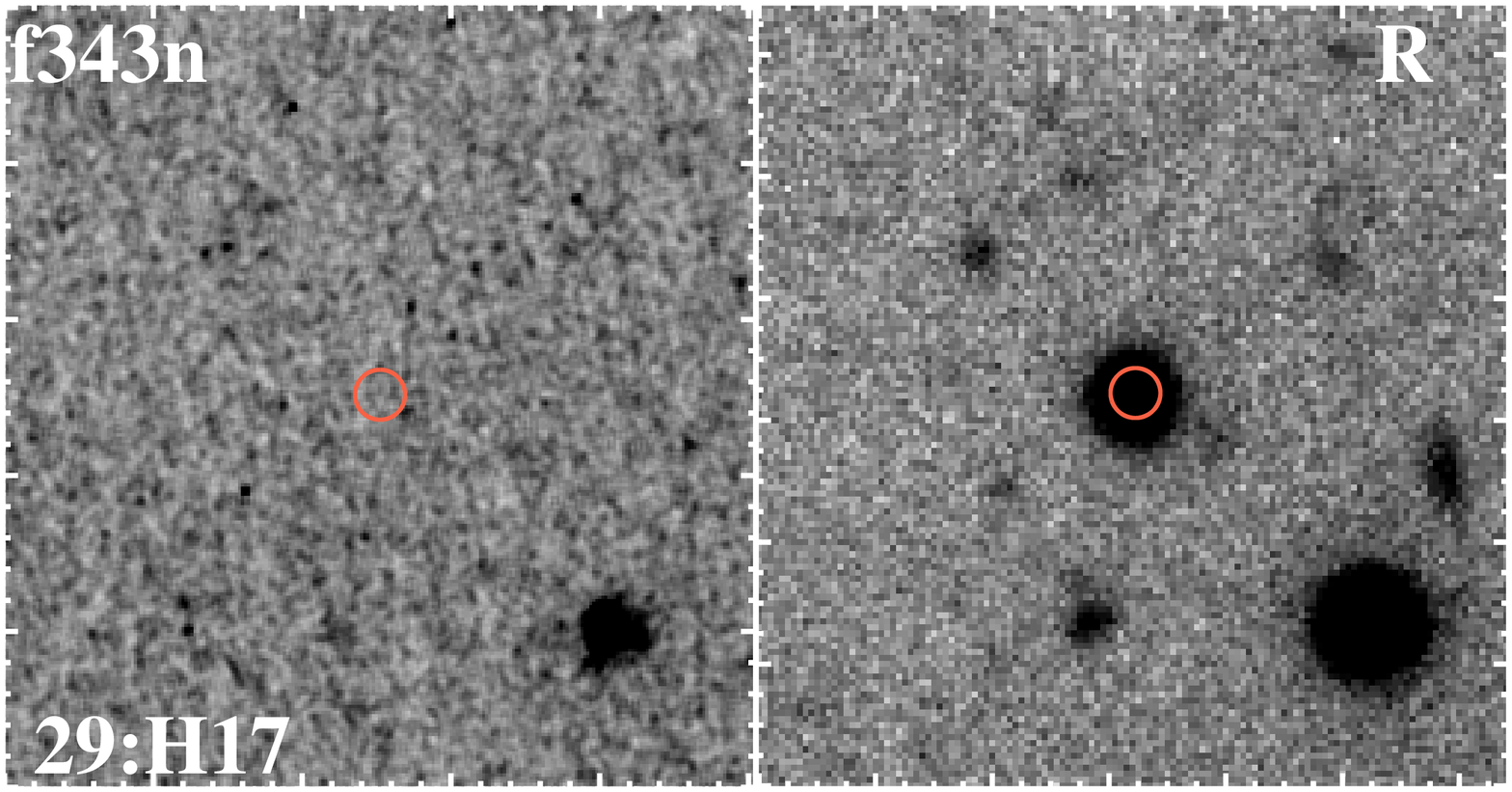}&
\includegraphics[scale=0.32]{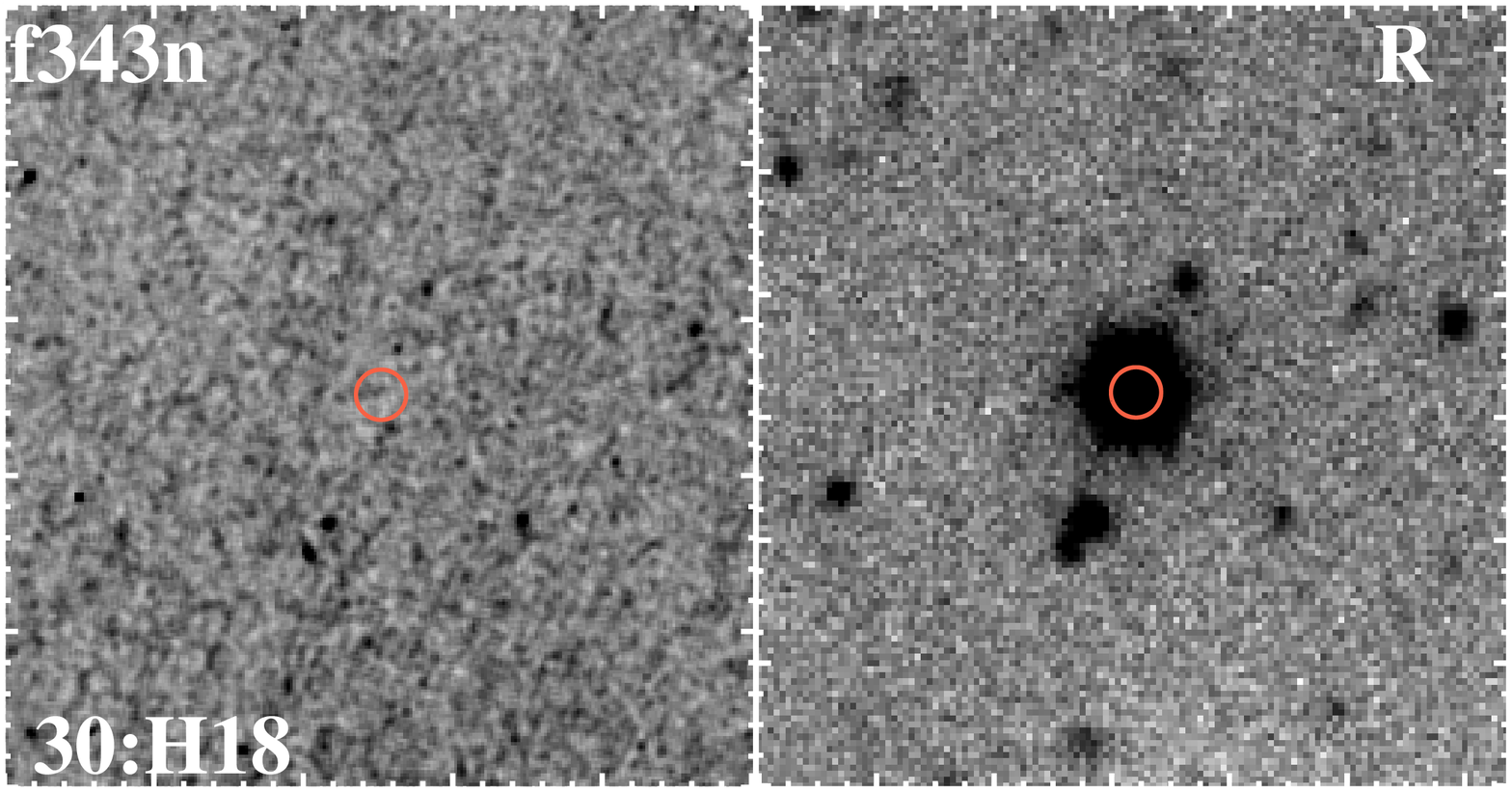}\\
\includegraphics[scale=0.32]{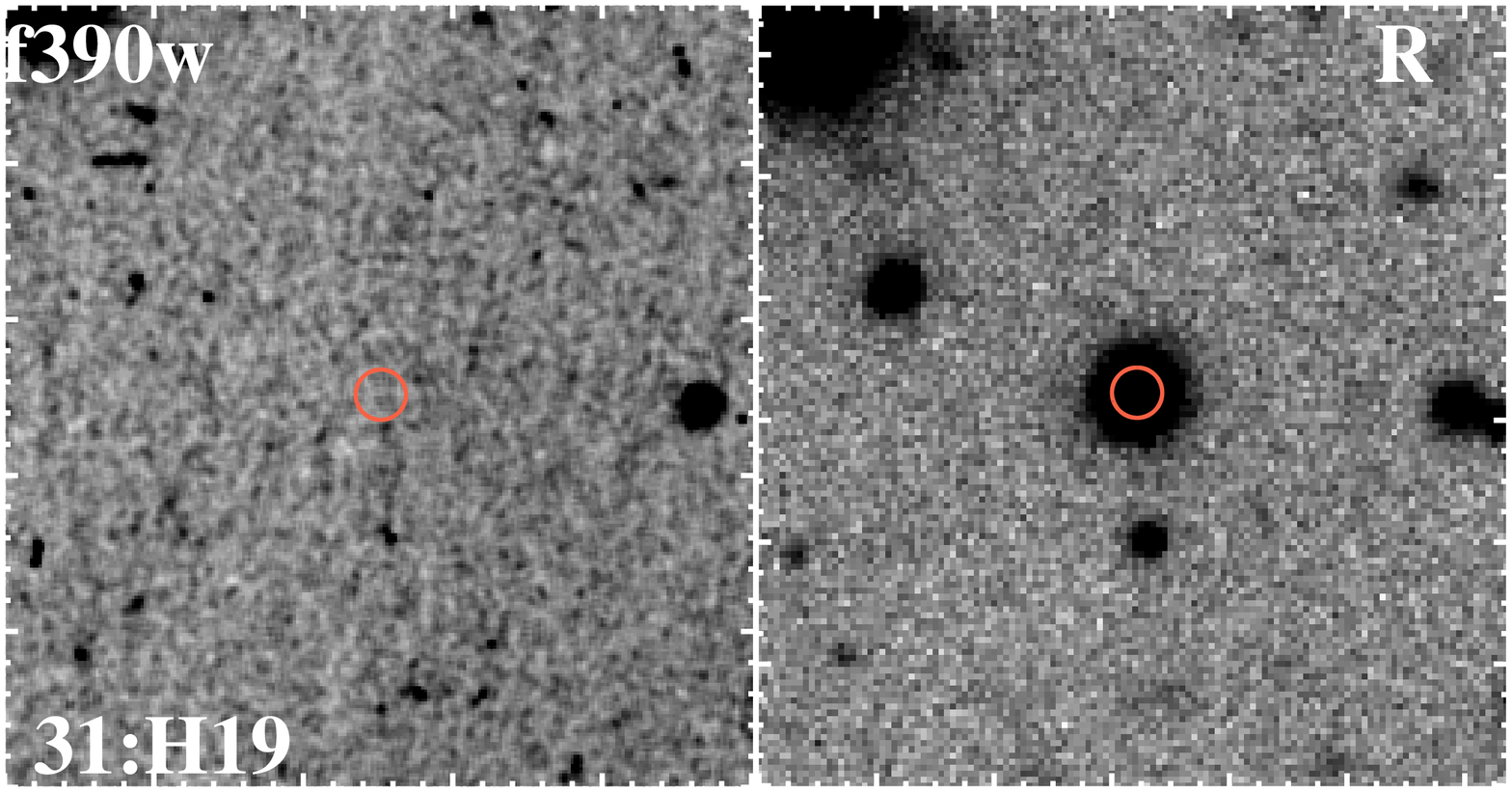}&
\includegraphics[scale=0.32]{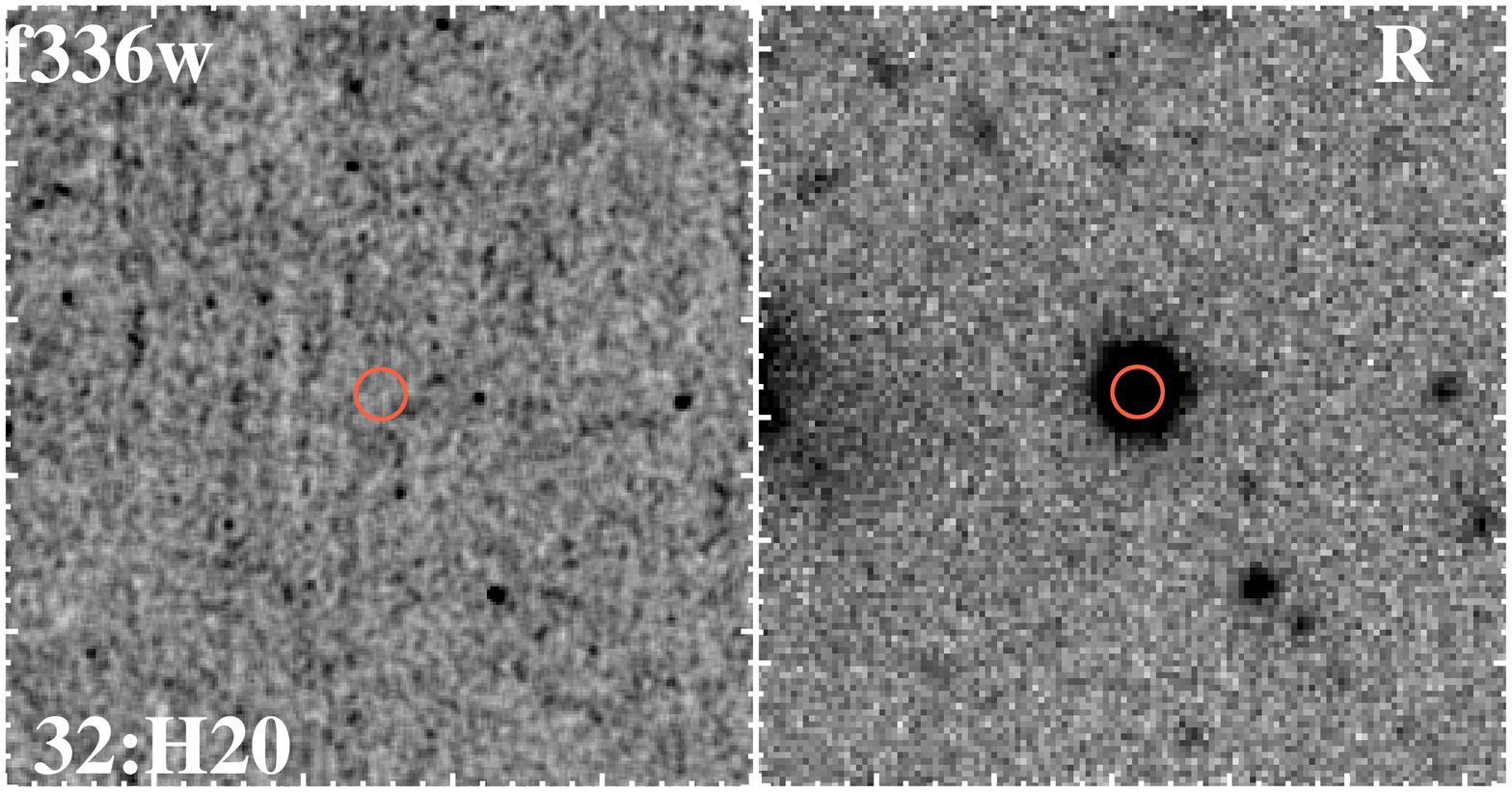}\\
\end{tabular}
\contcaption{Gallery of the HST and ground-based imaging. For each quasar field, we show on the left imaging in the bluest avilable filter and on the right imaging in the $R-$band filter. Each panel is $30''$ on a side, with North up and East to the left. The quasar position is marked by a red circle of $1''$ in radius.}
\end{figure*}

As shown by Figure \ref{design}, these conditions ensure that the quasar light and 
the light of the galaxy associated with the high-redshift Lyman limit system (LLS)
are fully blocked to avoid contamination in the selected blue filters.  
The above selection criteria restrict the number of quasars for which this experiment can be performed
in SDSS/DR5 (plus a few sightlines in the southern sky with known DLAs) to several tens of sightlines,
32 of which enter our final sample purely because of optimal scheduling of the imaging observations. The 
sample properties are summarized in Table \ref{tabprop}. Twenty of these quasar fields 
(hereafter the HST sample, labeled by the letter ``H'') have been selected for imaging with HST, while twelve 
additional fields have been imaged using ground-based facilities
(hereafter the ground-based sample, labeled by the letter ``G'').  

\begin{table*}
\caption{Log of the imaging observations for the ground-based sample.}\label{tabimg}
\centering
\begin{tabular}{l c c l c c c c c c c c c}
\hline
Field$^a$&R.A.$^b$&Dec.$^b$&UT Date$^c$&Filter$^d$&Instr.$^e$& Time$^f$& A.M.$^g$& Pixel$^h$ & FWHM$^i$&Depth$^l$& Compl.$^m$&$A_{\rm X}$$^n$\\
     &(J2000)&(J2000)&   & & &(s)& &($''$)&($''$)&(mag)& & (mag)\\
\hline
 1:G1 &21:14:43.95&-00:55:32.7&2008 Oct, 2$^{\rm nd}$	   &$u'$    & LRIS  & 5400 &  1.08&  0.135  & 0.67& 29.01&27.2/25.5 &0.24 \\		
      &      	  & 	      &  ''			   &$V'$    & LRIS  & 1320 &  1.08&  0.135  & 0.63& 28.09&	    &0.15 \\		
      &      	  & 	      &  ''			   &$R'$    & LRIS  & 1320 &  1.07&  0.135  & 0.67& 27.88&	    &0.13 \\			 
      &      	  & 	      &  ''			   &$I'$    & LRIS  & 1515 &  1.07&  0.135  & 0.79& 27.37&	    &0.10 \\			
 2:G2 &07:31:49.50&+28:54:48.6&2009 Jan, 28$^{\rm th}$     &$u'$    & LRIS  & 5400 &  1.10&  0.135  & 0.88& 28.61&26.6/25.6 &0.21 \\			
      &      	  & 	      &  ''			   &$V'$    & LRIS  & 2140 &  1.10&  0.135  & 0.89& 27.43&	    &0.14 \\			
      &      	  & 	      &  ''			   &$R'$    & LRIS  & 2420 &  1.08&  0.135  & 0.83& 27.43&	    &0.11 \\			
 3:G3 &09:56:04.43&+34:44:15.5&2009 Jan, 28$^{\rm th}$     &$u'$    & LRIS  & 5400 &  1.16&  0.135  & 1.00& 28.88&26.8/26.0 &0.05 \\			
      &      	  & 	      &  ''			   &$V'$    & LRIS  & 2280 &  1.15&  0.135  & 0.92& 27.82&	    &0.03 \\			
      &      	  & 	      &  ''			   &$R'$    & LRIS  & 2280 &  1.17&  0.135  & 0.90& 27.74&	    &0.03 \\			
 4:G4 &23:43:49.41&-10:47:42.0&2009 Jul, 23$^{\rm rd~1}$   &$u'$    & LRIS  & 6300 &  1.18&  0.135  & 1.31& 28.44&26.0/25.2 &0.13 \\			
      &      	  & 	      &  ''			   &$V$    & LRIS   & 1400 &  1.17&  0.135  & 0.99& 27.58&	    &0.08 \\					
      &      	  & 	      &  ''			   &$R$    & LRIS   & 1440 &  1.18&  0.135  & 1.27& 27.46&	    &0.07 \\			 
      &      	  & 	      &  ''			   &$I$    & LRIS   & 1540 &  1.17&  0.135  & 1.15& 27.22&	    &0.05 \\			 
 5:G5 &03:43:00.88&-06:22:29.9&2009 Sep, 20$^{\rm th}$     &$Us$   & LBC    & 3912 &  1.30&  0.224  & 1.15& 28.04&25.7/25.1 &0.23 \\	 
      &      	  & 	      &  ''			   &$V$    & LBC    &  900 &  1.30&  0.224  & 1.06& 27.62&	    &0.15 \\		 
      &      	  & 	      &  ''			   &$R$    & LBC    & 1200 &  1.29&  0.224  & 0.98& 27.71&     	    &0.12 \\		 
      &       	  & 	      &  ''			   &$I$    & LBC    & 1650 &  1.35&  0.224  & 0.98& 27.40&	    &0.09 \\		 
 6:G6 &23:51:52.80&+16:00:48.9&2009 Sep, 21$^{\rm st}$     &$B$    & LBC    & 1655 &  1.09&  0.224  & 1.19& 28.32&25.9/25.0 &0.12 \\	 
      &      	  & 	      &  ''			   &$V$    & LBC    &  450 &  1.09&  0.224  & 0.97& 27.32&	    &0.09 \\		 
      &      	  & 	      &  ''			   &$R$    & LBC    &  600 &  1.11&  0.224  & 0.99& 27.40&	    &0.08 \\		 
      &      	  & 	      &  ''			   &$I$    & LBC    &  750 &  1.13&  0.224  & 1.02& 27.08&	    &0.06 \\		 
 7:G7 &00:42:19.74&-10:20:09.4&2009 Dec, 17$^{\rm st}$     &$u'$    & LRIS  & 5400 &  1.16&  0.135  & 0.75& 28.64&26.4/25.4 &0.13 \\			   
      &      	  & 	      &  ''			   &$V$    & LRIS   & 1400 &  1.16&  0.135  & 0.83& 27.64&	    &0.08 \\			   
      &      	  & 	      &  ''			   &$R$    & LRIS   & 1400 &  1.16&  0.135  & 0.84& 27.38&	    &0.07 \\			   
      &      	  & 	      &  ''			   &$I$    & LRIS   & 1440 &  1.16&  0.135  & 0.83& 27.03&	    &0.05 \\			   
 8:G9 &09:49:27.88&+11:15:18.2&2009 Dec, 17$^{\rm st}$     &$u'$    & LRIS  & 5400 &  1.16&  0.135  & 0.77& 28.94&27.1/26.0 &0.09 \\			   
      &      	  & 	      &  ''			   &$V$    & LRIS   & 1200 &  1.14&  0.135  & 0.81& 27.80&	    &0.06 \\			   
      &      	  & 	      &  ''			   &$R$    & LRIS   & 1200 &  1.16&  0.135  & 0.73& 27.68&	    &0.05 \\			   
      &      	  & 	      &  ''			   &$I$    & LRIS   & 1440 &  1.13&  0.135  & 0.75& 27.38&	    &0.04 \\			   
 9:G10&10:18:06.28&+31:06:27.2&2009 Dec, 17$^{\rm st}$     &$u'$    & LRIS  & 3600 &  1.02&  0.135  & 0.92& 29.08&26.9/25.5 &0.12 \\			   
      &      	  & 	      &  ''			   &$V$    & LRIS   &  800 &  1.02&  0.135  & 0.85& 27.68&	    &0.08 \\			   
      &      	  & 	      &  ''			   &$R$    & LRIS   &  800 &  1.02&  0.135  & 0.86& 27.64&	    &0.06 \\			   
      &      	  & 	      &  ''			   &$I$    & LRIS   &  960 &  1.02&  0.135  & 0.96& 27.46&	    &0.05 \\			   
10:G11&08:51:43.72&+23:32:08.9&2009 Dec, 18$^{\rm st}$     &$B$    & LRIS   & 3600 &  1.19&  0.135  & 1.21& 28.69&26.7/25.0 &0.10 \\			   
      &      	  & 	      &  ''			   &$V$    & LRIS   &  800 &  1.02&  0.135  & 1.15& 27.42&	    &0.08 \\			   
      &      	  & 	      &  ''			   &$R$    & LRIS   &  800 &  1.02&  0.135  & 1.14& 27.32&	    &0.07 \\			   
      &      	  & 	      &  ''			   &$I$    & LRIS   &  960 &  1.02&  0.135  & 1.16& 26.97&	    &0.05 \\			   
11:G12&09:56:05.09&+14:48:54.7&2009 Dec, 18$^{\rm st}$     &$u'$    & LRIS  & 5400 &  1.19&  0.135  & 1.30& 28.60&26.0/25.4 &0.11 \\			   
      &      	  & 	      &  ''			   &$V$    & LRIS   & 1200 &  1.17&  0.135  & 1.20& 27.61&	    &0.07 \\			   
      &      	  & 	      &  ''			   &$R$    & LRIS   & 1200 &  1.19&  0.135  & 1.09& 27.53&	    &0.06 \\			   
      &      	  & 	      &  ''			   &$I$    & LRIS   & 1440 &  1.16&  0.135  & 1.14& 27.18&	    &0.04 \\			   
12:G13&11:51:30.48&+35:36:25.0&2009 Dec, 18$^{\rm st}$     &$u'$    & LRIS  & 3600 &  1.21&  0.135  & 1.21& 28.61&26.4/25.1 &0.08 \\	     
      &      	  & 	      &  ''			   &$V$    & LRIS   & 1200 &  1.21&  0.135  & 1.09& 27.88&	    &0.05 \\	     
      &      	  & 	      &  ''			   &$R$    & LRIS   & 1200 &  1.20&  0.135  & 1.14& 27.78&	    &0.04 \\	     
      &      	  & 	      &  ''			   &$I$    & LRIS   & 1440 &  1.18&  0.135  & 1.10& 27.45&	    &0.03 \\	     
\hline															   
\end{tabular}														   
\\\flushleft{$^a$ ID of the quasar field. $^b$ Right ascension and declination of the quasar. $^c$ UT date in which observations 
were conducted. $^d$ Adopted filter. $^e$ Instrument. $^f$ Total exposure time.  
$^g$ Typical airmass during observations. $^{h}$ Pixel size of the final reprojected image.
$^{i}$ Point source full-width at half-maximum. $^{l}$ Image depth at $2\sigma$ measured in a $1''$ aperture. $^m$ $90\%$ completeness limit
from simulated images (left) and as empirically determined based on the recovered number counts (right). 
$^{n}$ Galactic extinction in the adopted filter. $^1$ Additional observations from 2009 Sep 17$^{\rm th}$ were coadded in the preparation of the final images.} 										   
\end{table*}														   
\begin{table*}														   
\contcaption{Log of the imaging observations for the HST sample.}							   
\centering														   
\begin{tabular}{l c c l c c c c c c c c c}										   
\hline															   
Field&R.A.&Dec.&UT Date&Filter&Instr. & Time& A.M. &Pixel &FWHM&Depth &Compl.&$A_{\rm X}$\\ 			   
     &(J2000)&(J2000)&   & & &(s)& &($''$)&($''$)&(mag)& &(mag)\\ 							   
\hline															   
															   
13:H1 &21:23:57.56&-00:53:50.1&2009 Dec,  4$^{\rm th}$ &$F390M$& UVIS  & 5130 &     & 0.040& 0.08  &25.97&26.4/25.2 & 0.14 \\  
      &           &           &2013 Aug, 11$^{\rm th}$ &$V$    & ESI   &  360 & 1.25& 0.156& 0.94  &27.11&	    & 0.10 \\  
      &           &           &  ''                    &$R$    & ESI   &  360 & 1.27& 0.156& 0.93  &26.99&	    & 0.08 \\  
14:H2 &04:07:18.06&-44:10:14.0&2010 Feb, 13$^{\rm th}$ &$F275W$& UVIS  & 5361 &     & 0.040& 0.08  &26.60&27.2/26.3 & 0.07 \\  
      &           &           &2013 Oct, 06$^{\rm th}$ &$V$    & IMACS &  600 & 1.10& 0.111& 0.54  &26.67&	    & 0.03 \\      
      &           &           &  ''                    &$R$    & IMACS &  900 & 1.11& 0.111& 0.50  &26.84&	    & 0.03 \\      
15:H3 &02:55:18.58&+00:48:47.6&2010 Feb, 13$^{\rm th}$ &$F390W$& UVIS  & 5130 &     & 0.040& 0.08  &27.39&27.9/27.2 & 0.32 \\      
      &      	  & 	      &2009 Dec, 17$^{\rm st}$ &$u'$   & LRIS  & 4500 & 1.07& 0.135& 0.65  &28.36&26.5/25.8 & 0.34 \\	
      &      	  & 	      &  ''		       &$V$    & LRIS  & 1000 & 1.09& 0.135& 0.67  &27.50&          & 0.21 \\	
      &      	  & 	      &  ''		       &$R$    & LRIS  & 1000 & 1.08& 0.135& 0.75  &27.42&          & 0.18 \\	
      &      	  & 	      &  ''		       &$I$    & LRIS  & 1200 & 1.07& 0.135& 0.71  &26.97&          & 0.14 \\	
16:H4 &08:16:18.99&+48:23:28.4&2010 Feb, 24$^{\rm th}$ &$F390M$& UVIS  & 5456 &     & 0.040& 0.08  &25.97&26.6/26.1 & 0.19 \\      
      & 	  &	      &2012 Jan, 25$^{\rm th}$ &$V$    & ESI   &  360 & 1.17& 0.156& 1.06  &27.35&	    & 0.13 \\    
      & 	  &	      &  ''		       &$R$    & ESI   &  360 & 1.18& 0.156& 1.07  &26.87&	    & 0.10 \\    
17:H5 &09:30:51.93&+60:23:01.1&2010 Feb, 28$^{\rm th}$ &$F390M$& UVIS  & 5721 &     & 0.040& 0.08  &25.94&26.4/25.5 & 0.11 \\       
      & 	  & 	      &2012 Mar, 17$^{\rm th}$ &$V$    & ESI   &  360 & 1.36& 0.156& 0.84  &27.08&	    & 0.07 \\    
      & 	  & 	      &  ''		       &$R$    & ESI   &  360 & 1.37& 0.156& 0.86  &26.80&	    & 0.06 \\    
18:H6 &09:08:10.36&+02:38:18.7&2010 Mar, 20$^{\rm th}$ &$F390M$& UVIS  & 5641 &     & 0.040& 0.08  &26.18&26.8/26.1 & 0.10 \\   
      & 	  & 	      &2012 Jan, 25$^{\rm th}$ &$V$    & ESI   &  360 & 1.25& 0.156& 1.14  &27.24&	    & 0.07 \\     
      & 	  & 	      &  ''		       &$R$    & ESI   &  360 & 1.22& 0.156& 0.84  &26.69&	    & 0.05 \\     
19:H7 &12:20:21.39&+09:21:35.7&2010 Apr, 13$^{\rm th}$ &$F438W$& UVIS  & 5657 &     & 0.040& 0.08  &26.98&27.5/26.8 & 0.08 \\     
      & 	  & 	      &2012 Jan, 25$^{\rm th}$ &$V$    & ESI   &  360 & 1.02& 0.156& 0.83  &27.22&	    & 0.06 \\  
      & 	  & 	      &  ''		       &$R$    & ESI   &  360 & 1.02& 0.156& 0.66  &26.71&	    & 0.05 \\  
20:H8 &14:42:33.01&+49:52:42.6&2010 Apr, 16$^{\rm th}$ &$F343N$& UVIS  & 5955 &     & 0.040& 0.08  &26.50&27.2/26.3 & 0.11 \\    
      & 	  & 	      &2012 Mar, 17$^{\rm th}$ &$V$    & ESI   &  360 & 1.21& 0.156& 0.89  &27.17&	    & 0.07 \\   
      & 	  & 	      &  ''		       &$R$    & ESI   &  360 & 1.20& 0.156& 0.87  &27.08&	    & 0.06 \\   
21:H9 &08:44:24.24&+12:45:46.7&2010 Apr, 24$^{\rm th}$ &$F275W$& UVIS  & 5663 &     & 0.040& 0.08  &26.72&27.5/26.9 & 0.23 \\    
      & 	  & 	      &2012 Jan, 25$^{\rm th}$ &$V$    & ESI   &  360 & 1.10& 0.156& 0.85  &27.14&	    & 0.11 \\    
      & 	  & 	      &  ''		       &$R$    & ESI   &  360 & 1.09& 0.156& 0.84  &26.76&	    & 0.09 \\   
22:H10&07:51:55.10&+45:16:19.6&2010 May,  4$^{\rm th}$ &$F343N$& UVIS  & 5955 &     & 0.040& 0.08  &26.40&27.1/26.6 & 0.19 \\   
      & 	  & 	      &2012 Jan, 25$^{\rm th}$ &$V$    & ESI   &  360 & 1.13& 0.156& 1.03  &27.26&	    & 0.12 \\  
      & 	  & 	      &  ''		       &$R$    & ESI   &  360 & 1.14& 0.156& 0.91  &26.72&	    & 0.09 \\  
23:H11&08:18:13.14&+07:20:54.9&2010 May, 10$^{\rm th}$ &$F390W$& UVIS  & 5657 &     & 0.040& 0.08  &27.40&28.1/26.7 & 0.07 \\   
      & 	  & 	      &2012 Jan, 25$^{\rm th}$ &$V$    & ESI   &  360 & 1.11& 0.156& 0.83  &27.10&   	    & 0.05 \\   
      & 	  & 	      &  ''		       &$R$    & ESI   &  360 & 1.09& 0.156& 0.82  &26.70&	    & 0.04 \\   
24:H12&08:18:13.05&+26:31:36.9&2010 May, 13$^{\rm th}$ &$F438W$& UVIS  & 5687 &     & 0.040& 0.08  &26.82&27.4/26.6 & 0.11 \\    
      & 	  & 	      &2012 Jan, 25$^{\rm th}$ &$V$    & ESI   &  360 & 1.09& 0.156& 0.91  &27.15&	    & 0.08 \\  
      & 	  & 	      &  ''		       &$R$    & ESI   &  360 & 1.11& 0.156& 0.98  &26.68&	    & 0.06 \\  
25:H13&08:11:14.32&+39:36:33.2&2010 May, 14$^{\rm th}$ &$F343N$& UVIS  & 5775 &     & 0.040& 0.08  &26.43&27.1/26.0 & 0.19 \\  
      & 	  & 	      &2012 Jan, 25$^{\rm th}$ &$V$    & ESI   &  360 & 1.09& 0.156& 0.93  &27.27&	    & 0.12 \\  
      & 	  & 	      &  ''		       &$R$    & ESI   &  360 & 1.09& 0.156& 0.91  &26.73&	    & 0.10 \\  
26:H14&15:08:51.94&+51:56:27.7&2010 Jun,  1$^{\rm st}$ &$F390M$& UVIS  & 6063 &     & 0.040& 0.08  &26.16&26.8/26.3 & 0.07 \\ 
      & 	  & 	      &2012 Mar, 17$^{\rm th}$ &$V$    & ESI   &  360 & 1.22& 0.156& 0.99  &27.13&	    & 0.05 \\ 
      & 	  & 	      &  ''		       &$R$    & ESI   &  360 & 1.21& 0.156& 0.88  &27.02&	    & 0.04 \\ 
27:H15&10:54:30.07&+49:19:47.1&2010 Jun,  4$^{\rm th}$ &$F390M$& UVIS  & 5955 &     & 0.040& 0.08  &26.18&26.8/26.3 & 0.07 \\ 
      & 	  & 	      &2012 Mar, 17$^{\rm th}$ &$V$    & ESI   &  360 & 1.28& 0.156& 1.16  &27.19& 	    & 0.05 \\  
      & 	  & 	      &  ''		       &$R$    & ESI   &  360 & 1.29& 0.156& 0.99  &27.00&	    & 0.04 \\  
28:H16&09:56:25.16&+47:34:42.5&2010 Jun,  6$^{\rm th}$ &$F438W$& UVIS  & 5955 &     & 0.040& 0.08  &26.89&27.5/27.2 & 0.03 \\ 
      & 	  & 	      &2012 Mar, 17$^{\rm th}$ &$V$    & ESI   &  360 & 1.18& 0.156& 0.97  &27.27&	    & 0.03 \\ 
      & 	  & 	      &  ''		       &$R$    & ESI   &  360 & 1.19& 0.156& 0.83  &27.06&	    & 0.02 \\ 
29:H17&14:41:47.52&+54:15:38.1&2010 Jun,  9$^{\rm th}$ &$F343N$& UVIS  & 6063 &     & 0.040& 0.08  &26.54&27.3/26.7 & 0.05 \\ 
      & 	  & 	      &2012 Jan, 25$^{\rm th}$ &$V$    & ESI   &  360 & 1.26& 0.156& 1.42  &27.20&	    & 0.03 \\  
      & 	  & 	      &  ''		       &$R$    & ESI   &  360 & 1.27& 0.156& 1.17  &26.68&	    & 0.03 \\  
30:H18&11:55:38.60&+05:30:50.5&2010 Jul, 12$^{\rm th}$ &$F343N$& UVIS  & 5657 &     & 0.040& 0.08  &26.39&27.2/26.4 & 0.07 \\ 
      & 	  & 	      &2012 Jan, 25$^{\rm th}$ &$V$    & ESI   &  360 & 1.04& 0.156& 0.80  &27.22&	    & 0.04 \\ 
      & 	  & 	      &  ''		       &$R$    & ESI   &  360 & 1.04& 0.156& 0.75  &26.68&	    & 0.03 \\ 
31:H19&15:24:13.35&+43:05:37.4&2010 Jul, 14$^{\rm th}$ &$F390W$& UVIS  & 5855 &     & 0.040& 0.08  &27.68&28.3/27.3 & 0.09 \\ 
      & 	  & 	      &2012 Mar, 17$^{\rm th}$ &$V$    & ESI   &  360 & 1.13& 0.156& 0.83  &27.11&	    & 0.06 \\   
      & 	  & 	      &  ''		       &$R$    & ESI   &  360 & 1.12& 0.156& 0.84  &27.01&	    & 0.05 \\ 
32:H20&13:20:05.97&+13:10:15.3&2010 Nov, 23$^{\rm rd}$ &$F336W$& UVIS  & 5663 &     & 0.040& 0.08  &26.82&27.5/26.6 & 0.10 \\
      & 	  &	      &2012 Jan, 25$^{\rm th}$ &$V$    & ESI   &  360 & 1.01& 0.156& 0.88  &27.33&	    & 0.06 \\ 
      & 	  &	      &  ''		       &$R$    & ESI   &  360 & 1.01& 0.156& 1.07  &26.84&	    & 0.05 \\ 
\hline
\end{tabular}
\end{table*}

\section{Imaging Observations}\label{obserimg}

In this section we discuss new imaging observations and the data processing for both the 
HST and the ground-based samples.

\begin{figure}
\centering
\includegraphics[scale=0.4]{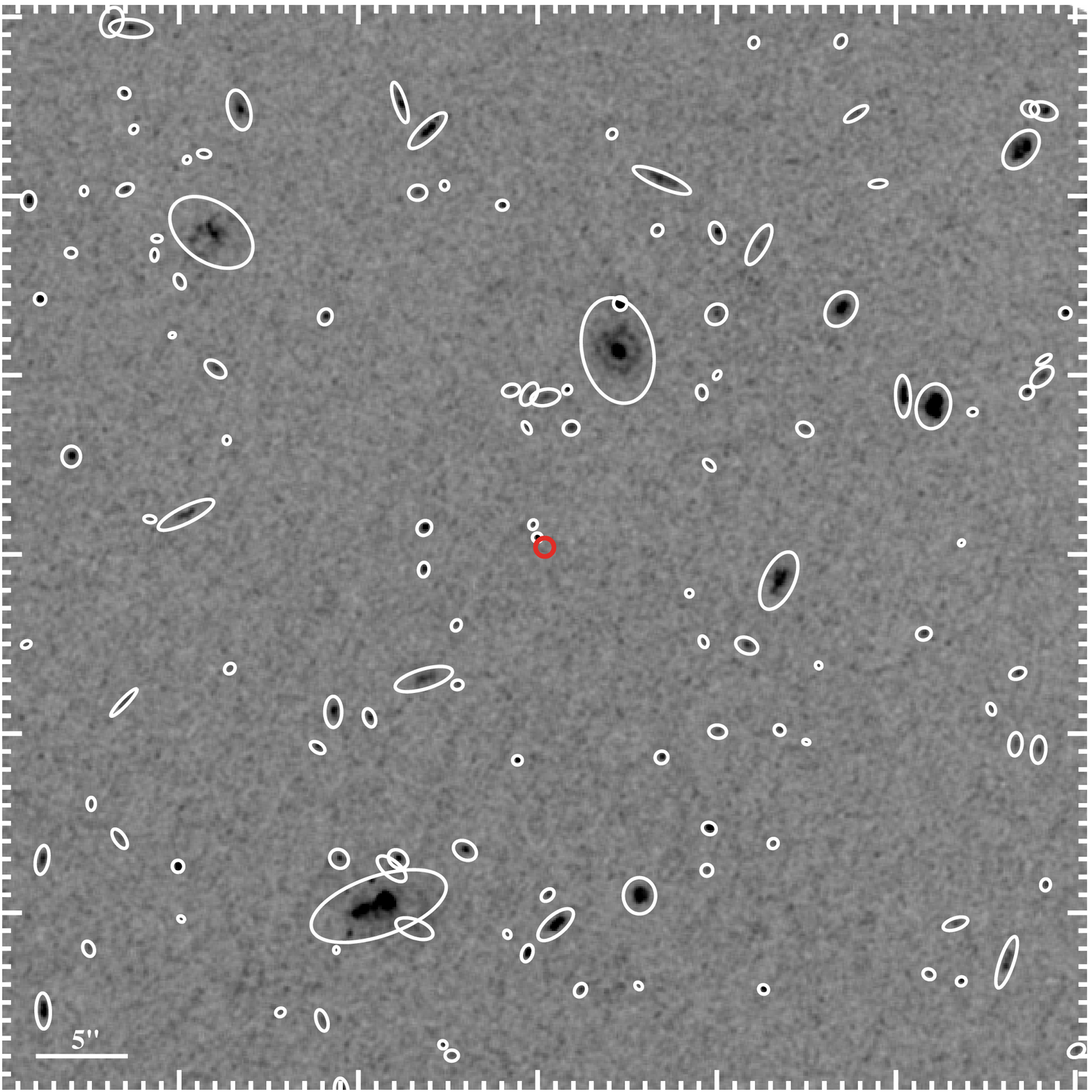}
\caption{A region of $1'\times 1'$ around the quasar position in the field 23:H11 is shown.
The red circle marks the quasar location, while the detected galaxies are identified by 
the Kron apertures used for photometry.}\label{catexam}
\end{figure}

\subsection{HST Imaging}

\subsubsection{Data Processing}

Imaging observations in the near-UV (NUV) for the HST sample were acquired during cycle 17 
(PI O'Meara, PID 11595), using the UVIS channel of WFC3. For each field, 
observations were conducted in two orbits using the filter that maximizes the overlap 
between the transmission curve and the wavelength interval defined by the DLA Lyman limit 
and the LLS Lyman limit (see Figure \ref{design}). The dates in which observations were conducted, 
the filter choices, and the effective exposure times for each quasar field are listed in Table \ref{tabimg}. 

The final images presented in this work were retrieved from the Hubble Legacy Archive (DR7)
which provides enhanced data products in the form of final co-added images in units of 
$e/s$. Data have been re-projected to a regular grid of pixel size $0.04''$ and 
cleaned from cosmic rays.  The inverse variance images that express the 
associated noise and the maps of the effective exposure time in each pixel are also retrieved 
from the archive. For photometric calibration, we utilize the zero-points published at the time of 
imaging retrieval, for which we assume a typical uncertainty of $2\%$. We further assume a point spread 
function (PSF) of $0.08''$ full width at half maximum (FWHM). A zoom-in 
of the processed images centered on the quasars is shown in Figure \ref{fig:imggallery}.

\begin{figure*}
\centering
\includegraphics[scale=0.5]{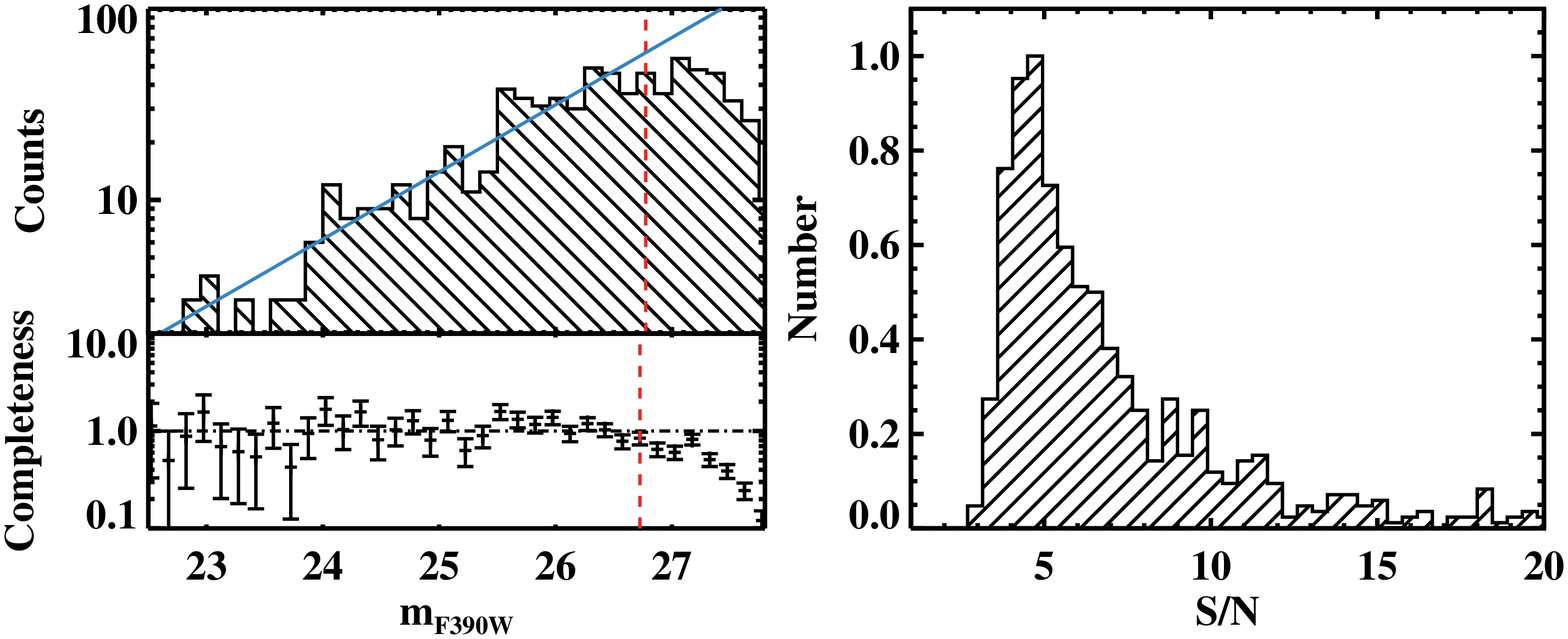}
\caption{ {\it (Left:)} Estimate of the empirical completeness for the HST field 23:H11. 
The dashed histogram in the top panel represents the observed number counts from the 
final source catalogue, while the blue solid line indicates a model fit to the
observed distribution for $m<26$ mag. The comparison between the observed and the modeled number counts 
is shown in the bottom panel. The magnitude at which the number counts deviate from the extrapolated 
model by $>10\%$ (shown by a red dashed line) is used as empirical estimator for the $90\%$ completeness limit.
{\it (Right:)} Histogram of the $S/N$ of the extracted sources which enter the final catalogue.}\label{catcompl}
\end{figure*}

\subsubsection{Source extraction and noise properties}  

To identify galaxies in the field, we use the {\sc SExtractor} 
software package \citep{ber96}, following a general procedure similar to the one employed in
the CANDELS survey \citep{gal13,guo13}, but adapted to our goals. 
We run {\sc SExtractor} with a set of parameters that maximizes the detection of faint 
and small objects, similar to the ``hot'' mode in \citet{guo13}. 

Specifically, we convolve the images with a Gaussian filter of 5 pixels in size, and we select sources 
with a minimum area of 10 pixels, using a detection threshold of 1.15, and an analysis threshold of 
3.0. At the depth of our imaging, blending of unrelated sources is a rare occurrence. Conversely, 
low-redshift and extended sources in the UV exhibit a clumpy structure, which is typical for 
star-forming galaxies. To prevent detecting these clumps as individual sources, we reduce the deblending 
efficiency during source extraction. Given the small width of the Gaussian filter, however, the very extended sources 
are still fragmented in multiple clumps, a limitation which does not affect significantly our analysis 
of galaxies at $z\gtrsim 2$. The background is computed locally in regions of 128 pixels and annuli of 48 pixels. 
Inverse variance maps are used as weight maps, and we further mask bright artifacts manually, including diffraction 
spikes from stars and ghost images. 

{\sc SExtractor} produces source catalogues with magnitudes and errors.
However, due to correlated noise in the reprojected images, the derived uncertainties underestimate 
the true errors. We therefore compute a noise model for each image, by measuring the flux standard
deviation in apertures that contain $n_{\rm pix}$ pixels within sky regions which do not 
overlap with sources according to the segmentation map. To avoid signal from residual large-scale 
fluctuations in the background, prior to this calculation, we subtract a sky model that has been smoothed 
on scales of 500 pixels, much larger than the largest aperture here considered with $n_{\rm pix}=100$. 
We then parametrize the size-dependent standard deviation on the flux with the functional form 
$\sigma(n_{\rm pix}) = \sigma_1 \alpha n_{\rm pix}^{\beta}$, where $\sigma_1$ is the flux standard deviation per pixel
across the entire image, and $\alpha$ and $\beta$ are the best-fit parameters \citep[cf.][]{gaw06}.
For each image, at fixed $n_{\rm pix}$, $\sigma(n_{\rm pix})$ is very well approximated by a Gaussian,
and typical parameters are $\sigma_1 \sim 0.001$, $\alpha \sim 1$, and $\beta \sim 0.6$,
which imply a modest but non-zero correlation in the noise. Using the measured noise properties 
of individual images, we compute limiting magnitudes at the $2\sigma$ confidence level (C.L.) within a 
circular aperture of $1''$ in diameter, as listed in Table \ref{tabimg}. These values can be used to 
quantify the depth of our images. 

\begin{figure}
\centering
\includegraphics[scale=0.32,angle=90]{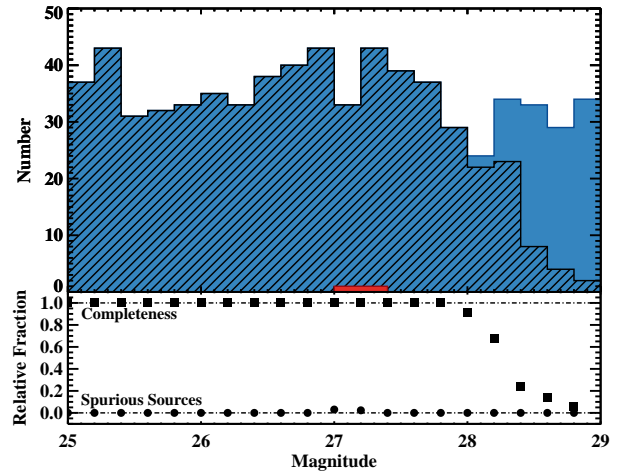}
\caption{{\it Top:} Number of the input sources in the mock image (blue filled histogram), 
compared to the number of recovered sources (black dashed histogram), and spurious 
sources we no real counterpart (red filled histogram).  
{\it Bottom:} Completeness (squares) and fraction of spurious sources (circles) as a function of 
magnitude for the final mock catalogue.}\label{hstmocks}
\end{figure}

\subsubsection{Galaxy Catalogues and Completeness}\label{sec:hstcatmw}

Having characterized the noise, we update the error on the fluxes computed within Kron apertures, 
and we include in the final catalogues only sources that have been detected with a $S/N\ge 3$.
While seemingly low, the inclusion of correlated noise translates this
limit into an effectively higher $S/N$ cut, once compared to the values returned by {\sc SExtractor}. 
Furthermore, the extraction parameters are chosen to minimize the inclusion of spurious sources (see below), 
and the resulting $S/N$ distribution peaks at $\sim 4-5$ (Figure \ref{catcompl}).
In the final catalogues, we further attempt to minimize the problem of excessive deblending of 
extended objects by removing sources that have more than 90\% of their area enclosed in another aperture. The 
fluxes of the larger apertures are also recomputed to account for the pixels that {\sc SExtractor} 
originally assigned to a different source. Figure \ref{catexam} shows an example of the sources included in 
the final catalogues for one of our quasar fields.

We assess the completeness and purity of these catalogues by means of 
simulated images, which we generate for each HST pointing.
First, we produce a noise model by filtering high frequencies from an otherwise Gaussian noise background 
so to introduce a small-scale correlation with $\beta \sim 0.6$. The final noise map is then normalized 
to the measured $\sigma_1$ in each HST image. A consistency check reveals that the noise properties of the 
mock images are in excellent agreement with the measured $\sigma(n_{\rm pix})$.
Next, we insert in these images 700 point sources uniformly drawn from the magnitude interval 
$m=25-29$ mag.

After running the same procedures used to generate the final catalogues for the science images,
we compare the input lists to the recovered mock catalogues in bins of 0.1 mag, and 
we determine the magnitude at which the completeness falls below $90\%$ as listed in Table \ref{tabimg}.
We also search for spurious detections, defined as extracted
sources with no input counterparts within a $0.4''$ search radius.
This exercise reveals that the purity of the final catalogues is 
$> 90\%$, with $\ge 95\%$ as typical value.
And while our mock images are idealized cases (with Gaussian point sources and theoretical noise models),
this test lets us conclude that the adopted extraction parameters ensure a good
compromise between depth and completeness. Using these mocks, we also confirm that our photometry is free 
from systematic errors unaccounted for in the magnitude uncertainties, given
that the input fluxes are recovered to within $2\sigma$. The results of this test for one of our
HST fields are shown in Figure \ref{hstmocks}.

We also offer a more empirical determination of the completeness limit, which also captures 
the presence of resolved (lower surface brightness) objects. 
These empirical limits are computed by modeling the number of detected sources $S$ in bins 
of 0.15 mag with a function $\log S \propto m^\chi$, where $\chi$ is a free parameter chosen 
to reproduce the linear portion of the observed number counts ($m<26.5$ in Figure \ref{catcompl}). 
The magnitudes at which the number of extracted sources fall below $90\%$ of the extrapolated number 
counts are listed in Table \ref{tabimg}. These empirical limits are 
clearly sensitive to the magnitude interval adopted to constrain $\chi$,
especially because of the variance due to the small field of view of each HST image.
For this reason, the quoted numbers are useful to gauge the location at which the number counts 
reach a maximum, but bear non-negligible uncertainties ($\sim 0.2$ mag).

As a final step in the production of photometric catalogues, we correct all the fluxes to account 
for Galactic extinction, using the \citet{fit99} extinction law and the \citet{sch98} dust map, which we 
re-calibrate as described in \citet{sch11}. Extinctions in each filter $A_{\rm X}$ are computed by 
convolving the wavelength-dependent extinction curve at 
each quasar position with the total instrument throughput (see Section \ref{secphotcal}) and a source spectral 
energy distribution (SED) of the form $f_{\lambda} \propto \lambda^{-2}$, as appropriate for high-redshift 
star-forming galaxies \citep[cf. Appendix B in][]{sch98}. Values are listed in Table \ref{tabimg}. 
In the following, we do not include the uncertainty on the Galactic extinction correction 
(at the level of $4-5$\%), which corresponds to a systematic error in each field.

\subsection{Ground-based Imaging}

\subsubsection{Observations}

Imaging observations for the ground-based sample were obtained in most part 
using the dual-arm Low Resolution Imaging Spectrometer \citep[LRIS;][]{oke95} at the 
Keck I telescope. In the blue camera, we consistently adopted either the $u'$ or $B$ 
filters, which were chosen to match the DLA redshifts, while we used the $V$, $R$, and $I$ 
filters in the red camera. On the blue arm, the two 2k $\times$ 4k Marconi (E2V) CCDs 
have been available throughout the entire duration of our imaging program, but the 
Tektronix/SITe 2k $\times$ 2k CCD on the red arm of LRIS was replaced with two 
2k $\times$ 4k LBNL CCD detectors in 2009 \citep{roc10}, after the observations of 
the first three fields (1:G1, 2:G2, and 3:G3). 
Additionally, two quasar fields (5:G5 and 6:G6) were observed with the Large Binocular Cameras 
\citep[LBC;][]{ped03} at the prime focus of the Large Binocular Telescope (LBT). Finally, in support 
of the HST imaging observations described in the previous section, we acquired from the ground 
imaging in the $V$ and $R$ filters, using either LRIS, or the Echellette 
Spectrograph and Imager \citep[ESI;][]{she02}, or the Inamori-Magellan Areal Camera 
and Spectrograph \citep[IMACS;][]{dre06}. A log book of the imaging observations 
is provided in Table \ref{tabimg}. Observations were conducted under a variety of 
weather conditions, mostly in clear and/or photometric skies, but sometimes in the presence of 
cirrus and patchy clouds. 

\subsubsection{Data Reduction}

All ground based images have been processed following standard reduction techniques, using 
a combination of in-house codes and the {\sc scamp},  {\sc SWarp}, and {\sc SExtractor} 
software packages \citep{ber96,ber02,ber06}. First, a bias level is subtracted from each frame 
according to the counts recorded in the overscan region. Next, we correct the pixel response 
across the image using twilight flats, or dome flats, or a combination of both. 
After applying the gain, we construct an inverse variance image which we also use to mask 
hot pixels and bad columns. Cosmic rays are identified and masked using the algorithm 
presented in \citet{van01}. Given the thickness of the upgraded CCDs on the red side of LRIS,
a much higher incidence of cosmic rays is present in these images. Further, because of the 
higher number of grazing events, trails of several arcseconds up to one arcminute are common 
in these images. In this case, we reject cosmic rays more aggressively by comparing pixels 
across multiple exposures of the same field to identify rare fluctuations in the number 
counts that are caused by transient events. 

\begin{figure}
\includegraphics[scale=0.4]{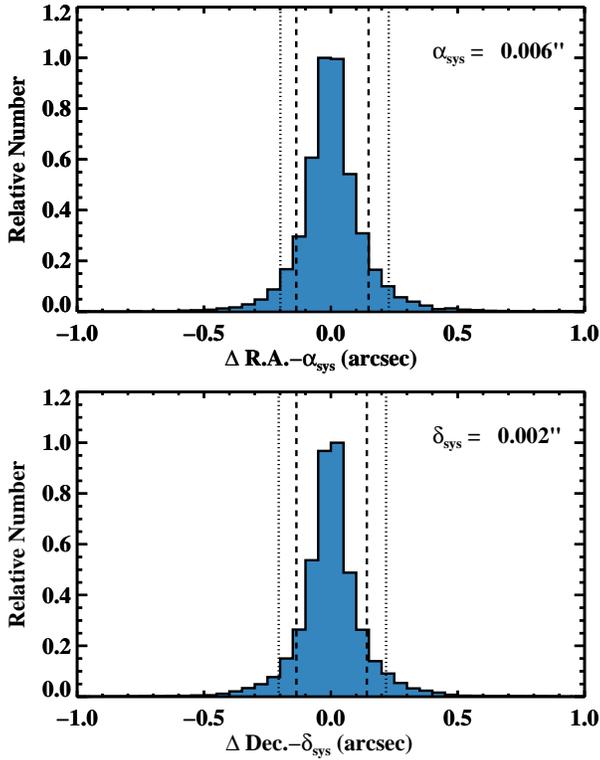}
\caption{Quality of the astrometric solutions for all imaged quasar fields.
Each panel shows the histogram of residual right ascensions (top) and declinations (bottom)
for all the detected sources with a match in SDSS-DR9 after a small systematic correction has been 
applied as indicated inside each panel. The dotted and dashed lines represent the 
5\%, 10\%, 90\%, and 95\% of the distributions. The final astrometric solutions 
have a precision of $\lesssim 1$ pixel and an accuracy of $\ll 1$ pixel.}\label{fig:astrores}
\end{figure}

Next, we fit an astrometric solution to each set of exposures of
individual quasar fields using the {\sc scamp} software. In all but one case (the 
southern field 14:H2), we use the SDSS-DR7 source catalogue as our reference coordinate system.
For the 14:H2 field instead, we use the USNO star catalogue.
Comparing all detected sources with a corresponding match in the reference catalogue, we 
find negligible systematic offsets for both the right ascension ($\alpha_{\rm sys}=0.006''$) and 
the declination ($\delta_{\rm sys}=0.002''$). Given that the native pixel sizes range from 
$0.111''$ for IMACS to $0.224''$ for LBC, our solutions thus achieve an accuracy 
of $\ll 1$ pixel. Figure \ref{fig:astrores} shows the residuals of these astrometric solutions 
relative to all the detected sources once the small systematic offsets have been subtracted. The 5\%, 10\%, 
90\%, and 95\% of the distributions are also shown, indicating that we achieve a precision 
in the astrometry across the imaged fields of $\lesssim 1$ pixel. 

Finally, using the {\sc SWarp} software, we reproject the background subtracted images
in $\rm e/s$ to a common grid of fixed pixel size. During this step, we employ 
a Lanczos resampling technique and, with the exception of the old CCD on the red side of LRIS, 
we preserve the native pixel size of each instrument ($0.156''$ for ESI, $0.224''$ for LBT, 
$0.135''$ for LRIS, and $0.111''$ for IMACS). 
As a last step, the reprojected images in the same filter 
are optimally combined, weighting by their inverse variance. A gallery of the processed images for 
each quasar field in the adopted blue filter and in the $R-$band is 
presented in Figure \ref{fig:imggallery}.

To assess the quality of the processed data, we generate a model for the PSF
by combining multiple sharp stars that are relatively isolated, i.e. without bright 
sources within a $2''$ radius. We then fit a Gaussian to the resultant light profiles 
in each filter. The inferred FWHMs are listed in Table \ref{tabimg}.

\begin{figure}
\centering
\includegraphics[scale=0.4]{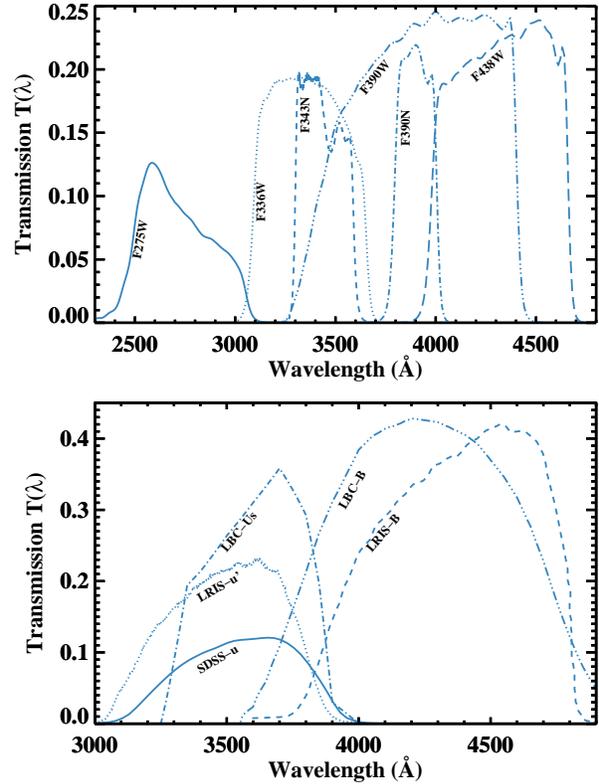}
\caption{Comparison of the effective filter transmission curves  
that have been used for this imaging survey blueward of $\sim 4500$\AA. 
The effects of the wavelength dependent quantum efficiency and, 
for ground based instruments, of the atmospheric extinction per 
unit airmass have been included. The SDSS $u$ filter is also shown 
for comparison.}\label{fig:filsetb}
\end{figure}

\begin{figure}
\centering
\includegraphics[scale=0.4]{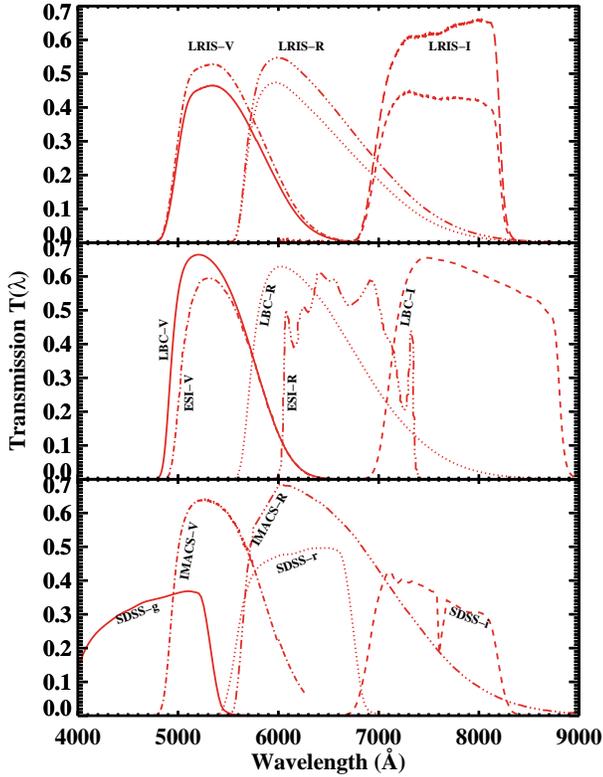}
\caption{Comparison of the effective filter transmission curves  
that have been used for this imaging survey redward of $\sim 4500$\AA. 
The effects of the wavelength dependent quantum efficiency and of the atmospheric 
extinction per unit airmass have been included. The SDSS $g$, $r$, and $i$ filters 
are also plotted for comparison. In the top panel, we display two transmission 
curves for each LRIS filter, one for the old Tektronix/SITe CCD and the other
for the new higher-throughput LBNL CCD.}\label{fig:filsetr}
\end{figure}

\begin{figure}
\centering
\includegraphics[scale=0.4]{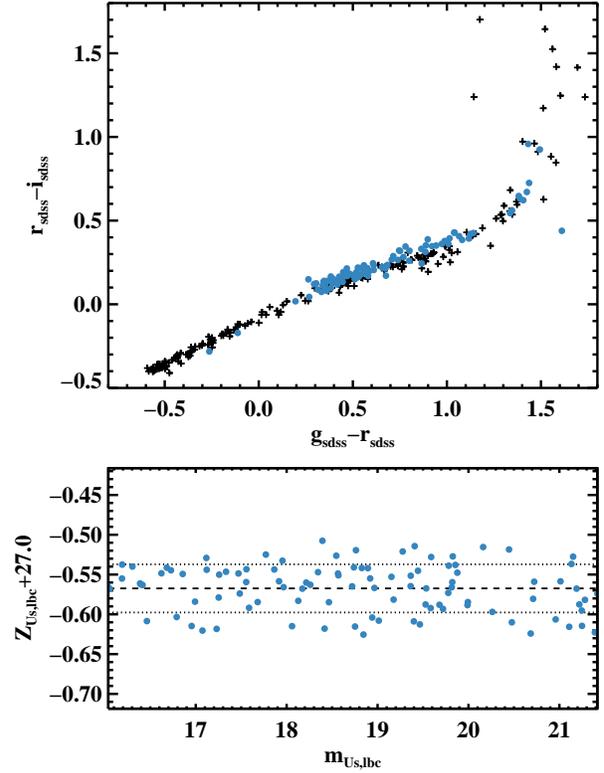}
\caption{Results of the photometric calibration for the 5:G5 field imaged with the $Us$ filter at LBC. 
In the top panel, the SDSS $g-r$ vs. $r-i$ colors are shown for the stellar templates (black crosses) and calibration stars 
(blue circles). The bottom panel shows the zero point derived for each calibrator (blue circles), together
with the mean (dashed line) and the associated standard deviation (dotted line).}\label{fig:zpcalib}
\end{figure}

\subsubsection{Photometric Calibrations}\label{secphotcal}

During the ground-based part of this imaging program, we made use of multiple instruments and filters sets:   
$Us$ and $B$ for the blue channel of LBC; $V$, $R$, and $I$ for the red channel of LBC; $V$ and $R$
for ESI; $u'$ and $B$ for the blue arm of LRIS;  $V$, $R$, and $I$ for the red arm of LRIS before 
and after the CCD upgrade\footnote{Filter transmission curves for the old LRIS detector are marked as 
$V'$, $R'$, and $I'$ }; $R$ and $V$ for IMACS at Magellan. 
For each filter, we compute the effective transmission curve
$T(\lambda)$, including the 
measured detector response $S(\lambda)$ and the best estimate of atmospheric extinction 
$\exp(-\alpha(\lambda)X)$ at airmass $X=1$ for each site. 
LRIS filter transmission curves are further corrected
for the D460 dichroic, in use during the observations.
These filter transmission curves are shown in Figures \ref{fig:filsetb} and \ref{fig:filsetr}, 
together with the adopted HST/WFC3 filters and, for comparison,  he SDSS $u$, $g$, $r$, $i$ filters.

To homogenize photometry across many different filters, instruments, 
and observing nights, we use stars within the SDSS footprint as calibrators, after applying 
color transformations and conversions to Pogson magnitudes in the AB system \citep{fuk96,lup99,smi02}.
To this purpose, we first need to compute transformations from the SDSS filter set to the 
filters available at each instrument. Both empirical and theoretical transformations from the Gunn to 
Johnson filters exist in the literature \citep[e.g.]{ken85,win91,fuk96,smi02}. However, because our filter 
set includes non-standard filters (e.g. $Us_{\rm lbc}$ and $R_{\rm esi}$), we recompute these 
transformations following common procedures \citep[e.g.][]{smi02}.  In this calculation,  
we employ stellar SEDs from the ``CALSPEC'' library of spectrophotometric stars available through 
the Hubble Space Telescope Calibration Database System, which also includes spectrophotometric stars 
from \citet{oke90}. Although this library is of high quality, it does not encompass all spectral types, 
which we recover by adding the stellar atlas by \citet{gun83}.

We test our procedures by convolving the BD+17~4708 spectrum with the SDSS filter transmission curves 
shown in Figure \ref{fig:filsetb} and \ref{fig:filsetr}, recovering the magnitudes 
$u_{\rm sdss}=10.54$, $g_{\rm sdss}=9.65$, $r_{\rm sdss}=9.35$, $i_{\rm sdss}=9.25$. These values are
in excellent agreement with those reported by \citet{fuk96} and \citet{smi02}. 
Further, for the $V_{\rm lris}$ filter which is similar to a standard Johnson $V$ filter, we 
recover the transformation $V_{\rm lris} = g_{\rm sdss}-0.025-0.60(g_{\rm sdss}-r_{\rm sdss})$, again 
in good agreement with \citet{fuk96} and \citet{smi02}. Coefficients for filter transformations 
in the form $Y_{1} = Y_{2} + a_{1} + a_{2} (Y_{3}-Y_{4})$ are listed in Table \ref{filtcoeff}, together 
with the color intervals within which these transformations hold. Typical uncertainties on the 
resulting magnitudes are $\sim 0.02-0.03$ \citep{fuk96}.

\begin{table}
\caption{Coefficients for the filter transformations used for photometric calibrations. Symbols 
are defined in the text.}\label{filtcoeff}
\centering
\begin{tabular}{c c c c c c}
\hline
$Y_{1}$         &$Y_{2}$       &$(Y_{3}-Y_{4})$          &  $a_{1}$    &   $a_{2}$    &   Interval   \\
\hline
$u'_{\rm lris}$    &$u_{\rm sdss}$ &$u_{\rm sdss}-g_{\rm sdss}$ &  $+0.007$    &   $+0.08$  &  $(-0.6,+0.8)$ \\
                  &               &                            &  $-0.008$    &   $+0.04$  &  $(+1.3,+3.8)$ \\
$Us_{\rm lbc}$    &$u_{\rm sdss}$ &$u_{\rm sdss}-g_{\rm sdss}$ &  $+0.007$    &   $-0.06$  &  $(-0.6,+1.3)$ \\
                  &               &                            &  $-0.014$    &   $-0.03$  &  $(+1.3,+3.8)$ \\
$B_{\rm lris}$    &$g_{\rm sdss}$ &$g_{\rm sdss}-r_{\rm sdss}$ &  $+0.052$    &   $+0.35$  &  $(-0.4,+1.4)$ \\
$B_{\rm lbc}$     &$g_{\rm sdss}$ &$g_{\rm sdss}-r_{\rm sdss}$ &  $+0.116$    &   $+0.50$  &  $(-0.4,+1.4)$ \\
$V_{\rm lris}$    &$g_{\rm sdss}$ &$g_{\rm sdss}-r_{\rm sdss}$ &  $-0.025$    &   $-0.60$  &  $(-0.4,+1.9)$ \\
$V'_{\rm lris}$   &$g_{\rm sdss}$ &$g_{\rm sdss}-r_{\rm sdss}$ &  $-0.025$    &   $-0.60$  &  $(-0.4,+1.9)$ \\
$V_{\rm lbc}$     &$g_{\rm sdss}$ &$g_{\rm sdss}-r_{\rm sdss}$ &  $-0.026$    &   $-0.54$  &  $(-0.4,+1.9)$ \\
$V_{\rm esi}$     &$g_{\rm sdss}$ &$g_{\rm sdss}-r_{\rm sdss}$ &  $-0.025$    &   $-0.58$  &  $(-0.4,+1.9)$ \\
$V_{\rm imacs}$   &$g_{\rm sdss}$ &$g_{\rm sdss}-r_{\rm sdss}$ &  $-0.026$    &   $-0.57$  &  $(-0.4,+1.9)$ \\
$R_{\rm lris}$    &$r_{\rm sdss}$ &$g_{\rm sdss}-r_{\rm sdss}$ &  $+0.014$    &   $-0.09$  &  $(-0.4,+1.3)$ \\
$R'_{\rm lris}$   &$r_{\rm sdss}$ &$g_{\rm sdss}-r_{\rm sdss}$ &  $+0.012$    &   $-0.08$  &  $(-0.4,+1.3)$ \\
$R_{\rm lbc}$     &$r_{\rm sdss}$ &$g_{\rm sdss}-r_{\rm sdss}$ &  $+0.015$    &   $-0.10$  &  $(-0.4,+1.3)$ \\
$R_{\rm esi}$     &$r_{\rm sdss}$ &$g_{\rm sdss}-r_{\rm sdss}$ &  $+0.035$    &   $-0.20$  &  $(-0.4,+1.3)$ \\
$R_{\rm imacs}$   &$r_{\rm sdss}$ &$g_{\rm sdss}-r_{\rm sdss}$ &  $+0.023$    &   $-0.14$  &  $(-0.4,+1.3)$ \\
$I_{\rm lris}$    &$i_{\rm sdss}$ &$r_{\rm sdss}-i_{\rm sdss}$ &  $+0.004$    &   $-0.07$  &  $(-0.4,+2.8)$ \\
$I'_{\rm lris}$   &$i_{\rm sdss}$ &$r_{\rm sdss}-i_{\rm sdss}$ &  $+0.003$    &   $-0.05$  &  $(-0.4,+2.8)$ \\
$I_{\rm lbc}$     &$i_{\rm sdss}$ &$r_{\rm sdss}-i_{\rm sdss}$ &  $+0.011$    &   $-0.22$  &  $(-0.4,+2.8)$ \\
\hline
\end{tabular}
\end{table}

Once the SDSS photometry has been transferred to the new system defined by the adopted filter set, 
we compute zero-points $Z_Y$ by comparing the observed instrumental magnitudes $Y_{\rm inst}$ to the 
intrinsic extra-atmospheric magnitudes $Y_0$, allowing for an additional offset due to atmospheric 
extinction $k_Y$ as a function of airmass X: 
$Y_{\rm inst} = Y_0 + Z_Y + k_Y X$. Since we do not aim to 
a precision of better than 0.02 mag, we neglect second-order color terms \citep[cf.][]{fuk96}.
Lacking observations across a wide range of airmasses in each filter, we avoid substantial 
degeneracy between $Z_Y$ and $k_Y$ by fixing the extinction coefficients to the following values:
$k_{Us}=0.47$, $k_{B}=0.22$, $k_{V}=0.16$, $k_{R}=0.13$, $k_{I}=0.04$ for LBC 
as reported on the instrument 
manual\footnote{http://abell.as.arizona.edu/$\sim$lbtsci/Instruments/LBC/lbc\_description.html}; 
$k_{u'}=0.41$, $k_{B}=0.19$, $k_{V}=0.12$, $k_{R}=0.11$, $k_{I}=0.07$ for LRIS \citep{coo05},
$k_{V}=0.12$, $k_{R}=0.09$ for ESI as inferred by scaling the LRIS coefficients to the 
ESI effective wavelengths;  and
$k_{V}=0.15$, $k_{R}=0.11$ for IMACS \citep{win00,kri13}.

For all the fields imaged with LRIS and LBC, we compute zero-points in each image 
(e.g. Figure \ref{fig:zpcalib}).
Conversely, only a handful of calibrators are found within the small field of view imaged by ESI, 
preventing us from computing individual zero-point reliably. However, this imaging has been 
acquired in three clear nights during which the SkyProbe at CFHT recorded an attenuation 
of $\lesssim 0.05$ mag. We therefore combine calibrators from each field and fit for a single 
zero-point in the $R_{\rm esi}$ and $V_{\rm esi}$ filters. Uncertainties on the zero-points are computed 
combining the standard errors on the mean  $Z_Y$ with an error of $0.05$ mag, which accounts for 
residuals in the filter transformations and errors in the aperture photometry for the 
calibrator stars.

For the southern field 14:H2, instead, zero-points are computed with photometric standard 
stars in the \citet{lan92} SA95 field, which we observed during the same night in 
which science data were taken, under photometric conditions.

\begin{figure}
\centering
\includegraphics[scale=0.4]{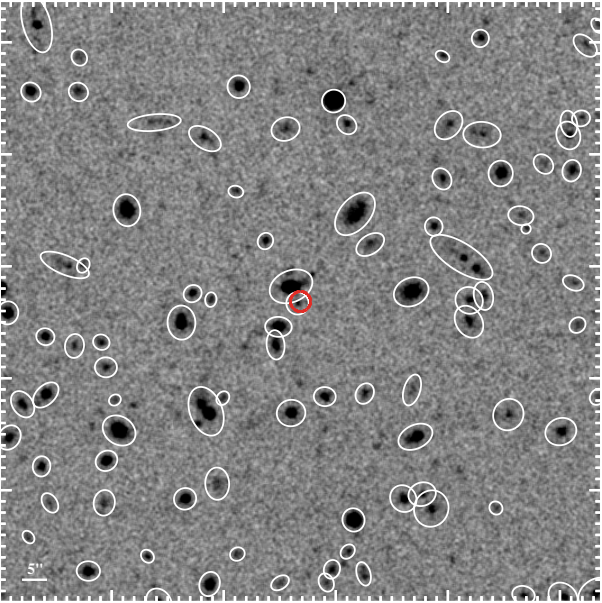}
\caption{A region of $1'\times 1'$, imaged in the $Us$-band, is shown
around the quasar position in the field 5:G5. The red circle marks the quasar location, 
while the detected galaxies are identified by the Kron apertures used for photometry.}\label{fig:grosou}
\end{figure}

\begin{figure*}
\centering
\includegraphics[scale=0.5]{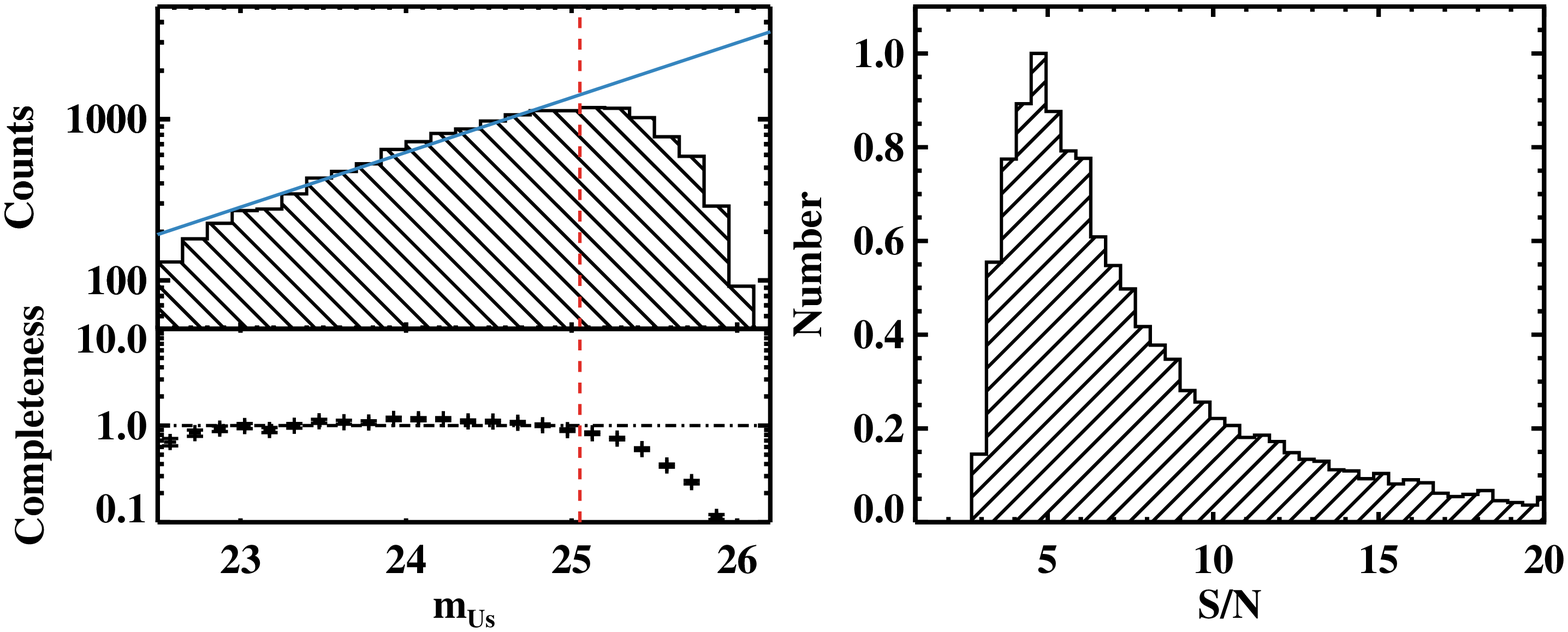}
\caption{ {\it (Left:)} Estimate of the empirical completeness for the field 5:G5
imaged in the $Us$-band with LBC. 
The dashed histogram in the top panel represents the observed number counts from the 
final source catalogue, while the blue solid line indicates a model fit to the
observed distribution for $m<25$ mag. The comparison between the observed and the modeled number counts 
is shown in the bottom panel. The magnitude at which the number counts deviate from the extrapolated 
model by $>10\%$ (shown by a red dashed line) is used as empirical estimator for the $90\%$ completeness limit.
{\it (Right:)} Histogram of the $S/N$ of the extracted sources which enter the final catalogue.}\label{fig:grocom}
\end{figure*}

\subsubsection{Source Extraction, Noise Properties and Completeness}

For the ground-based imaging, we generate source catalogues and we characterize the image noise and
completeness following similar procedures to those described in the previous section 
for the HST imaging. First, we generate source catalogues by running {\sc SExtractor} 
after convolving images with a Gaussian filter of $\sim 4-6$ pixels in size, and by 
selecting sources with a minimum area of $\sim 20-30$ pixels, using a detection threshold of $\sim 0.6-0.8$, 
and an analysis threshold of $\sim 3.0$ (e.g. Figure \ref{fig:grosou}). These parameters are adjusted 
for specific combinations of filters and instruments for optimization.  The background is computed locally 
in regions of 64 pixels and annuli of 28 pixels. Inverse variance maps are again used as weight maps, 
through which we also mask bright artifacts, such as diffraction spikes from bright stars and, 
occasionally, poorly-masked cosmic rays. Differently from the HST imaging, we enable deblending of 
overlapping sources. 

Following the above procedures, we model the noise in each image as
$\sigma(n_{\rm pix}) = \sigma_1 \alpha n_{\rm pix}^{\beta}$. We find a range of 
coefficient across different instruments and filters, in the interval $\beta \sim 0.55-0.70$. 
The noise estimates are used to measure limiting magnitudes at the $2\sigma$ C.L.
within a circular aperture of $1''$ in diameter (listed in Table \ref{tabimg}), and 
to recompute the errors on the fluxes within Kron apertures.
For the catalogues in the blue filters ($U,B$), which are the most relevant for our following analysis, 
we also compute completeness limits (e.g. Figure \ref{fig:grocom}) with the same procedures used for the 
HST imaging (i.e. both from measured number counts and from mocks with artificial point sources 
matched to the seeing). Magnitudes at which completeness falls below $90\%$ are listed in Table \ref{tabimg}.
With mock images we also test the quality of the photometry, concluding that input values are recovered
for all sources within 2$\sigma$ errors. The number of spurious detections is found to be $\lesssim 5-10\%$
above and at the completeness limits, and increasing between $10-20\%$ at the faintest magnitudes, where only
a handful of sources are typically detected. Finally, we correct all fluxes to account for Galactic 
extinction as described in Section \ref{sec:hstcatmw}.

\begin{table}
\caption{Log book of the quasar spectroscopic observations.}\label{logspec}
\begin{center}
\begin{tabular}{l l c c c c c c c c c c c}
\hline
Field$^a$&UT Date$^b$&Instr.$^c$&Slit$^d$  & Exp. Time$^e$ \\
     &       &      &($''$)&(s)        \\
\hline										  
    1:G1  &2010 Jan, 5$^{\rm th}$   &ESI&   0.5 & 1800 \\
    2:G2  &2010 Jan, 5$^{\rm th}$   &ESI&   0.5 & 1800 \\
    3:G3  &2010 Jan, 5$^{\rm th}$   &ESI&   0.5 & 2400 \\
    4:G4  &2010 Jan, 5$^{\rm th}$   &ESI&   0.5 & 3600 \\
    5:G5  &2012 Jan, 25$^{\rm th}$  &ESI&   0.75& 3600 \\
    6:G6  &2010 Jan, 5$^{\rm th}$   &ESI&   0.5 & 3600 \\
    7:G7  &2010 Jan, 5$^{\rm th}$   &ESI&   0.5 & 1800 \\
    8:G9  &2012 Jan, 25$^{\rm th}$  &ESI&   0.75& 2400 \\
    9:G10 &2010 Jan, 5$^{\rm th}$   &ESI&   0.5 & 1800 \\
    10:G11&2010 Jan, 5$^{\rm th}$   &ESI&   0.5 & 3600 \\
    11:G12&2010 Jan, 5$^{\rm th}$   &ESI&   0.5 & 1800 \\
    12:G13&2010 Jan, 5$^{\rm th}$   &ESI&   0.5 & 2400 \\
    13:H1 &2010 Jan, 5$^{\rm th}$   &ESI&   0.5 & 1800 \\
    14:H2 &2009 Dec, 22$^{\rm th}$  &MagE&  0.7 & 1800 \\
    15:H3 &2010 Jan, 5$^{\rm th}$   &ESI&   0.5 & 2400 \\
    16:H4 &2010 Jan, 5$^{\rm th}$   &ESI&   0.5 & 2400 \\
    17:H5 &-      		    &SDSS&    - &  -   \\
    18:H6 &2013 Mar, 2$^{\rm nd}$   &MagE&  0.7 & 3600 \\
    19:H7 &-       		    &[1]&     - &  -   \\
    20:H8 &-      		    &SDSS&    - &  -   \\
    21:H9 &2013 Jan,  4$^{\rm th}$  &ESI&   0.75&  960 \\
    22:H10&2010 Jan,  5$^{\rm th}$  &ESI&   0.5 & 2400 \\
    23:H11&2012 Jan, 25$^{\rm th}$  &ESI&   0.75& 3600 \\
    24:H12&2013 Jan,  4$^{\rm th}$  &ESI&   0.75& 3600 \\
    25:H13&2012 Jan, 25$^{\rm th}$  &ESI&   0.75& 3600 \\
    26:H14&-      		    &SDSS&   -  &  -   \\
    27:H15&-      		    &SDSS&   -  &  -   \\
    28:H16&                         &[2] &   -  &  -   \\
    29:H17&-      		    &SDSS&   -  &  -   \\
    30:H18&-                        &[3] &   -  &  -   \\
    31:H19&                         &SDSS&   -  &  -   \\
    32:H20&                         &[4] &   -  &  -   \\
\hline 
\end{tabular}
\end{center}
$^a$ ID of the quasar field. $^b$ UT date during which observations were conducted.
$^c$ Instrument/survey, or reference to the source of archival data. $^d$ Slit width.
$^e$ Exposure time. [1] MagE data from \citet{jor13}; [2] ESI data from \citet{pro03b}; [3] ESI data from \citet{wol08};
[4] X-shooter data from Cooke et al. (private communication).
\end{table}

\section{Spectroscopic Observations}\label{obserspe}

To characterize the absorption properties of both the DLAs and the higher redshift LLSs, 
we collect and analyze spectroscopic data for the targeted quasars in each field, as 
detailed in Table \ref{logspec}. For 21 quasars, we acquired new observations using ESI 
in echellette mode or, for two sightlines, the Magellan Echellette
Spectrograph \citep[MagE;][]{mar08}. For the remaining 11 quasars, we rely on spectra published 
in the literature \citep{jor13,pro03b,wol08} or on SDSS spectroscopy, and, for 32:H20, 
on archival X-shooter \citep{ver11} data from programme ID 087.A-0022 (PI R. Cooke).

\subsection{New Observations and Data Reduction}

New spectroscopic observations for 20 quasars have been obtained with ESI 
under good to moderate weather and seeing conditions. ESI spectra were acquired with two 
different choices of slit width, $0.5''$ or $0.75''$, 
matched to the seeing conditions. The corresponding velocity resolution of ESI is $\sim 37~\rm km~s^{-1}$
for the $0.5''$ slit and $\sim 56~\rm km~s^{-1}$ for the 
$0.75''$ slit. Observations for two additional quasars were instead obtained with
MagE using a $0.7''$ slit, under good weather conditions. In this configuration, MagE yields 
spectra at a resolution of $\sim 70~\rm km~s^{-1}$. A log book of the spectroscopic observations
with the corresponding exposure times is provided in Table \ref{logspec}. 

ESI data were reduced using the {\sc ESIRedux} software 
package\footnote{https://www2.keck.hawaii.edu/realpublic/realpublic/inst/esi/ESIRedux}.
The pipeline processes the 2D frames by performing bias subtraction and flat fielding. Next, it 
creates a wavelength solution using arc lines, and it extracts and coadds the spectra in each 
order, producing wavelength- and flux-calibrated 1D spectra, together with associated errors.
Flux calibration is achieved with repeated observations of spectrophotometric standard stars 
that were observed throughout the night. MagE spectra are processed following
similar procedures, but using the {\sc mase} pipeline \citep{boc09}.

\begin{figure}
\centering
\includegraphics[scale=0.32]{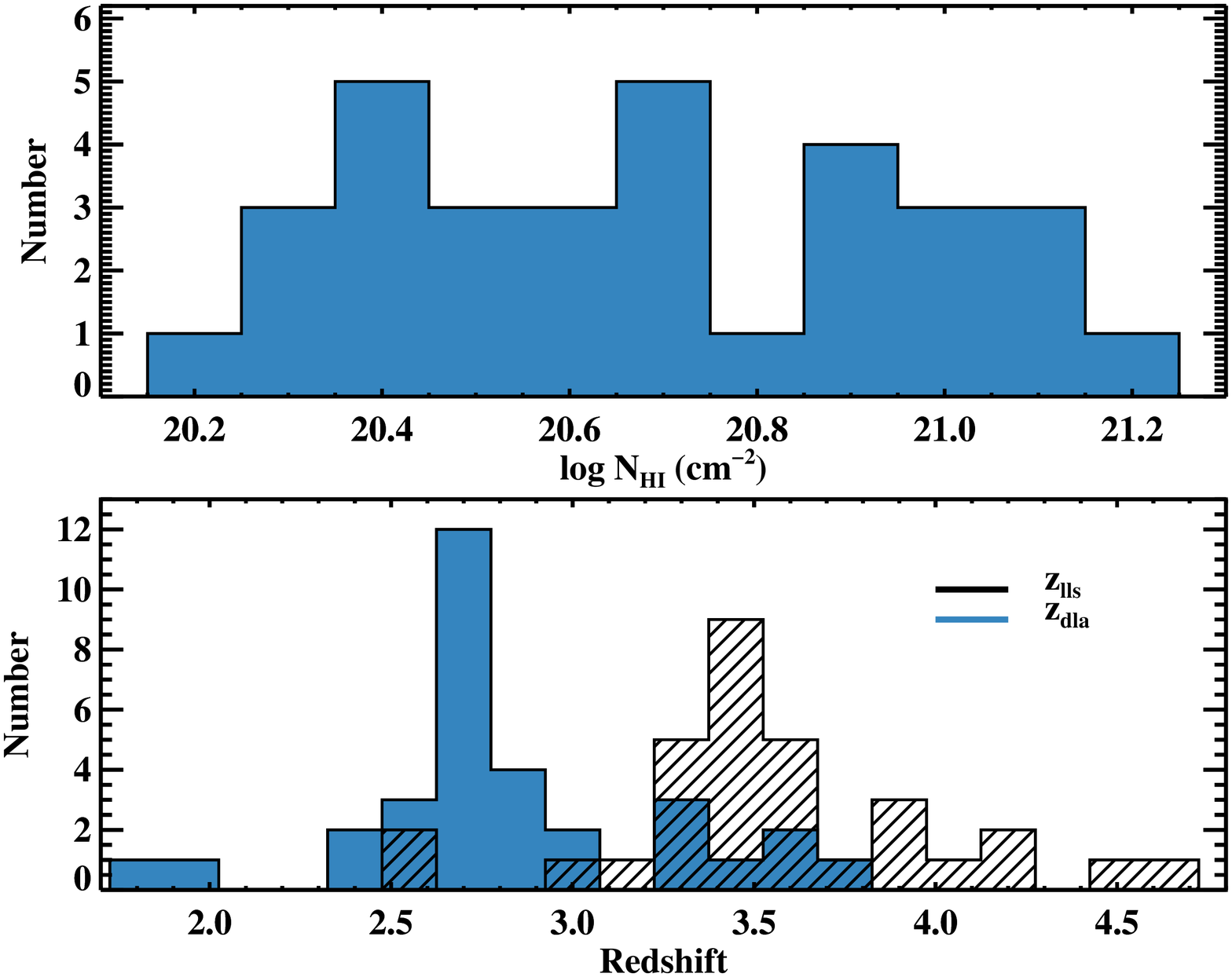}
\caption{{\it Top:} The \ion{H}{I} column density distribution of the targeted 
DLAs. {\it Bottom:} The redshift distributions of the targeted DLAs and of the 
higher redshift LLSs that act as blocking filters. Our sample includes absorbers 
that are representative of the general DLA populations at $z\sim 2.5-3$, although 
they are not a statistical selection from the column density distribution
function.}\label{fig:specprop}
\end{figure}

\begin{figure*}
  \begin{tabular}{cccc}
  \includegraphics[scale=0.15,angle=90]{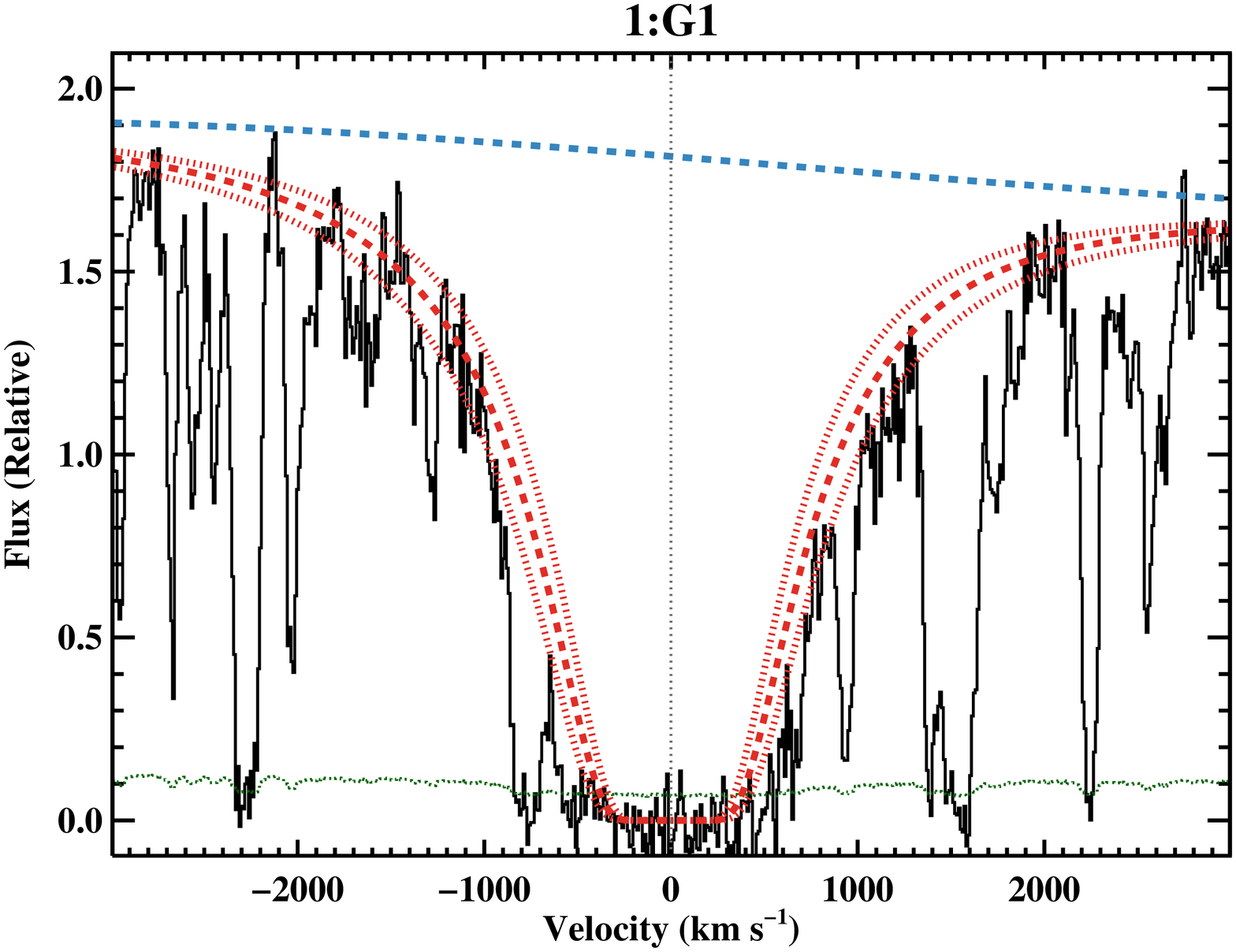}&
  \includegraphics[scale=0.15,angle=90]{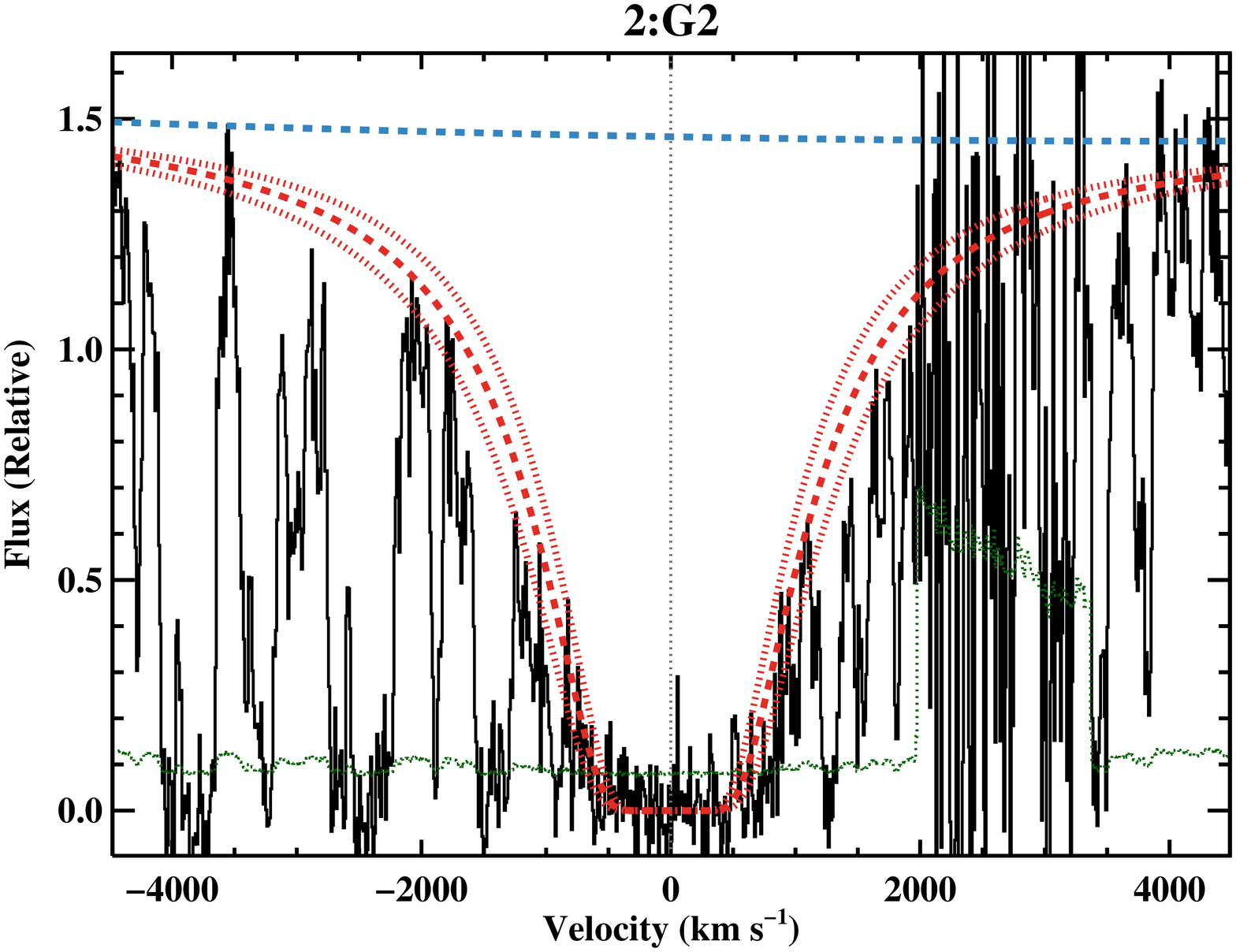}&
  \includegraphics[scale=0.15,angle=90]{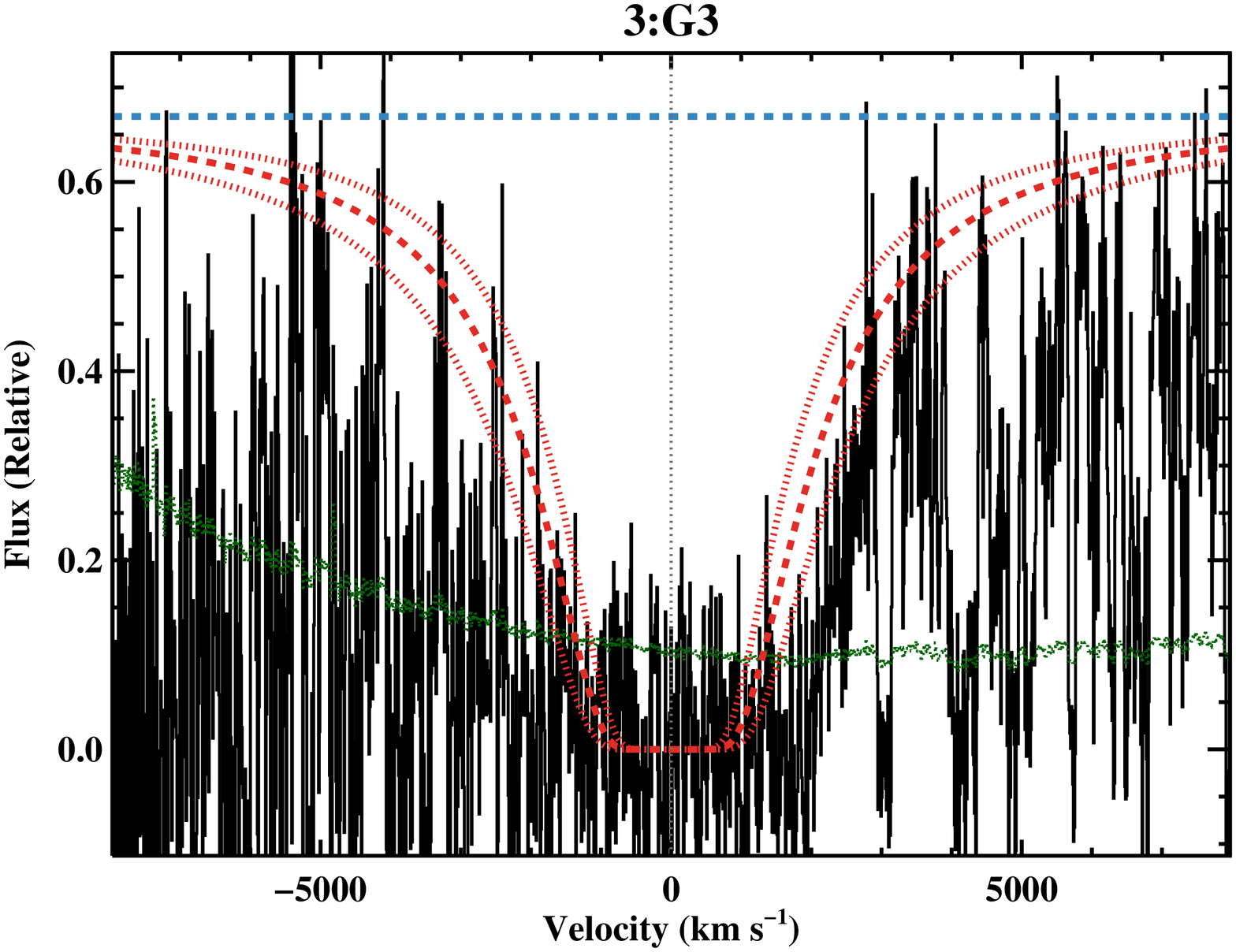}&
  \includegraphics[scale=0.15,angle=90]{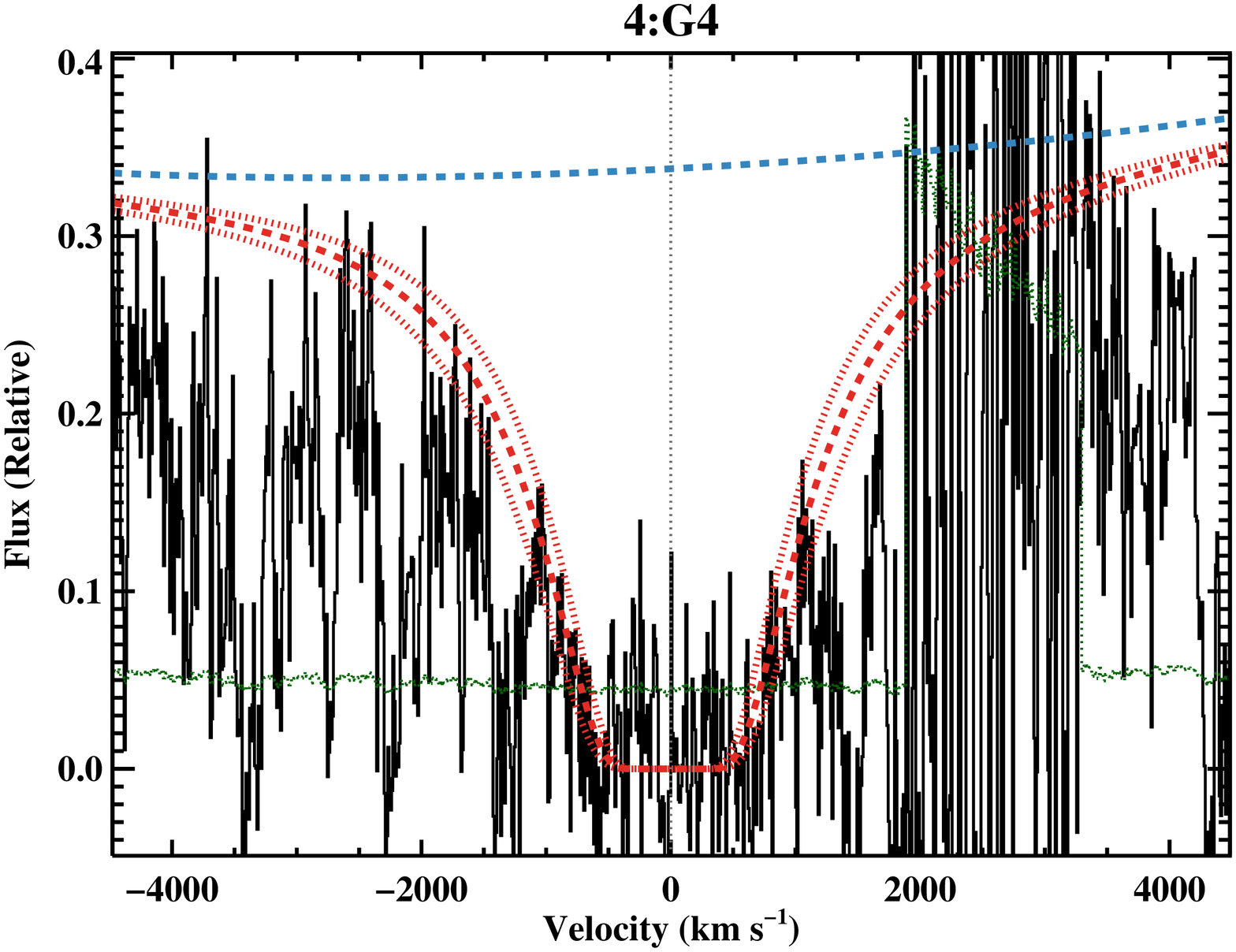}\\
  \includegraphics[scale=0.15,angle=90]{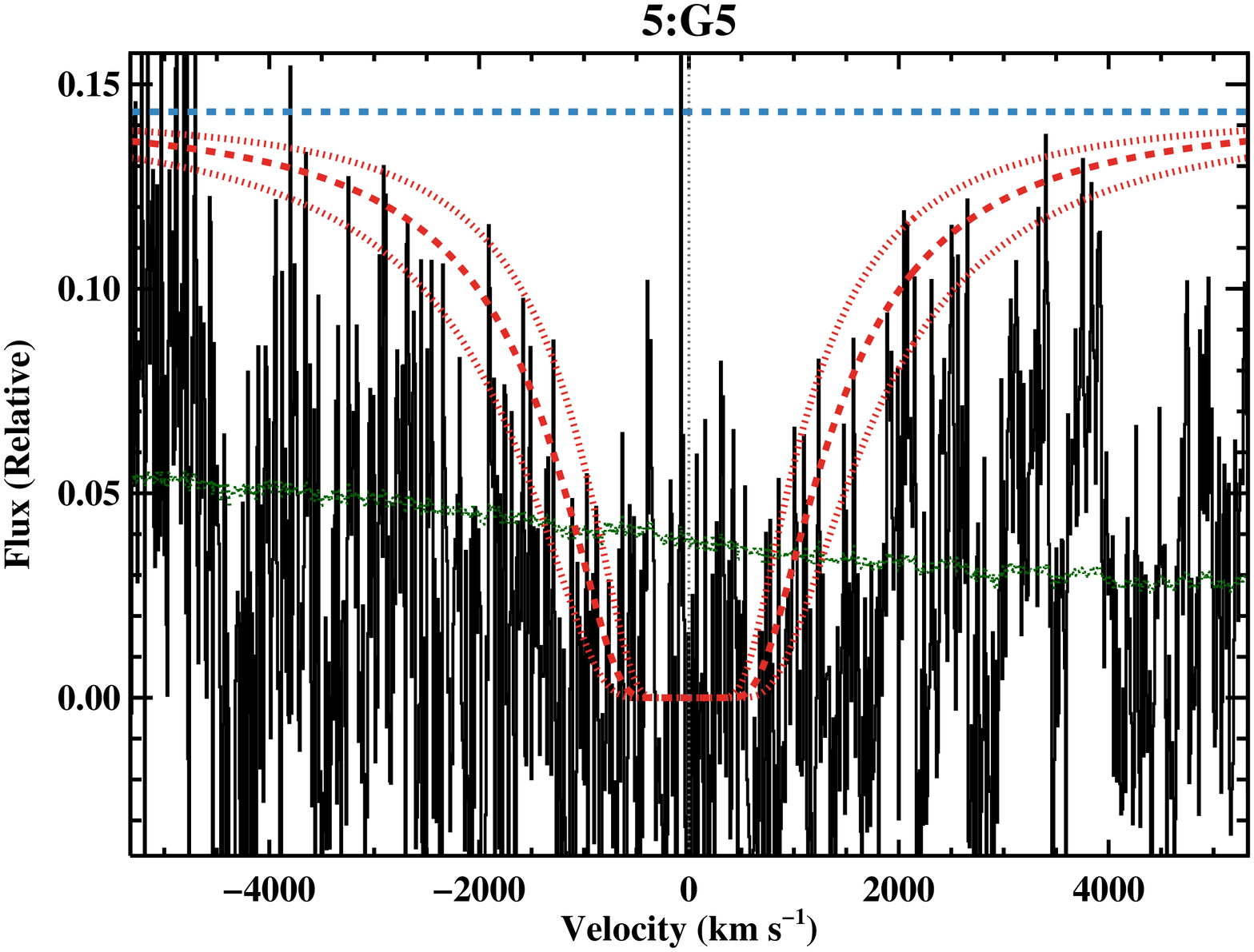}&
  \includegraphics[scale=0.15,angle=90]{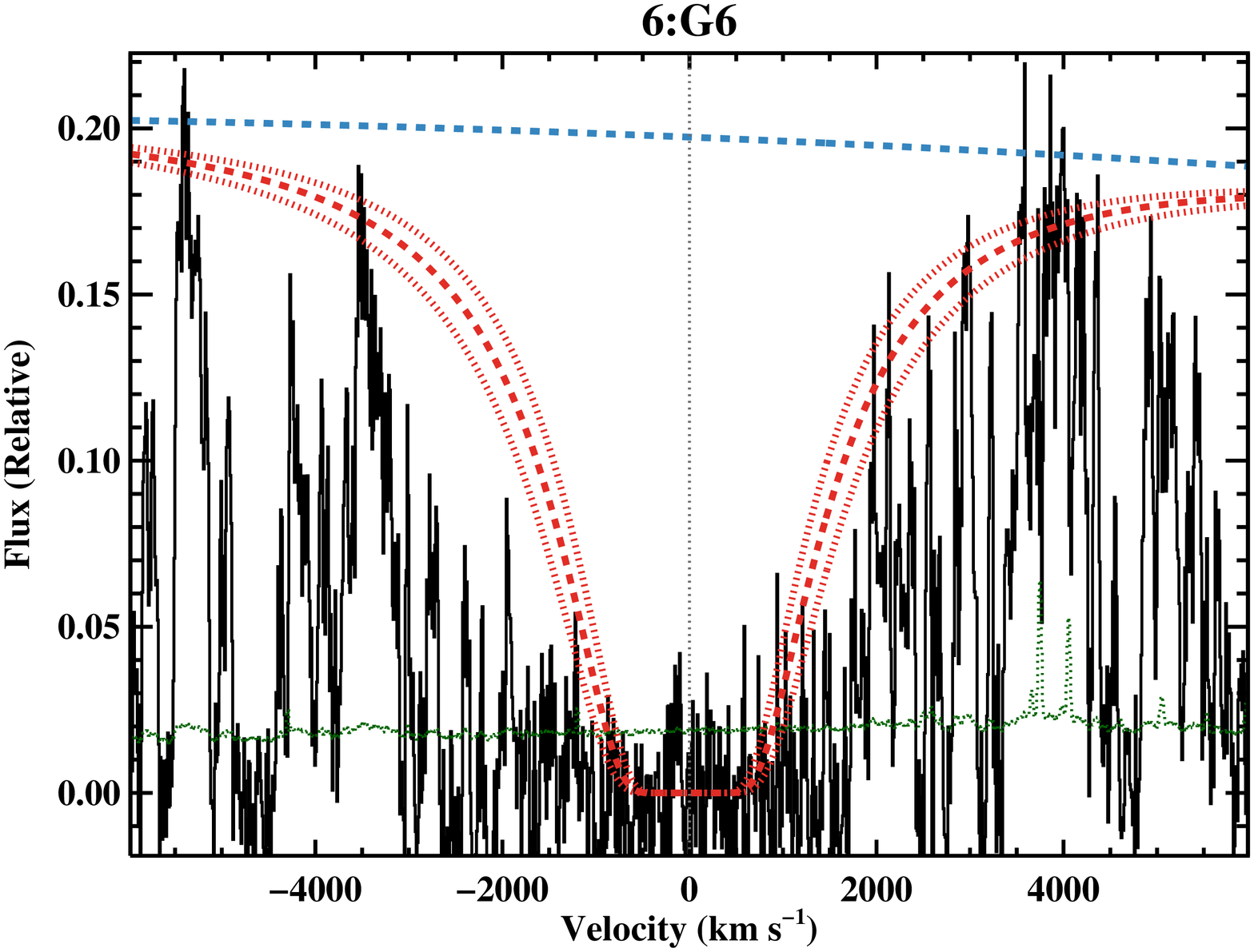}&
  \includegraphics[scale=0.15,angle=90]{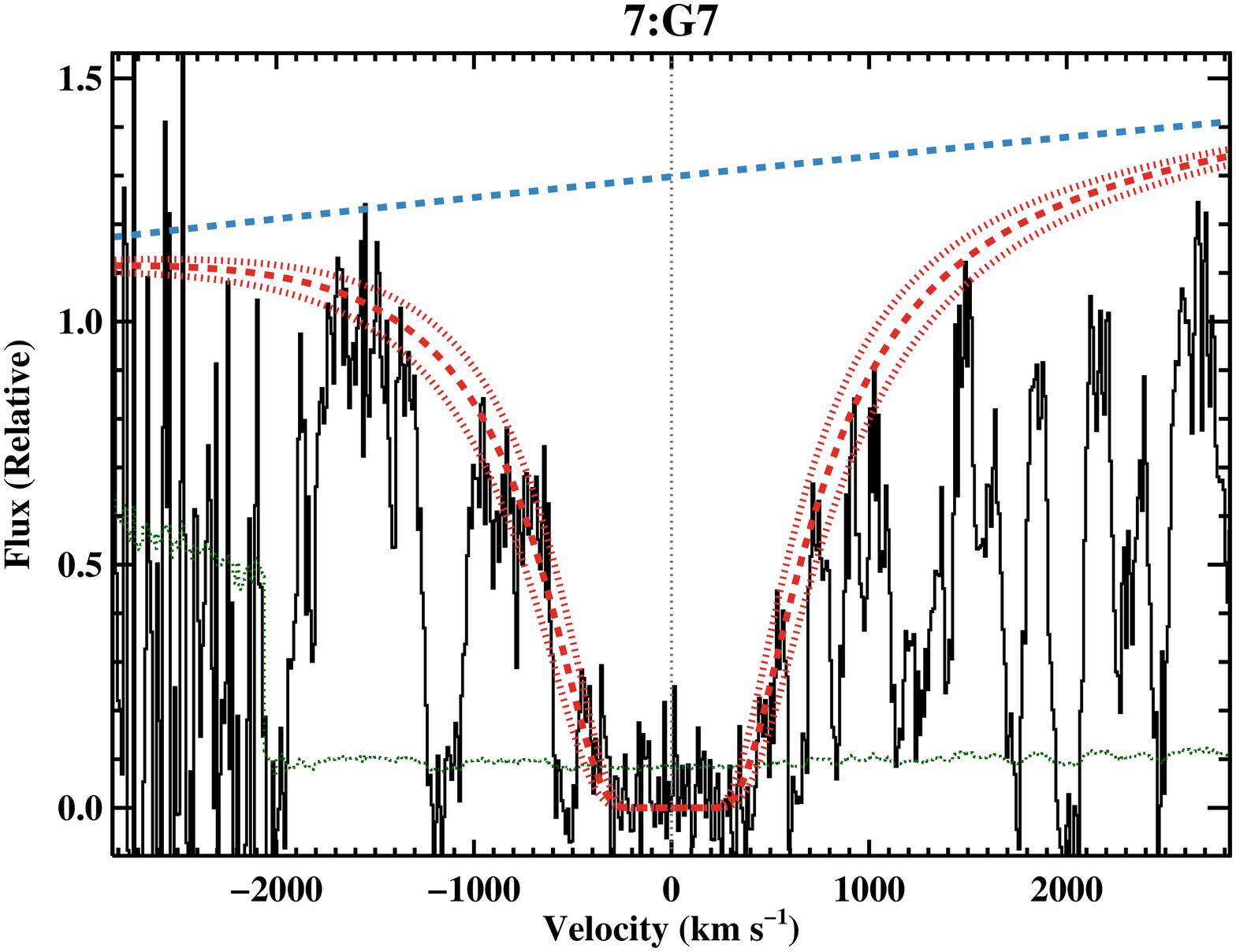}&
  \includegraphics[scale=0.15,angle=90]{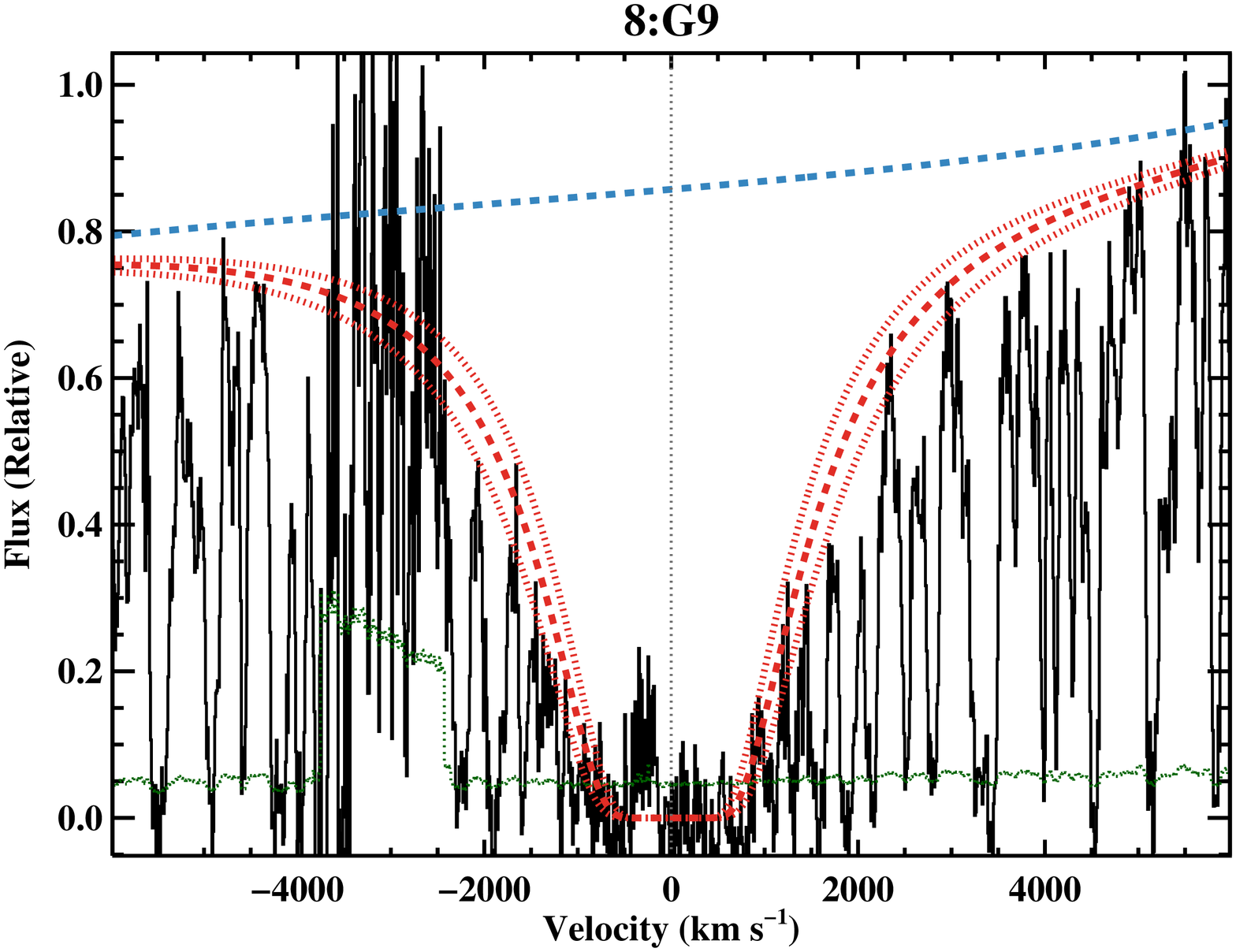}\\
  \includegraphics[scale=0.15,angle=90]{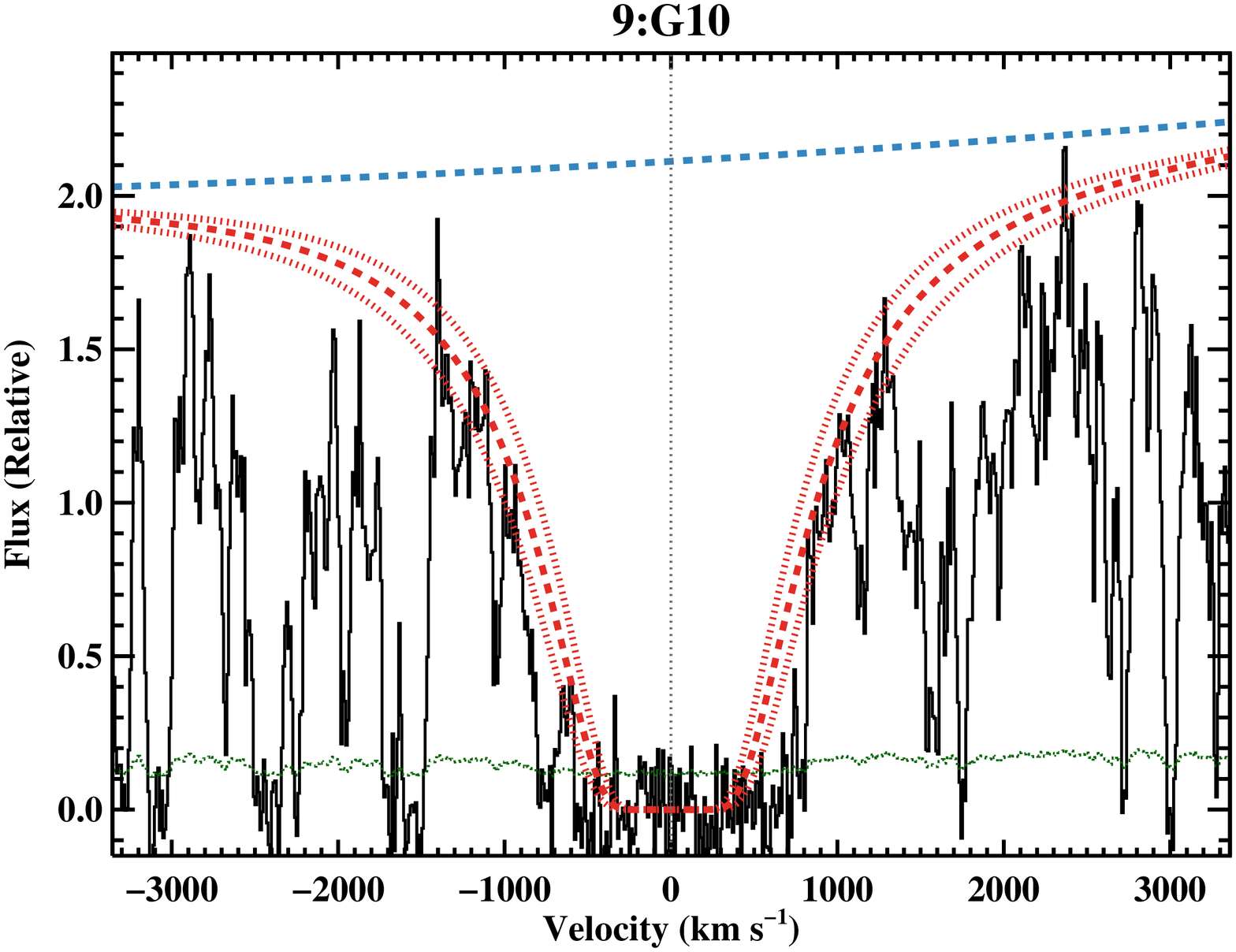}&
  \includegraphics[scale=0.15,angle=90]{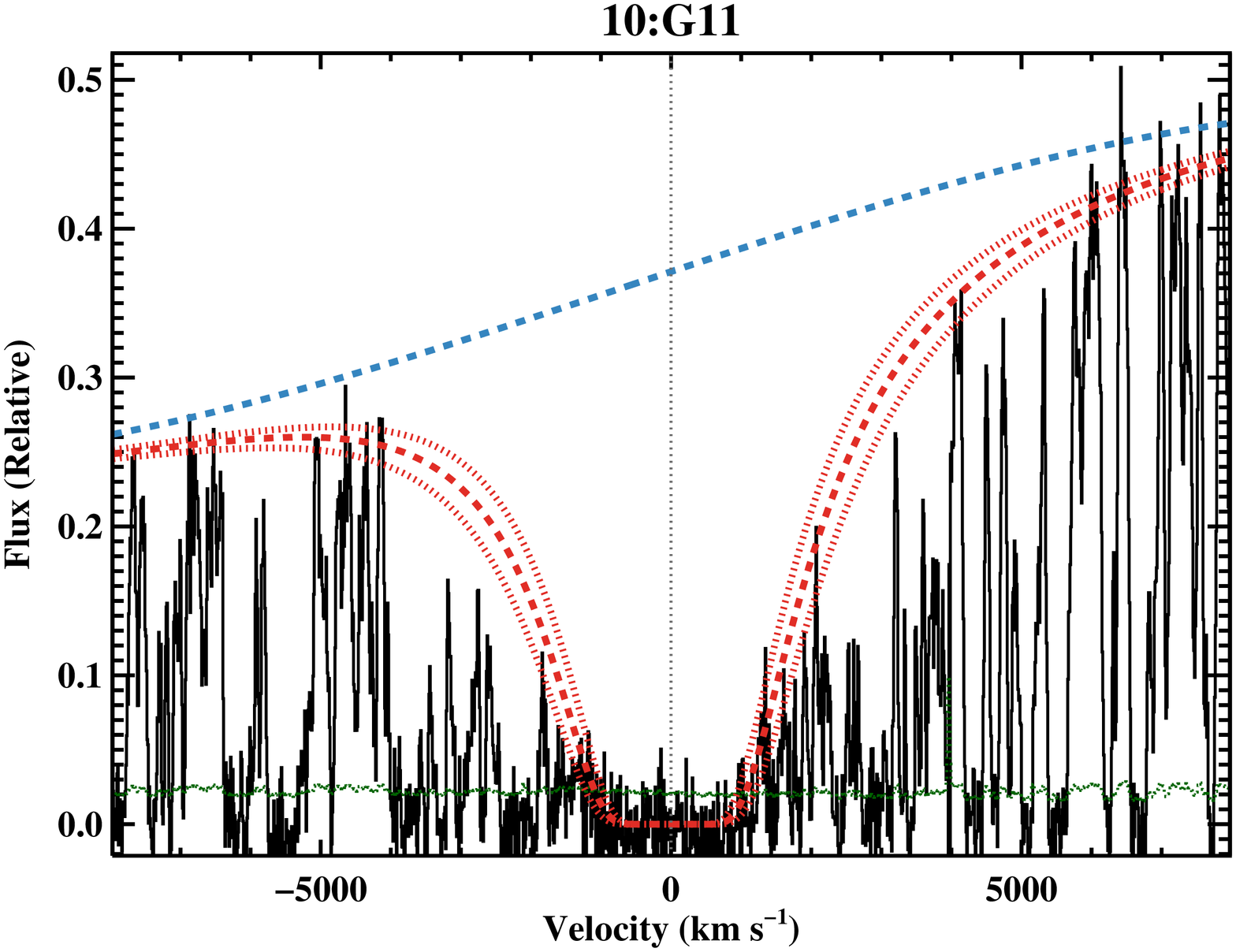}&
  \includegraphics[scale=0.15,angle=90]{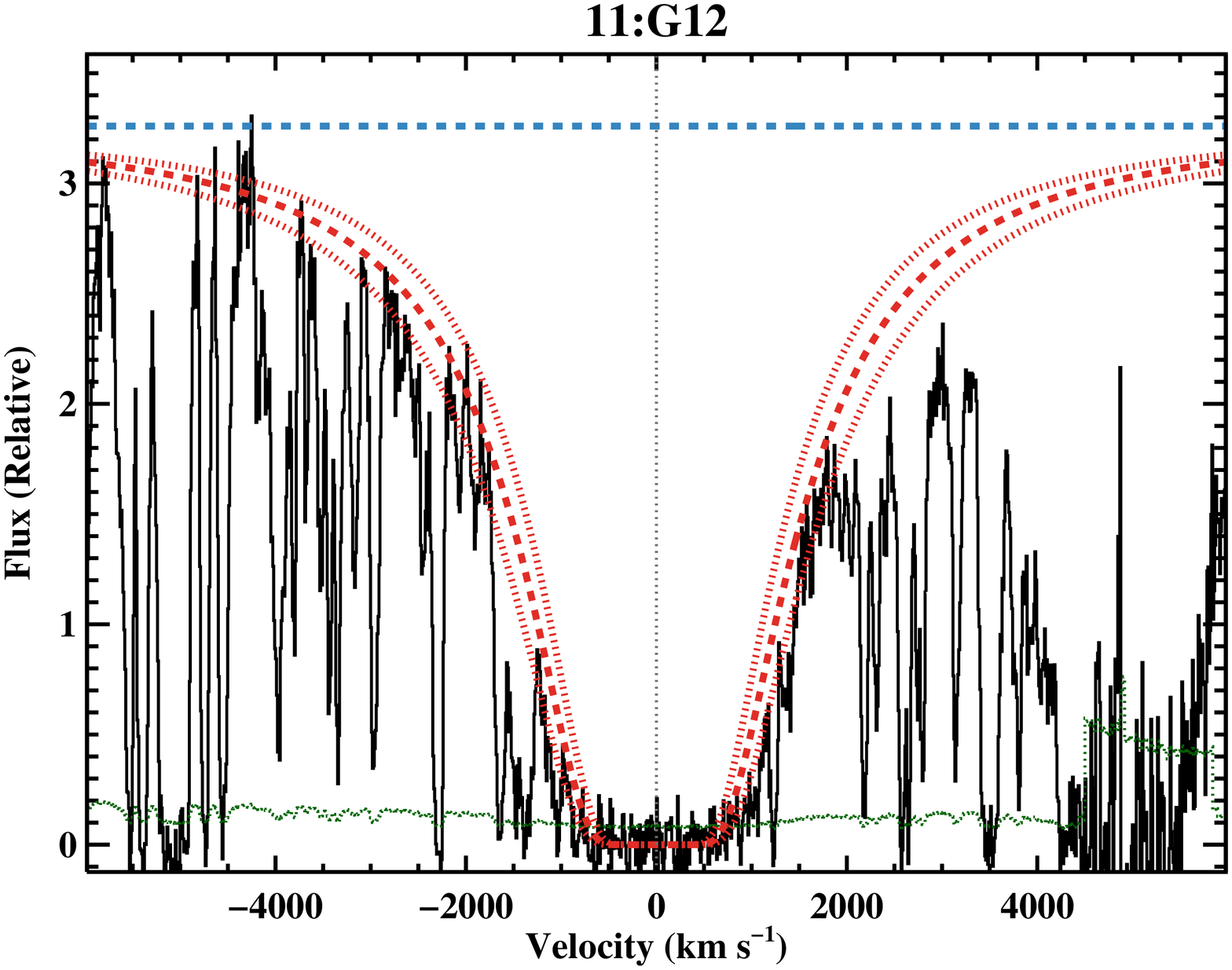}&
  \includegraphics[scale=0.15,angle=90]{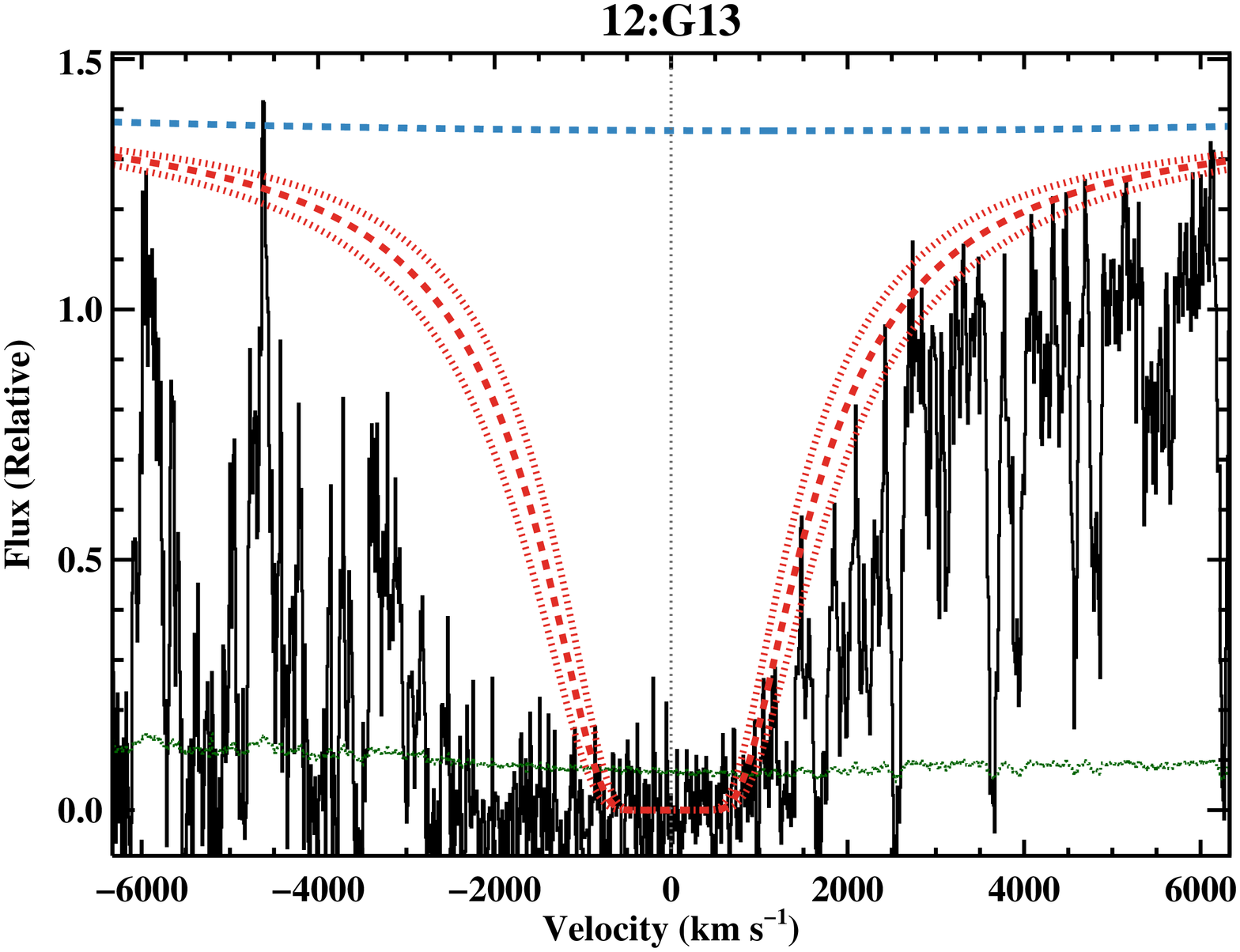}\\
  \includegraphics[scale=0.15,angle=90]{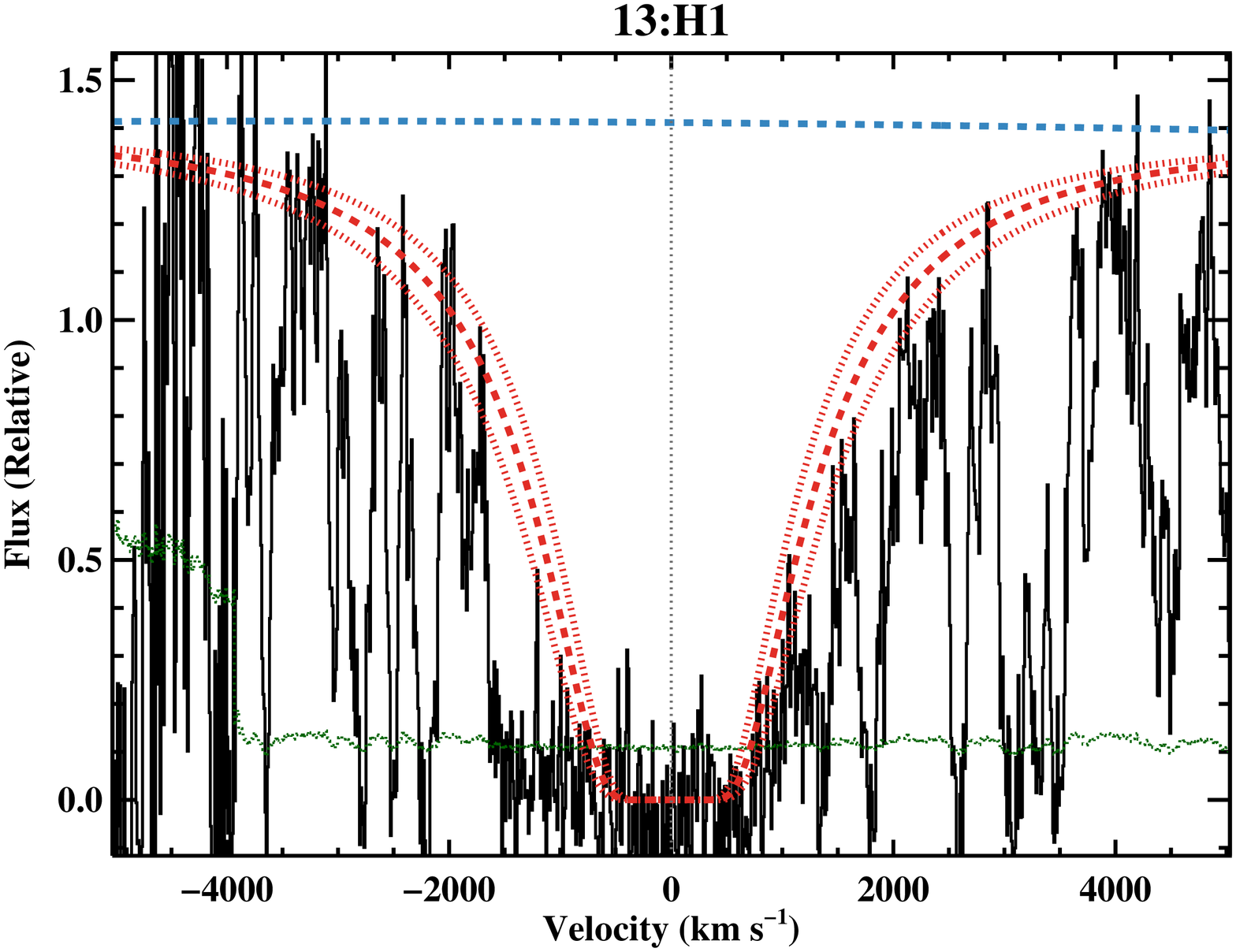}&
  \includegraphics[scale=0.15,angle=90]{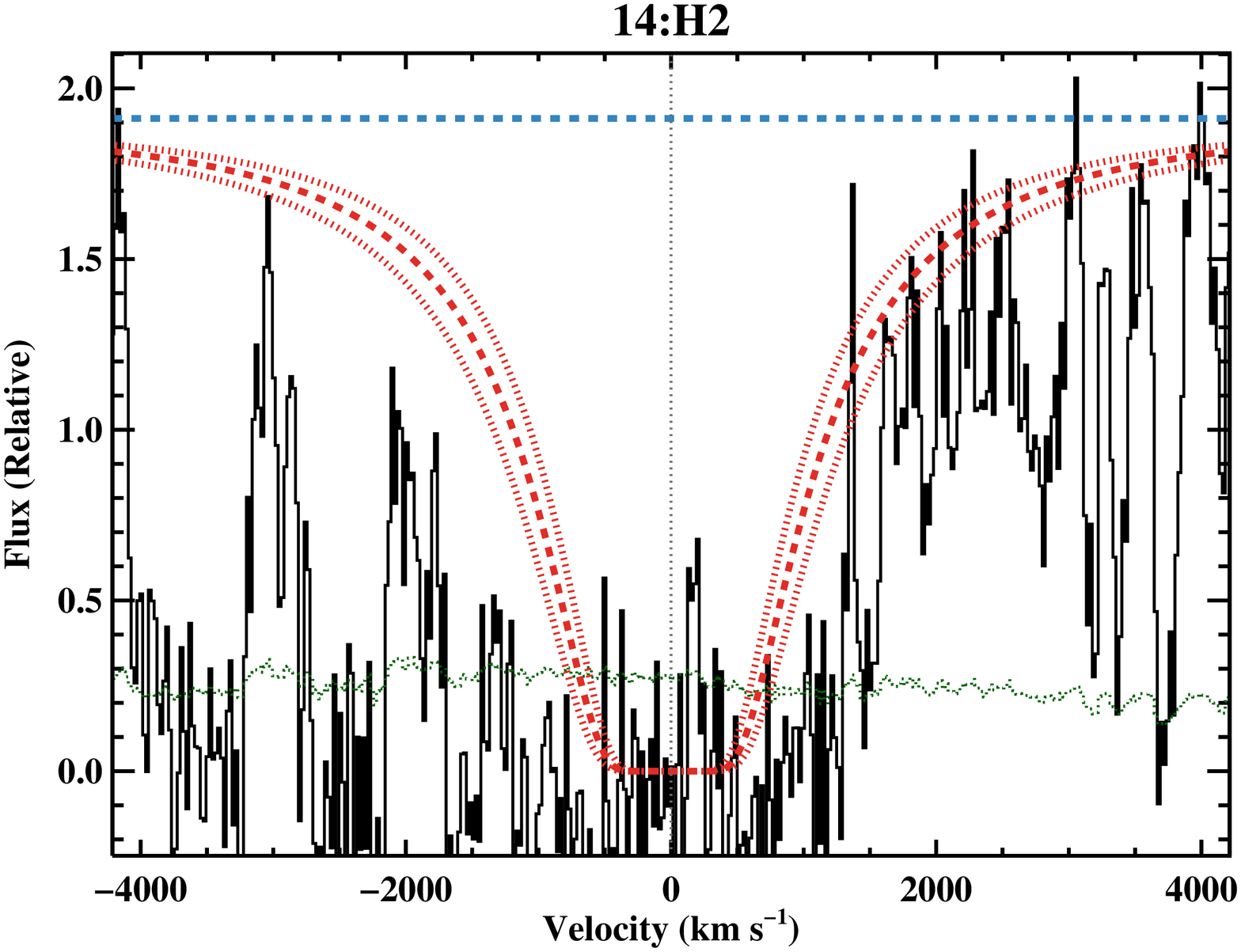}&
  \includegraphics[scale=0.15,angle=90]{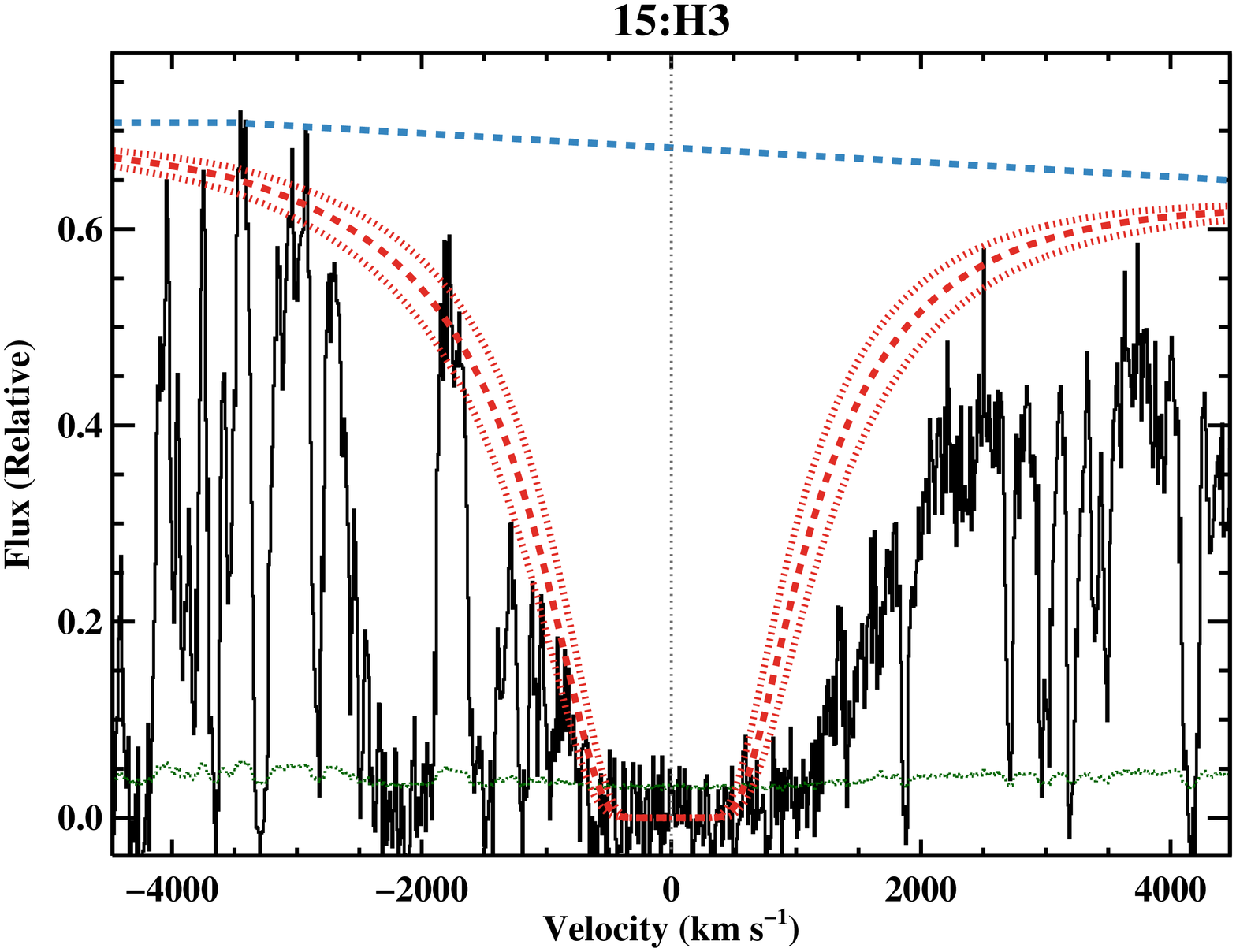}&
  \includegraphics[scale=0.15,angle=90]{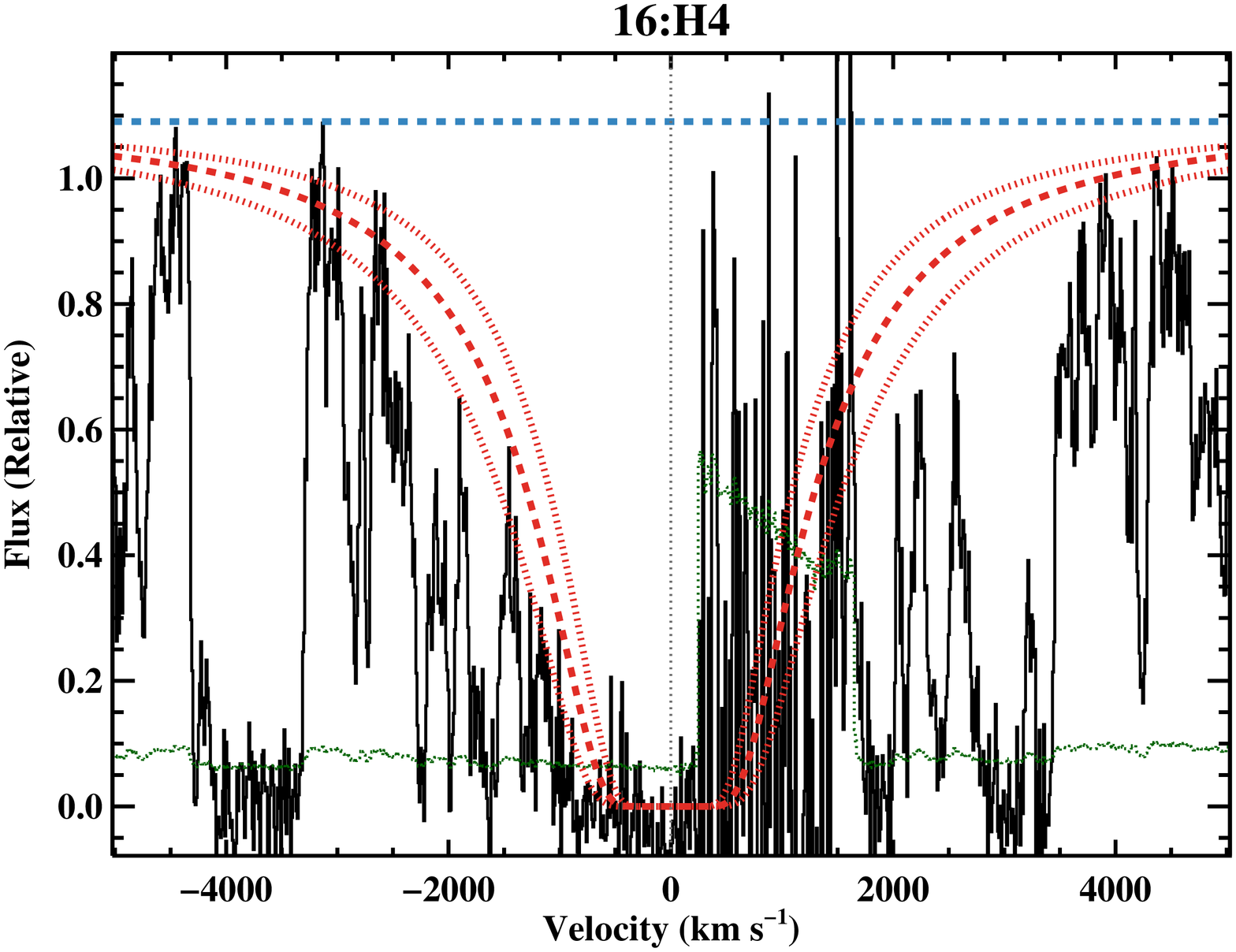}\\
  \end{tabular}
  \caption{Voigt profiles of the Ly$\alpha$ 
    absorption line of the targeted DLAs. In each panel, we superimpose on the data (black histograms) 
    the quasar continuum level (blue dashed line), the absorption line models (red long-dashed line), and the
    corresponding $1\sigma$ errors (red dotted lines) for the main hydrogen component only. Uncertainties 
    on the flux are shown by a green dotted line, while the systemic redshift of each DLA is marked by vertical gray 
    dotted line.\label{fig:dlafit}}
\end{figure*}

\begin{figure*}
  \begin{tabular}{cccc}
  \includegraphics[scale=0.15,angle=90]{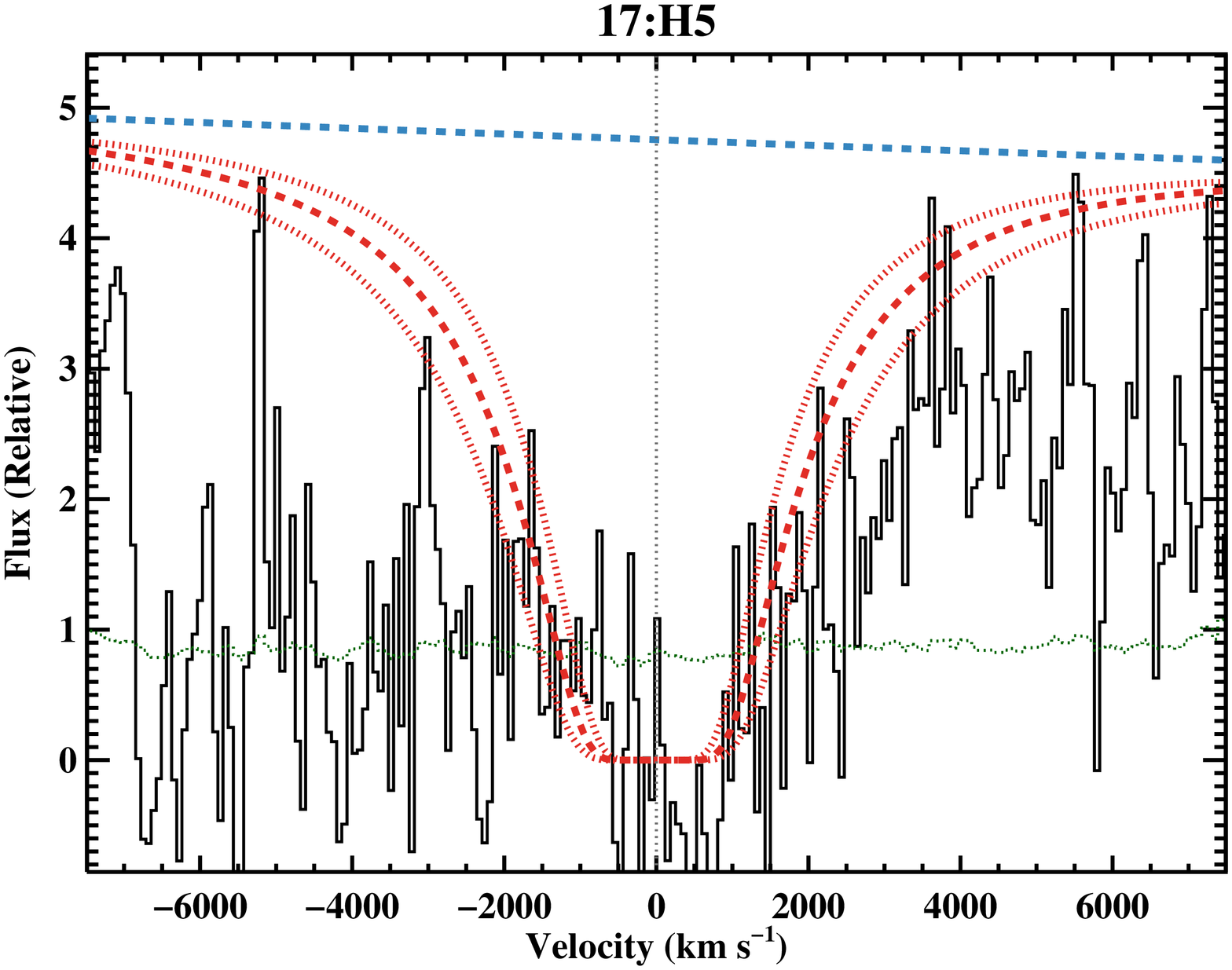}&
  \includegraphics[scale=0.15,angle=90]{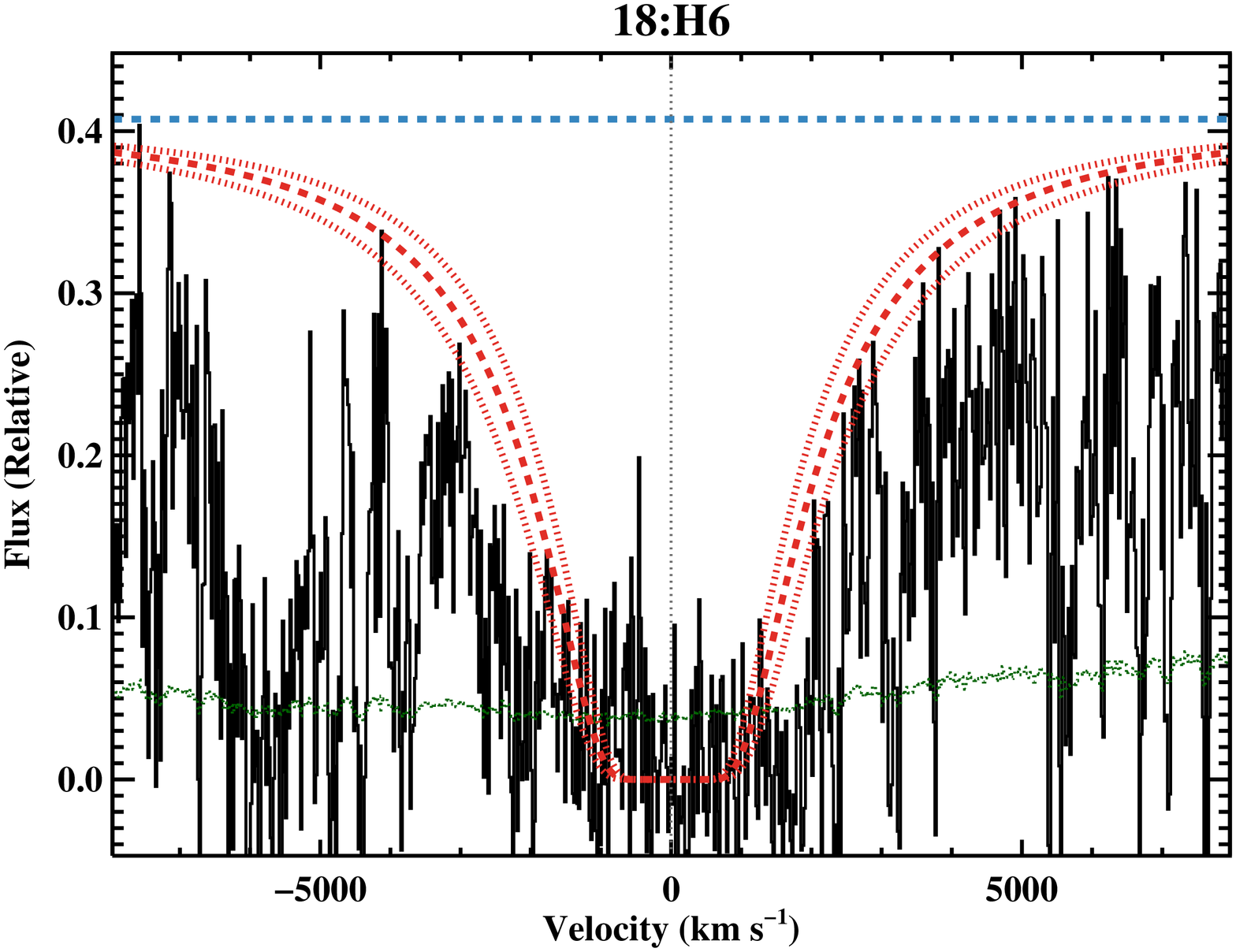}&
  \includegraphics[scale=0.15,angle=90]{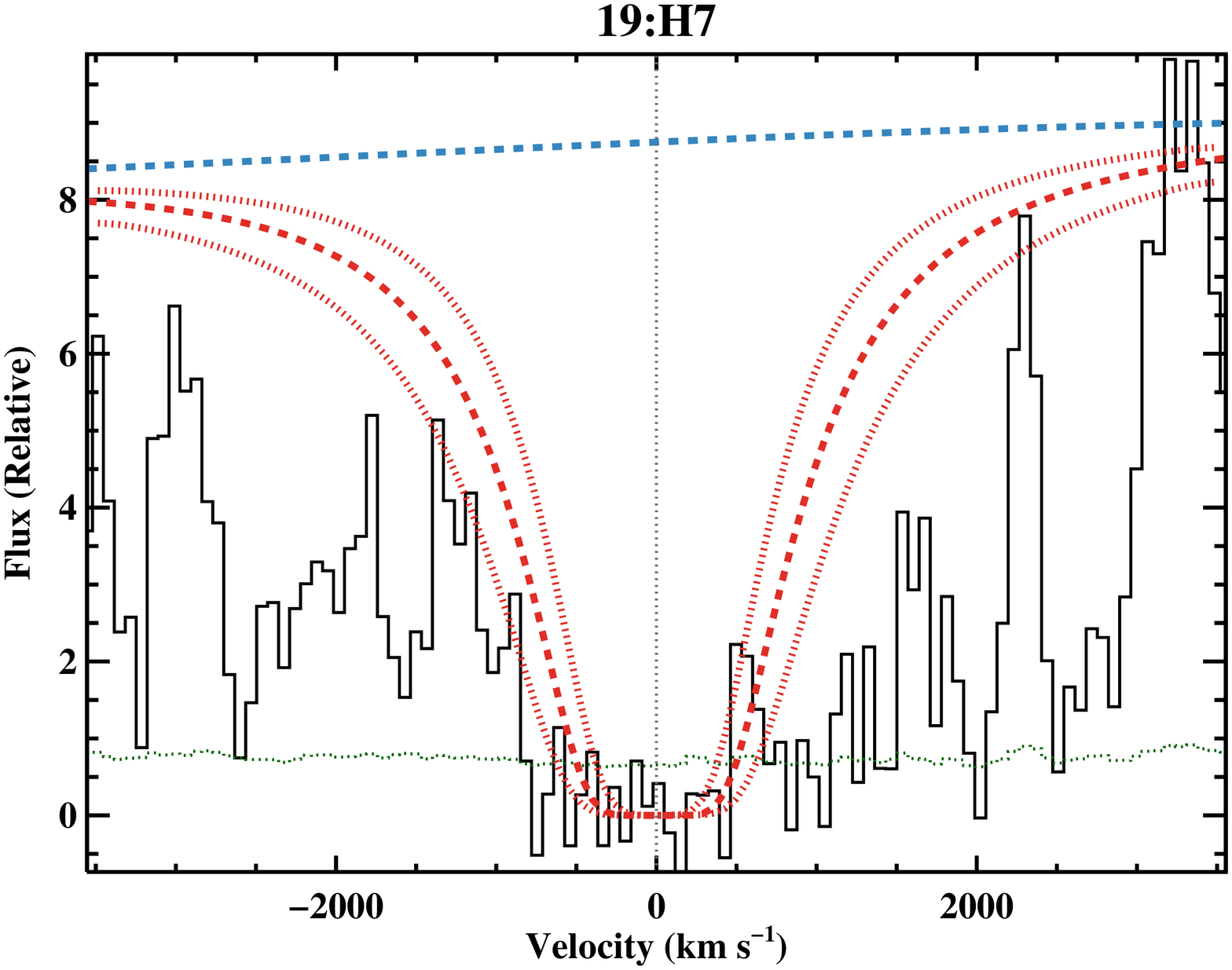}&
  \includegraphics[scale=0.15,angle=90]{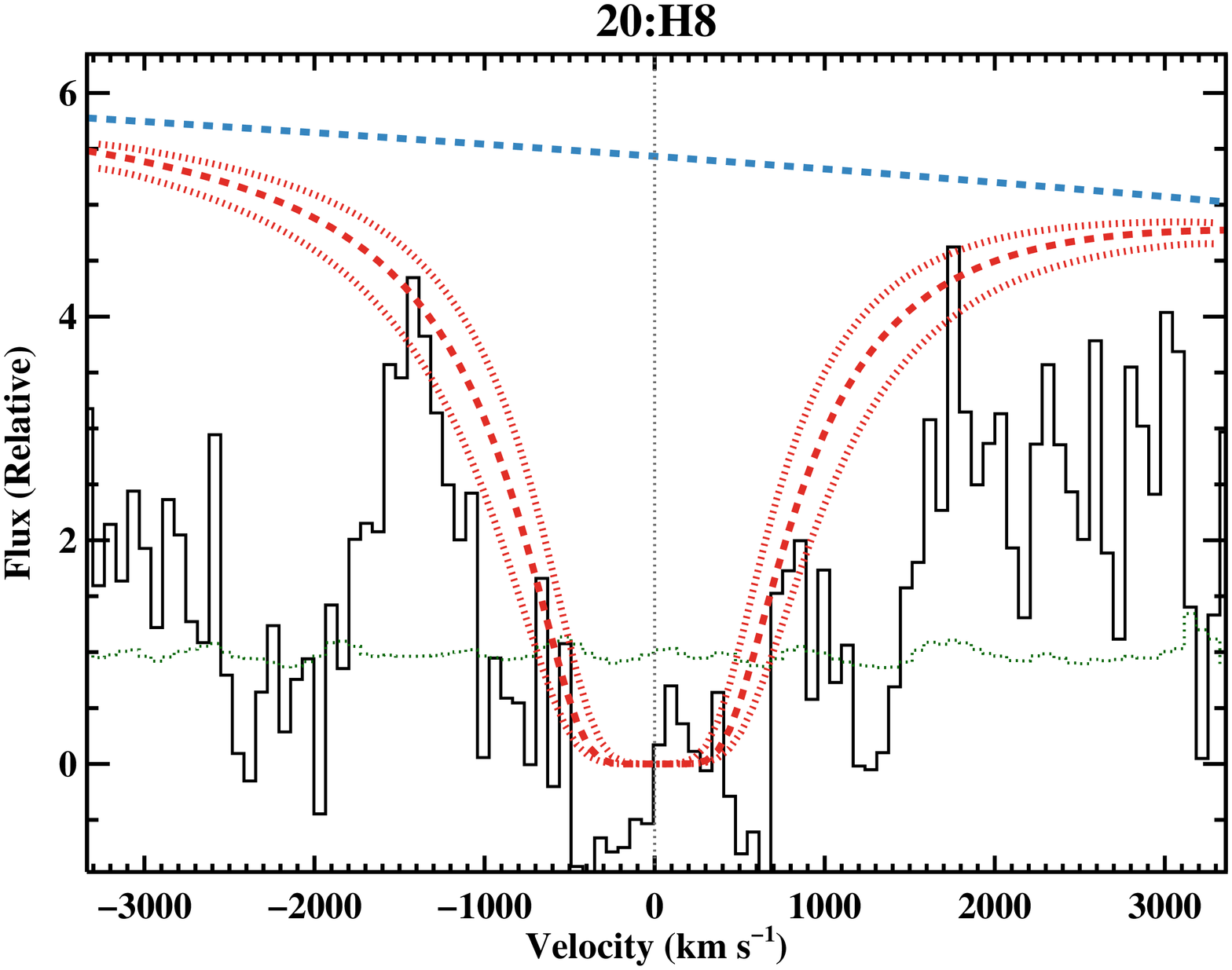}\\
  \includegraphics[scale=0.15,angle=90]{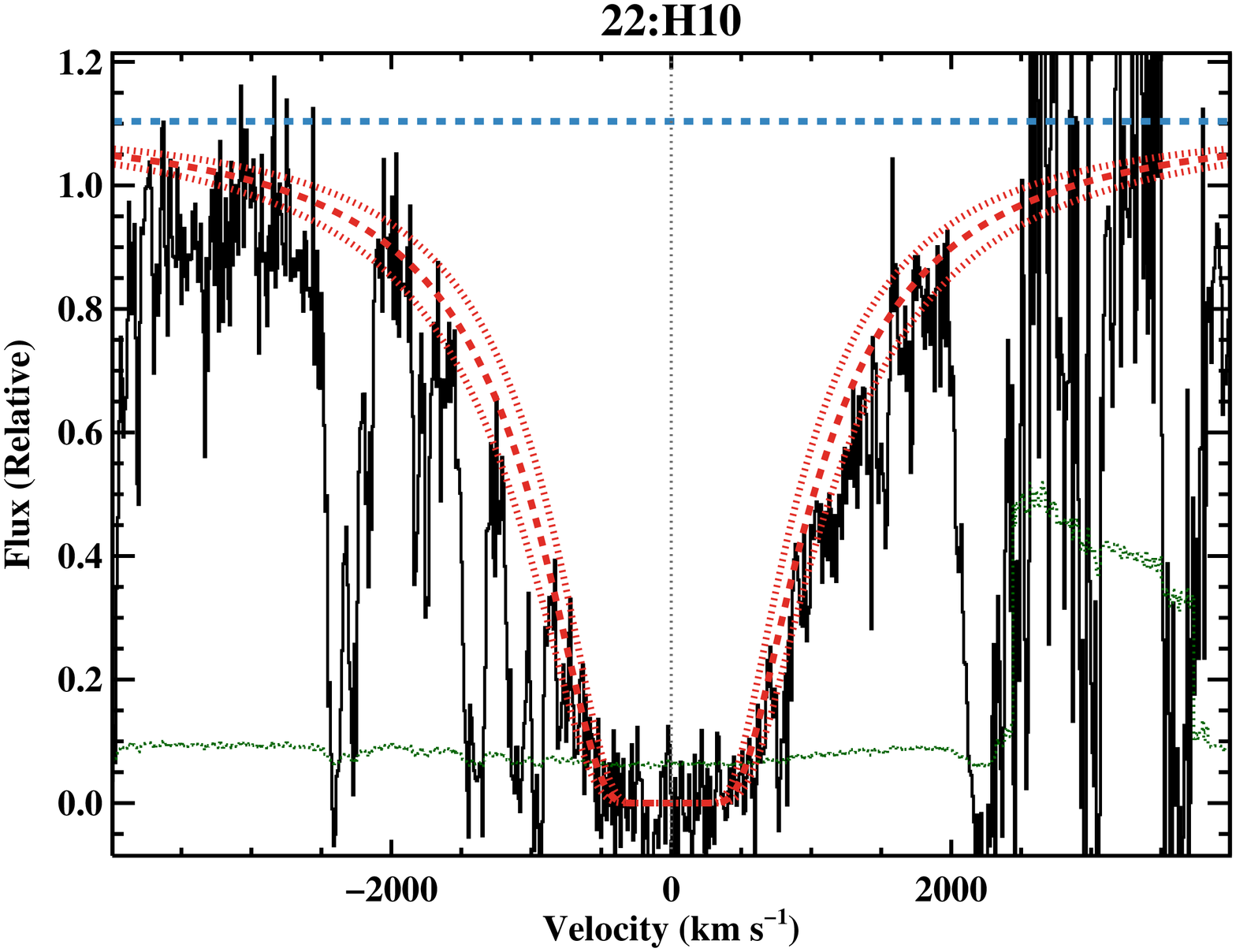}&
  \includegraphics[scale=0.15,angle=90]{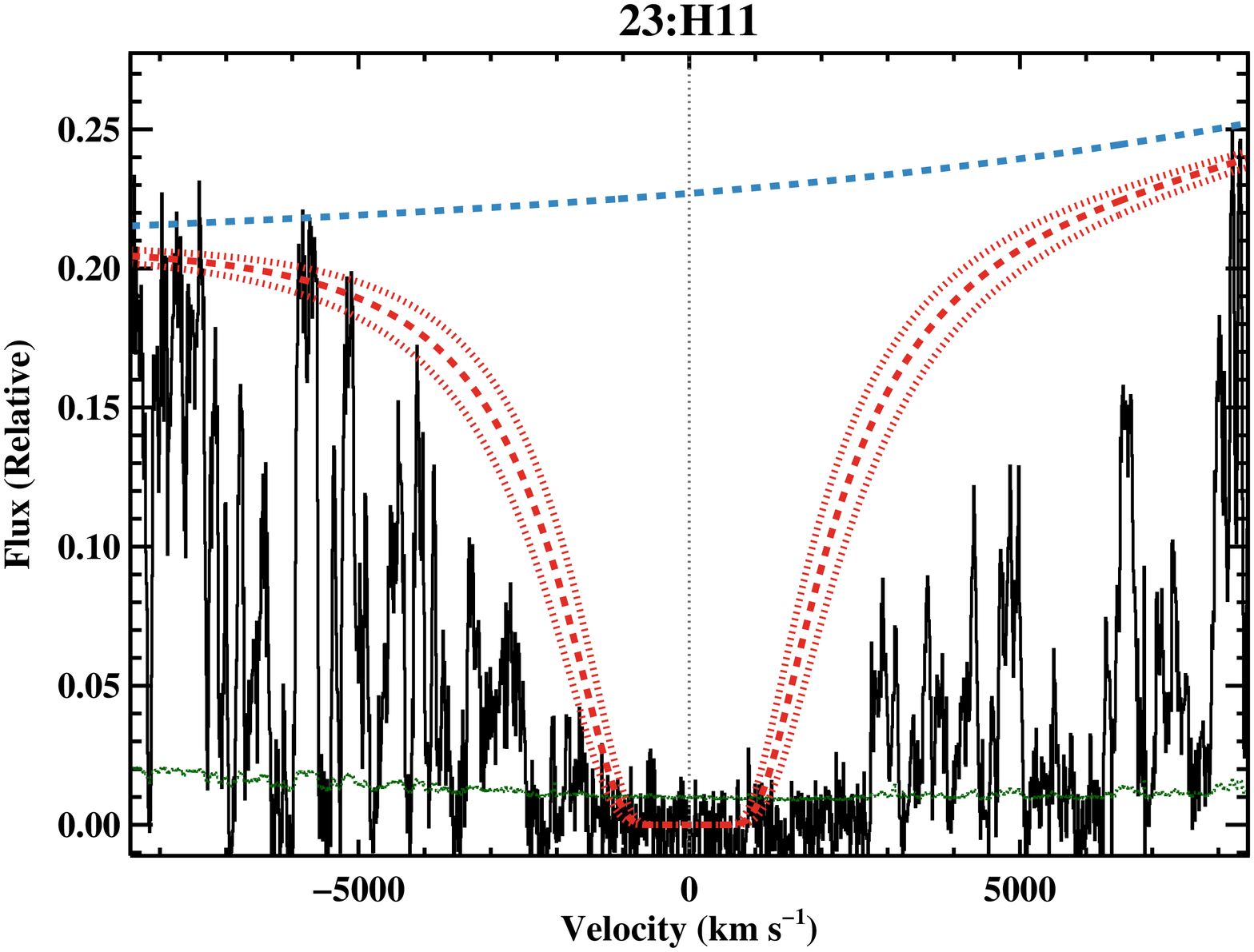}&
  \includegraphics[scale=0.15,angle=90]{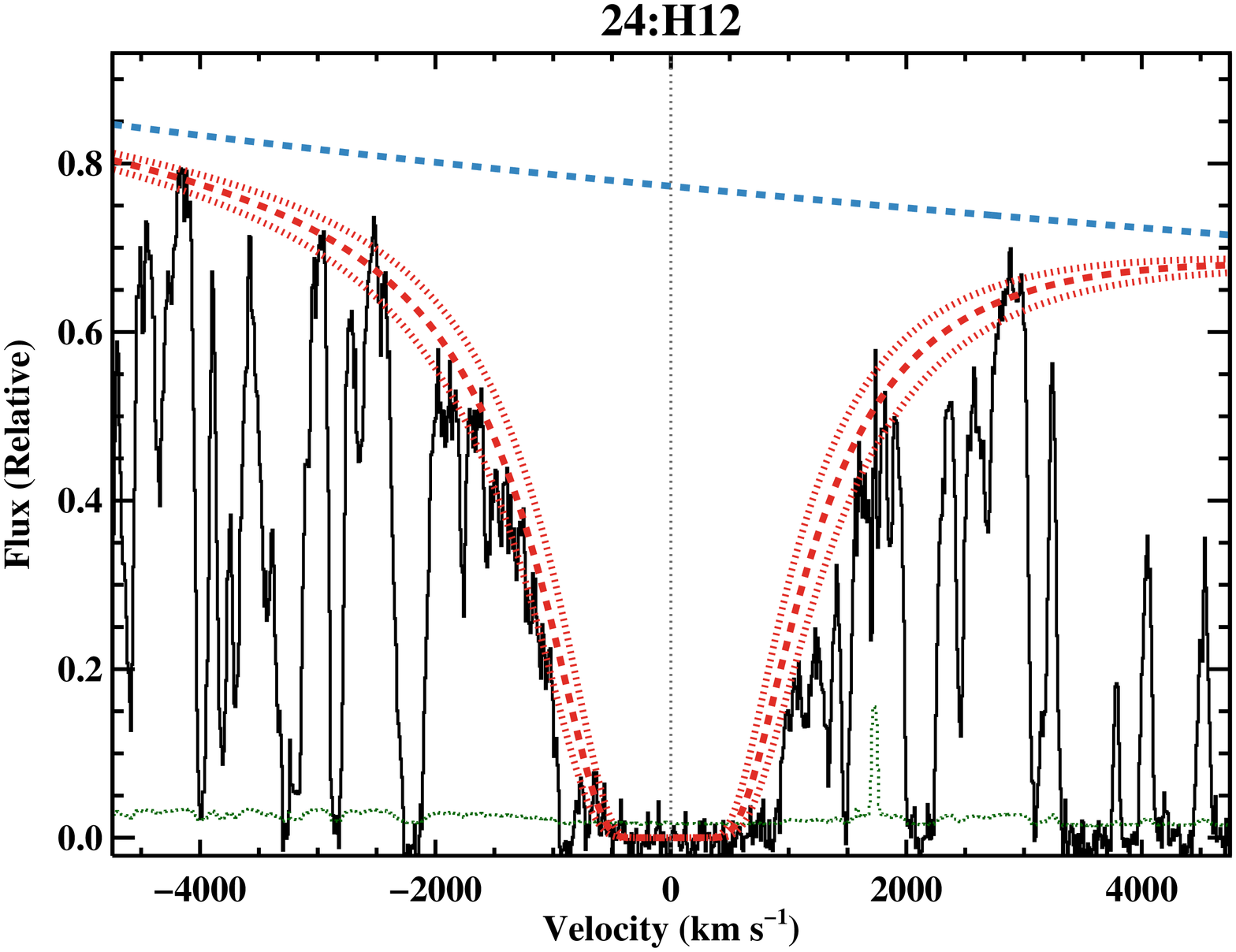}&
  \includegraphics[scale=0.15,angle=90]{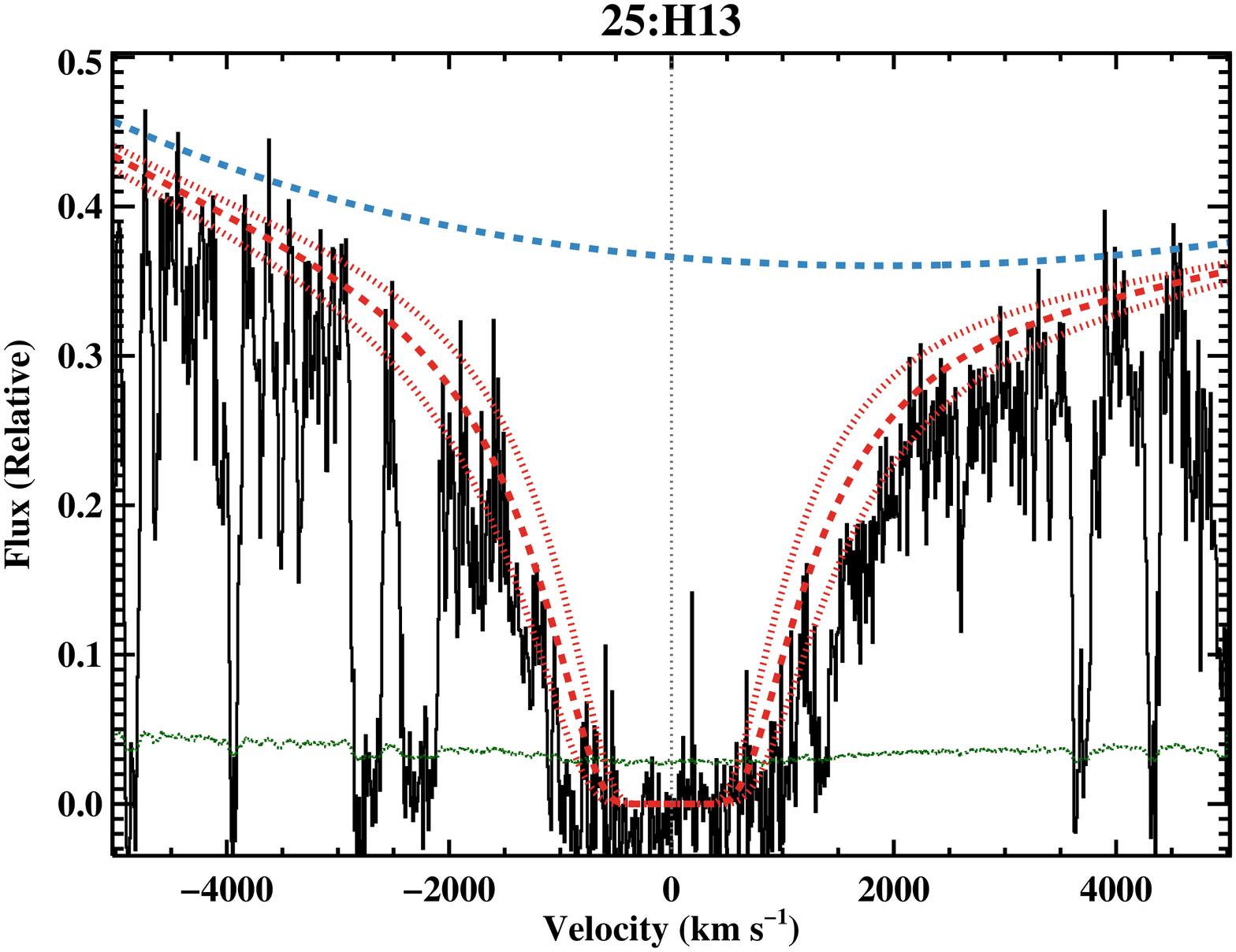}\\
  \includegraphics[scale=0.15,angle=90]{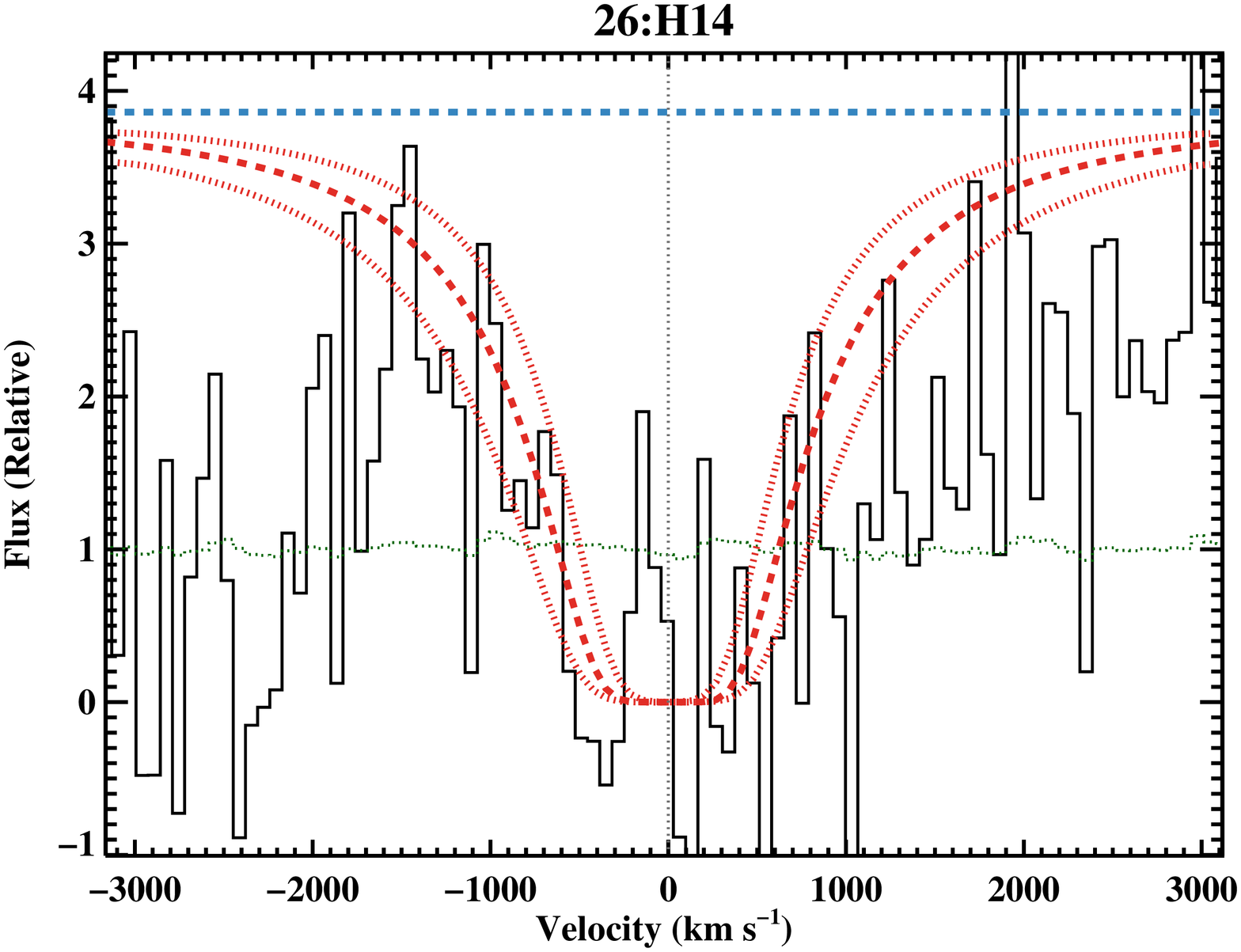}&
  \includegraphics[scale=0.15,angle=90]{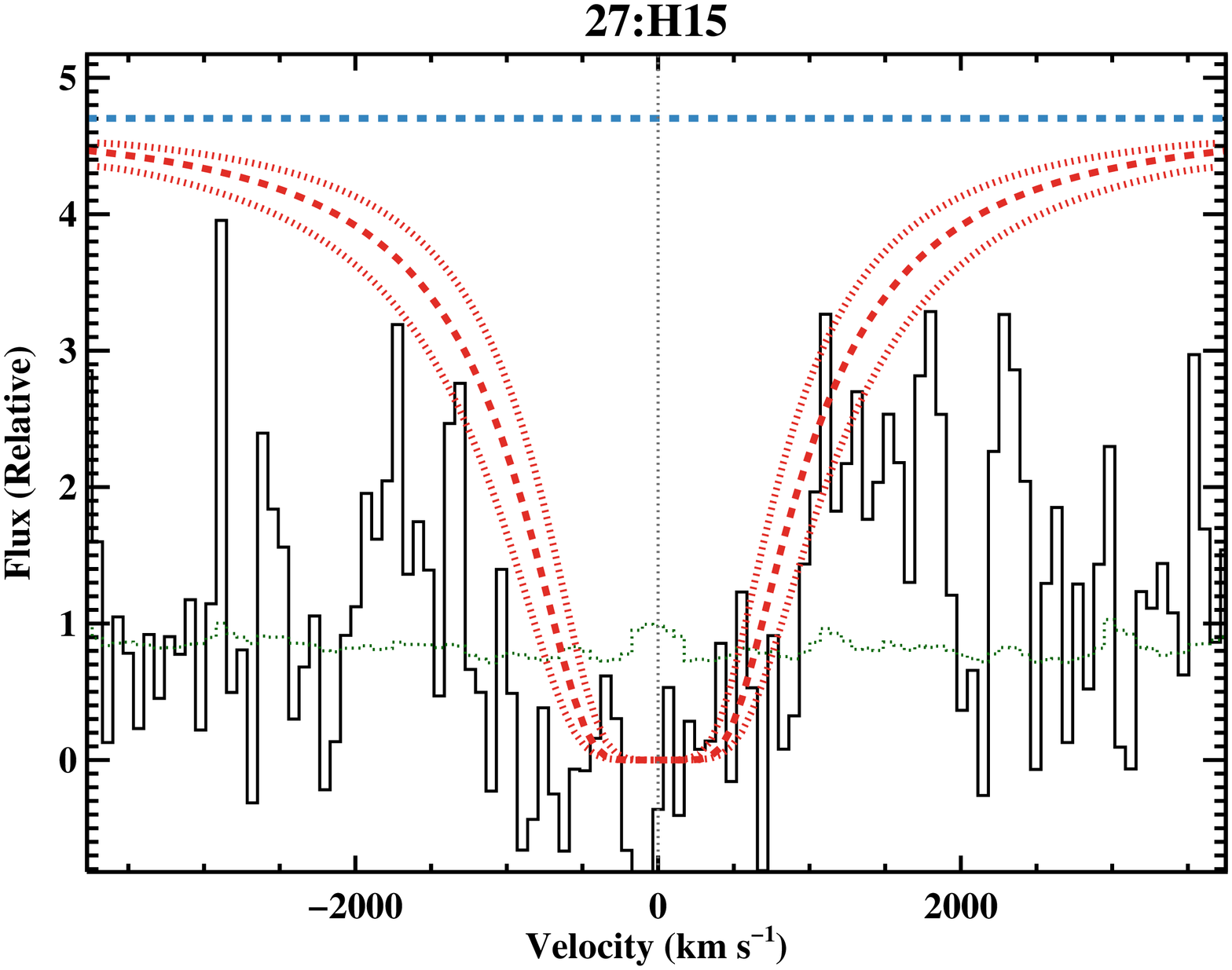}&
  \includegraphics[scale=0.15,angle=90]{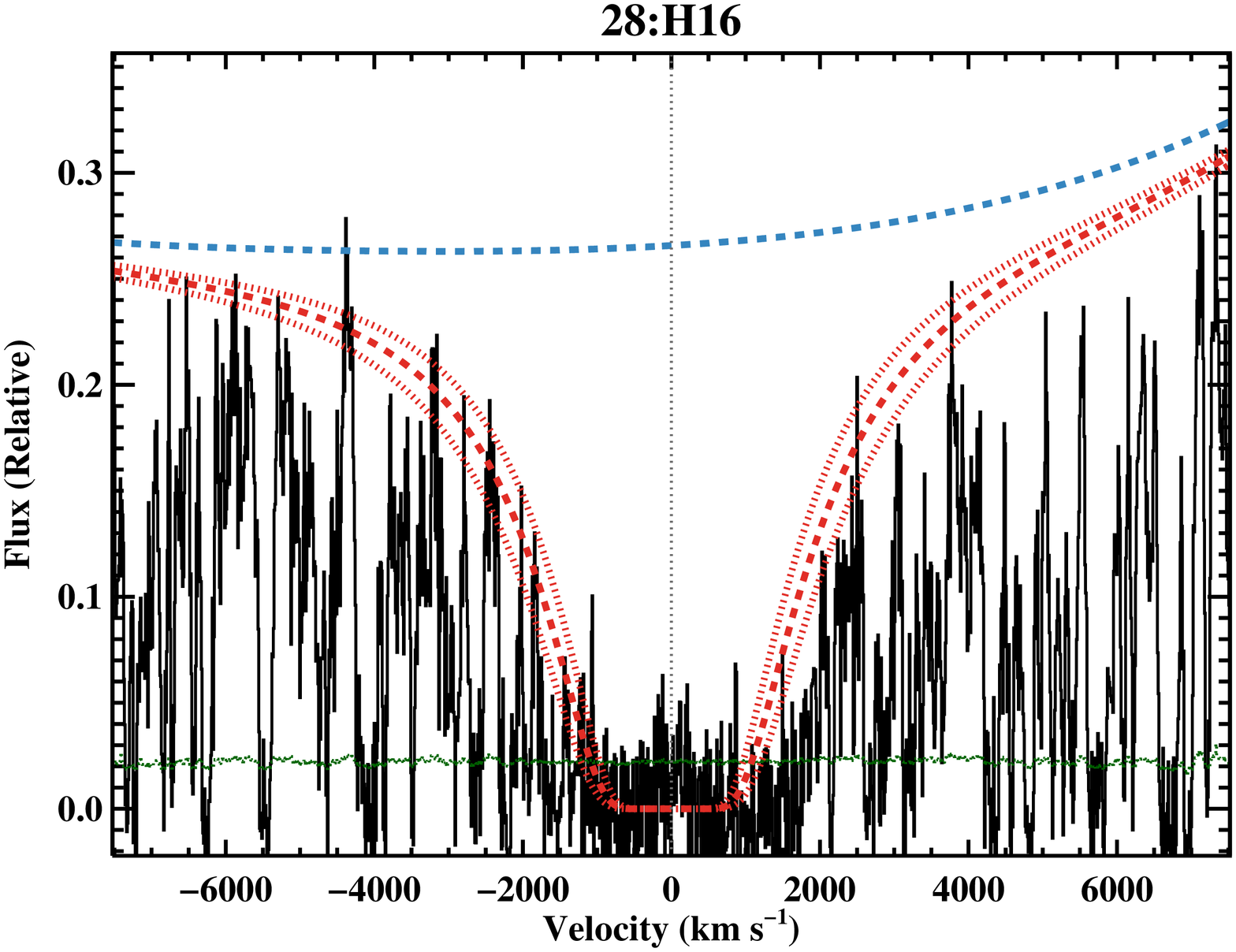}&
  \includegraphics[scale=0.15,angle=90]{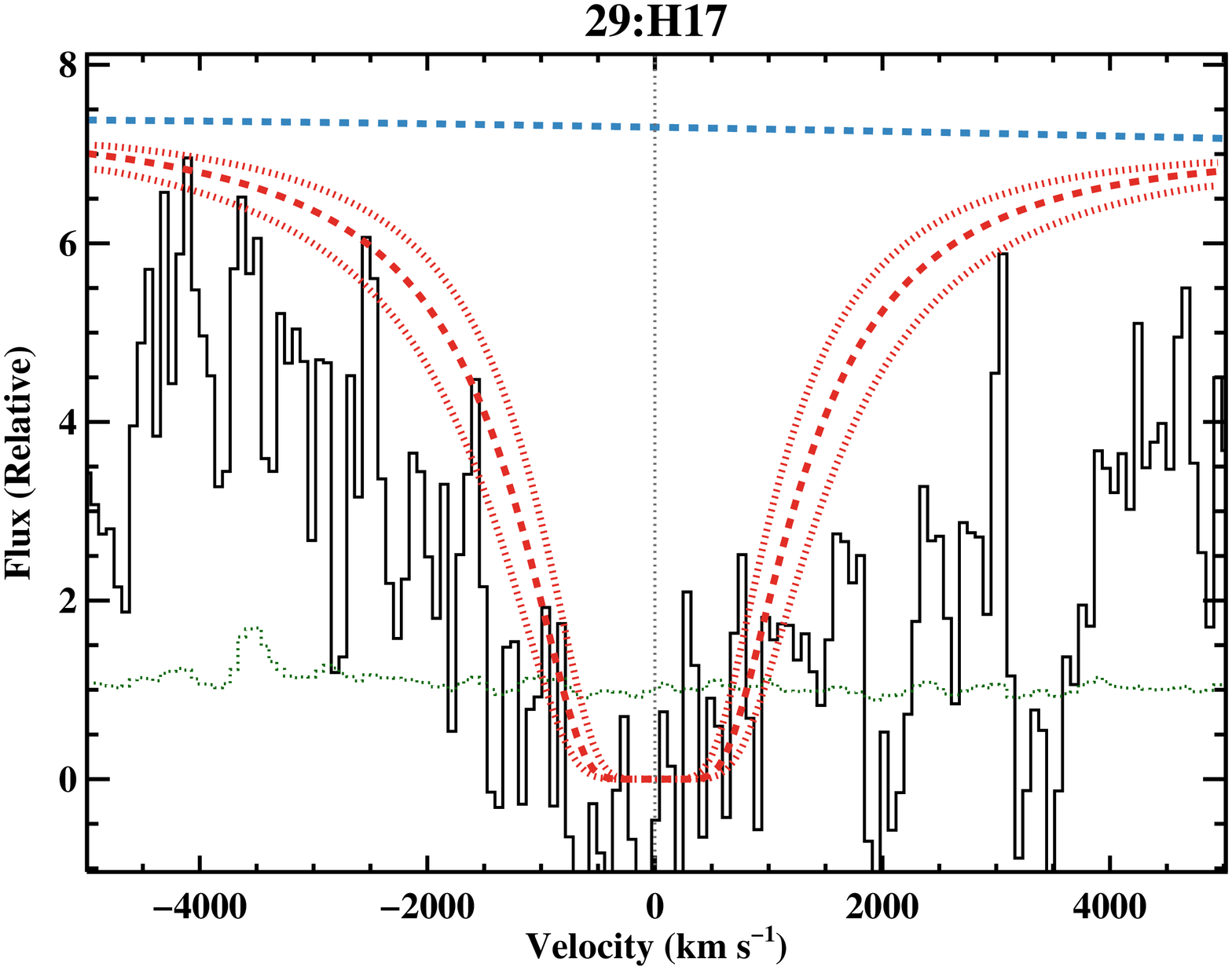}\\
  \includegraphics[scale=0.15,angle=90]{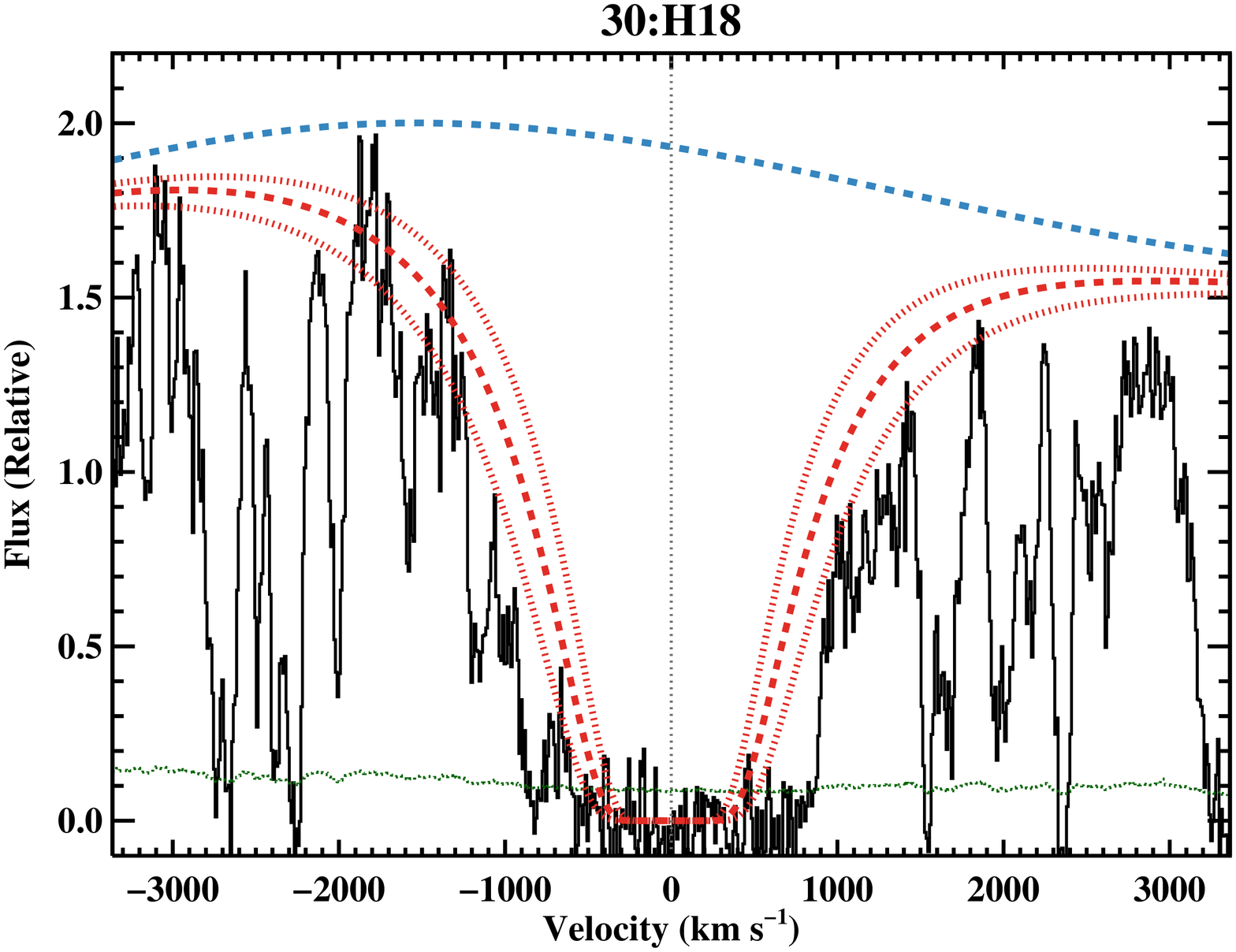}&
  \includegraphics[scale=0.15,angle=90]{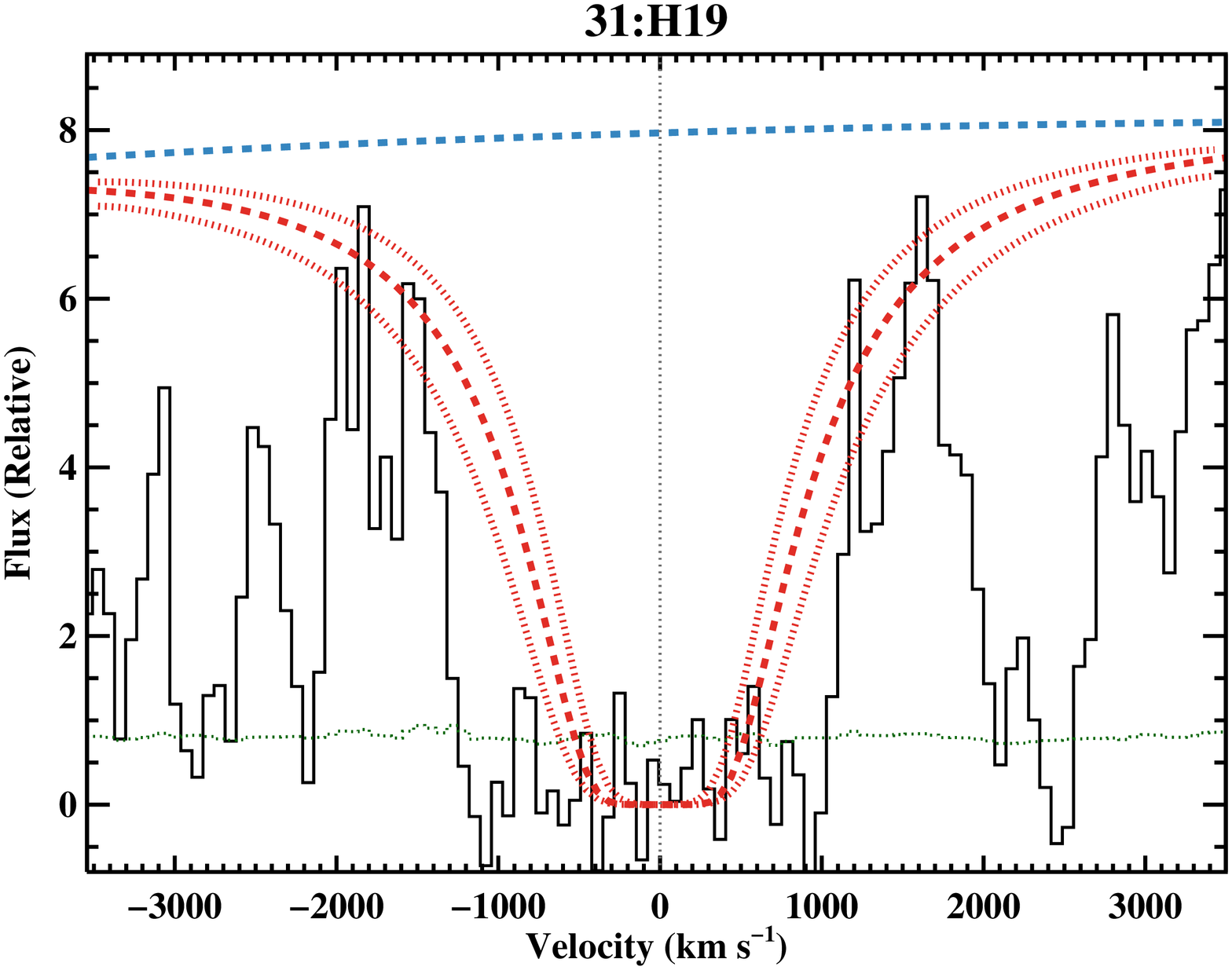}&
  \includegraphics[scale=0.15,angle=90]{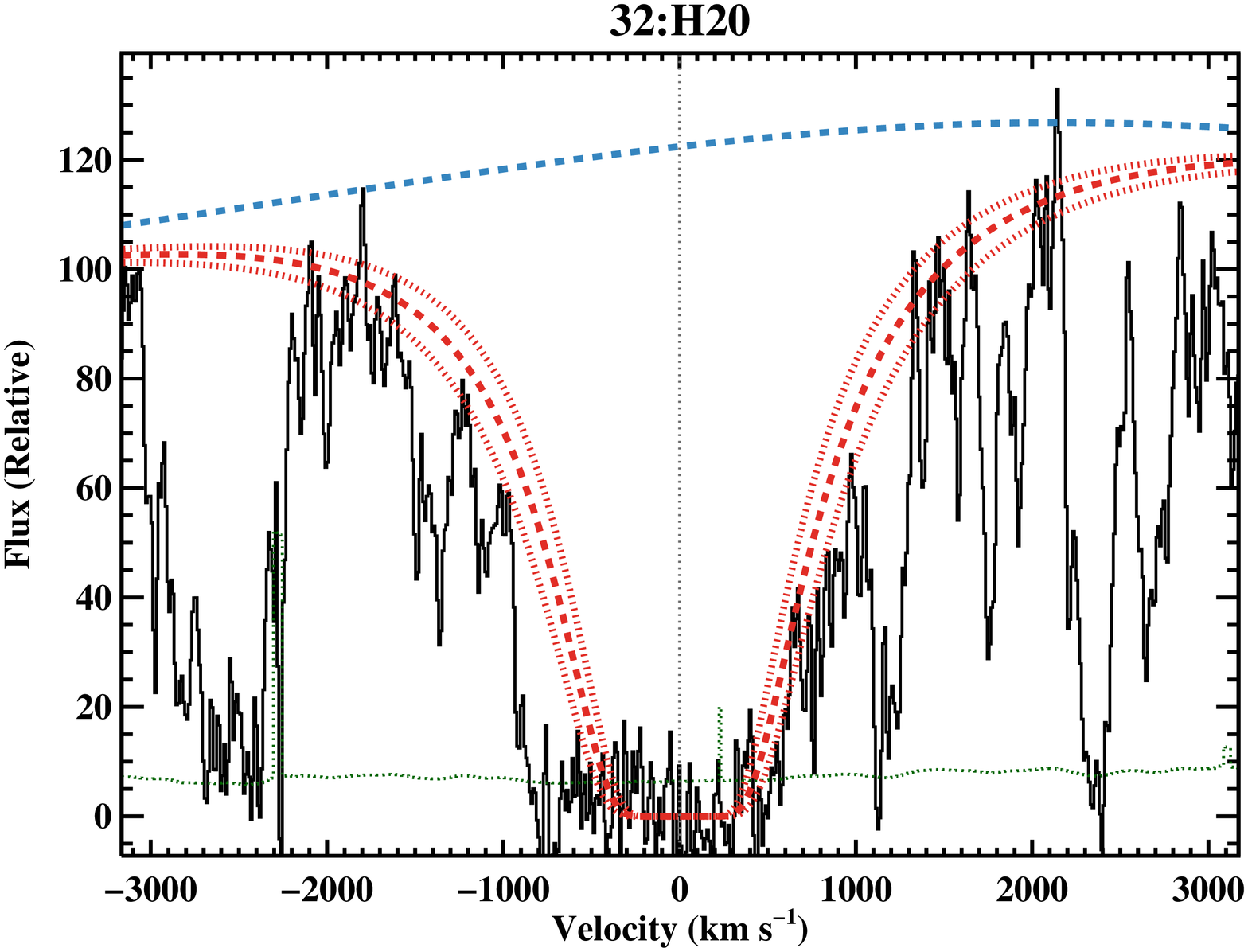}&\\
  \end{tabular}
  \contcaption{Voigt profiles of the Ly$\alpha$ 
    absorption line of the targeted DLAs. In each panel, we superimpose on the data (black histograms) 
    the quasar continuum level (blue dashed line), the absorption line models (red long-dashed line), and the
    corresponding $1\sigma$ errors (red dotted lines) for the main hydrogen component only. Uncertainties 
    on the flux are shown by a green dotted line, while the systemic redshift of each DLA is marked by vertical gray 
    dotted line.}
\end{figure*}

\begin{figure}
\centering
\includegraphics[scale=0.34]{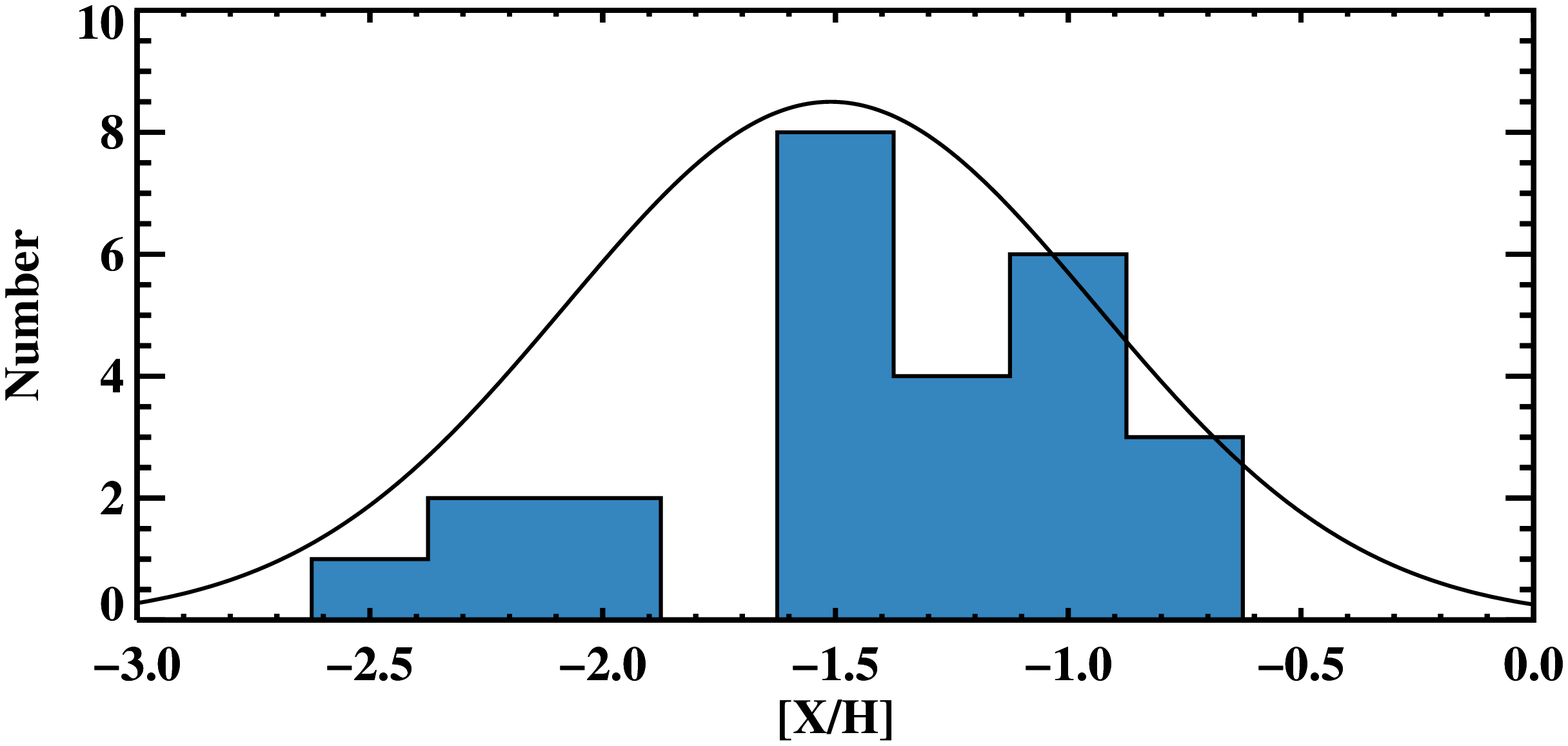}
\caption{Metallicity distribution for the sub-sample of DLAs with 
echellette spectroscopy (blue histogram). Superimposed, a Gaussian with 
mean $-1.51$ and dispersion $0.57$ that represents the observed distribution in a large sample 
of DLAs from \citet{raf12}. Our sample includes DLAs over a large range of 
metallicities and it is representative of the general DLA 
population.}\label{fig:metdla}
\end{figure}

\subsection{Redshifts and \ion{H}{I} Column Densities}

In each spectrum, we search for metal lines that are associated with the 
targeted DLAs. These transitions, and preferentially those at 
low ionization states (e.g. \ion{O}{I}, \ion{Si}{II}, \ion{C}{II}, or \ion{Al}{II}), 
are used to pinpoint the redshifts of the absorbers. Similarly, we use both metal lines
and high-order transitions in the Lyman series to measure the redshift of the LLSs
that act as blocking filters. Typical errors on the redshifts are $\delta_{\rm z} \sim 0.0001$ 
for ESI and MagE spectra, and $\delta_{\rm z} \sim 0.0005$ for SDSS spectra.
Values are listed in Table \ref{tabprop} and shown in the bottom panel of 
Figure \ref{fig:specprop}. The majority of the targeted DLAs lie in the  
redshift window $z\sim 2.5-3.0$, with a handful of outliers at higher and lower redshifts. 
The majority of the high-redshift blocking systems lie instead around redshifts $z\sim 3.5$, 
by construction of our experiment.

For the DLAs, we also measure the \ion{H}{I} column densities by fitting 
Voigt profiles to the Ly$\alpha$ lines in the flux-calibrated spectra, 
as shown in Figure \ref{fig:dlafit}. The amplitude of the 
errors is set according to the results of the simulations described in 
\citet{raf14}. In this previous work, after inserting $z\gtrsim 4$ DLAs into 
actual sightlines and into mock spectra at the resolution of ESI, 
we recovered column densities with mean offsets of $0.02-0.06$ dex compared to the 
input values, and with standard deviations of $0.07-0.12$ dex. Following this analysis,
for the quasars with ESI and MagE spectra, errors on the column densities 
are estimated to be $\sim 0.10$ dex and $\sim 0.15$ dex at lower $S/N$. 
At the lower redshift of our DLAs ($z\sim 3$), contamination from the Ly$\alpha$
forest is less important and the simulations in \citet{raf14} suggest that systematic errors are 
smaller than the quoted errors. Column densities are more uncertain at the lower resolutions 
and lower $S/N$ of the SDSS spectra, with estimated errors between $\sim 0.15-0.20$ dex. 
These errors are also consistent with the dispersion found by \citet{jor13} when comparing 
the \ion{H}{I} column densities measured in MagE and SDSS spectra. Given that our ESI spectrum 
does not cover the Ly$\alpha$ transition for the DLA at $z_{\rm dla} = 1.8639$ in the 
field 21:H9, we take the column density value from \citet{led06}.
The top panel of Figure \ref{fig:specprop} shows the column density distribution of 
the targeted DLAs. Our selection is dictated purely by redshift separations
between the intervening DLAs and LLSs, and thus our sample includes a 
wide range of column densities, which can be considered typical 
of the general DLA populations. We note, however, that this sample is not a strict 
statistical selection from the observed column density distribution function.

In Table \ref{tabprop}, we also list the \ion{H}{I} column densities of the 
higher redshift blocking systems. For DLAs or sub-DLAs which exhibit a clear 
Voigt profile, column densities are measured as done for the lower redshift DLAs. 
Some complications arise from systems which are proximate to the quasar (flagged by an asterisk in 
Table \ref{tabprop}), as accurate modeling of the quasar continuum near the Ly$\alpha$ emission 
line is required for a reliable estimate of $N_{\rm HI}$. For the most difficult cases, 
we therefore quote only a conservative lower limit. For systems with $N_{\rm HI} \lesssim 10^{19}~\rm cm^{-2}$
instead, we quote lower limits on the column density as inferred from the transmitted flux at the Lyman 
limit and the properties of the Lyman series, following \citet{fum13}.

\subsection{Metallicities}\label{metal}

We compute the gas-phase metallicity for the targeted DLAs for which we have new echellette 
data, following the procedures discussed in 
\citet{pro03b} and \citet{raf12}. The basics are only briefly summarized here. 
We continuum normalize each quasar either with a polynomial function, which we fit to 
the data redward to the quasar Ly$\alpha$ emission line, or with a spline function 
constrained by pixels without evident IGM absorption \citep[see e.g.][]{fum13}. 
Metallicities are then computed using the apparent optical depth method \citep{sav91} 
that we apply, when possible, to multiple transitions of the same ions. 
Ionization corrections are not included, being negligible at the high
\ion{H}{I} column densities of DLAs. The derived metallicities $\rm [X/H]_{\rm dla}$, together with 
the ions used as tracers, are summarized in Table \ref{tabprop}. Measurements of individual metal transitions 
in each system are further listed in the Appendix. We do not attempt to derive metallicities from the 
low resolution SDSS spectroscopy, as hidden saturation and noise would prevent a reliable determination 
of $\rm [X/H]_{\rm dla}$ \citep[e.g.][]{jor13}. Finally, for the few systems with archival data, we adopt published 
values, as detailed in Table \ref{logspec}. The resultant distribution of metallicities is
shown in Figure \ref{fig:metdla}, which we compare against a model for the observed distribution in a 
large sample of $z>1.5$ DLAs from \citet{raf12}. Our 
selection does not bias our sample towards a particular class of DLAs, also because 
metallicities for most of the selected systems were unknown at the beginning of our survey.
Similarly to the hydrogen column densities, the metal content of the selected DLAs is 
therefore consistent with that of the general DLA population at comparable redshifts, although our sample 
is not a strict statistical selection from the parent sample. 

\section{Summary}\label{conclusion}

In this second paper of a series, we have presented new imaging and spectroscopic data which we collected as
part of a survey that aims to characterize the {\it in-situ} star formation rates of DLAs and the connection 
between the DLA gas seen in absorption and galaxies seen in emission. 

We have collected optical and near-UV multiwavelength imaging of 32 quasar fields, 
using HST, Keck, LBT, and the Magellan telescopes. These sightlines have been selected because of the presence
of both high redshift ($z\sim 3.5$)
optically-thick absorption systems and lower redshift ($z\sim 2.7$) DLAs. Given this fortuitous alignment, 
the high-redshift absorber acts as a natural blocking filter for the quasar light, which is completely absorbed 
below $\sim 4000-4500$\AA. In turn, this ``Lyman limit technique'' allows us to image, 
blueward of the above blocking wavelengths, candidate DLA host galaxies at all impact 
parameters, including small separations ($\lesssim 1''$) from the quasars.  

In this manuscript, we discussed the observations and data reduction for the HST and ground-based 
imaging, including details on the procedures adopted for photometric calibration and the preparation 
of the galaxy catalogues. We also discussed and analyzed new spectroscopic observations for 22 quasars, 
presenting new measurements of the hydrogen column densities and gas-phase metallicities of the 
intervening DLAs. Further, we compared the properties of the targeted DLAs to the properties of 
a larger sample of DLAs at comparable redshifts, concluding that our sample represents an unbiased selection 
with respect to metallicity and column density, although it is not a strict statistical selection from the 
 general population of $z\gtrsim 2$ DLAs. Therefore, this sample is useful to
characterize the connection between neutral gas and galaxies and the rate with which stars are formed 
in DLAs, simply defined as systems with hydrogen column densities above $N_{\rm HI} \ge 10^{20.3}~\rm cm^{-2}$
without any prior information on the associated metal content or the presence of candidate host galaxies. 
This analysis will be the subject of the third paper in the series.

\section*{Acknowledgments}

This work is dedicated to the memory of our co-author and long-time
collaborator, Arthur M. Wolfe. His passion for DLAs 
has accompanied us for many years, and it will continue to inspire our work.
We thank Marc Rafelski, Marcel Neeleman, and Joseph Hennawi for their 
help during observations and for useful discussions. We thank Ryan Cooke 
for sharing the reduced spectrum of 32:H20. This paper has benefited 
from a careful review of an anonymous referee. This work has been partially supported by 
NASA grants HST-GO-10878.05-A. MF acknowledges support by the Science and Technology 
Facilities Council [grant numbers ST/I001573/1, ST/F001166/1]. JMO thanks the office of the VPAA 
at St. Michael’s College for travel support. JXP acknowledges support from NSF grant AST-1109447.
NK acknowledges the Department of Science and Technology for support via a Ramanujan Fellowship.
This work is based on observations made with the NASA/ESA Hubble Space Telescope (PID 11595), 
and obtained from the Hubble Legacy Archive, which is a collaboration between the Space Telescope Science Institute
(STScI/NASA), the Space Telescope European Coordinating Facility (ST-ECF/ESA) and the Canadian 
Astronomy Data Centre (CADC/NRC/CSA). Some of the data presented herein were 
obtained at the W.M. Keck Observatory, which is operated as a scientific partnership among the California Institute of 
Technology, the University of California and the National Aeronautics and Space Administration. 
The Observatory was made possible by the generous financial support of the W.M. Keck Foundation. 
Part of the data presented in this manuscript have been collected through time awarded 
by NOAO (Proposal ID \#2009A-0184). This paper includes data gathered with the 6.5 meter Magellan 
Telescopes located at Las Campanas Observatory, Chile, and data gathered at the Large Binocular Telescope (LBT). 
The LBT is an international collaboration among institutions in the United States, Italy and Germany. LBT Corporation 
partners are: The University of Arizona on behalf of the Arizona university system; Istituto Nazionale di Astrofisica, Italy 
LBT Beteiligungsgesellschaft, Germany, representing the Max-Planck Society, the Astrophysical Institute Potsdam, 
and Heidelberg University; The Ohio State University, and The Research Corporation, on behalf of The University 
of Notre Dame, University of Minnesota and University of Virginia. Imaging and spectroscopy data from the Sloan 
Digital Sky Survey have been a key element for the design and  completion of this survey. This manuscript includes 
observations made with ESO Telescopes at the La Silla Paranal Observatory under programme ID 087.A-0022. For 
access to the data used in this paper, please contact the authors. 

\appendix

\section{New metallicity measurements}
\label{lastpage}

\begin{table*}
\caption{Summary of all metallicity measurements for the DLAs with new echellette data.}\label{tabmetsumm}
\begin{tabular}{lcccccccccccc}
\hline
Name$^a$ & f$_M^b$ & [M/H]$^{c}$ & $\sigma(M)^c$ & f$_\alpha^d$ & [$\alpha$/H]$^e$ & $\sigma(\alpha)^e$ & f$_{Zn}^f$ & [Zn/H]$^g$ & $\sigma$(Zn)$^g$ & f$_{Fe}^h$ & [Fe/H]$^i$ & $\sigma$(Fe)$^i$ \\
\hline
J211443.94-005532 (1:G1)  & 4&$-0.63$&0.11& 4&$-0.63$&0.04& 3&$-0.78$&    & 4&$-1.11$&0.06\\
J073149.50+285448 (2:G2)  & 1&$-1.45$&0.17& 1&$-1.45$&0.14& 3&$-1.16$&    & 1&$-1.46$&0.11\\
J095604.43+344415 (3:G3)  & 2&$-1.00$&0.17& 2&$-1.48$&    & 1&$-1.00$&0.09& 1&$-1.46$&0.11\\
J234349.41-104742 (4:G4)  &13&$-1.27$&0.20&13&$-1.27$&0.20& 3&$-0.63$&    & 1&$-1.77$&0.03\\
J034300.88-062229 (5:G5)  &14&$-2.02$&0.26& 3&$-1.24$&    & 3&$-0.72$&    & 1&$-2.32$&0.05\\
J235152.80+160048 (6:G6)  &14&$-2.03$&0.20& 2&$-2.26$&    & 0&$     $&    & 1&$-2.33$&0.06\\
J004219.74-102009 (7:G7)  &14&$-0.96$&0.16& 3&$-0.19$&    & 0&$     $&    & 1&$-1.26$&0.01\\
J094927.88+111518 (8:G9)  & 1&$-0.95$&0.10& 1&$-0.95$&0.02& 0&$     $&    & 1&$-1.20$&0.05\\
J101806.28+310627 (9:G10) &13&$-1.19$&0.28& 2&$-1.48$&    & 3&$-0.91$&    & 1&$-1.50$&0.03\\
J085143.72+233208 (10:G11)& 2&$-1.05$&0.15& 2&$-1.04$&    & 1&$-1.05$&0.11& 4&$-1.51$&0.05\\
J095605.09+144854 (11:G12)& 1&$-1.46$&0.12& 1&$-1.46$&0.07& 1&$-1.17$&0.11& 1&$-1.63$&0.08\\
J115130.48+353625 (12:G13)& 1&$-1.28$&0.11& 1&$-1.28$&0.05& 0&$     $&    & 1&$-1.74$&0.01\\
J212357.56-005350 (13:H1) &13&$-1.59$&0.15&13&$-1.59$&0.15& 0&$     $&    & 1&$-1.88$&0.02\\
J040718.06-441014 (14:H2) & 1&$-0.77$&0.11& 1&$-0.77$&0.05& 1&$-0.41$&0.09& 1&$-0.59$&0.03\\
J025518.58+004847 (15:H3) & 1&$-0.80$&0.11& 1&$-0.80$&0.05& 3&$-0.68$&    & 1&$-1.28$&0.02\\
J081618.99+482328 (16:H4) & 1&$-2.36$&0.15& 1&$-2.36$&0.02& 3&$-1.17$&    & 1&$-2.41$&0.03\\
J090810.36+023818 (18:H6) & 1&$-0.93$&0.12& 1&$-0.93$&0.07& 3&$-0.41$&    &25&$-1.31$&0.29\\
J084424.24+124546 (21:H9) & 1&$-1.54$&0.12& 1&$-1.54$&0.06& 1&$-1.37$&0.12& 1&$-1.66$&0.06\\
J075155.10+451619 (22:H10)& 1&$-1.16$&0.13& 1&$-1.16$&0.09& 3&$-0.81$&    & 1&$-1.80$&0.03\\
J081813.14+072054 (23:H11)&13&$-1.41$&0.25& 2&$-1.67$&    & 3&$-1.16$&    & 4&$-1.36$&0.07\\
J081813.05+263136 (24:H12)&13&$-0.93$&0.24& 2&$-1.17$&    & 3&$-0.68$&    & 4&$-1.10$&0.05\\
J081114.32+393633 (25:H13)&13&$-1.44$&0.15&13&$-1.44$&0.15& 3&$-0.67$&    & 1&$-1.70$&0.03\\
J132005.97+131015 (32:H20)& 1&$-2.30$&0.10& 1&$-2.30$&0.02& 3&$-0.46$&    & 1&$-2.81$&0.09\\
\hline
\end{tabular}
 
\flushleft{
$^a$ Quasar name. $^{b}$ Flag indicating how
the metallicity is computed: 1=[Si/H]; 2=[Zn/H]; 
4=[S/H] ; 13=Bracked by Si,Zn limits; 14=[Fe/H]+0.3\,dex.
$^c$ Adopted metallicity with associated error. 
$^{d}$ Flag indicating how the abundance of 
$\alpha-$elements is computed: 0=No measurement; 1=Si measurement; 2=Si lower limit; 
3=Si upper limit; 4=[S/H] ; 5=[O/H] ; 13=S+Si limits.
$^e$ Abundance of  $\alpha-$elements with associated error.
$^{f}$ Flag indicating how the abundance of zinc is computed: 
0=No measurement; 1=Zn measurement; 2=Zn lower limit; 3=Zn upper limit.
$^g$ Abundance of  zinc with associated error.
$^{h}$ Flag indicating how the abundance of iron is computed: 
0=No measurement; 1=Fe measurement; 2=Fe lower limit;
3=Fe upper limit; 4=[Ni/H]-0.1dex;
5=[Cr/H] - 0.2dex;
6=[Al/H]; 11-16=Fe, Ni, Cr, Al limits; 25=Mean of Fe limits.
$^{i}$ Abundance of iron with associated error.
Note that none of the limits reported take into account the
uncertainty in the \nhi\ value.}
\end{table*}

\begin{figure*}
\centering
\includegraphics[scale=0.8]{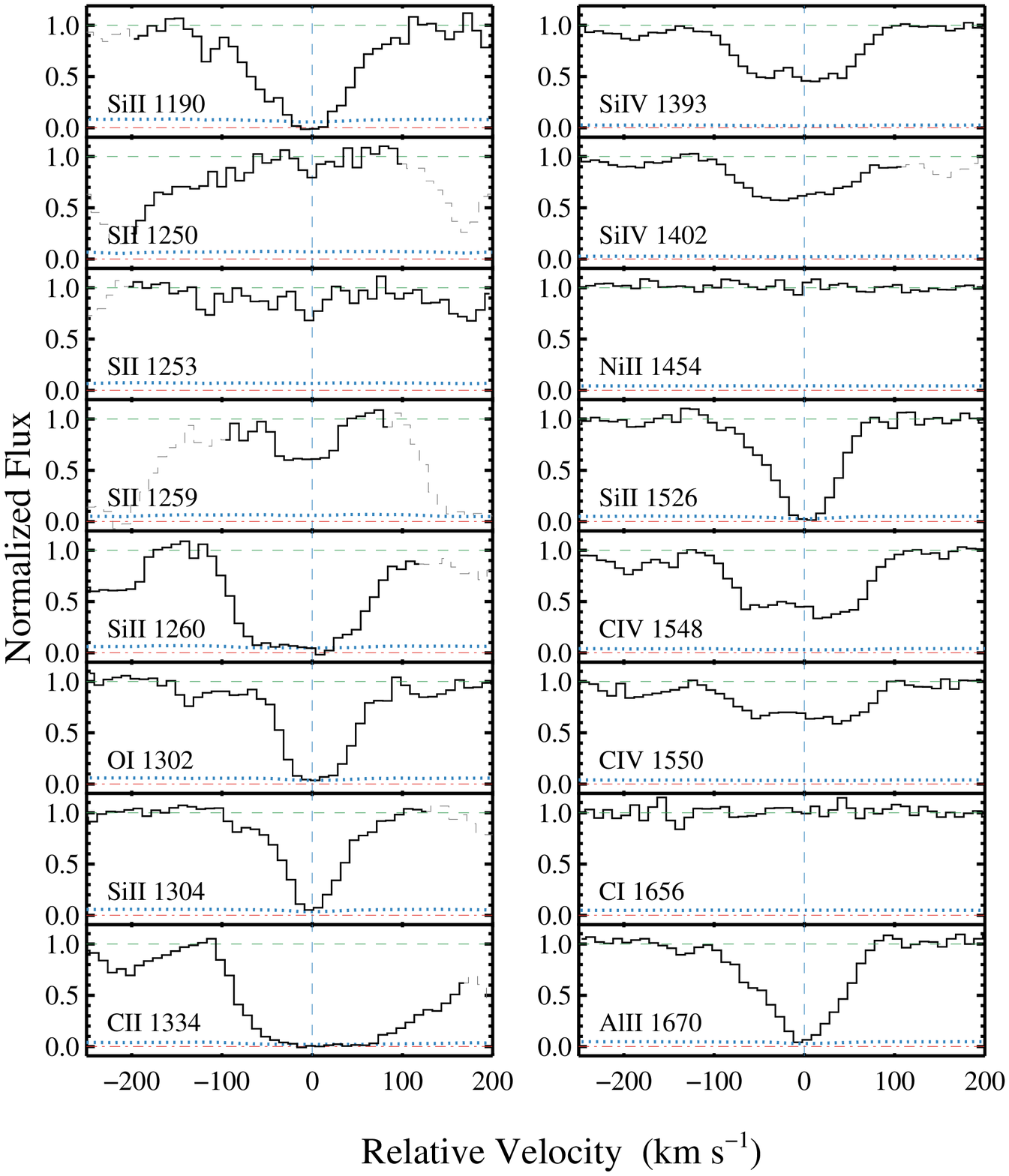}
\caption{Velocity plot of the ion transitions associated to the DLA J2114-0055 (1:G1) at $z_{\rm dla}=2.9181$.
Data are shown as black histograms, with grey dashed lines highlighting regions affected by 
unrelated absorption. The normalized continuum levels are marked by a green dashed line, while 
the red dash-dotted lines mark the zero level. The adopted systemic velocities are indicated by vertical blue dashed lines. Errors on the flux are shown with dotted blue lines.}\label{fig:vpdla01a}
\end{figure*}

\begin{figure*}
\centering
\includegraphics[scale=0.8]{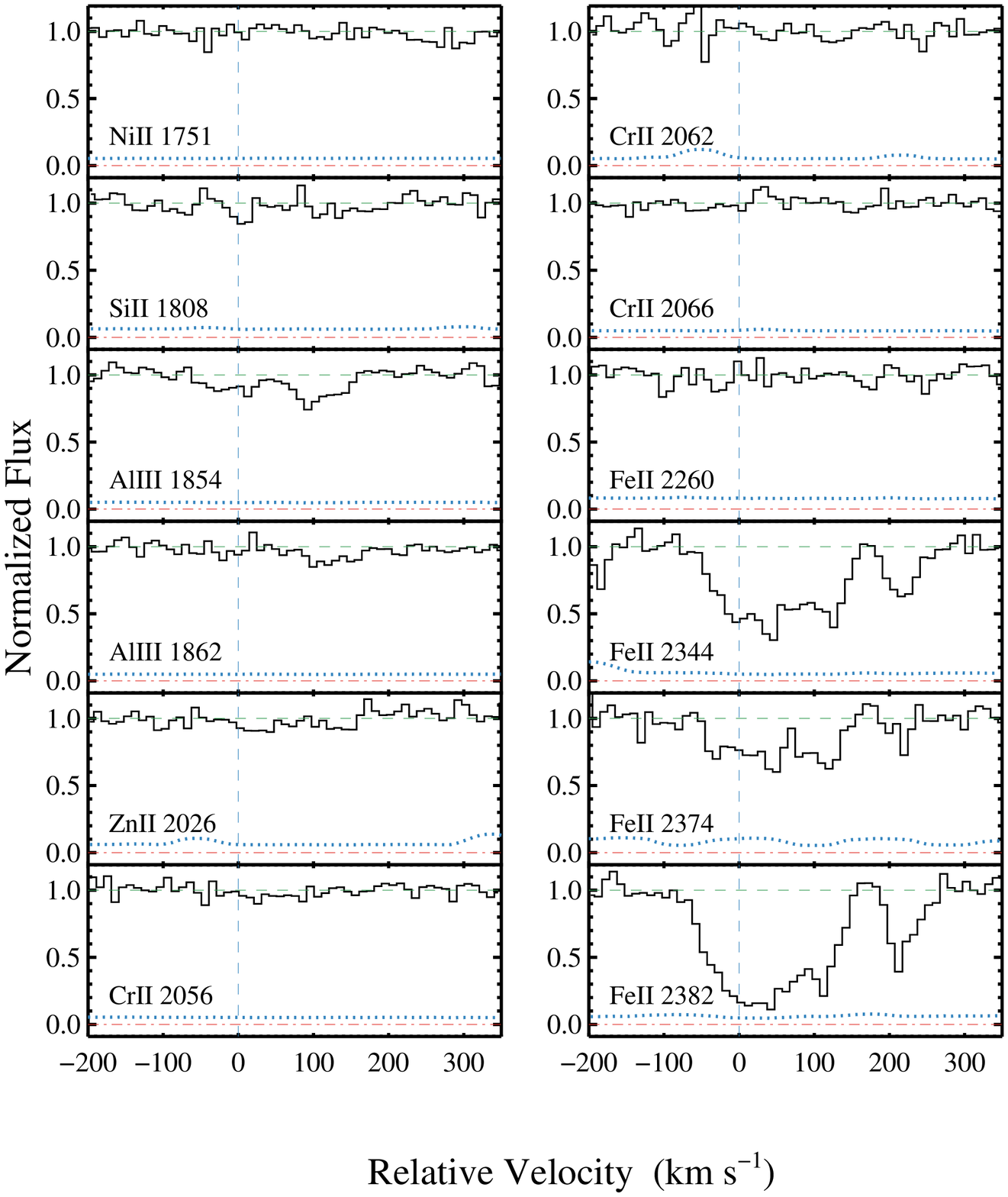}
\caption{Velocity plot of the ion transitions associated to the DLA J0811+3936 (25:H13) at  $z_{\rm dla}=2.6500$ (continued).
See Figure \ref{fig:vpdla01a} for an explanation of the different line colors.}\label{fig:vphst13b}
\end{figure*}

In this Appendix, we present details of the metallicity measurements for the intervening 
DLAs with new echellette spectra. A summary of the metallicities together with the
$\alpha-$ element, zinc, and iron abundances is provided in Table \ref{tabmetsumm}.
The best estimate for the DLA metallicity $\rm [X/H]_{\rm dla}$, which we infer assuming 
solar abundance pattern \citep{asp09}, is also reported in Table \ref{tabprop}.
In Tables \ref{tab:dla01}-\ref{tab:hst20}, we list instead all the metal transitions 
measured in each DLA. For each ionic transition, we provide information on 
the adopted wavelength $\lambda$, the oscillator 
strength $f$ from \citet{mor03}, and the velocity window $v_{\rm int}$ over which we measure equivalent 
widths $W_\lambda$ and column densities $N$. The adopted column density for each ion $N_{\rm adopt}$ is 
also indicated. Velocity plots of all analyzed transitions are shown in Figures 
\ref{fig:vpdla01a}-\ref{fig:vphst13b}. Additional information on the methodology here adopted 
can be found in \citet{pro03b} and \citet{raf12}.

\clearpage
\begin{table*}
\caption{Ionic column densities for J2114-0055 (1:G1) at $z_{\rm dla}=2.9181$\label{tab:dla01}}
\begin{center}
\begin{tabular}{lcccccccc}
\hline
Ion & $\lambda$ & $\log f$ & Instr. & $v_{\rm int}^a$ & $W_\lambda^b$ & $\log N$ & $\log N_{\rm adopt}$ \\
& (\AA) & & & (\kms) & (m\AA) & \\
\hline
\ion{C}{I}\\
&1560.3092 &$ -0.8808$&ESI&$[-100, 100]$&$<  18.0$&$< 12.98$&$< 12.98$\\
&1656.9283 &$ -0.8273$&ESI&$[-100, 100]$&$<  25.5$&$< 13.02$& \\
\ion{C}{II}\\
&1334.5323 &$ -0.8935$&ESI&$[ -50,  50]$&$ 476.6 \pm   3.3$&$> 14.85$&$> 14.85$\\
\ion{C}{IV}\\
&1548.1950 &$ -0.7194$&ESI&$[-120, 120]$&$ 482.5 \pm   9.0$&$  14.23 \pm 0.01$&$  14.23 \pm 0.01$\\
&1550.7700 &$ -1.0213$&ESI&$[-120, 100]$&$ 290.3 \pm   9.1$&$  14.24 \pm 0.02$& \\
\ion{O}{I}\\
&1302.1685 &$ -1.3110$&ESI&$[ -70,  90]$&$ 380.8 \pm   8.7$&$> 15.07$&$> 15.07$\\
\ion{Al}{II}\\
&1670.7874 &$  0.2742$&ESI&$[-100, 100]$&$ 504.7 \pm  10.5$&$> 13.33$&$> 13.33$\\
\ion{Al}{III}\\
&1854.7164 &$ -0.2684$&ESI&$[-150,  80]$&$ 272.6 \pm  11.0$&$  13.34 \pm 0.02$&$  13.33 \pm 0.02$\\
&1862.7895 &$ -0.5719$&ESI&$[-100, 100]$&$ 124.2 \pm  14.4$&$  13.27 \pm 0.05$& \\
\ion{Si}{II}\\
&1190.4158 &$ -0.6017$&ESI&$[-120, 120]$&$ 487.1 \pm  14.4$&$> 14.51$&$  15.00 \pm 0.06$\\
&1260.4221 &$  0.0030$&ESI&$[-100, 100]$&$ 626.0 \pm  10.5$&$> 14.03$& \\
&1304.3702 &$ -1.0269$&ESI&$[-100, 100]$&$ 336.2 \pm   9.9$&$> 14.66$& \\
&1526.7066 &$ -0.8962$&ESI&$[-120,  90]$&$ 426.0 \pm  10.3$&$> 14.55$& \\
&1808.0130 &$ -2.6603$&ESI&$[ -60,  60]$&$  58.1 \pm   7.8$&$  15.00 \pm 0.06$& \\
\ion{Si}{IV}\\
&1393.7550 &$ -0.2774$&ESI&$[-140, 100]$&$ 351.6 \pm   5.5$&$> 13.71$&$  13.85 \pm 0.01$\\
&1402.7700 &$ -0.5817$&ESI&$[-120,  90]$&$ 260.8 \pm   5.6$&$  13.85 \pm 0.01$& \\
\ion{S}{II}\\
&1250.5840 &$ -2.2634$&ESI&$[ -50,  50]$&$<  20.3$&$< 14.63$&$  14.77 \pm 0.04$\\
&1253.8110 &$ -1.9634$&ESI&$[ -50, 110]$&$  72.8 \pm  12.2$&$  14.70 \pm 0.08$& \\
&1259.5190 &$ -1.7894$&ESI&$[ -60,  60]$&$ 115.2 \pm   9.8$&$  14.80 \pm 0.04$& \\
\ion{Cr}{II}\\
&2056.2539 &$ -0.9788$&ESI&$[ -70,  70]$&$  32.0 \pm  10.3$&$  12.93 \pm 0.14$&$  12.93 \pm 0.14$\\
&2066.1610 &$ -1.2882$&ESI&$[ -70,  70]$&$<  22.7$&$< 13.25$& \\
\ion{Fe}{II}\\
&2249.8768 &$ -2.7397$&ESI&$[ -60,  60]$&$<  18.9$&$< 14.55$&$> 14.44$\\
&2260.7805 &$ -2.6126$&ESI&$[ -60,  60]$&$<  19.4$&$< 14.43$& \\
&2344.2140 &$ -0.9431$&ESI&$[-100, 100]$&$ 488.0 \pm  15.1$&$> 14.19$& \\
&2374.4612 &$ -1.5045$&ESI&$[ -80,  80]$&$ 319.4 \pm  19.1$&$> 14.44$& \\
&2382.7650 &$ -0.4949$&ESI&$[-100, 100]$&$ 659.0 \pm  27.1$&$> 13.94$& \\
\ion{Ni}{II}\\
&1454.8420 &$ -1.4908$&ESI&$[ -50,  50]$&$<  13.8$&$< 13.53$&$  13.44 \pm 0.06$\\
&1709.6042 &$ -1.4895$&ESI&$[ -50,  50]$&$  30.9 \pm   6.8$&$  13.60 \pm 0.09$& \\
&1741.5531 &$ -1.3696$&ESI&$[ -50,  50]$&$  28.3 \pm   4.4$&$  13.40 \pm 0.07$& \\
&1751.9157 &$ -1.5575$&ESI&$[ -50,  50]$&$<  10.2$&$< 13.31$& \\
\ion{Zn}{II}\\
&2026.1360 &$ -0.3107$&ESI&$[ -40,  40]$&$<  14.5$&$< 12.10$&$< 12.10$\\
\hline
\end{tabular}
\end{center}
 
$^{a}$ Velocity interval over which the equivalent width and column density are measured.
$^{b}$ Rest equivalent width.
\end{table*}

\clearpage
\begin{table*}
\caption{Ionic column densities for J0731+2854 (2:G2) at  $z_{\rm dla}=2.6878$\label{tab:dla02}}
\begin{center}
\begin{tabular}{lcccccccc}
\hline
Ion & $\lambda$ & $\log f$ & Instr. & $v_{\rm int}^a$ & $W_\lambda^b$ & $\log N$ & $\log N_{\rm adopt}$ \\
& (\AA) & & & (\kms) & (m\AA) & \\
\hline
\ion{C}{I}\\
&1560.3092 &$ -0.8808$&ESI&$[-100, 100]$&$<  13.1$&$< 12.84$&$< 12.84$\\
\ion{C}{II}\\
&1334.5323 &$ -0.8935$&ESI&$[ -70,  70]$&$ 304.9 \pm  10.1$&$> 14.49$&$> 14.49$\\
\ion{C}{IV}\\
&1548.1950 &$ -0.7194$&ESI&$[-100, 100]$&$ 116.9 \pm   5.6$&$  13.50 \pm 0.02$&$  13.50 \pm 0.02$\\
&1550.7700 &$ -1.0213$&ESI&$[-100,  50]$&$ 111.8 \pm   4.9$&$< 13.80$& \\
\ion{Al}{II}\\
&1670.7874 &$  0.2742$&ESI&$[ -70,  70]$&$ 210.7 \pm   7.8$&$> 12.87$&$> 12.87$\\
\ion{Al}{III}\\
&1854.7164 &$ -0.2684$&ESI&$[ -70,  70]$&$<  18.4$&$< 12.23$&$< 12.23$\\
\ion{Si}{II}\\
&1193.2897 &$ -0.3018$&ESI&$[ -70,  70]$&$ 337.6 \pm  11.7$&$> 14.08$&$  14.66 \pm 0.14$\\
&1808.0130 &$ -2.6603$&ESI&$[ -50,  50]$&$  26.9 \pm   8.6$&$  14.66 \pm 0.14$& \\
\ion{Si}{IV}\\
&1402.7700 &$ -0.5817$&ESI&$[-100, 200]$&$ 257.3 \pm  13.0$&$< 13.82$&$< 13.82$\\
\ion{Mn}{II}\\
&2594.4990 &$ -0.5670$&ESI&$[ -50,  50]$&$<  57.2$&$< 12.72$&$< 12.72$\\
&2606.4620 &$ -0.7151$&ESI&$[ -50,  50]$&$<  61.2$&$< 12.91$& \\
\ion{Fe}{II}\\
&1608.4511 &$ -1.2366$&ESI&$[ -50,  50]$&$ 155.1 \pm   7.1$&$> 14.23$&$  14.59 \pm 0.11$\\
&1611.2005 &$ -2.8665$&ESI&$[ -50,  50]$&$<  15.8$&$< 14.89$& \\
&2249.8768 &$ -2.7397$&ESI&$[ -40,  40]$&$<  22.7$&$< 14.64$& \\
&2260.7805 &$ -2.6126$&ESI&$[ -40,  40]$&$  40.6 \pm  10.5$&$  14.59 \pm 0.11$& \\
&2344.2140 &$ -0.9431$&ESI&$[ -70,  60]$&$ 350.7 \pm   9.1$&$> 14.06$& \\
&2374.4612 &$ -1.5045$&ESI&$[ -50,  50]$&$ 198.1 \pm   7.9$&$> 14.23$& \\
&2382.7650 &$ -0.4949$&ESI&$[ -60,  60]$&$ 396.3 \pm   7.9$&$> 13.71$& \\
&2586.6500 &$ -1.1605$&ESI&$[ -50,  50]$&$ 309.6 \pm  25.7$&$> 14.04$& \\
&2600.1729 &$ -0.6216$&ESI&$[ -60,  60]$&$ 417.3 \pm  23.5$&$> 13.73$& \\
\ion{Ni}{II}\\
&1709.6042 &$ -1.4895$&ESI&$[ -40,  40]$&$<  14.7$&$< 13.43$&$  13.88 \pm 0.04$\\
&1741.5531 &$ -1.3696$&ESI&$[ -30,  80]$&$  89.3 \pm   8.7$&$  13.98 \pm 0.04$& \\
&1751.9157 &$ -1.5575$&ESI&$[ -50,  50]$&$  29.4 \pm   8.4$&$  13.64 \pm 0.12$& \\
\ion{Zn}{II}\\
&2026.1360 &$ -0.3107$&ESI&$[ -30,  30]$&$<  13.8$&$< 12.07$&$< 12.07$\\
\hline
\end{tabular}
\end{center}
 
$^{a}$ Velocity interval over which the equivalent width and column density are measured.
$^{b}$ Rest equivalent width.
\end{table*}

\clearpage
\begin{table*}
\caption{Ionic column densities for J0956+3444 (3:G3) at  $z_{\rm dla}=2.3887$\label{tab:dla03}}
\begin{center}
\begin{tabular}{lcccccccc}
\hline
Ion & $\lambda$ & $\log f$ & Instr. & $v_{\rm int}^a$ & $W_\lambda^b$ & $\log N$ & $\log N_{\rm adopt}$ \\
& (\AA) & & & (\kms) & (m\AA) & \\
\hline
\ion{C}{I}\\
&1656.9283 &$ -0.8273$&ESI&$[ -50,  50]$&$<  20.3$&$< 12.91$&$< 12.91$\\
\ion{C}{IV}\\
&1550.7700 &$ -1.0213$&ESI&$[-140, 150]$&$ 765.3 \pm  13.8$&$> 14.78$&$> 14.78$\\
\ion{O}{I}\\
&1302.1685 &$ -1.3110$&ESI&$[-150, 150]$&$ 886.4 \pm  23.6$&$> 15.45$&$> 15.45$\\
\ion{Al}{II}\\
&1670.7874 &$  0.2742$&ESI&$[-130, 130]$&$ 778.6 \pm  15.1$&$> 13.62$&$> 13.62$\\
\ion{Al}{III}\\
&1854.7164 &$ -0.2684$&ESI&$[ -80,  80]$&$ 240.7 \pm  14.1$&$  13.25 \pm 0.03$&$  13.25 \pm 0.03$\\
\ion{Si}{II}\\
&1304.3702 &$ -1.0269$&ESI&$[-120, 120]$&$ 780.9 \pm  20.4$&$> 15.13$&$> 15.13$\\
&1526.7066 &$ -0.8962$&ESI&$[-180, 180]$&$ 877.5 \pm  15.9$&$> 14.89$& \\
\ion{Cr}{II}\\
&2056.2539 &$ -0.9788$&ESI&$[-120, 120]$&$ 129.5 \pm  17.7$&$  13.56 \pm 0.06$&$  13.57 \pm 0.05$\\
&2062.2340 &$ -1.1079$&ESI&$[-120, 120]$&$ 130.6 \pm  18.0$&$< 13.69$& \\
&2066.1610 &$ -1.2882$&ESI&$[-120, 120]$&$  72.7 \pm  18.4$&$  13.62 \pm 0.10$& \\
\ion{Mn}{II}\\
&2594.4990 &$ -0.5670$&ESI&$[ -60,  60]$&$ 116.5 \pm  18.6$&$  12.91 \pm 0.07$&$  12.91 \pm 0.07$\\
\ion{Fe}{II}\\
&1608.4511 &$ -1.2366$&ESI&$[-150, 150]$&$ 625.6 \pm  11.3$&$> 15.03$&$  15.09 \pm 0.11$\\
&1611.2005 &$ -2.8665$&ESI&$[-100,  60]$&$  35.9 \pm   9.6$&$  15.09 \pm 0.11$& \\
&2344.2140 &$ -0.9431$&ESI&$[-150, 150]$&$1043.0 \pm  20.5$&$> 14.67$& \\
&2374.4612 &$ -1.5045$&ESI&$[-150, 150]$&$ 756.2 \pm  23.5$&$> 15.07$& \\
&2382.7650 &$ -0.4949$&ESI&$[-150, 150]$&$1234.8 \pm  22.6$&$> 14.30$& \\
&2586.6500 &$ -1.1605$&ESI&$[-150, 150]$&$1116.5 \pm  34.9$&$> 14.84$& \\
&2600.1729 &$ -0.6216$&ESI&$[-150, 150]$&$1334.0 \pm  21.8$&$> 14.38$& \\
\ion{Ni}{II}\\
&1709.6042 &$ -1.4895$&ESI&$[-100, 100]$&$  86.9 \pm  18.1$&$  14.06 \pm 0.09$&$  14.02 \pm 0.05$\\
&1741.5531 &$ -1.3696$&ESI&$[-100, 100]$&$  97.5 \pm  18.9$&$  13.99 \pm 0.08$& \\
&1751.9157 &$ -1.5575$&ESI&$[-100, 100]$&$  66.6 \pm  18.7$&$  14.00 \pm 0.11$& \\
\ion{Zn}{II}\\
&2026.1360 &$ -0.3107$&ESI&$[-100, 100]$&$  69.7 \pm  15.6$&$  12.73 \pm 0.09$&$  12.73 \pm 0.09$\\
\hline
\end{tabular}
\end{center}
 
$^{a}$ Velocity interval over which the equivalent width and column density are measured.
$^{b}$ Rest equivalent width.
\end{table*}

\clearpage
\begin{table*}
\caption{Ionic column densities for J2343-1047 (4:G4) at  $z_{\rm dla}=2.6880$\label{tab:dla04}}
\begin{center}
\begin{tabular}{lcccccccc}
\hline
Ion & $\lambda$ & $\log f$ & Instr. & $v_{\rm int}^a$ & $W_\lambda^b$ & $\log N$ & $\log N_{\rm adopt}$ \\
& (\AA) & & & (\kms) & (m\AA) & \\
\hline
\ion{C}{I}\\
&1560.3092 &$ -0.8808$&ESI&$[ -80,  80]$&$<  19.8$&$< 13.03$&$< 13.03$\\
&1656.9283 &$ -0.8273$&ESI&$[ -80,  80]$&$<  35.1$&$< 13.16$& \\
\ion{C}{II}\\
&1334.5323 &$ -0.8935$&ESI&$[ -80,  50]$&$ 527.4 \pm  17.2$&$> 14.82$&$> 14.82$\\
\ion{C}{IV}\\
&1548.1950 &$ -0.7194$&ESI&$[-150, 150]$&$ 490.6 \pm  10.1$&$< 14.19$&$< 14.19$\\
\ion{O}{I}\\
&1302.1685 &$ -1.3110$&ESI&$[ -80,  60]$&$ 504.8 \pm  11.6$&$> 15.24$&$> 15.24$\\
\ion{Al}{II}\\
&1670.7874 &$  0.2742$&ESI&$[-120,  90]$&$ 328.9 \pm  16.6$&$> 13.03$&$> 13.03$\\
\ion{Si}{II}\\
&1260.4221 &$  0.0030$&ESI&$[-130,  80]$&$ 534.2 \pm  31.6$&$> 13.90$&$> 14.64$\\
&1304.3702 &$ -1.0269$&ESI&$[ -50,  80]$&$ 329.9 \pm  13.3$&$> 14.64$& \\
&1526.7066 &$ -0.8962$&ESI&$[-100, 130]$&$ 425.5 \pm   5.5$&$> 14.47$& \\
&1808.0130 &$ -2.6603$&ESI&$[-100, 100]$&$<  45.5$&$< 15.04$& \\
\ion{Fe}{II}\\
&1608.4511 &$ -1.2366$&ESI&$[-100, 100]$&$ 198.4 \pm  23.4$&$  14.29 \pm 0.05$&$  14.27 \pm 0.03$\\
&1611.2005 &$ -2.8665$&ESI&$[ -80,  80]$&$<  40.5$&$< 15.30$& \\
&2260.7805 &$ -2.6126$&ESI&$[-100, 100]$&$<  63.9$&$< 14.94$& \\
&2344.2140 &$ -0.9431$&ESI&$[-130, 130]$&$ 480.6 \pm  27.3$&$> 14.13$& \\
&2374.4612 &$ -1.5045$&ESI&$[-100,  80]$&$ 238.1 \pm  15.7$&$  14.27 \pm 0.03$& \\
&2382.7650 &$ -0.4949$&ESI&$[-150, 150]$&$ 711.2 \pm  18.4$&$> 13.99$& \\
&2600.1729 &$ -0.6216$&ESI&$[-150, 150]$&$ 814.2 \pm  77.7$&$> 14.07$& \\
\ion{Ni}{II}\\
&1741.5531 &$ -1.3696$&ESI&$[-100, 100]$&$<  39.4$&$< 13.71$&$< 13.71$\\
&1751.9157 &$ -1.5575$&ESI&$[-100, 100]$&$<  36.3$&$< 13.88$& \\
\ion{Zn}{II}\\
&2026.1360 &$ -0.3107$&ESI&$[-100, 100]$&$<  47.5$&$< 12.60$&$< 12.60$\\
\hline
\end{tabular}
\end{center}
 
$^{a}$ Velocity interval over which the equivalent width and column density are measured.
$^{b}$ Rest equivalent width.
\end{table*}

\clearpage
\begin{table*}
\caption{Ionic column densities for J0343-0622 (5:G5) at  $z_{\rm dla}=2.5713$\label{tab:dla05}}
\begin{center}
\begin{tabular}{lcccccccc}
\hline
Ion & $\lambda$ & $\log f$ & Instr. & $v_{\rm int}^a$ & $W_\lambda^b$ & $\log N$ & $\log N_{\rm adopt}$ \\
& (\AA) & & & (\kms) & (m\AA) & \\
\hline
\ion{C}{I}\\
&1656.9283 &$ -0.8273$&ESI&$[ -60,  60]$&$<  65.6$&$< 13.40$&$< 13.40$\\
\ion{Al}{II}\\
&1670.7874 &$  0.2742$&ESI&$[ -80,  60]$&$ 154.4 \pm  28.3$&$  12.62 \pm 0.08$&$  12.62 \pm 0.08$\\
\ion{Al}{III}\\
&1854.7164 &$ -0.2684$&ESI&$[ -80,  80]$&$<  65.2$&$< 12.77$&$< 12.77$\\
\ion{Si}{II}\\
&1808.0130 &$ -2.6603$&ESI&$[ -80,  80]$&$<  43.9$&$< 15.02$&$< 15.02$\\
\ion{Cr}{II}\\
&2056.2539 &$ -0.9788$&ESI&$[ -80,  80]$&$< 137.9$&$< 13.74$&$< 13.56$\\
&2062.2340 &$ -1.1079$&ESI&$[ -80,  80]$&$<  69.2$&$< 13.56$& \\
&2066.1610 &$ -1.2882$&ESI&$[ -80,  80]$&$<  63.8$&$< 13.70$& \\
\ion{Mn}{II}\\
&2594.4990 &$ -0.5670$&ESI&$[ -80,  80]$&$< 112.3$&$< 13.04$&$< 13.04$\\
\ion{Fe}{II}\\
&1608.4511 &$ -1.2366$&ESI&$[ -50,  70]$&$  96.3 \pm  13.0$&$  13.93 \pm 0.06$&$  13.88 \pm 0.05$\\
&1611.2005 &$ -2.8665$&ESI&$[ -80,  80]$&$<  33.6$&$< 15.22$& \\
&2260.7805 &$ -2.6126$&ESI&$[ -80,  80]$&$<  61.3$&$< 14.93$& \\
&2344.2140 &$ -0.9431$&ESI&$[ -60,  60]$&$ 242.9 \pm  28.6$&$> 13.75$& \\
&2374.4612 &$ -1.5045$&ESI&$[ -40,  40]$&$  74.9 \pm  22.3$&$  13.75 \pm 0.13$& \\
&2382.7650 &$ -0.4949$&ESI&$[ -50,  50]$&$ 260.5 \pm  22.4$&$> 13.35$& \\
&2586.6500 &$ -1.1605$&ESI&$[ -50,  40]$&$ 203.2 \pm  37.8$&$> 13.81$& \\
\ion{Ni}{II}\\
&1741.5531 &$ -1.3696$&ESI&$[ -80,  80]$&$<  49.9$&$< 13.83$&$< 13.83$\\
\ion{Zn}{II}\\
&2026.1360 &$ -0.3107$&ESI&$[ -80,  80]$&$<  52.8$&$< 12.66$&$< 12.66$\\
\hline
\end{tabular}
\end{center}
 
$^{a}$ Velocity interval over which the equivalent width and column density are measured.
$^{b}$ Rest equivalent width.
\end{table*}

\clearpage
\begin{table*}
\caption{Ionic column densities for J2351+1600 (6:G6) at  $z_{\rm dla}=3.7861$\label{tab:dla06}}
\begin{center}
\begin{tabular}{lcccccccc}
\hline
Ion & $\lambda$ & $\log f$ & Instr. & $v_{\rm int}^a$ & $W_\lambda^b$ & $\log N$ & $\log N_{\rm adopt}$ \\
& (\AA) & & & (\kms) & (m\AA) & \\
\hline
\ion{C}{I}\\
&1656.9283 &$ -0.8273$&ESI&$[-100, 100]$&$<  47.3$&$< 13.32$&$< 13.32$\\
\ion{C}{IV}\\
&1548.1950 &$ -0.7194$&ESI&$[-100, 130]$&$ 105.9 \pm  20.4$&$  13.47 \pm 0.08$&$  13.47 \pm 0.08$\\
\ion{O}{I}\\
&1302.1685 &$ -1.3110$&ESI&$[-100,  60]$&$ 363.7 \pm  14.2$&$> 14.96$&$> 14.96$\\
\ion{Al}{II}\\
&1670.7874 &$  0.2742$&ESI&$[ -40,  40]$&$ 108.8 \pm  12.6$&$  12.49 \pm 0.05$&$  12.49 \pm 0.05$\\
\ion{Al}{III}\\
&1862.7895 &$ -0.5719$&ESI&$[ -80,  80]$&$<  40.6$&$< 12.88$&$< 12.88$\\
\ion{Si}{II}\\
&1526.7066 &$ -0.8962$&ESI&$[ -60,  70]$&$ 190.5 \pm  22.6$&$> 14.10$&$> 14.10$\\
&1808.0130 &$ -2.6603$&ESI&$[-100,  70]$&$<  50.7$&$< 15.10$& \\
\ion{Fe}{II}\\
&1608.4511 &$ -1.2366$&ESI&$[ -60,  60]$&$ 100.4 \pm  13.8$&$  13.97 \pm 0.06$&$  13.97 \pm 0.06$\\
\ion{Ni}{II}\\
&1454.8420 &$ -1.4908$&ESI&$[ -80,  80]$&$<  16.4$&$< 13.61$&$< 13.61$\\
\hline
\end{tabular}
\end{center}
 
$^{a}$ Velocity interval over which the equivalent width and column density are measured.
$^{b}$ Rest equivalent width.
\end{table*}

\clearpage
\begin{table*}
\caption{Ionic column densities for J0042-1020 (7:G7) at  $z_{\rm dla}=2.7544$\label{tab:dla07}}
\begin{center}
\begin{tabular}{lcccccccc}
\hline
Ion & $\lambda$ & $\log f$ & Instr. & $v_{\rm int}^a$ & $W_\lambda^b$ & $\log N$ & $\log N_{\rm adopt}$ \\
& (\AA) & & & (\kms) & (m\AA) & \\
\hline
\ion{C}{I}\\
&1656.9283 &$ -0.8273$&ESI&$[ -80,  80]$&$<  20.2$&$< 12.93$&$< 12.93$\\
\ion{C}{II}\\
&1334.5323 &$ -0.8935$&ESI&$[-100, 220]$&$ 981.0 \pm  11.0$&$> 15.07$&$> 15.07$\\
\ion{O}{I}\\
&1302.1685 &$ -1.3110$&ESI&$[ -80, 160]$&$ 729.9 \pm  12.5$&$> 15.32$&$> 15.31$\\
\ion{Al}{II}\\
&1670.7874 &$  0.2742$&ESI&$[-180, 200]$&$ 732.2 \pm  14.4$&$> 13.42$&$> 13.42$\\
\ion{Al}{III}\\
&1854.7164 &$ -0.2684$&ESI&$[-100, 100]$&$<  23.9$&$< 12.35$&$< 12.35$\\
&1862.7895 &$ -0.5719$&ESI&$[-100, 100]$&$<  24.8$&$< 12.65$& \\
\ion{Si}{II}\\
&1808.0130 &$ -2.6603$&ESI&$[-120, 180]$&$ 186.8 \pm  11.7$&$< 15.52$&$< 15.52$\\
\ion{Cr}{II}\\
&2056.2539 &$ -0.9788$&ESI&$[-100, 100]$&$<  34.5$&$< 13.15$&$< 13.15$\\
&2066.1610 &$ -1.2882$&ESI&$[-100, 100]$&$<  24.7$&$< 13.29$& \\
\ion{Fe}{II}\\
&1608.4511 &$ -1.2366$&ESI&$[-100, 190]$&$ 309.9 \pm   9.2$&$  14.44 \pm 0.01$&$  14.39 \pm 0.01$\\
&1611.2005 &$ -2.8665$&ESI&$[-100, 190]$&$<  20.3$&$< 14.99$& \\
&2249.8768 &$ -2.7397$&ESI&$[-100, 190]$&$<  39.0$&$< 14.85$& \\
&2260.7805 &$ -2.6126$&ESI&$[-100, 190]$&$<  37.1$&$< 14.70$& \\
&2344.2140 &$ -0.9431$&ESI&$[-100, 190]$&$ 627.1 \pm  16.2$&$> 14.17$& \\
&2374.4612 &$ -1.5045$&ESI&$[-100, 190]$&$ 236.4 \pm  19.7$&$  14.23 \pm 0.04$& \\
&2600.1729 &$ -0.6216$&ESI&$[-100, 190]$&$ 925.4 \pm  43.9$&$> 13.99$& \\
\ion{Ni}{II}\\
&1709.6042 &$ -1.4895$&ESI&$[-100, 100]$&$<  27.4$&$< 13.69$&$< 13.56$\\
&1741.5531 &$ -1.3696$&ESI&$[-100, 100]$&$<  27.6$&$< 13.57$& \\
&1751.9157 &$ -1.5575$&ESI&$[-100, 100]$&$<  34.4$&$< 13.83$& \\
\hline
\end{tabular}
\end{center}
 
$^{a}$ Velocity interval over which the equivalent width and column density are measured.
$^{b}$ Rest equivalent width.
\end{table*}

\clearpage
\begin{table*}
\caption{Ionic column densities for J0949+1115 (8:G9) at  $z_{\rm dla}=2.7584$\label{tab:dla09}}
\begin{center}
\begin{tabular}{lcccccccc}
\hline
Ion & $\lambda$ & $\log f$ & Instr. & $v_{\rm int}^a$ & $W_\lambda^b$ & $\log N$ & $\log N_{\rm adopt}$ \\
& (\AA) & & & (\kms) & (m\AA) & \\
\hline
\ion{C}{I}\\
&1656.9283 &$ -0.8273$&ESI&$[-100, 100]$&$<  15.8$&$< 12.82$&$< 12.82$\\
\ion{Al}{II}\\
&1670.7874 &$  0.2742$&ESI&$[-130, 130]$&$ 777.7 \pm   7.4$&$> 13.60$&$> 13.60$\\
\ion{Al}{III}\\
&1854.7164 &$ -0.2684$&ESI&$[-100, 100]$&$ 178.4 \pm   8.7$&$  13.09 \pm 0.02$&$  13.05 \pm 0.02$\\
&1862.7895 &$ -0.5719$&ESI&$[-100, 100]$&$  61.4 \pm   9.0$&$  12.90 \pm 0.06$& \\
\ion{Si}{II}\\
&1304.3702 &$ -1.0269$&ESI&$[-100, 100]$&$ 533.5 \pm   9.0$&$> 14.95$&$  15.41 \pm 0.02$\\
&1808.0130 &$ -2.6603$&ESI&$[-100, 100]$&$ 145.4 \pm   7.7$&$  15.41 \pm 0.02$& \\
\ion{Cr}{II}\\
&2056.2539 &$ -0.9788$&ESI&$[-100, 100]$&$  80.2 \pm   9.7$&$  13.34 \pm 0.05$&$  13.31 \pm 0.05$\\
&2062.2340 &$ -1.1079$&ESI&$[-100,  70]$&$  57.8 \pm  10.1$&$< 13.32$& \\
&2066.1610 &$ -1.2882$&ESI&$[-100, 100]$&$  28.4 \pm   9.0$&$  13.19 \pm 0.13$& \\
\ion{Fe}{II}\\
&1608.4511 &$ -1.2366$&ESI&$[-120, 120]$&$ 522.3 \pm   6.4$&$> 14.88$&$  15.10 \pm 0.05$\\
&2260.7805 &$ -2.6126$&ESI&$[-100, 100]$&$ 128.4 \pm  13.2$&$  15.10 \pm 0.05$& \\
&2344.2140 &$ -0.9431$&ESI&$[-100, 100]$&$ 981.7 \pm   8.2$&$> 14.62$& \\
&2374.4612 &$ -1.5045$&ESI&$[-100, 100]$&$ 916.8 \pm  20.5$&$> 15.06$& \\
&2382.7650 &$ -0.4949$&ESI&$[-100, 100]$&$1185.8 \pm   7.5$&$> 14.26$& \\
&2586.6500 &$ -1.1605$&ESI&$[-100, 100]$&$ 804.1 \pm  35.8$&$> 14.58$& \\
&2600.1729 &$ -0.6216$&ESI&$[-180, 180]$&$1420.0 \pm  26.2$&$> 14.35$& \\
\ion{Ni}{II}\\
&1709.6042 &$ -1.4895$&ESI&$[-100, 100]$&$  50.5 \pm   9.3$&$  13.80 \pm 0.08$&$  13.72 \pm 0.06$\\
&1741.5531 &$ -1.3696$&ESI&$[-100, 100]$&$  45.0 \pm  10.4$&$  13.63 \pm 0.10$& \\
&1751.9157 &$ -1.5575$&ESI&$[-100, 100]$&$  40.3 \pm  11.3$&$  13.75 \pm 0.12$& \\
\hline
\end{tabular}
\end{center}
 
$^{a}$ Velocity interval over which the equivalent width and column density are measured.
$^{b}$ Rest equivalent width.
\end{table*}

\clearpage
\begin{table*}
\caption{Ionic column densities for J1018+3106 (9:G10) at  $z_{\rm dla}=2.4592$\label{tab:dla10}}
\begin{center}
\begin{tabular}{lcccccccc}
\hline
Ion & $\lambda$ & $\log f$ & Instr. & $v_{\rm int}^a$ & $W_\lambda^b$ & $\log N$ & $\log N_{\rm adopt}$ \\
& (\AA) & & & (\kms) & (m\AA) & \\
\hline
\ion{C}{II}\\
&1334.5323 &$ -0.8935$&ESI&$[-170,  60]$&$ 455.3 \pm  10.8$&$> 14.70$&$> 14.70$\\
\ion{O}{I}\\
&1302.1685 &$ -1.3110$&ESI&$[-120,  80]$&$ 480.0 \pm   8.8$&$> 15.18$&$> 15.18$\\
\ion{Al}{II}\\
&1670.7874 &$  0.2742$&ESI&$[ -80,  80]$&$ 296.4 \pm   5.7$&$> 13.10$&$> 13.10$\\
\ion{Al}{III}\\
&1854.7164 &$ -0.2684$&ESI&$[ -70,  70]$&$ 102.3 \pm   8.0$&$  12.86 \pm 0.03$&$  12.86 \pm 0.03$\\
&1862.7895 &$ -0.5719$&ESI&$[ -70,  70]$&$  53.1 \pm   7.9$&$  12.85 \pm 0.06$& \\
\ion{Si}{II}\\
&1526.7066 &$ -0.8962$&ESI&$[ -70,  70]$&$ 304.6 \pm   7.0$&$> 14.38$&$> 14.38$\\
&1808.0130 &$ -2.6603$&ESI&$[ -70,  70]$&$ 109.6 \pm   7.3$&$< 15.30$& \\
\ion{Si}{IV}\\
&1393.7550 &$ -0.2774$&ESI&$[ -70,  70]$&$ 302.6 \pm   6.2$&$> 13.83$&$> 13.83$\\
\ion{S}{II}\\
&1250.5840 &$ -2.2634$&ESI&$[ -50,  50]$&$<  22.7$&$< 14.66$&$< 14.65$\\
\ion{Cr}{II}\\
&2062.2340 &$ -1.1079$&ESI&$[ -50,  50]$&$<  12.0$&$< 12.80$&$< 12.80$\\
&2066.1610 &$ -1.2882$&ESI&$[ -50,  50]$&$<  12.1$&$< 12.96$& \\
\ion{Mn}{II}\\
&2594.4990 &$ -0.5670$&ESI&$[ -50,  50]$&$  33.8 \pm   9.9$&$  12.34 \pm 0.13$&$  12.34 \pm 0.13$\\
&2606.4620 &$ -0.7151$&ESI&$[ -50,  50]$&$<  21.0$&$< 12.44$& \\
\ion{Fe}{II}\\
&2249.8768 &$ -2.7397$&ESI&$[ -50,  50]$&$<  16.2$&$< 14.48$&$  14.30 \pm 0.03$\\
&2260.7805 &$ -2.6126$&ESI&$[ -50,  50]$&$<  15.4$&$< 14.33$& \\
&2344.2140 &$ -0.9431$&ESI&$[ -50,  50]$&$ 361.4 \pm   7.8$&$> 14.09$& \\
&2374.4612 &$ -1.5045$&ESI&$[ -70,  50]$&$ 240.6 \pm  13.2$&$  14.30 \pm 0.03$& \\
&2382.7650 &$ -0.4949$&ESI&$[ -80,  60]$&$ 500.6 \pm  12.0$&$> 13.80$& \\
&2586.6500 &$ -1.1605$&ESI&$[ -50,  50]$&$ 367.1 \pm  10.3$&$> 14.19$& \\
&2600.1729 &$ -0.6216$&ESI&$[ -50,  50]$&$ 510.8 \pm   7.8$&$> 13.89$& \\
\ion{Ni}{II}\\
&1709.6042 &$ -1.4895$&ESI&$[ -50,  50]$&$<  13.8$&$< 13.40$&$< 13.26$\\
&1741.5531 &$ -1.3696$&ESI&$[ -50,  50]$&$<  13.8$&$< 13.26$& \\
&1751.9157 &$ -1.5575$&ESI&$[ -50,  50]$&$<  13.3$&$< 13.42$& \\
\ion{Zn}{II}\\
&2026.1360 &$ -0.3107$&ESI&$[ -50,  50]$&$<  13.7$&$< 12.07$&$< 12.07$\\
\hline
\end{tabular}
\end{center}
 
$^{a}$ Velocity interval over which the equivalent width and column density are measured.
$^{b}$ Rest equivalent width.
\end{table*}

\clearpage
\begin{table*}
\caption{Ionic column densities for J0851+2332 (10:G11) at  $z_{\rm dla}=3.5297$\label{tab:dla11}}
\begin{center}
\begin{tabular}{lcccccccc}
\hline
Ion & $\lambda$ & $\log f$ & Instr. & $v_{\rm int}^a$ & $W_\lambda^b$ & $\log N$ & $\log N_{\rm adopt}$ \\
& (\AA) & & & (\kms) & (m\AA) & \\
\hline
\ion{C}{I}\\
&1560.3092 &$ -0.8808$&ESI&$[-100, 100]$&$<  24.9$&$< 13.12$&$< 13.04$\\
&1656.9283 &$ -0.8273$&ESI&$[-100, 100]$&$<  25.6$&$< 13.04$& \\
\ion{C}{II}\\
&1334.5323 &$ -0.8935$&ESI&$[-200, 210]$&$1826.6 \pm  12.5$&$> 15.42$&$> 15.42$\\
\ion{C}{IV}\\
&1548.1950 &$ -0.7194$&ESI&$[-270, 250]$&$1144.3 \pm  16.2$&$> 14.72$&$> 14.76$\\
&1550.7700 &$ -1.0213$&ESI&$[-250, 250]$&$ 811.7 \pm  16.5$&$> 14.76$& \\
\ion{O}{I}\\
&1302.1685 &$ -1.3110$&ESI&$[-190, 260]$&$1768.4 \pm  19.0$&$> 15.82$&$> 15.82$\\
\ion{Al}{II}\\
&1670.7874 &$  0.2742$&ESI&$[-200, 250]$&$1864.2 \pm  13.5$&$> 14.00$&$> 14.00$\\
\ion{Si}{II}\\
&1260.4221 &$  0.0030$&ESI&$[-190, 220]$&$1670.7 \pm  10.9$&$> 14.52$&$> 15.57$\\
&1304.3702 &$ -1.0269$&ESI&$[-230, 260]$&$1958.1 \pm  17.1$&$> 15.57$& \\
&1526.7066 &$ -0.8962$&ESI&$[-220, 250]$&$1765.4 \pm  11.0$&$> 15.22$& \\
\ion{Fe}{II}\\
&1608.4511 &$ -1.2366$&ESI&$[-200, 250]$&$1149.1 \pm  18.4$&$> 15.16$&$> 15.16$\\
&1611.2005 &$ -2.8665$&ESI&$[ -80,  80]$&$<  33.2$&$< 15.23$& \\
\ion{Ni}{II}\\
&1709.6042 &$ -1.4895$&ESI&$[-100, 100]$&$  58.2 \pm  11.5$&$  13.86 \pm 0.09$&$  13.89 \pm 0.05$\\
&1741.5531 &$ -1.3696$&ESI&$[-100, 100]$&$  87.8 \pm  13.3$&$  13.91 \pm 0.07$& \\
&1751.9157 &$ -1.5575$&ESI&$[-100, 100]$&$<  33.5$&$< 13.84$& \\
\ion{Zn}{II}\\
&2026.1360 &$ -0.3107$&ESI&$[ -80, 100]$&$  77.2 \pm  19.5$&$  12.68 \pm 0.11$&$  12.68 \pm 0.11$\\
\hline
\end{tabular}
\end{center}
 
$^{a}$ Velocity interval over which the equivalent width and column density are measured.
$^{b}$ Rest equivalent width.
\end{table*}

\clearpage
\begin{table*}
\caption{Ionic column densities for J0956+1448 (11:G12) at  $z_{\rm dla}=2.6606$\label{tab:dla12}}
\begin{center}
\begin{tabular}{lcccccccc}
\hline
Ion & $\lambda$ & $\log f$ & Instr. & $v_{\rm int}^a$ & $W_\lambda^b$ & $\log N$ & $\log N_{\rm adopt}$ \\
& (\AA) & & & (\kms) & (m\AA) & \\
\hline
\ion{C}{I}\\
&1560.3092 &$ -0.8808$&ESI&$[-100, 100]$&$<  15.7$&$< 12.92$&$< 12.85$\\
&1656.9283 &$ -0.8273$&ESI&$[-100, 100]$&$<  16.9$&$< 12.85$& \\
\ion{C}{IV}\\
&1548.1950 &$ -0.7194$&ESI&$[-150,  60]$&$ 181.5 \pm   6.8$&$  13.74 \pm 0.02$&$  13.72 \pm 0.01$\\
&1550.7700 &$ -1.0213$&ESI&$[-150,  50]$&$  78.6 \pm   7.0$&$  13.64 \pm 0.04$& \\
\ion{N}{I}\\
&1200.2233 &$ -1.0645$&ESI&$[ -50,  50]$&$ 191.1 \pm   6.4$&$< 14.41$&$  14.36 \pm 0.03$\\
&1200.7098 &$ -1.3665$&ESI&$[ -50,  50]$&$ 101.1 \pm   7.1$&$  14.36 \pm 0.03$& \\
\ion{Al}{II}\\
&1670.7874 &$  0.2742$&ESI&$[-120,  50]$&$ 325.1 \pm   6.4$&$> 13.09$&$> 13.09$\\
\ion{Al}{III}\\
&1854.7164 &$ -0.2684$&ESI&$[ -50,  60]$&$  46.1 \pm   5.5$&$  12.48 \pm 0.05$&$  12.46 \pm 0.05$\\
&1862.7895 &$ -0.5719$&ESI&$[ -50,  60]$&$  20.8 \pm   5.4$&$  12.41 \pm 0.11$& \\
\ion{Si}{II}\\
&1526.7066 &$ -0.8962$&ESI&$[-100, 100]$&$ 324.0 \pm   5.5$&$> 14.40$&$  14.90 \pm 0.07$\\
&1808.0130 &$ -2.6603$&ESI&$[ -60,  50]$&$  46.8 \pm   7.4$&$  14.90 \pm 0.07$& \\
\ion{Si}{IV}\\
&1393.7550 &$ -0.2774$&ESI&$[-150,  50]$&$ 319.6 \pm   7.4$&$  13.67 \pm 0.01$&$  13.67 \pm 0.01$\\
\ion{Cr}{II}\\
&2056.2539 &$ -0.9788$&ESI&$[ -70,  80]$&$  28.9 \pm   9.0$&$  12.89 \pm 0.13$&$  12.95 \pm 0.10$\\
&2062.2340 &$ -1.1079$&ESI&$[ -80,  80]$&$<  16.5$&$< 12.94$& \\
&2066.1610 &$ -1.2882$&ESI&$[ -80,  80]$&$<  16.4$&$  13.11 \pm 0.14$& \\
\ion{Fe}{II}\\
&1608.4511 &$ -1.2366$&ESI&$[ -70,  60]$&$ 209.9 \pm   6.5$&$> 14.38$&$  14.67 \pm 0.08$\\
&2249.8768 &$ -2.7397$&ESI&$[ -60,  20]$&$<  19.9$&$< 14.57$& \\
&2260.7805 &$ -2.6126$&ESI&$[ -50,  20]$&$  48.4 \pm   8.3$&$  14.67 \pm 0.08$& \\
&2344.2140 &$ -0.9431$&ESI&$[ -80,  50]$&$ 418.4 \pm   6.5$&$> 14.13$& \\
&2382.7650 &$ -0.4949$&ESI&$[-100,  70]$&$ 743.8 \pm   7.3$&$> 14.03$& \\
&2586.6500 &$ -1.1605$&ESI&$[ -80,  50]$&$ 422.0 \pm  22.2$&$> 14.24$& \\
&2600.1729 &$ -0.6216$&ESI&$[ -80,  50]$&$ 584.4 \pm  19.9$&$> 13.93$& \\
\ion{Ni}{II}\\
&1751.9157 &$ -1.5575$&ESI&$[ -40,  70]$&$  31.1 \pm   7.1$&$  13.64 \pm 0.10$&$  13.64 \pm 0.10$\\
\ion{Zn}{II}\\
&2026.1360 &$ -0.3107$&ESI&$[ -80,  80]$&$  35.2 \pm   9.1$&$  12.31 \pm 0.11$&$  12.31 \pm 0.11$\\
\hline
\end{tabular}
\end{center}
 
$^{a}$ Velocity interval over which the equivalent width and column density are measured.
$^{b}$ Rest equivalent width.
\end{table*}

\clearpage
\begin{table*}
\caption{Ionic column densities for J1151+3536 (12:G13) at $z_{\rm dla}=2.5978$\label{tab:dla13}}
\begin{center}
\begin{tabular}{lcccccccc}
\hline
Ion & $\lambda$ & $\log f$ & Instr. & $v_{\rm int}^a$ & $W_\lambda^b$ & $\log N$ & $\log N_{\rm adopt}$ \\
& (\AA) & & & (\kms) & (m\AA) & \\
\hline
\ion{C}{I}\\
&1656.9283 &$ -0.8273$&ESI&$[-100, 100]$&$<  19.5$&$< 12.92$&$< 12.92$\\
\ion{Al}{II}\\
&1670.7874 &$  0.2742$&ESI&$[-110, 110]$&$ 619.9 \pm   8.4$&$> 13.38$&$> 13.38$\\
\ion{Al}{III}\\
&1854.7164 &$ -0.2684$&ESI&$[-100, 100]$&$<  22.1$&$< 12.31$&$< 12.31$\\
&1862.7895 &$ -0.5719$&ESI&$[-100, 100]$&$<  21.1$&$< 12.59$& \\
\ion{Si}{II}\\
&1808.0130 &$ -2.6603$&ESI&$[ -70,  80]$&$  80.3 \pm   9.3$&$  15.13 \pm 0.05$&$  15.13 \pm 0.05$\\
\ion{Cr}{II}\\
&2066.1610 &$ -1.2882$&ESI&$[ -80,  80]$&$<  20.4$&$< 13.20$&$< 13.20$\\
\ion{Fe}{II}\\
&1608.4511 &$ -1.2366$&ESI&$[-180, 180]$&$ 401.1 \pm  10.9$&$  14.62 \pm 0.01$&$  14.61 \pm 0.01$\\
&1611.2005 &$ -2.8665$&ESI&$[-100, 100]$&$<  18.1$&$< 14.95$& \\
&2260.7805 &$ -2.6126$&ESI&$[-180, 200]$&$<  34.8$&$< 14.68$& \\
&2344.2140 &$ -0.9431$&ESI&$[-180, 200]$&$ 954.3 \pm  18.3$&$> 14.54$& \\
&2374.4612 &$ -1.5045$&ESI&$[-110, 120]$&$ 492.3 \pm  11.3$&$  14.61 \pm 0.01$& \\
&2382.7650 &$ -0.4949$&ESI&$[-180, 180]$&$1252.3 \pm  11.3$&$> 14.27$& \\
&2600.1729 &$ -0.6216$&ESI&$[-120, 130]$&$1297.0 \pm  20.2$&$> 14.35$& \\
\hline
\end{tabular}
\end{center}
 
$^{a}$ Velocity interval over which the equivalent width and column density are measured.
$^{b}$ Rest equivalent width.
\end{table*}

\clearpage
\begin{table*}
\caption{Ionic column densities for J2123-0053 (13:H1) at  $z_{\rm dla}=2.7803$\label{tab:hst01}}
\begin{center}
\begin{tabular}{lcccccccc}
\hline
Ion & $\lambda$ & $\log f$ & Instr. & $v_{\rm int}^a$ & $W_\lambda^b$ & $\log N$ & $\log N_{\rm adopt}$ \\
& (\AA) & & & (\kms) & (m\AA) & \\
\hline
\ion{C}{I}\\
&1560.3092 &$ -0.8808$&ESI&$[-100, 100]$&$<  33.4$&$< 13.26$&$< 13.00$\\
&1656.9283 &$ -0.8273$&ESI&$[-100, 100]$&$<  23.6$&$< 13.00$& \\
\ion{Al}{II}\\
&1670.7874 &$  0.2742$&ESI&$[-150, 150]$&$ 543.3 \pm  13.1$&$> 13.29$&$> 13.29$\\
\ion{Al}{III}\\
&1854.7164 &$ -0.2684$&ESI&$[ -50,  50]$&$  66.5 \pm   7.7$&$  12.64 \pm 0.05$&$  12.65 \pm 0.05$\\
&1862.7895 &$ -0.5719$&ESI&$[ -50,  50]$&$  35.4 \pm   7.5$&$  12.66 \pm 0.09$& \\
\ion{Si}{II}\\
&1526.7066 &$ -0.8962$&ESI&$[-100, 160]$&$ 509.4 \pm  11.2$&$> 14.54$&$> 14.54$\\
&1808.0130 &$ -2.6603$&ESI&$[ -60,  60]$&$<  20.9$&$< 14.71$& \\
\ion{Cr}{II}\\
&2056.2539 &$ -0.9788$&ESI&$[ -50,  50]$&$<  18.9$&$< 12.87$&$< 12.87$\\
&2062.2340 &$ -1.1079$&ESI&$[ -50,  50]$&$<  28.7$&$< 13.16$& \\
&2066.1610 &$ -1.2882$&ESI&$[ -50,  50]$&$<  22.8$&$< 13.27$& \\
\ion{Fe}{II}\\
&1608.4511 &$ -1.2366$&ESI&$[-100, 120]$&$ 200.5 \pm  11.9$&$  14.28 \pm 0.03$&$  14.27 \pm 0.02$\\
&2260.7805 &$ -2.6126$&ESI&$[ -80, 120]$&$<  29.0$&$< 14.60$& \\
&2344.2140 &$ -0.9431$&ESI&$[ -90, 140]$&$ 492.3 \pm  13.5$&$> 14.13$& \\
&2374.4612 &$ -1.5045$&ESI&$[ -80, 100]$&$ 223.7 \pm  14.7$&$  14.26 \pm 0.03$& \\
&2382.7650 &$ -0.4949$&ESI&$[-100, 120]$&$ 763.2 \pm  15.0$&$> 13.97$& \\
&2586.6500 &$ -1.1605$&ESI&$[-100, 100]$&$ 319.0 \pm  39.2$&$> 14.08$& \\
&2600.1729 &$ -0.6216$&ESI&$[-100, 120]$&$ 808.0 \pm  43.3$&$> 13.97$& \\
\ion{Ni}{II}\\
&1709.6042 &$ -1.4895$&ESI&$[-100, 100]$&$<  26.1$&$< 13.66$&$< 13.61$\\
&1741.5531 &$ -1.3696$&ESI&$[-100, 100]$&$<  31.6$&$< 13.61$& \\
&1751.9157 &$ -1.5575$&ESI&$[-100, 100]$&$<  28.7$&$< 13.76$& \\
\hline
\end{tabular}
\end{center}
 
$^{a}$ Velocity interval over which the equivalent width and column density are measured.
$^{b}$ Rest equivalent width.
\end{table*}

\clearpage
\begin{table*}
\caption{Ionic column densities for J0407-4410 (14:H2) at   $z_{\rm dla}=1.9127$\label{tab:hst02}}
\begin{center}
\begin{tabular}{lcccccccc}
\hline
Ion & $\lambda$ & $\log f$ & Instr. & $v_{\rm int}^a$ & $W_\lambda^b$ & $\log N$ & $\log N_{\rm adopt}$ \\
& (\AA) & & & (\kms) & (m\AA) & \\
\hline
\ion{Mg}{II}\\
&2796.3520 &$ -0.2130$&MagE&$[-200, 200]$&$1710.3 \pm  31.2$&$> 14.01$&$> 14.27$\\
&2803.5310 &$ -0.5151$&MagE&$[-200, 200]$&$1569.2 \pm  32.7$&$> 14.26$& \\
\ion{Al}{III}\\
&1854.7164 &$ -0.2684$&MagE&$[-300, 150]$&$ 240.8 \pm  21.8$&$  13.22 \pm 0.04$&$  13.22 \pm 0.04$\\
\ion{Si}{II}\\
&1808.0130 &$ -2.6603$&MagE&$[-150, 150]$&$ 112.3 \pm  14.0$&$  15.29 \pm 0.05$&$  15.29 \pm 0.05$\\
\ion{Cr}{II}\\
&2056.2539 &$ -0.9788$&MagE&$[-150, 150]$&$ 133.8 \pm  14.9$&$  13.56 \pm 0.05$&$  13.60 \pm 0.04$\\
&2066.1610 &$ -1.2882$&MagE&$[-150, 150]$&$  95.3 \pm  13.3$&$  13.70 \pm 0.06$& \\
\ion{Mn}{II}\\
&2576.8770 &$ -0.4549$&MagE&$[-150, 150]$&$ 118.8 \pm  15.0$&$  12.79 \pm 0.05$&$  12.85 \pm 0.04$\\
&2594.4990 &$ -0.5670$&MagE&$[-150, 150]$&$ 125.8 \pm  14.9$&$  12.92 \pm 0.05$& \\
\ion{Fe}{II}\\
&2249.8768 &$ -2.7397$&MagE&$[-150, 150]$&$ 134.0 \pm  19.7$&$  15.24 \pm 0.06$&$  15.41 \pm 0.03$\\
&2260.7805 &$ -2.6126$&MagE&$[-150, 150]$&$ 288.5 \pm  17.4$&$  15.47 \pm 0.03$& \\
&2344.2140 &$ -0.9431$&MagE&$[-150, 150]$&$1039.6 \pm  11.9$&$> 14.66$& \\
&2374.4612 &$ -1.5045$&MagE&$[-150, 150]$&$ 846.7 \pm  12.4$&$> 14.99$& \\
&2382.7650 &$ -0.4949$&MagE&$[-150, 150]$&$1273.6 \pm  11.2$&$> 14.27$& \\
&2586.6500 &$ -1.1605$&MagE&$[-150, 150]$&$1131.6 \pm  12.7$&$> 14.82$& \\
&2600.1729 &$ -0.6216$&MagE&$[-150, 150]$&$1272.4 \pm  12.3$&$> 14.34$& \\
\ion{Ni}{II}\\
&1741.5531 &$ -1.3696$&MagE&$[-150, 150]$&$ 111.8 \pm  10.8$&$  14.01 \pm 0.04$&$  14.02 \pm 0.04$\\
\ion{Zn}{II}\\
&2026.1360 &$ -0.3107$&MagE&$[-150, 150]$&$  99.9 \pm  20.4$&$  12.77 \pm 0.09$&$  12.77 \pm 0.09$\\
\hline
\end{tabular}
\end{center}
 
$^{a}$ Velocity interval over which the equivalent width and column density are measured.
$^{b}$ Rest equivalent width.
\end{table*}

\clearpage
\begin{table*}
\caption{Ionic column densities for J0255+0048 (15:H3) at   $z_{\rm dla}=3.2530$\label{tab:hst03}}
\begin{center}
\begin{tabular}{lcccccccc}
\hline
Ion & $\lambda$ & $\log f$ & Instr. & $v_{\rm int}^a$ & $W_\lambda^b$ & $\log N$ & $\log N_{\rm adopt}$ \\
& (\AA) & & & (\kms) & (m\AA) & \\
\hline
\ion{C}{I}\\
&1560.3092 &$ -0.8808$&ESI&$[-100, 100]$&$<  34.6$&$< 13.26$&$< 13.09$\\
&1656.9283 &$ -0.8273$&ESI&$[-100, 100]$&$<  29.0$&$< 13.09$& \\
\ion{C}{IV}\\
&1548.1950 &$ -0.7194$&ESI&$[-150, 300]$&$ 683.9 \pm  26.8$&$  14.39 \pm 0.02$&$  14.41 \pm 0.02$\\
&1550.7700 &$ -1.0213$&ESI&$[-150, 300]$&$ 480.9 \pm  27.0$&$  14.46 \pm 0.03$& \\
\ion{Al}{II}\\
&1670.7874 &$  0.2742$&ESI&$[-200, 300]$&$1212.7 \pm  22.7$&$> 13.77$&$> 13.77$\\
\ion{Al}{III}\\
&1854.7164 &$ -0.2684$&ESI&$[-100, 100]$&$ 104.1 \pm  16.8$&$  12.86 \pm 0.07$&$  12.86 \pm 0.07$\\
&1862.7895 &$ -0.5719$&ESI&$[-100, 100]$&$  75.2 \pm  24.9$&$< 13.01$& \\
\ion{Si}{II}\\
&1526.7066 &$ -0.8962$&ESI&$[-280, 300]$&$1056.9 \pm  27.2$&$> 14.94$&$  15.31 \pm 0.05$\\
&1808.0130 &$ -2.6603$&ESI&$[-100, 200]$&$ 122.4 \pm  12.5$&$  15.31 \pm 0.04$& \\
\ion{Cr}{II}\\
&2056.2539 &$ -0.9788$&ESI&$[-100, 100]$&$<  37.5$&$< 13.14$&$< 13.14$\\
&2066.1610 &$ -1.2882$&ESI&$[-100, 100]$&$<  37.1$&$< 13.46$& \\
\ion{Fe}{II}\\
&1608.4511 &$ -1.2366$&ESI&$[-280, 300]$&$ 561.2 \pm  24.2$&$  14.77 \pm 0.02$&$  14.77 \pm 0.02$\\
\ion{Ni}{II}\\
&1741.5531 &$ -1.3696$&ESI&$[-100, 230]$&$  69.9 \pm  23.1$&$  13.82 \pm 0.14$&$  13.82 \pm 0.14$\\
\ion{Zn}{II}\\
&2026.1360 &$ -0.3107$&ESI&$[-100, 100]$&$<  41.3$&$< 12.55$&$< 12.55$\\
\hline
\end{tabular}
\end{center}
 
$^{a}$ Velocity interval over which the equivalent width and column density are measured.
$^{b}$ Rest equivalent width.
\end{table*}

\clearpage
\begin{table*}
\caption{Ionic column densities for J0816+4823 (16:H4) at  $z_{\rm dla}=2.7067$\label{tab:hst04}}
\begin{center}
\begin{tabular}{lcccccccc}
\hline
Ion & $\lambda$ & $\log f$ & Instr. & $v_{\rm int}^a$ & $W_\lambda^b$ & $\log N$ & $\log N_{\rm adopt}$ \\
& (\AA) & & & (\kms) & (m\AA) & \\
\hline
\ion{C}{I}\\
&1656.9283 &$ -0.8273$&ESI&$[-100, 100]$&$<  20.3$&$< 12.93$&$< 12.93$\\
\ion{O}{I}\\
&1302.1685 &$ -1.3110$&ESI&$[ -60,  60]$&$ 236.0 \pm   8.5$&$> 14.70$&$> 14.70$\\
\ion{Al}{II}\\
&1670.7874 &$  0.2742$&ESI&$[ -60,  60]$&$ 103.6 \pm   7.7$&$  12.44 \pm 0.03$&$  12.44 \pm 0.03$\\
\ion{Si}{II}\\
&1526.7066 &$ -0.8962$&ESI&$[ -60,  60]$&$ 133.7 \pm   5.0$&$  13.85 \pm 0.02$&$  13.85 \pm 0.02$\\
&1808.0130 &$ -2.6603$&ESI&$[ -60,  50]$&$<  17.4$&$< 14.62$& \\
\ion{Mn}{II}\\
&2594.4990 &$ -0.5670$&ESI&$[ -60,  60]$&$<  53.3$&$< 12.68$&$< 12.68$\\
&2606.4620 &$ -0.7151$&ESI&$[ -60,  60]$&$<  54.0$&$< 12.82$& \\
\ion{Fe}{II}\\
&1611.2005 &$ -2.8665$&ESI&$[ -50,  50]$&$<  16.3$&$< 14.88$&$  13.74 \pm 0.03$\\
&2249.8768 &$ -2.7397$&ESI&$[ -50,  50]$&$<  23.1$&$< 14.64$& \\
&2260.7805 &$ -2.6126$&ESI&$[ -50,  50]$&$<  21.9$&$< 14.48$& \\
&2344.2140 &$ -0.9431$&ESI&$[ -60,  50]$&$ 178.2 \pm   7.9$&$> 13.63$& \\
&2374.4612 &$ -1.5045$&ESI&$[ -40,  40]$&$  84.5 \pm   7.3$&$  13.78 \pm 0.04$& \\
&2382.7650 &$ -0.4949$&ESI&$[ -40,  60]$&$ 221.3 \pm   8.7$&$> 13.31$& \\
&2586.6500 &$ -1.1605$&ESI&$[ -60,  60]$&$ 155.3 \pm  25.0$&$  13.65 \pm 0.07$& \\
&2600.1729 &$ -0.6216$&ESI&$[ -50,  50]$&$ 233.2 \pm  22.0$&$> 13.33$& \\
\ion{Ni}{II}\\
&1709.6042 &$ -1.4895$&ESI&$[ -60,  60]$&$<  16.9$&$< 13.49$&$< 13.49$\\
&1751.9157 &$ -1.5575$&ESI&$[ -60,  60]$&$<  20.4$&$< 13.62$& \\
\ion{Zn}{II}\\
&2026.1360 &$ -0.3107$&ESI&$[ -60,  60]$&$<  17.3$&$< 12.16$&$< 12.16$\\
\hline
\end{tabular}
\end{center}
 
$^{a}$ Velocity interval over which the equivalent width and column density are measured.
$^{b}$ Rest equivalent width.
\end{table*}

\clearpage
\begin{table*}
\caption{Ionic column densities for J0908+0238 (18:H6) at  $z_{\rm dla}=2.9586$\label{tab:hst06}}
\begin{center}
\begin{tabular}{lcccccccc}
\hline
Ion & $\lambda$ & $\log f$ & Instr. & $v_{\rm int}^a$ & $W_\lambda^b$ & $\log N$ & $\log N_{\rm adopt}$ \\
& (\AA) & & & (\kms) & (m\AA) & \\
\hline
\ion{C}{IV}\\
&1550.7700 &$ -1.0213$&MagE&$[ -70,  70]$&$<  52.2$&$< 13.58$&$< 13.58$\\
\ion{Al}{II}\\
&1670.7874 &$  0.2742$&MagE&$[-100, 100]$&$ 637.4 \pm  31.3$&$> 13.46$&$> 13.47$\\
\ion{Al}{III}\\
&1854.7164 &$ -0.2684$&MagE&$[-100, 100]$&$< 154.6$&$< 13.16$&$< 13.16$\\
&1862.7895 &$ -0.5719$&MagE&$[-100, 100]$&$<  90.4$&$< 13.25$& \\
\ion{Si}{II}\\
&1526.7066 &$ -0.8962$&MagE&$[-100, 100]$&$ 533.4 \pm  27.6$&$> 14.65$&$  15.68 \pm 0.07$\\
&1808.0130 &$ -2.6603$&MagE&$[-100, 100]$&$ 234.1 \pm  37.1$&$  15.68 \pm 0.07$& \\
\ion{Cr}{II}\\
&2056.2539 &$ -0.9788$&MagE&$[-100, 100]$&$< 167.5$&$< 13.92$&$< 13.92$\\
\ion{Fe}{II}\\
&1608.4511 &$ -1.2366$&MagE&$[-100, 100]$&$ 601.1 \pm  96.3$&$> 14.95$&$> 14.95$\\
&1611.2005 &$ -2.8665$&MagE&$[-100, 100]$&$<  83.1$&$< 15.62$& \\
\ion{Zn}{II}\\
&2026.1360 &$ -0.3107$&MagE&$[-100, 100]$&$ 298.6 \pm  65.7$&$< 13.32$&$< 13.32$\\
\hline
\end{tabular}
\end{center}
 
$^{a}$ Velocity interval over which the equivalent width and column density are measured.
$^{b}$ Rest equivalent width.
\end{table*}

\clearpage
\begin{table*}
\caption{Ionic column densities for J0844+1245 (21:H9) at   $z_{\rm dla}=1.8639$\label{tab:hst09}}
\begin{center}
\begin{tabular}{lcccccccc}
\hline
Ion & $\lambda$ & $\log f$ & Instr. & $v_{\rm int}^a$ & $W_\lambda^b$ & $\log N$ & $\log N_{\rm adopt}$ \\
& (\AA) & & & (\kms) & (m\AA) & \\
\hline
\ion{C}{IV}\\
&1548.1950 &$ -0.7194$&ESI&$[-200, 100]$&$ 262.8 \pm  14.3$&$  13.90 \pm 0.02$&$  13.91 \pm 0.02$\\
&1550.7700 &$ -1.0213$&ESI&$[-200, 100]$&$ 161.3 \pm  14.7$&$  13.96 \pm 0.04$& \\
\ion{Mg}{II}\\
&2796.3520 &$ -0.2130$&ESI&$[-100, 100]$&$ 484.0 \pm  13.5$&$> 13.33$&$> 13.58$\\
&2803.5310 &$ -0.5151$&ESI&$[-100, 100]$&$ 452.0 \pm  16.0$&$> 13.58$& \\
\ion{Al}{III}\\
&1854.7164 &$ -0.2684$&ESI&$[-100, 100]$&$  96.2 \pm  11.2$&$  12.82 \pm 0.05$&$  12.83 \pm 0.04$\\
&1862.7895 &$ -0.5719$&ESI&$[-100, 100]$&$  54.4 \pm  11.3$&$  12.85 \pm 0.09$& \\
\ion{Si}{II}\\
&1526.7066 &$ -0.8962$&ESI&$[-100, 100]$&$ 202.7 \pm  12.4$&$> 14.05$&$  14.97 \pm 0.06$\\
&1808.0130 &$ -2.6603$&ESI&$[ -50,  50]$&$  54.1 \pm   7.7$&$  14.97 \pm 0.06$& \\
\ion{Cr}{II}\\
&2066.1610 &$ -1.2882$&ESI&$[-100, 100]$&$<  24.9$&$< 13.29$&$< 13.29$\\
\ion{Mn}{II}\\
&2576.8770 &$ -0.4549$&ESI&$[ -50,  50]$&$  46.8 \pm  10.5$&$  12.38 \pm 0.10$&$  12.38 \pm 0.10$\\
&2594.4990 &$ -0.5670$&ESI&$[-100, 100]$&$  85.7 \pm  14.0$&$< 12.75$& \\
\ion{Fe}{II}\\
&1608.4511 &$ -1.2366$&ESI&$[-100, 100]$&$ 174.0 \pm  13.9$&$> 14.24$&$  14.79 \pm 0.06$\\
&1611.2005 &$ -2.8665$&ESI&$[-100, 100]$&$<  28.6$&$< 15.15$& \\
&2249.8768 &$ -2.7397$&ESI&$[ -50,  50]$&$  41.6 \pm   8.7$&$  14.74 \pm 0.09$& \\
&2260.7805 &$ -2.6126$&ESI&$[-100, 100]$&$  71.0 \pm  12.3$&$  14.84 \pm 0.07$& \\
&2344.2140 &$ -0.9431$&ESI&$[-100, 100]$&$ 295.8 \pm  11.0$&$> 13.90$& \\
&2374.4612 &$ -1.5045$&ESI&$[-100, 100]$&$ 223.0 \pm  11.1$&$> 14.29$& \\
&2382.7650 &$ -0.4949$&ESI&$[-100, 100]$&$ 332.8 \pm  10.8$&$> 13.54$& \\
&2586.6500 &$ -1.1605$&ESI&$[-100, 100]$&$ 388.2 \pm  13.0$&$> 14.14$& \\
&2600.1729 &$ -0.6216$&ESI&$[-100, 100]$&$ 356.8 \pm  12.6$&$> 13.60$& \\
\ion{Ni}{II}\\
&1741.5531 &$ -1.3696$&ESI&$[-100, 100]$&$<  26.0$&$  13.55 \pm 0.14$&$  13.55 \pm 0.14$\\
\ion{Zn}{II}\\
&2026.1360 &$ -0.3107$&ESI&$[ -50,  50]$&$  31.0 \pm   8.4$&$  12.26 \pm 0.12$&$  12.26 \pm 0.12$\\
\hline
\end{tabular}
\end{center}
 
$^{a}$ Velocity interval over which the equivalent width and column density are measured.
$^{b}$ Rest equivalent width.
\end{table*}

\clearpage
\begin{table*}
\caption{Ionic column densities for J0751+4516 (22:H10) at   $z_{\rm dla}=2.6826$\label{tab:hst10}}
\begin{center}
\begin{tabular}{lcccccccc}
\hline
Ion & $\lambda$ & $\log f$ & Instr. & $v_{\rm int}^a$ & $W_\lambda^b$ & $\log N$ & $\log N_{\rm adopt}$ \\
& (\AA) & & & (\kms) & (m\AA) & \\
\hline
\ion{C}{I}\\
&1560.3092 &$ -0.8808$&ESI&$[-100, 100]$&$<  32.4$&$< 13.23$&$< 12.99$\\
&1656.9283 &$ -0.8273$&ESI&$[-100, 100]$&$<  23.3$&$< 12.99$& \\
\ion{C}{IV}\\
&1548.1950 &$ -0.7194$&ESI&$[ -50,  70]$&$  99.7 \pm  10.5$&$  13.45 \pm 0.05$&$  13.44 \pm 0.04$\\
&1550.7700 &$ -1.0213$&ESI&$[ -50,  70]$&$  44.0 \pm  11.0$&$  13.37 \pm 0.11$& \\
\ion{O}{I}\\
&1302.1685 &$ -1.3110$&ESI&$[ -40,  30]$&$ 236.2 \pm   8.2$&$> 14.81$&$> 14.81$\\
\ion{Al}{II}\\
&1670.7874 &$  0.2742$&ESI&$[ -70,  60]$&$ 169.9 \pm  10.3$&$> 12.72$&$> 12.72$\\
\ion{Al}{III}\\
&1854.7164 &$ -0.2684$&ESI&$[ -50,  40]$&$  36.9 \pm  10.2$&$  12.38 \pm 0.12$&$  12.38 \pm 0.12$\\
&1862.7895 &$ -0.5719$&ESI&$[ -50,  40]$&$<  18.4$&$< 12.53$& \\
\ion{Si}{II}\\
&1304.3702 &$ -1.0269$&ESI&$[ -60,  50]$&$ 246.8 \pm  11.5$&$> 14.42$&$  14.85 \pm 0.09$\\
&1526.7066 &$ -0.8962$&ESI&$[ -60,  50]$&$ 152.9 \pm   9.6$&$> 13.98$& \\
&1808.0130 &$ -2.6603$&ESI&$[ -60,  50]$&$  42.2 \pm   8.6$&$  14.85 \pm 0.09$& \\
\ion{Si}{IV}\\
&1393.7550 &$ -0.2774$&ESI&$[ -40,  40]$&$ 123.0 \pm   7.8$&$  13.23 \pm 0.03$&$  13.23 \pm 0.03$\\
\ion{Cr}{II}\\
&2056.2539 &$ -0.9788$&ESI&$[ -50,  50]$&$<  24.0$&$< 12.97$&$< 12.97$\\
\ion{Fe}{II}\\
&1608.4511 &$ -1.2366$&ESI&$[ -50,  50]$&$ 135.2 \pm  12.1$&$  14.14 \pm 0.04$&$  14.15 \pm 0.03$\\
&1611.2005 &$ -2.8665$&ESI&$[ -50,  50]$&$<  25.2$&$< 15.06$& \\
&2249.8768 &$ -2.7397$&ESI&$[ -50,  50]$&$<  39.5$&$< 14.91$& \\
&2260.7805 &$ -2.6126$&ESI&$[ -50,  50]$&$<  27.2$&$< 14.58$& \\
&2344.2140 &$ -0.9431$&ESI&$[ -60,  50]$&$ 261.8 \pm  17.3$&$> 13.88$& \\
&2374.4612 &$ -1.5045$&ESI&$[ -60,  50]$&$ 174.0 \pm  14.0$&$  14.15 \pm 0.04$& \\
&2382.7650 &$ -0.4949$&ESI&$[ -60,  50]$&$ 285.9 \pm  13.6$&$> 13.47$& \\
&2600.1729 &$ -0.6216$&ESI&$[ -60,  50]$&$ 244.8 \pm  40.6$&$> 13.57$& \\
\ion{Ni}{II}\\
&1454.8420 &$ -1.4908$&ESI&$[ -50,  50]$&$<  12.6$&$< 13.49$&$< 13.49$\\
&1709.6042 &$ -1.4895$&ESI&$[ -50,  50]$&$<  22.5$&$< 13.63$& \\
&1741.5531 &$ -1.3696$&ESI&$[ -50,  50]$&$<  25.9$&$< 13.55$& \\
&1751.9157 &$ -1.5575$&ESI&$[ -50,  50]$&$<  26.3$&$< 13.73$& \\
\ion{Zn}{II}\\
&2026.1360 &$ -0.3107$&ESI&$[ -50,  50]$&$<  24.8$&$< 12.33$&$< 12.32$\\
\hline
\end{tabular}
\end{center}
 
$^{a}$ Velocity interval over which the equivalent width and column density are measured.
$^{b}$ Rest equivalent width.
\end{table*}

\clearpage
\begin{table*}
\caption{Ionic column densities for J0818+0720 (23:H11) at   $z_{\rm dla}=3.2332$\label{tab:hst11}}
\begin{center}
\begin{tabular}{lcccccccc}
\hline
Ion & $\lambda$ & $\log f$ & Instr. & $v_{\rm int}^a$ & $W_\lambda^b$ & $\log N$ & $\log N_{\rm adopt}$ \\
& (\AA) & & & (\kms) & (m\AA) & \\
\hline
\ion{C}{I}\\
&1560.3092 &$ -0.8808$&ESI&$[-200, 300]$&$<  48.8$&$< 13.41$&$< 13.26$\\
&1656.9283 &$ -0.8273$&ESI&$[-200, 300]$&$<  43.8$&$< 13.26$& \\
\ion{C}{II}\\
&1334.5323 &$ -0.8935$&ESI&$[-250, 300]$&$2192.2 \pm  14.6$&$> 15.47$&$> 15.47$\\
\ion{C}{IV}\\
&1548.1950 &$ -0.7194$&ESI&$[-160, 320]$&$ 749.9 \pm  19.1$&$> 14.57$&$> 14.82$\\
&1550.7700 &$ -1.0213$&ESI&$[-160, 320]$&$ 657.9 \pm  21.6$&$> 14.82$& \\
\ion{Al}{II}\\
&1670.7874 &$  0.2742$&ESI&$[-200, 320]$&$1178.8 \pm  19.7$&$> 13.69$&$> 13.69$\\
\ion{Si}{II}\\
&1260.4221 &$  0.0030$&ESI&$[-100, 320]$&$1423.2 \pm  12.4$&$> 14.41$&$> 14.99$\\
&1526.7066 &$ -0.8962$&ESI&$[-200, 320]$&$1214.6 \pm  14.4$&$> 14.99$& \\
\ion{Cr}{II}\\
&2056.2539 &$ -0.9788$&ESI&$[-200, 300]$&$<  55.0$&$< 13.33$&$< 13.33$\\
&2062.2340 &$ -1.1079$&ESI&$[-200, 300]$&$  99.5 \pm  26.0$&$< 13.59$& \\
&2066.1610 &$ -1.2882$&ESI&$[-200, 300]$&$<  53.4$&$< 13.62$& \\
\ion{Fe}{II}\\
&1608.4511 &$ -1.2366$&ESI&$[-200, 320]$&$ 598.1 \pm  18.7$&$> 14.82$&$> 15.20$\\
&1611.2005 &$ -2.8665$&ESI&$[-100, 100]$&$<  25.1$&$< 15.09$& \\
&2249.8768 &$ -2.7397$&ESI&$[-200, 300]$&$< 315.7$&$< 15.61$& \\
&2374.4612 &$ -1.5045$&ESI&$[-150, 150]$&$1023.0 \pm 116.1$&$> 15.20$& \\
\ion{Ni}{II}\\
&1709.6042 &$ -1.4895$&ESI&$[-150, 280]$&$  96.6 \pm  23.3$&$  14.09 \pm 0.10$&$  14.09 \pm 0.07$\\
&1741.5531 &$ -1.3696$&ESI&$[-150, 220]$&$ 123.6 \pm  30.2$&$  14.08 \pm 0.10$& \\
\ion{Zn}{II}\\
&2026.1360 &$ -0.3107$&ESI&$[-180, 300]$&$<  49.3$&$< 12.62$&$< 12.62$\\
\hline
\end{tabular}
\end{center}
 
$^{a}$ Velocity interval over which the equivalent width and column density are measured.
$^{b}$ Rest equivalent width.
\end{table*}

\clearpage
\begin{table*}
\caption{Ionic column densities for J0818+2631 (24:H12) at   $z_{\rm dla}=3.5629$\label{tab:hst12}}
\begin{center}
\begin{tabular}{lcccccccc}
\hline
Ion & $\lambda$ & $\log f$ & Instr. & $v_{\rm int}^a$ & $W_\lambda^b$ & $\log N$ & $\log N_{\rm adopt}$ \\
& (\AA) & & & (\kms) & (m\AA) & \\
\hline
\ion{C}{I}\\
&1560.3092 &$ -0.8808$&ESI&$[-200, 300]$&$  93.3 \pm  11.8$&$  13.55 \pm 0.05$&$  13.55 \pm 0.05$\\
&1656.9283 &$ -0.8273$&ESI&$[-200, 300]$&$<  26.4$&$< 13.03$& \\
\ion{C}{II}\\
&1334.5323 &$ -0.8935$&ESI&$[-250, 300]$&$1759.4 \pm   8.1$&$> 15.35$&$> 15.35$\\
\ion{C}{IV}\\
&1548.1950 &$ -0.7194$&ESI&$[-160, 320]$&$ 529.7 \pm  11.0$&$  14.25 \pm 0.01$&$  14.26 \pm 0.01$\\
&1550.7700 &$ -1.0213$&ESI&$[-160, 320]$&$ 329.8 \pm  11.4$&$  14.28 \pm 0.02$& \\
\ion{Al}{II}\\
&1670.7874 &$  0.2742$&ESI&$[-200, 320]$&$1577.5 \pm  10.7$&$> 13.88$&$> 13.88$\\
\ion{Si}{II}\\
&1526.7066 &$ -0.8962$&ESI&$[-200, 320]$&$1170.6 \pm   9.9$&$> 14.99$&$> 14.99$\\
\ion{Si}{IV}\\
&1393.7550 &$ -0.2774$&ESI&$[-100, 230]$&$ 601.3 \pm   6.8$&$> 14.02$&$> 14.02$\\
\ion{Fe}{II}\\
&1608.4511 &$ -1.2366$&ESI&$[-200, 320]$&$ 684.5 \pm  16.2$&$> 14.97$&$> 14.97$\\
&1611.2005 &$ -2.8665$&ESI&$[-200, 300]$&$  78.5 \pm  14.9$&$< 15.42$& \\
\ion{Ni}{II}\\
&1709.6042 &$ -1.4895$&ESI&$[-150, 280]$&$  55.3 \pm  11.0$&$  13.85 \pm 0.08$&$  13.85 \pm 0.05$\\
&1741.5531 &$ -1.3696$&ESI&$[-150, 220]$&$  77.6 \pm  10.4$&$  13.86 \pm 0.06$& \\
\ion{Zn}{II}\\
&2026.1360 &$ -0.3107$&ESI&$[-180, 300]$&$<  47.5$&$< 12.60$&$< 12.60$\\
\hline
\end{tabular}
\end{center}
 
$^{a}$ Velocity interval over which the equivalent width and column density are measured.
$^{b}$ Rest equivalent width.
\end{table*}

\clearpage
\begin{table*}
\caption{Ionic column densities for J0811+3936 (25:H13) at  $z_{\rm dla}=2.6500$\label{tab:hst13}}
\begin{center}
\begin{tabular}{lcccccccc}
\hline
Ion & $\lambda$ & $\log f$ & Instr. & $v_{\rm int}^a$ & $W_\lambda^b$ & $\log N$ & $\log N_{\rm adopt}$ \\
& (\AA) & & & (\kms) & (m\AA) & \\
\hline
\ion{C}{I}\\
&1560.3092 &$ -0.8808$&ESI&$[-100, 200]$&$<  27.0$&$< 13.15$&$< 13.15$\\
&1656.9283 &$ -0.8273$&ESI&$[-100, 200]$&$<  36.4$&$< 13.18$& \\
\ion{C}{II}\\
&1334.5323 &$ -0.8935$&ESI&$[-100, 300]$&$1074.3 \pm  25.2$&$> 15.11$&$> 15.11$\\
\ion{C}{IV}\\
&1548.1950 &$ -0.7194$&ESI&$[-100, 250]$&$ 553.1 \pm  13.7$&$> 14.43$&$> 14.51$\\
&1550.7700 &$ -1.0213$&ESI&$[-100, 250]$&$ 428.6 \pm  13.7$&$> 14.51$& \\
\ion{O}{I}\\
&1302.1685 &$ -1.3110$&ESI&$[-100, 300]$&$1118.4 \pm  15.8$&$> 15.53$&$> 15.53$\\
\ion{Al}{II}\\
&1670.7874 &$  0.2742$&ESI&$[-100, 300]$&$ 807.6 \pm  17.2$&$> 13.41$&$> 13.41$\\
\ion{Al}{III}\\
&1854.7164 &$ -0.2684$&ESI&$[-100, 300]$&$ 132.4 \pm  19.7$&$  12.95 \pm 0.06$&$  12.98 \pm 0.05$\\
&1862.7895 &$ -0.5719$&ESI&$[-100, 300]$&$  98.0 \pm  19.7$&$  13.10 \pm 0.09$& \\
\ion{Si}{II}\\
&1190.4158 &$ -0.6017$&ESI&$[-200, 200]$&$ 728.5 \pm  29.2$&$> 14.68$&$> 14.68$\\
&1260.4221 &$  0.0030$&ESI&$[-100, 300]$&$1392.0 \pm  24.3$&$> 14.39$& \\
&1526.7066 &$ -0.8962$&ESI&$[-100, 300]$&$ 803.6 \pm  17.5$&$> 14.68$& \\
&1808.0130 &$ -2.6603$&ESI&$[ -40, 100]$&$<  29.6$&$< 14.86$& \\
\ion{Si}{IV}\\
&1393.7550 &$ -0.2774$&ESI&$[-120, 200]$&$ 434.4 \pm  12.6$&$> 13.87$&$  13.95 \pm 0.02$\\
&1402.7700 &$ -0.5817$&ESI&$[-120, 200]$&$ 295.7 \pm  14.3$&$  13.95 \pm 0.02$& \\
\ion{Cr}{II}\\
&2056.2539 &$ -0.9788$&ESI&$[-100, 200]$&$<  39.7$&$< 13.19$&$< 13.19$\\
&2062.2340 &$ -1.1079$&ESI&$[-100, 200]$&$<  54.5$&$< 13.45$& \\
&2066.1610 &$ -1.2882$&ESI&$[-100, 200]$&$<  39.1$&$< 13.48$& \\
\ion{Fe}{II}\\
&1608.4511 &$ -1.2366$&ESI&$[-120, 280]$&$ 304.1 \pm  23.6$&$  14.42 \pm 0.03$&$  14.45 \pm 0.03$\\
&1611.2005 &$ -2.8665$&ESI&$[-100, 200]$&$<  42.0$&$< 15.31$& \\
&2260.7805 &$ -2.6126$&ESI&$[-100, 250]$&$<  73.5$&$< 15.01$& \\
&2344.2140 &$ -0.9431$&ESI&$[-100, 300]$&$ 865.2 \pm  27.6$&$> 14.33$& \\
&2374.4612 &$ -1.5045$&ESI&$[ -60, 250]$&$ 414.7 \pm  39.6$&$  14.50 \pm 0.04$& \\
&2382.7650 &$ -0.4949$&ESI&$[-100, 300]$&$1268.8 \pm  31.7$&$> 14.14$& \\
\ion{Ni}{II}\\
&1709.6042 &$ -1.4895$&ESI&$[-100, 200]$&$<  28.4$&$< 13.71$&$< 13.62$\\
&1741.5531 &$ -1.3696$&ESI&$[-100, 200]$&$<  31.8$&$< 13.62$& \\
&1751.9157 &$ -1.5575$&ESI&$[-100, 200]$&$<  35.1$&$< 13.85$& \\
\ion{Zn}{II}\\
&2026.1360 &$ -0.3107$&ESI&$[-100, 200]$&$<  52.7$&$< 12.66$&$< 12.66$\\
\hline
\end{tabular}
\end{center}
 
$^{a}$ Velocity interval over which the equivalent width and column density are measured.
$^{b}$ Rest equivalent width.
\end{table*}

\clearpage
\begin{table*}
\caption{Ionic column densities for J1320+1310 (32:H20) at  $z_{\rm dla}=2.6722$\label{tab:hst20}}
\begin{center}
\begin{tabular}{lcccccccc}
\hline
Ion & $\lambda$ & $\log f$ & Instr. & $v_{\rm int}^a$ & $W_\lambda^b$ & $\log N$ & $\log N_{\rm adopt}$ \\
& (\AA) & & & (\kms) & (m\AA) & \\
\hline
\ion{C}{IV}\\
&1548.1950 &$ -0.7194$&X-shooter&$[-100, 100]$&$<  56.6$&$< 13.33$&$< 13.32$\\
&1550.7700 &$ -1.0213$&X-shooter&$[-100, 100]$&$<  58.2$&$< 13.66$& \\
\ion{Al}{II}\\
&1670.7874 &$  0.2742$&X-shooter&$[ -50,  50]$&$<  39.0$&$< 12.15$&$< 12.15$\\
\ion{Al}{III}\\
&1854.7164 &$ -0.2684$&X-shooter&$[ -50,  50]$&$<  26.4$&$< 12.36$&$< 12.36$\\
&1862.7895 &$ -0.5719$&X-shooter&$[ -50,  50]$&$<  29.9$&$< 12.77$& \\
\ion{Si}{II}\\
&1260.4221 &$  0.0030$&X-shooter&$[-100, 100]$&$ 313.2 \pm  11.2$&$  13.51 \pm 0.02$&$  13.51 \pm 0.02$\\
&1526.7066 &$ -0.8962$&X-shooter&$[-100, 100]$&$ 125.9 \pm  38.7$&$< 13.86$& \\
&1808.0130 &$ -2.6603$&X-shooter&$[ -50,  50]$&$<  26.0$&$< 14.77$& \\
\ion{S}{II}\\
&1259.5190 &$ -1.7894$&X-shooter&$[ -50,  50]$&$  64.4 \pm   8.5$&$< 14.49$&$< 14.49$\\
\ion{Cr}{II}\\
&2056.2539 &$ -0.9788$&X-shooter&$[ -50,  50]$&$<  31.1$&$< 13.08$&$< 13.08$\\
\ion{Fe}{II}\\
&1608.4511 &$ -1.2366$&X-shooter&$[-100, 100]$&$<  76.2$&$< 14.03$&$  12.94 \pm 0.09$\\
&2249.8768 &$ -2.7397$&X-shooter&$[ -50,  50]$&$<  26.5$&$< 14.68$& \\
&2260.7805 &$ -2.6126$&X-shooter&$[ -50,  50]$&$<  36.7$&$< 14.70$& \\
&2344.2140 &$ -0.9431$&X-shooter&$[ -70,  70]$&$<  37.7$&$< 13.04$& \\
&2382.7650 &$ -0.4949$&X-shooter&$[ -70,  70]$&$ 116.1 \pm  24.8$&$  12.94 \pm 0.09$& \\
\ion{Ni}{II}\\
&1741.5531 &$ -1.3696$&X-shooter&$[ -50,  50]$&$<  40.1$&$< 13.72$&$< 13.72$\\
\ion{Zn}{II}\\
&2026.1360 &$ -0.3107$&X-shooter&$[ -50,  50]$&$<  33.6$&$< 12.47$&$< 12.47$\\
\hline
\end{tabular}
\end{center}
 
$^{a}$ Velocity interval over which the equivalent width and column density are measured.
$^{b}$ Rest equivalent width.
\end{table*}

\clearpage

\clearpage
\begin{figure*}
\centering
\includegraphics[scale=0.8]{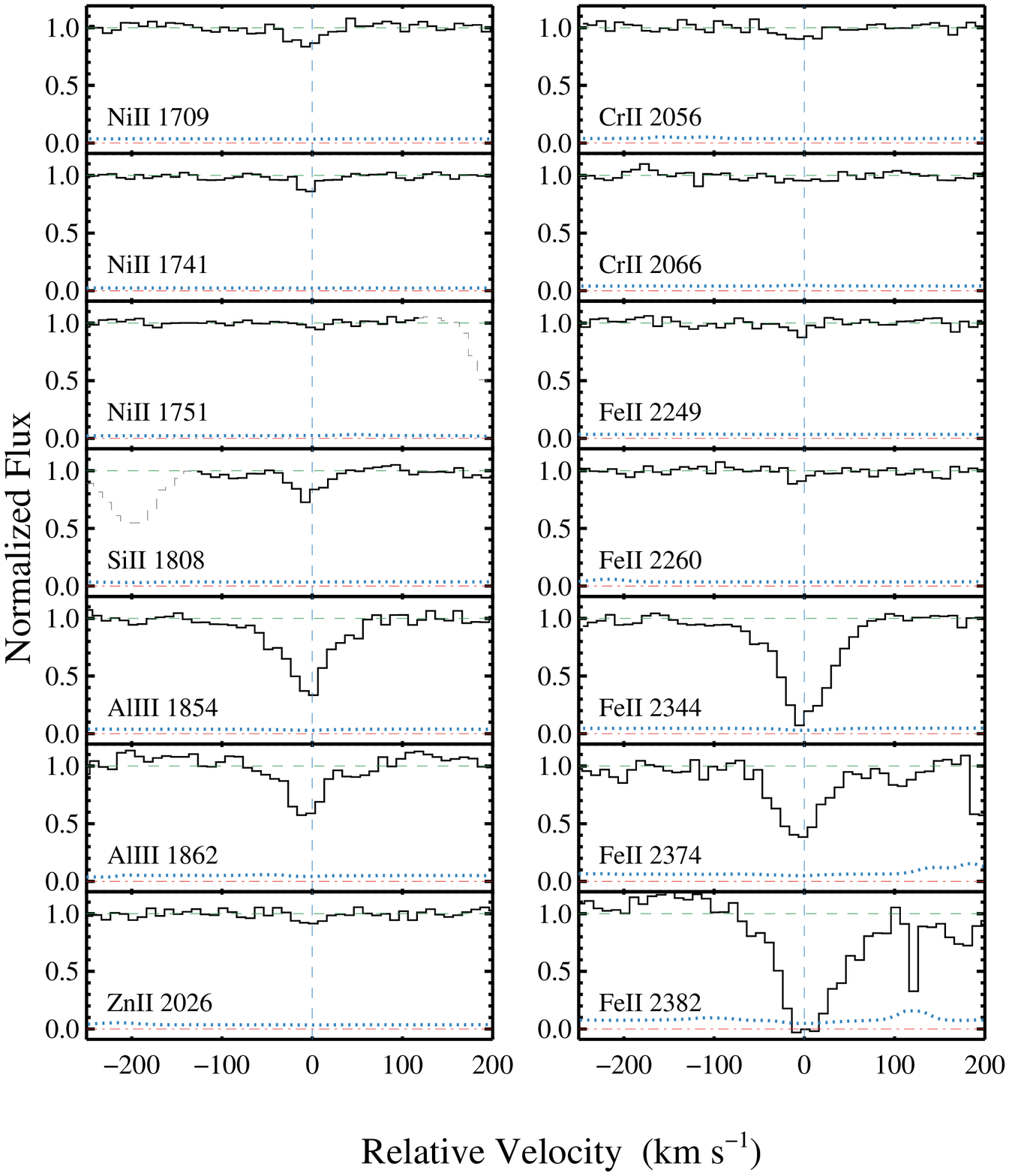}
\caption{Velocity plot of the ion transitions associated to the DLA J2114-0055 (1:G1) at $z_{\rm dla}=2.9181$ (continued).
See Figure \ref{fig:vpdla01a} for an explanation of the different line colors.}
\end{figure*}

\clearpage
\begin{figure*}
\centering
\includegraphics[scale=0.8]{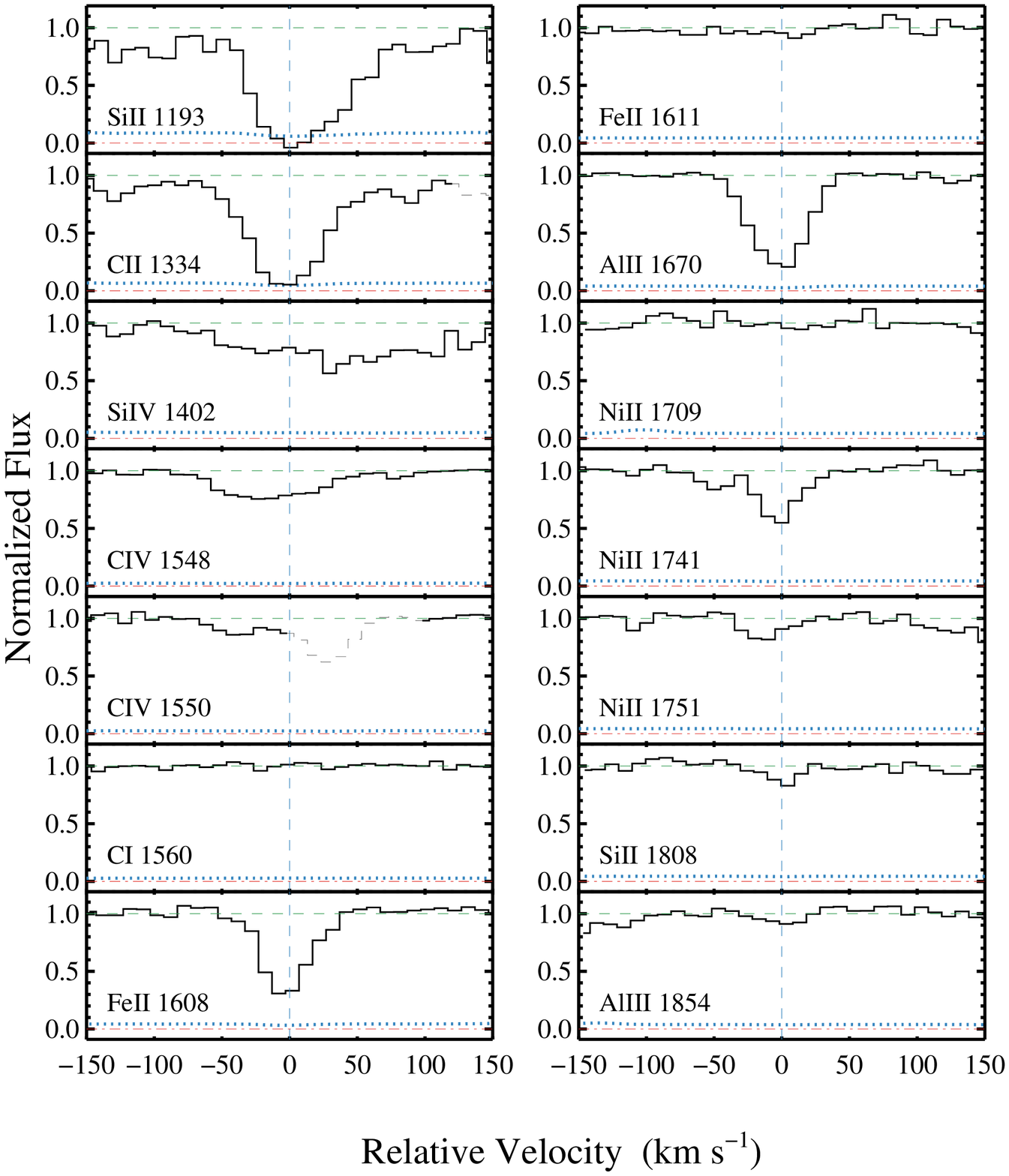}
\caption{Velocity plot of the ion transitions associated to the DLA J0731+2854 (2:G2) at  $z_{\rm dla}=2.6878$.
See Figure \ref{fig:vpdla01a} for an explanation of the different line colors.}
\end{figure*}

\clearpage
\begin{figure*}
\centering
\includegraphics[scale=0.8]{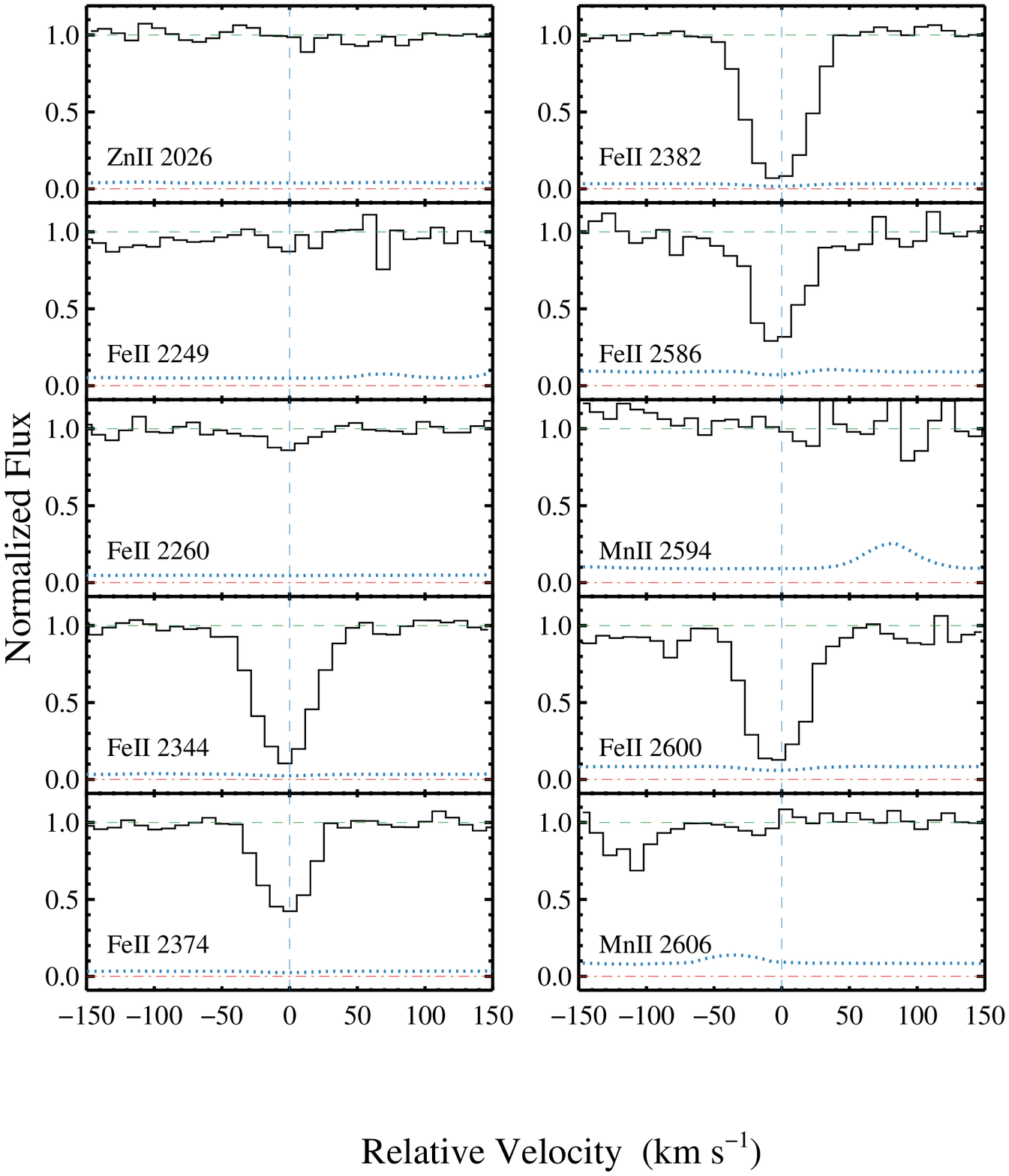}
\caption{Velocity plot of the ion transitions associated to the DLA J0731+2854 (2:G2) at  $z_{\rm dla}=2.6878$ (continued).
See Figure \ref{fig:vpdla01a} for an explanation of the different line colors.}
\end{figure*}

\clearpage
\begin{figure*}
\centering
\includegraphics[scale=0.8]{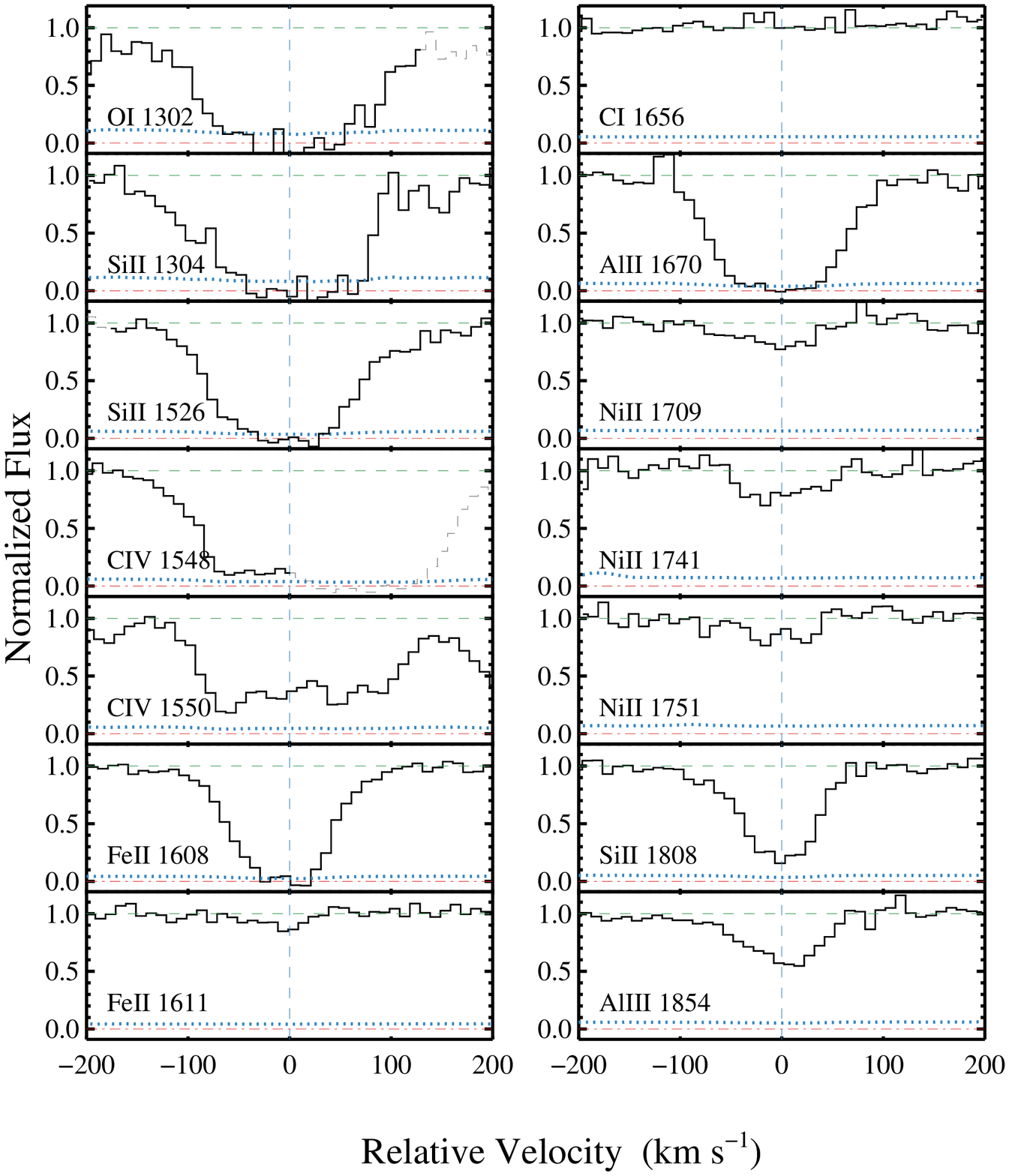}
\caption{Velocity plot of the ion transitions associated to the DLA J0956+3444 (3:G3) at  $z_{\rm dla}=2.3887$.
See Figure \ref{fig:vpdla01a} for an explanation of the different line colors. The \ion{Si}{II} 1808 transition
is blended with \ion{Si}{IV} 1393 from the higher redshift LLS and it is excluded from the analysis.}
\end{figure*}

\clearpage
\begin{figure*}
\centering
\includegraphics[scale=0.8]{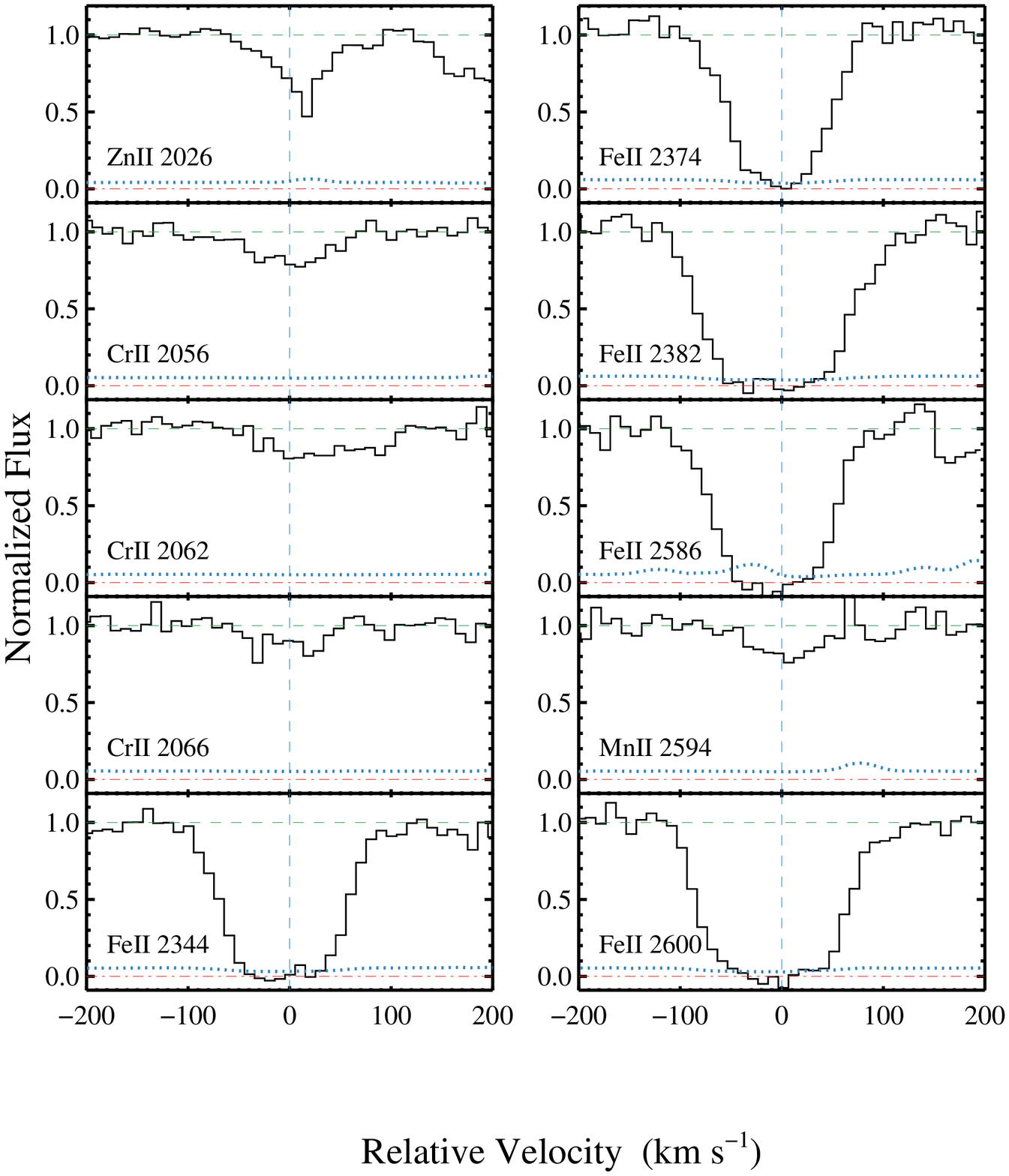}
\caption{Velocity plot of the ion transitions associated to the DLA J0956+3444 (3:G3) at  $z_{\rm dla}=2.3887$ (continued).
See Figure \ref{fig:vpdla01a} for an explanation of the different line colors.}
\end{figure*}

\clearpage
\begin{figure*}
\centering
\includegraphics[scale=0.8]{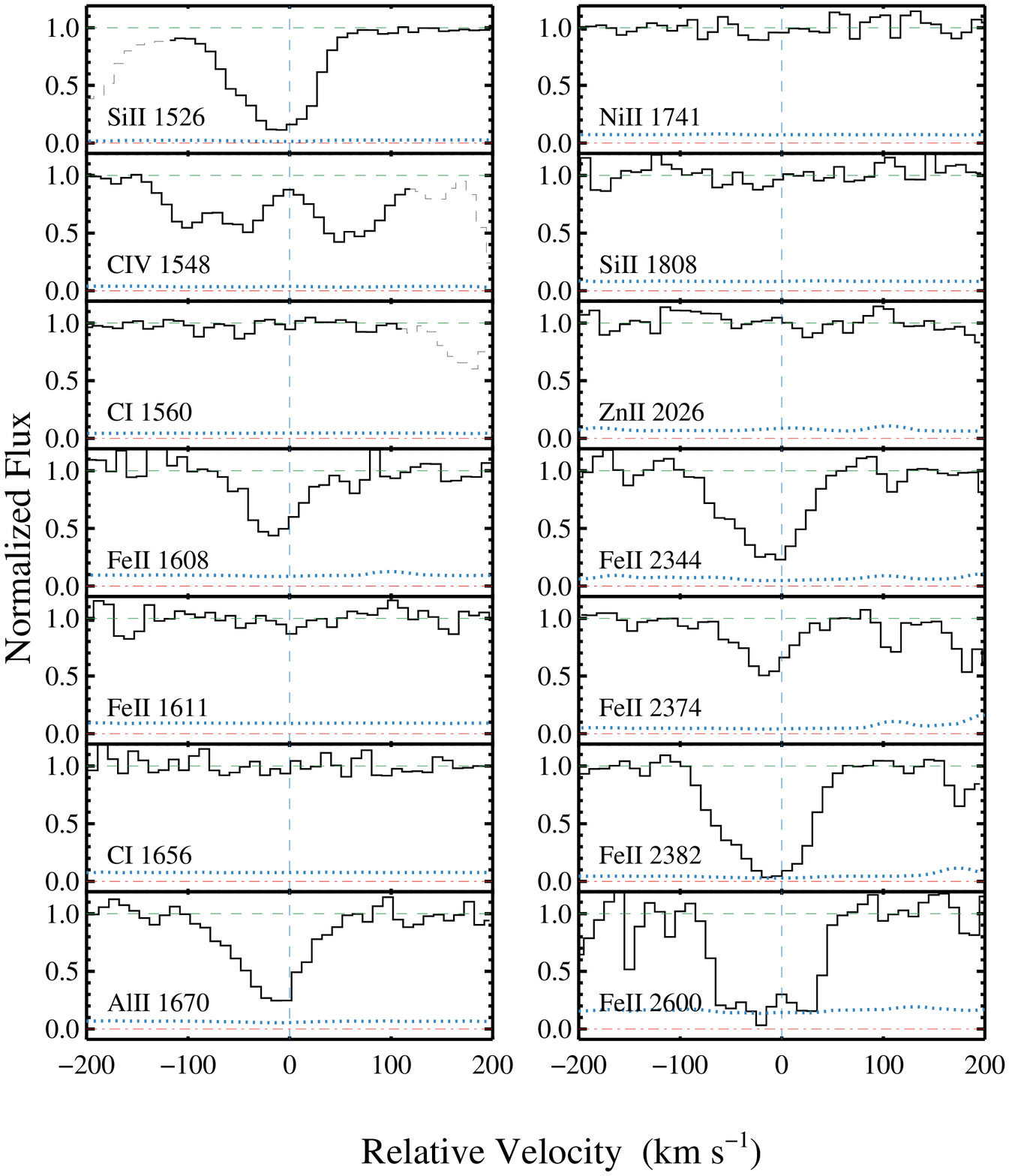}
\caption{Velocity plot of the ion transitions associated to the DLA J2343-1047 (4:G4) at  $z_{\rm dla}=2.6880$.
See Figure \ref{fig:vpdla01a} for an explanation of the different line colors.}
\end{figure*}

\clearpage
\begin{figure*}
\centering
\includegraphics[scale=0.8]{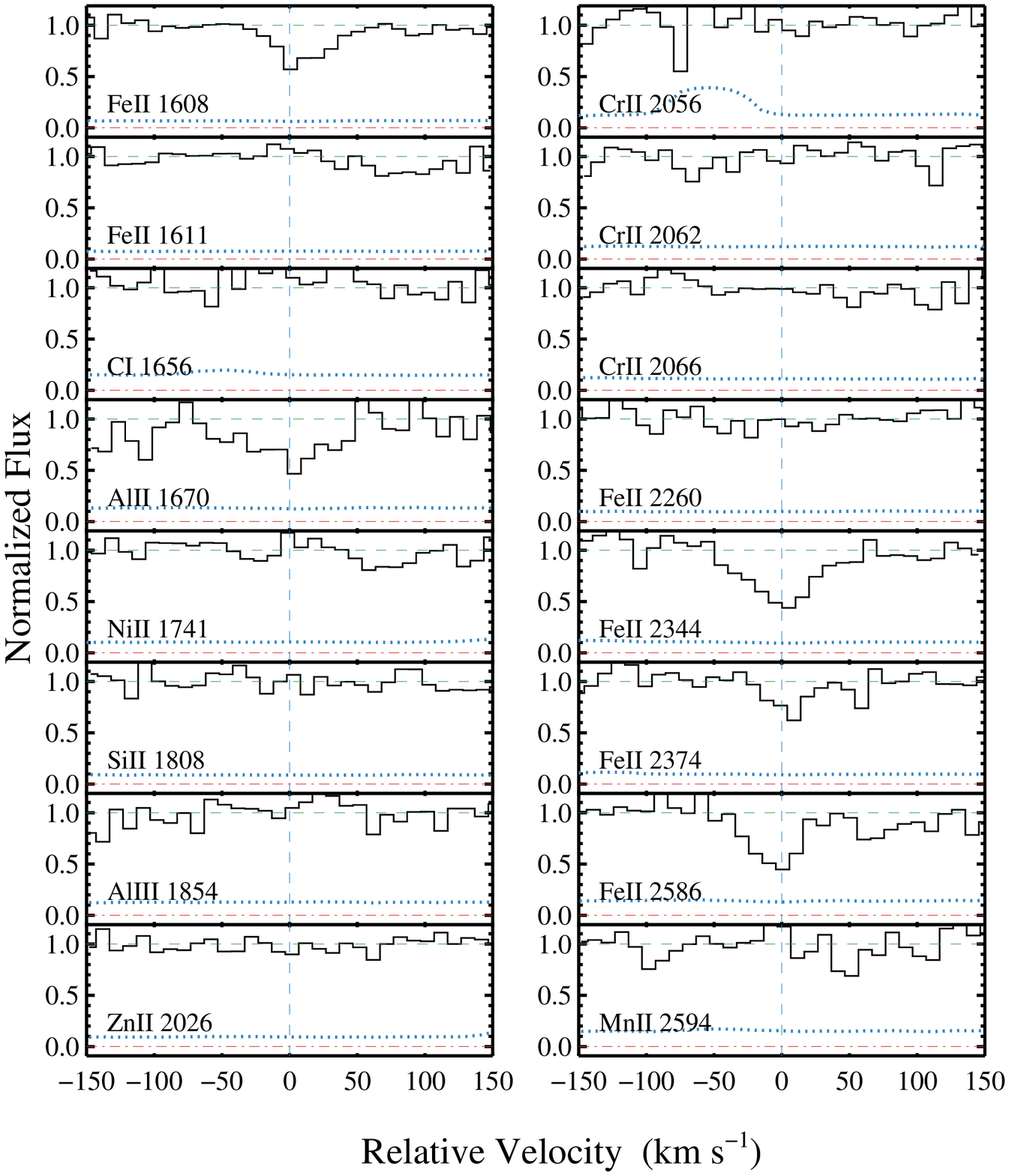}
\caption{Velocity plot of the ion transitions associated to the DLA J0343-0622 (5:G5) at  $z_{\rm dla}=2.5713$.
See Figure \ref{fig:vpdla01a} for an explanation of the different line colors.}
\end{figure*}

\clearpage
\begin{figure*}
\centering
\includegraphics[scale=0.8]{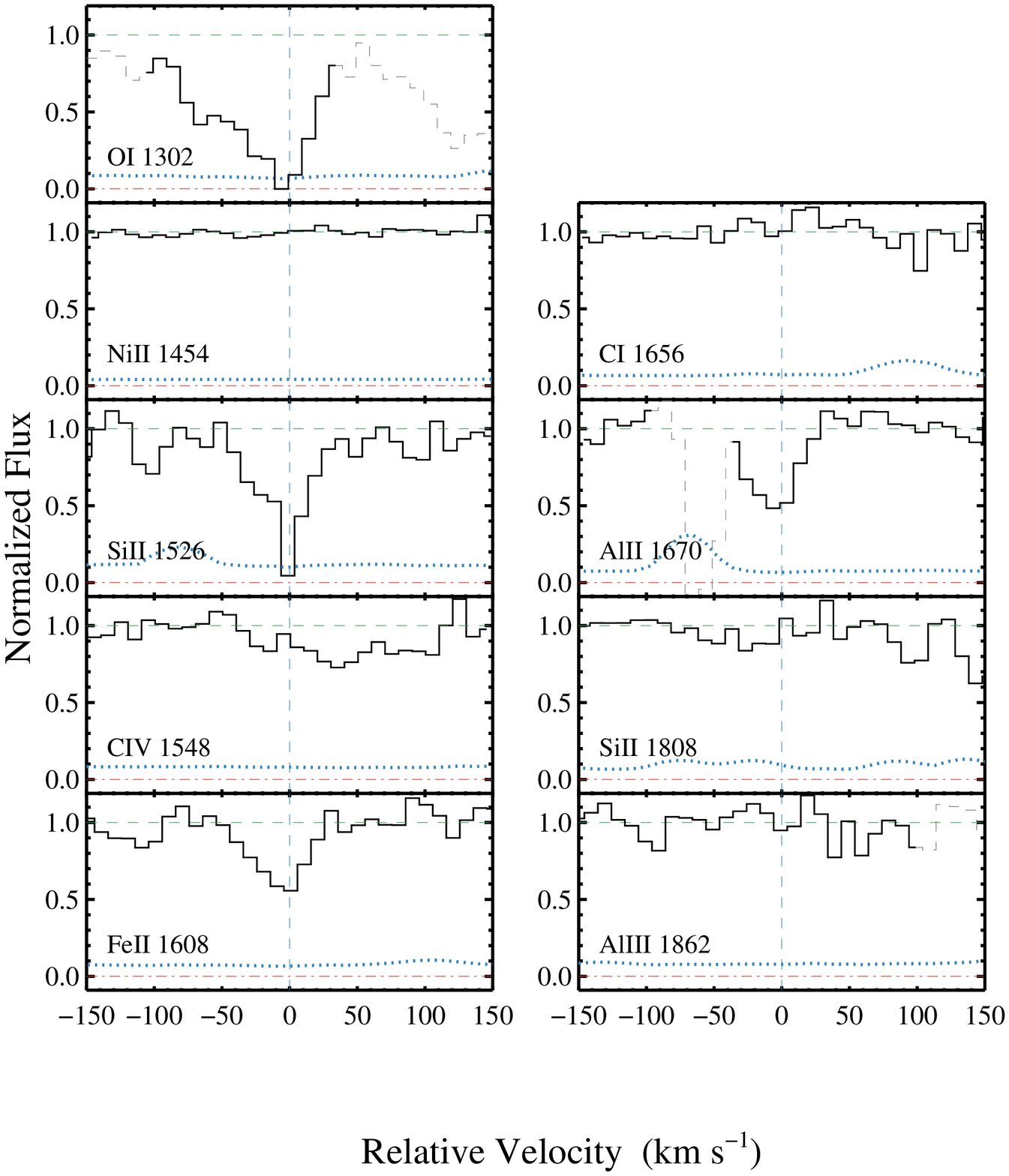}
\caption{Velocity plot of the ion transitions associated to the DLA J2351+1600 (6:G6) at  $z_{\rm dla}=3.7861$.
See Figure \ref{fig:vpdla01a} for an explanation of the different line colors.}
\end{figure*}

\clearpage
\begin{figure*}
\centering
\includegraphics[scale=0.8]{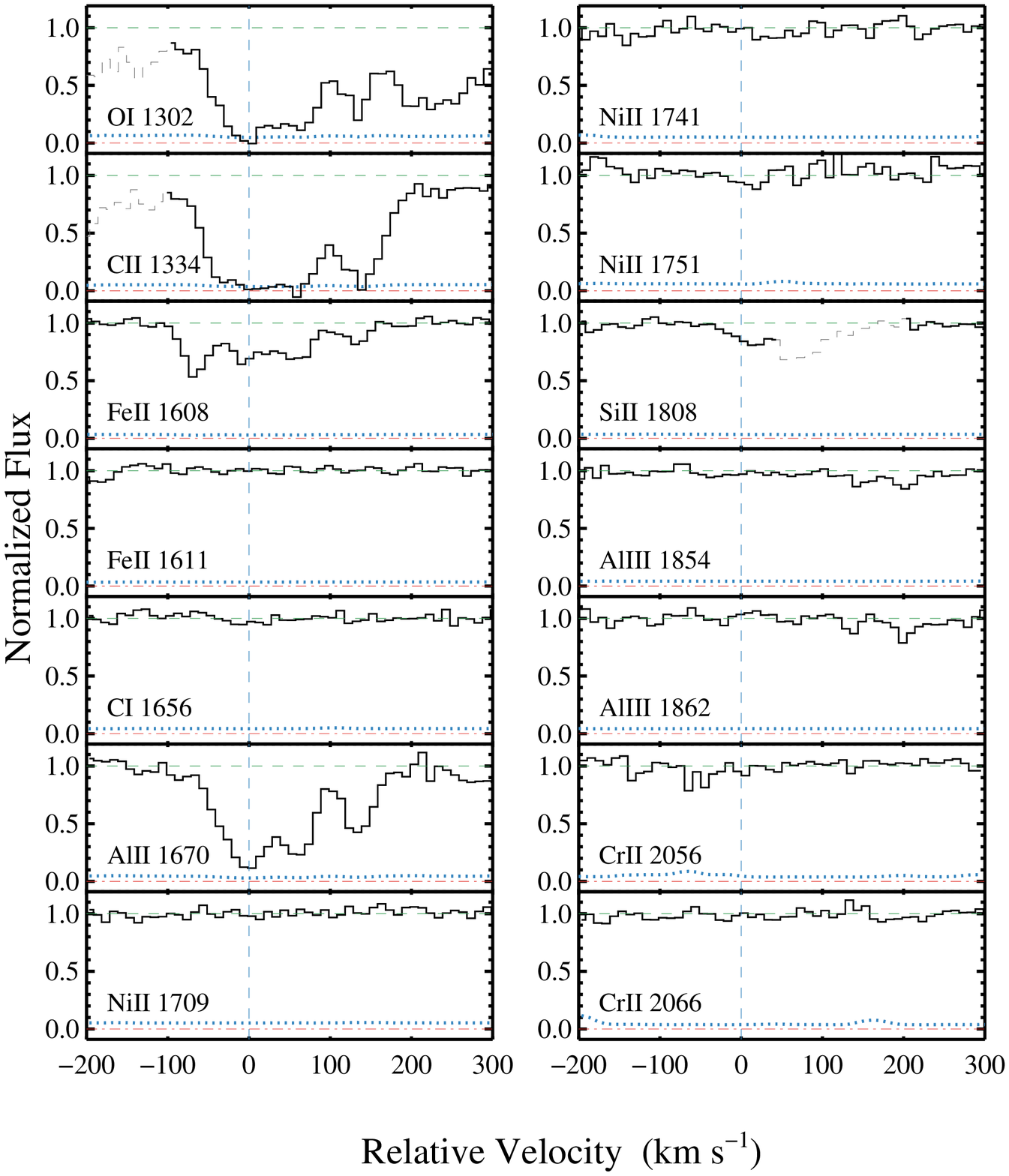}
\caption{Velocity plot of the ion transitions associated to the DLA J0042-1020 (7:G7) at  $z_{\rm dla}=2.7544$.
See Figure \ref{fig:vpdla01a} for an explanation of the different line colors.}
\end{figure*}

\clearpage
\begin{figure*}
\centering
\includegraphics[scale=0.8]{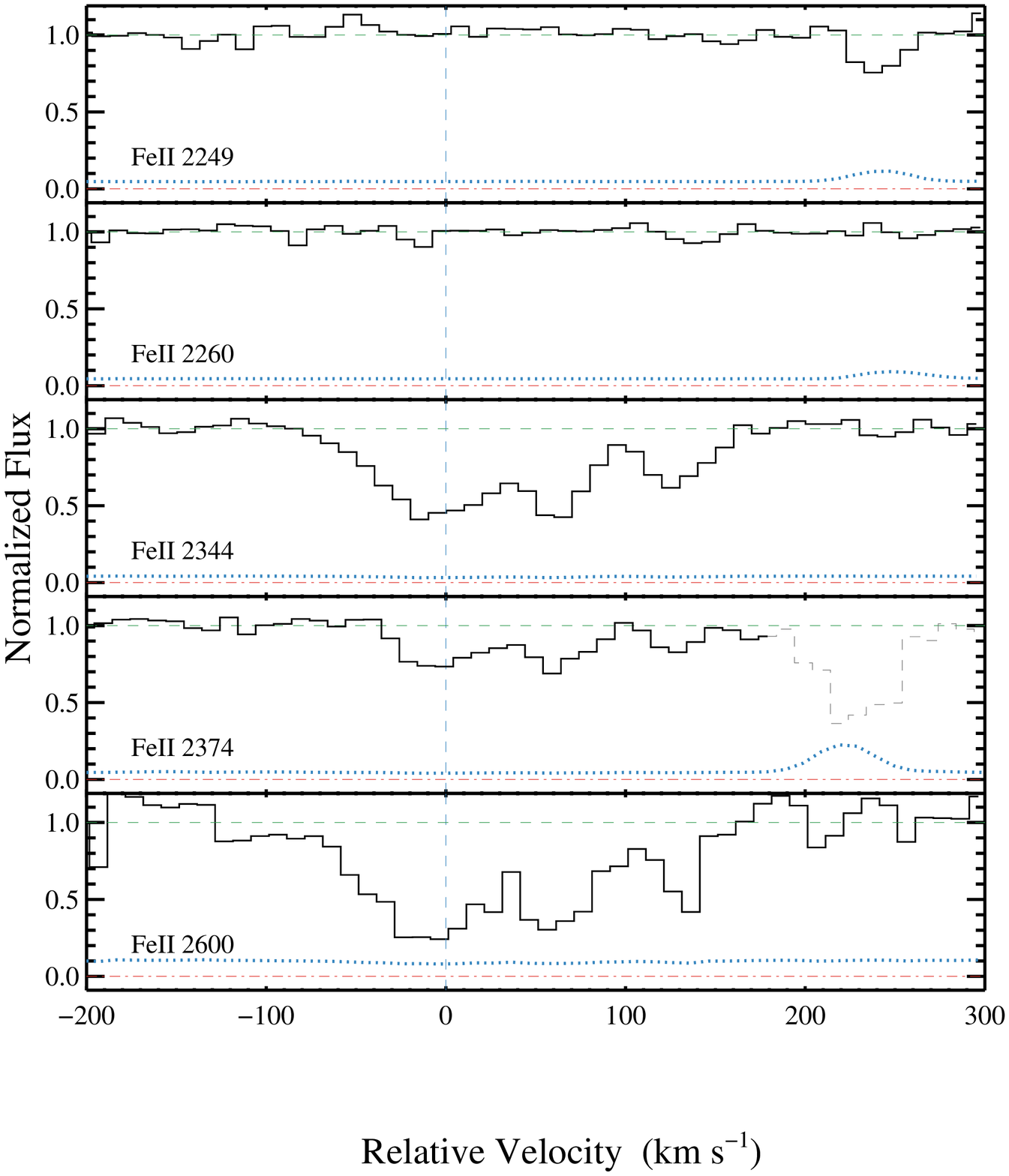}
\caption{Velocity plot of the ion transitions associated to the DLA J0042-1020 (7:G7) at  $z_{\rm dla}=2.7544$ (continued).
See Figure \ref{fig:vpdla01a} for an explanation of the different line colors.}
\end{figure*}

\clearpage
\begin{figure*}
\centering
\includegraphics[scale=0.8]{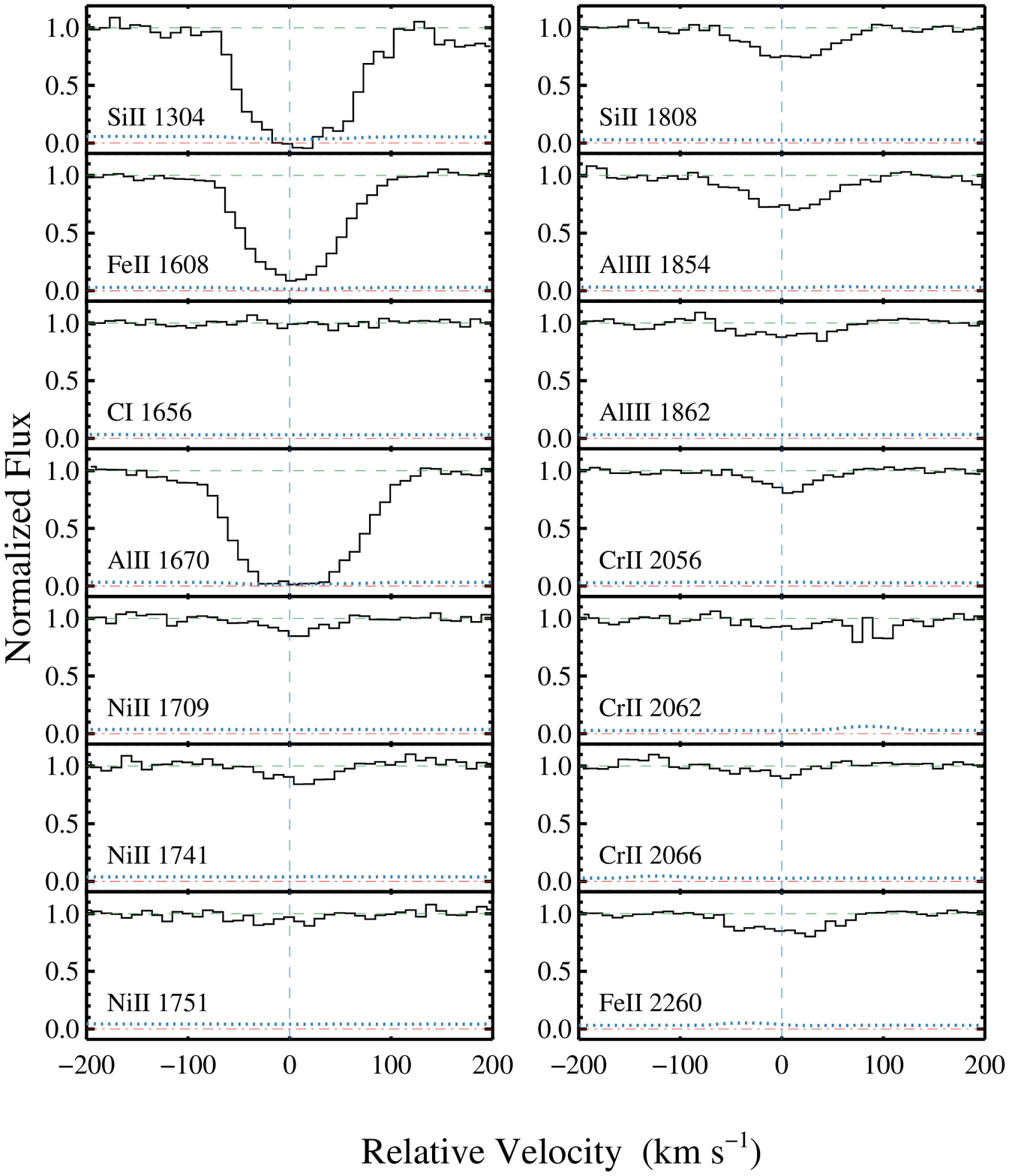}
\caption{Velocity plot of the ion transitions associated to the DLA J0949+1115 (8:G9) at  $z_{\rm dla}=2.7584$.
See Figure \ref{fig:vpdla01a} for an explanation of the different line colors.}
\end{figure*}

\clearpage
\begin{figure*}
\centering
\includegraphics[scale=0.8]{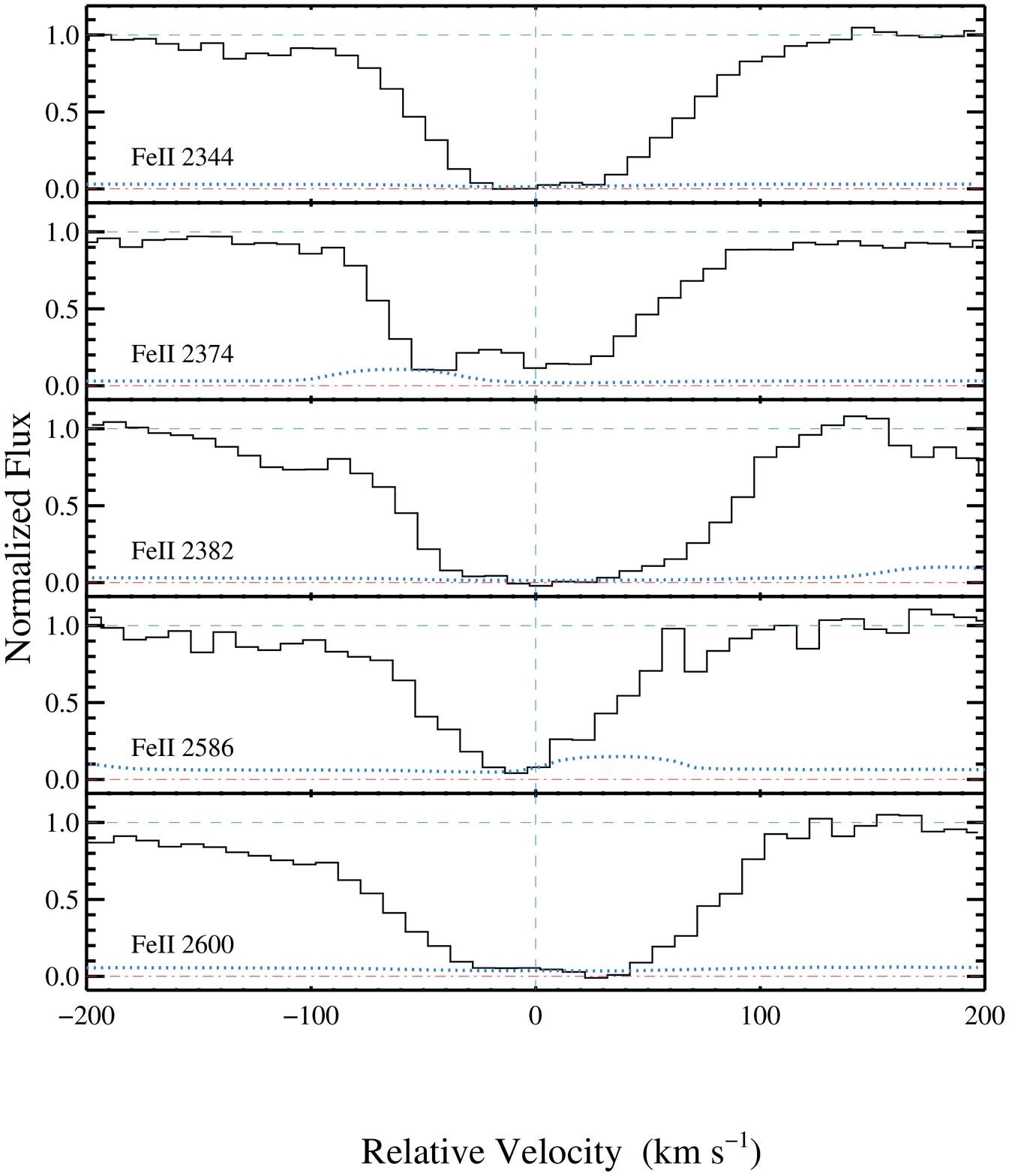}
\caption{Velocity plot of the ion transitions associated to the DLA J0949+1115 (8:G9) at  $z_{\rm dla}=2.7584$ (continued).
See Figure \ref{fig:vpdla01a} for an explanation of the different line colors.}
\end{figure*}

\clearpage
\begin{figure*}
\centering
\includegraphics[scale=0.8]{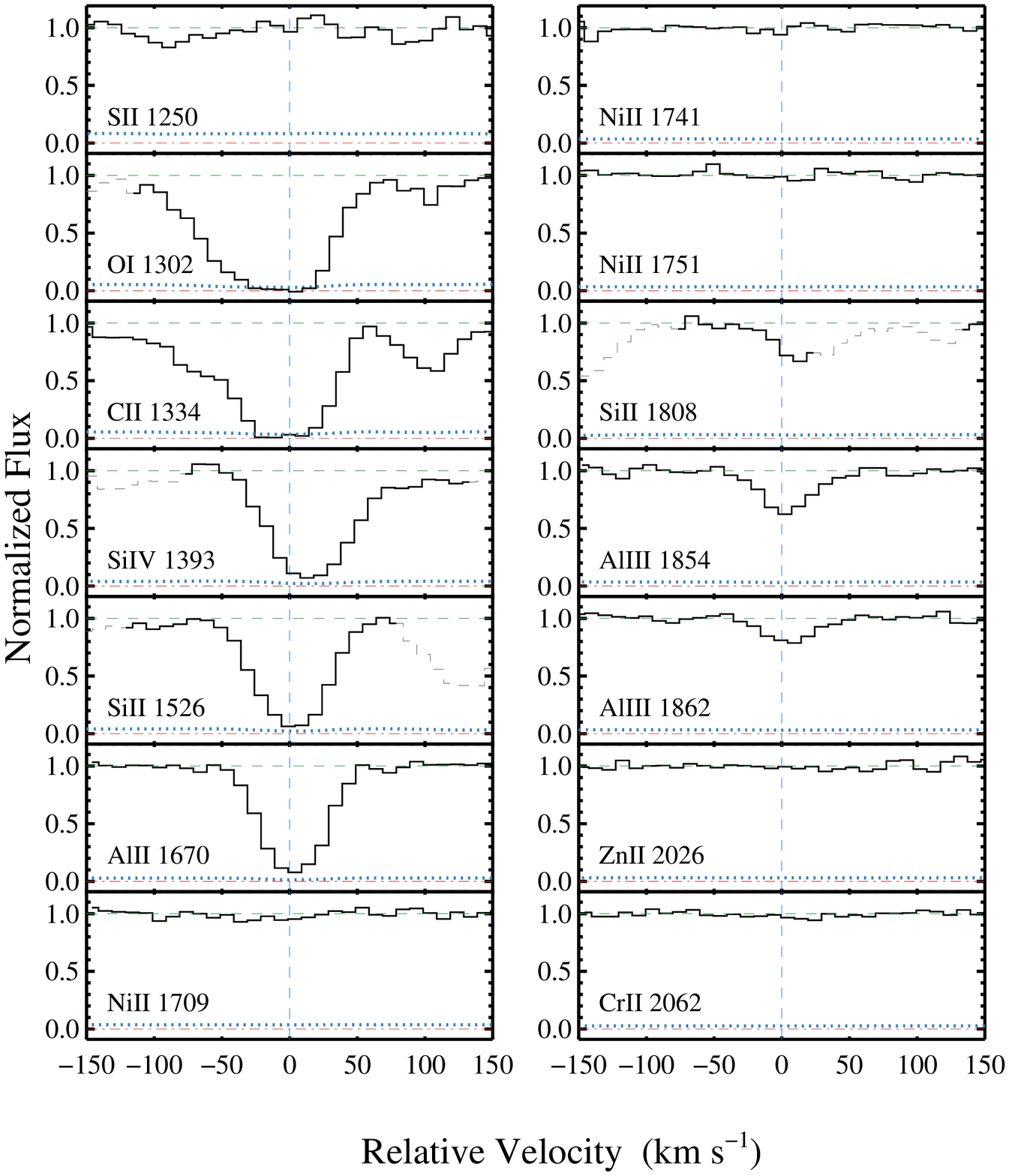}
\caption{Velocity plot of the ion transitions associated to the DLA J1018+3106 (9:G10) at  $z_{\rm dla}=2.4592$.
See Figure \ref{fig:vpdla01a} for an explanation of the different line colors.}
\end{figure*}

\clearpage
\begin{figure*}
\centering
\includegraphics[scale=0.8]{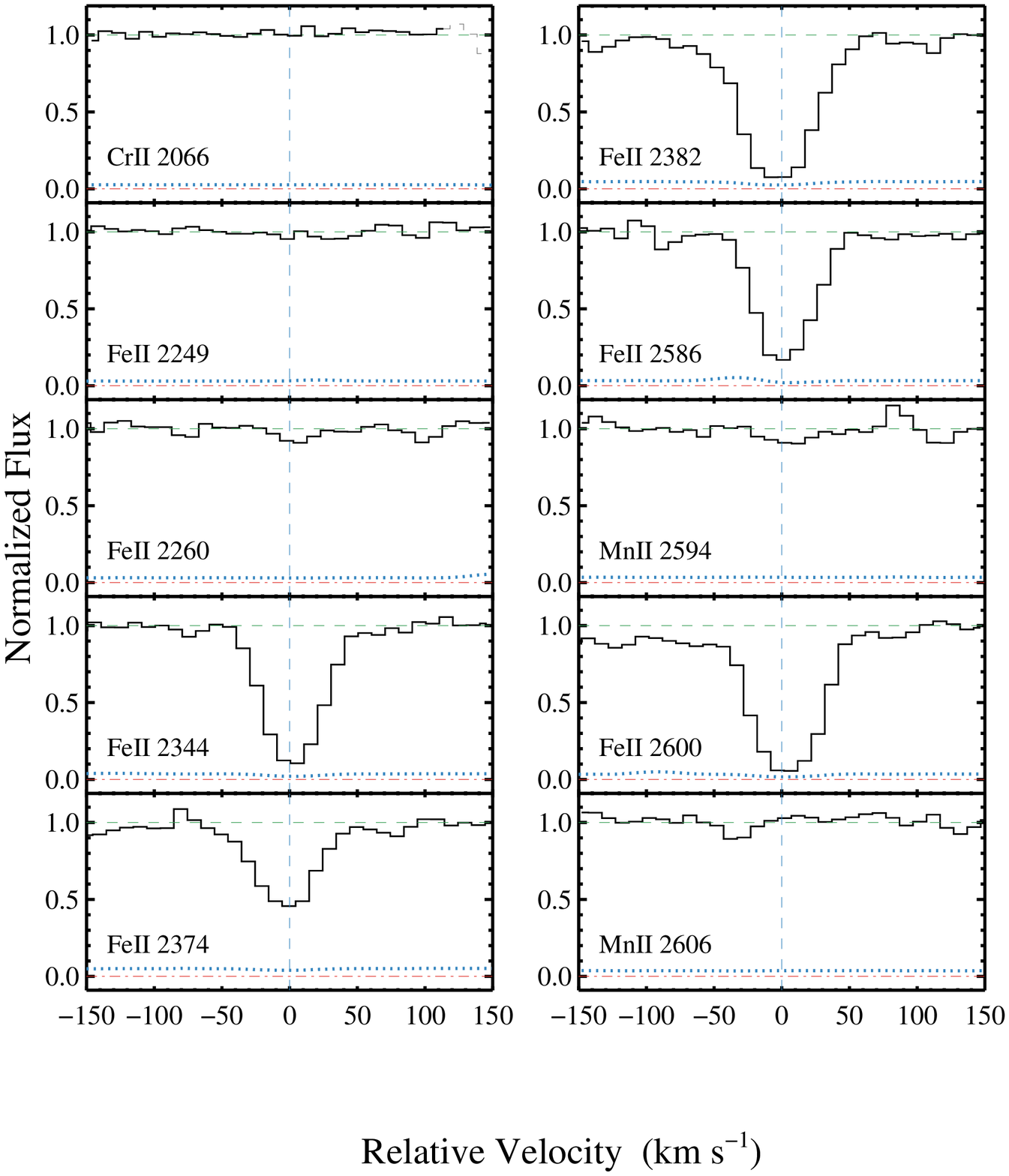}
\caption{Velocity plot of the ion transitions associated to the DLA J1018+3106 (9:G10) at  $z_{\rm dla}=2.4592$ (continued).
See Figure \ref{fig:vpdla01a} for an explanation of the different line colors.}
\end{figure*}

\clearpage
\begin{figure*}
\centering
\includegraphics[scale=0.8]{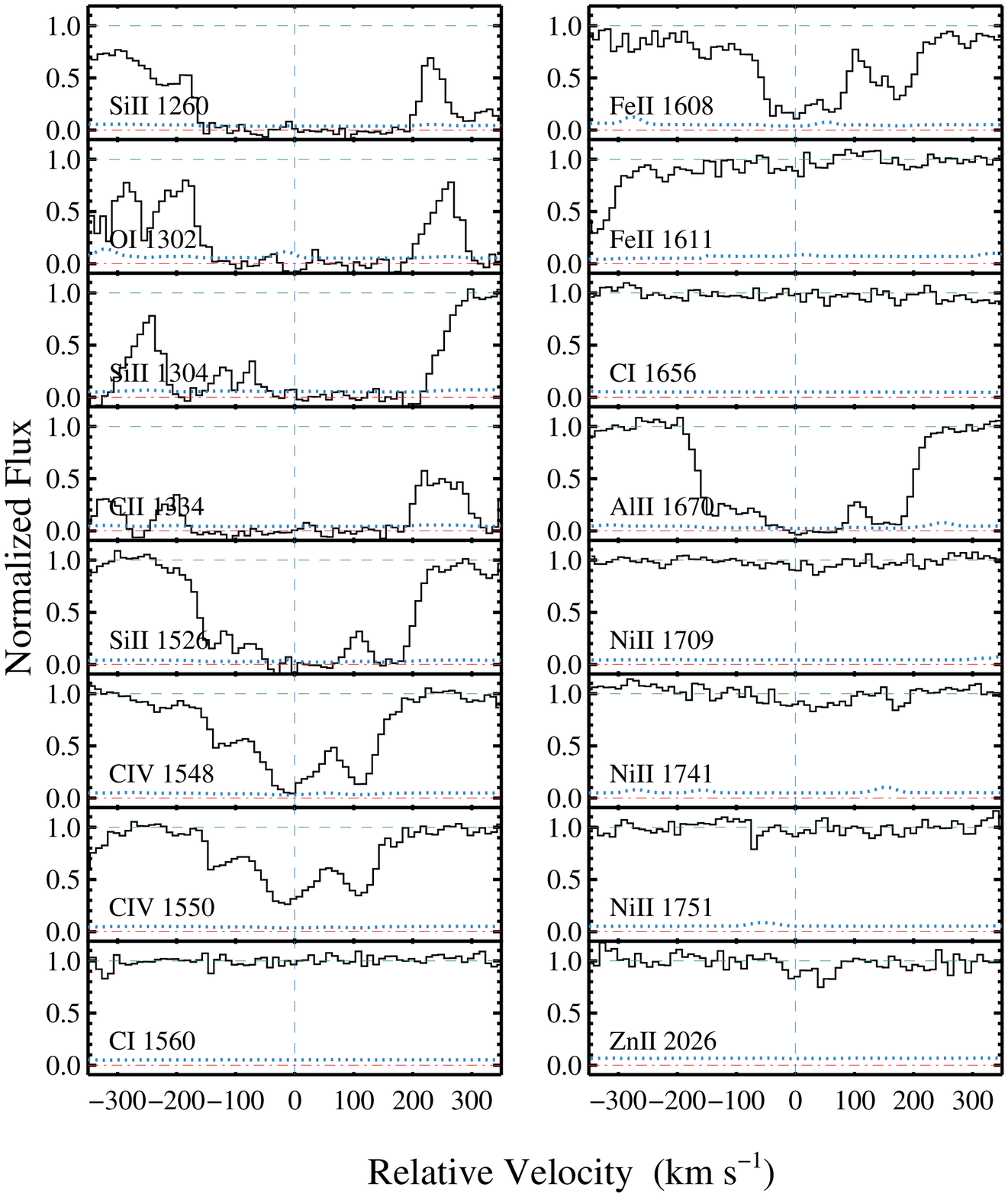}
\caption{Velocity plot of the ion transitions associated to the DLA J0851+2332 (10:G11) at  $z_{\rm dla}=3.5297$.
See Figure \ref{fig:vpdla01a} for an explanation of the different line colors.}
\end{figure*}

\clearpage
\begin{figure*}
\centering
\includegraphics[scale=0.8]{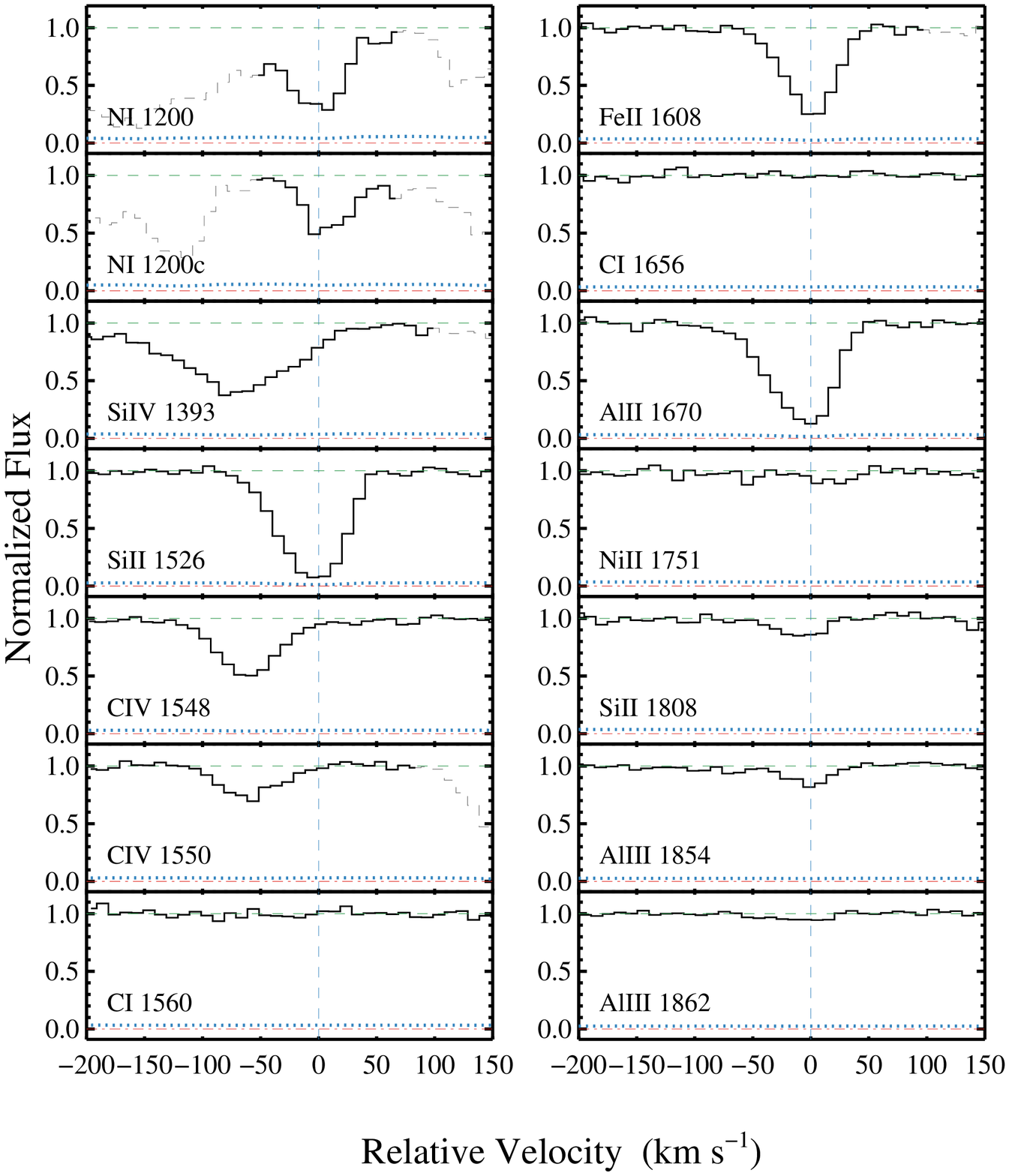}
\caption{Velocity plot of the ion transitions associated to the DLA J0956+1448 (11:G12) at  $z_{\rm dla}=2.6606$.
See Figure \ref{fig:vpdla01a} for an explanation of the different line colors.}
\end{figure*}

\clearpage
\begin{figure*}
\centering
\includegraphics[scale=0.8]{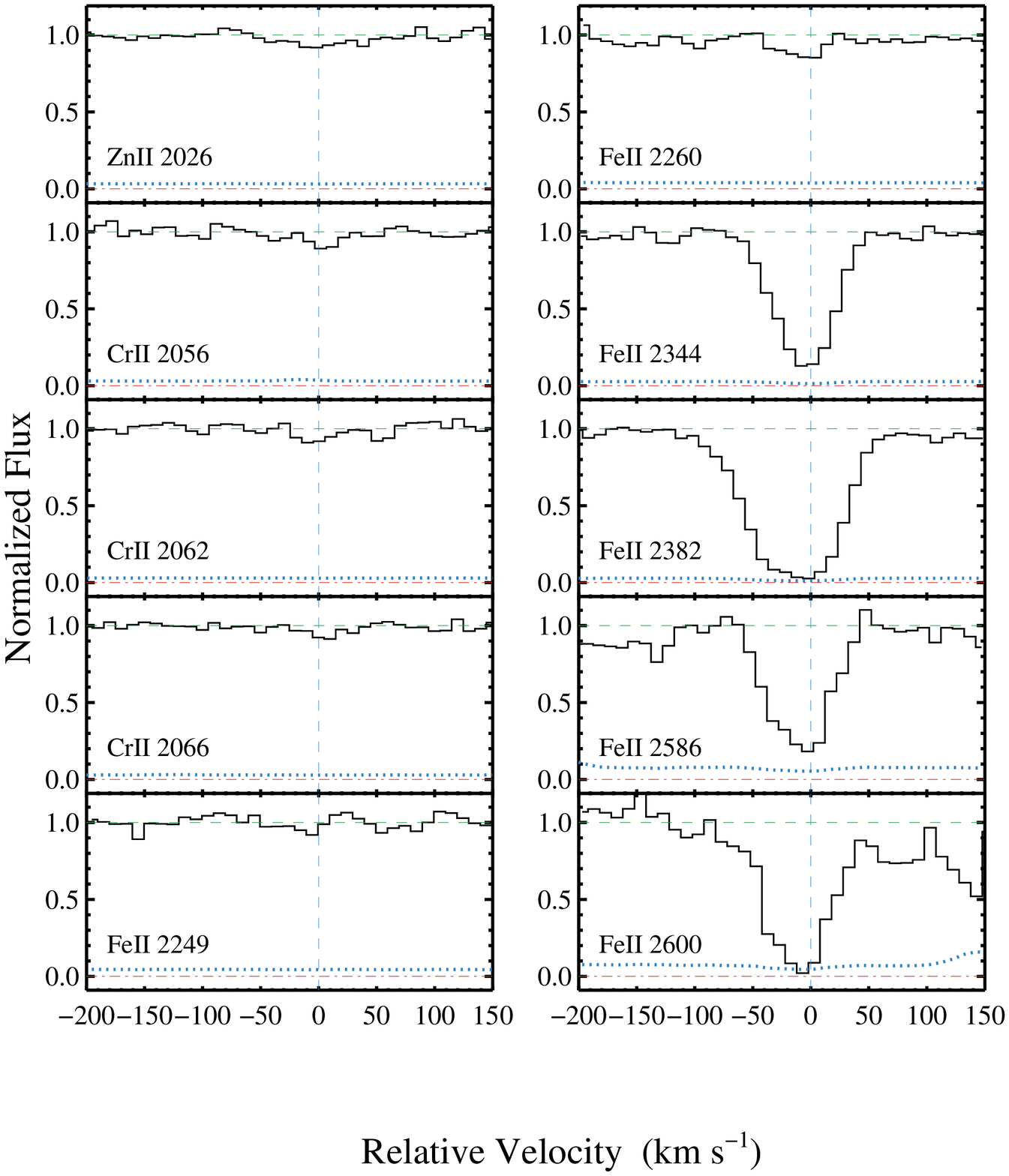}
\caption{Velocity plot of the ion transitions associated to the DLA J0956+1448 (11:G12) at  $z_{\rm dla}=2.6606$ (continued).
See Figure \ref{fig:vpdla01a} for an explanation of the different line colors.}
\end{figure*}

\clearpage
\begin{figure*}
\centering
\includegraphics[scale=0.8]{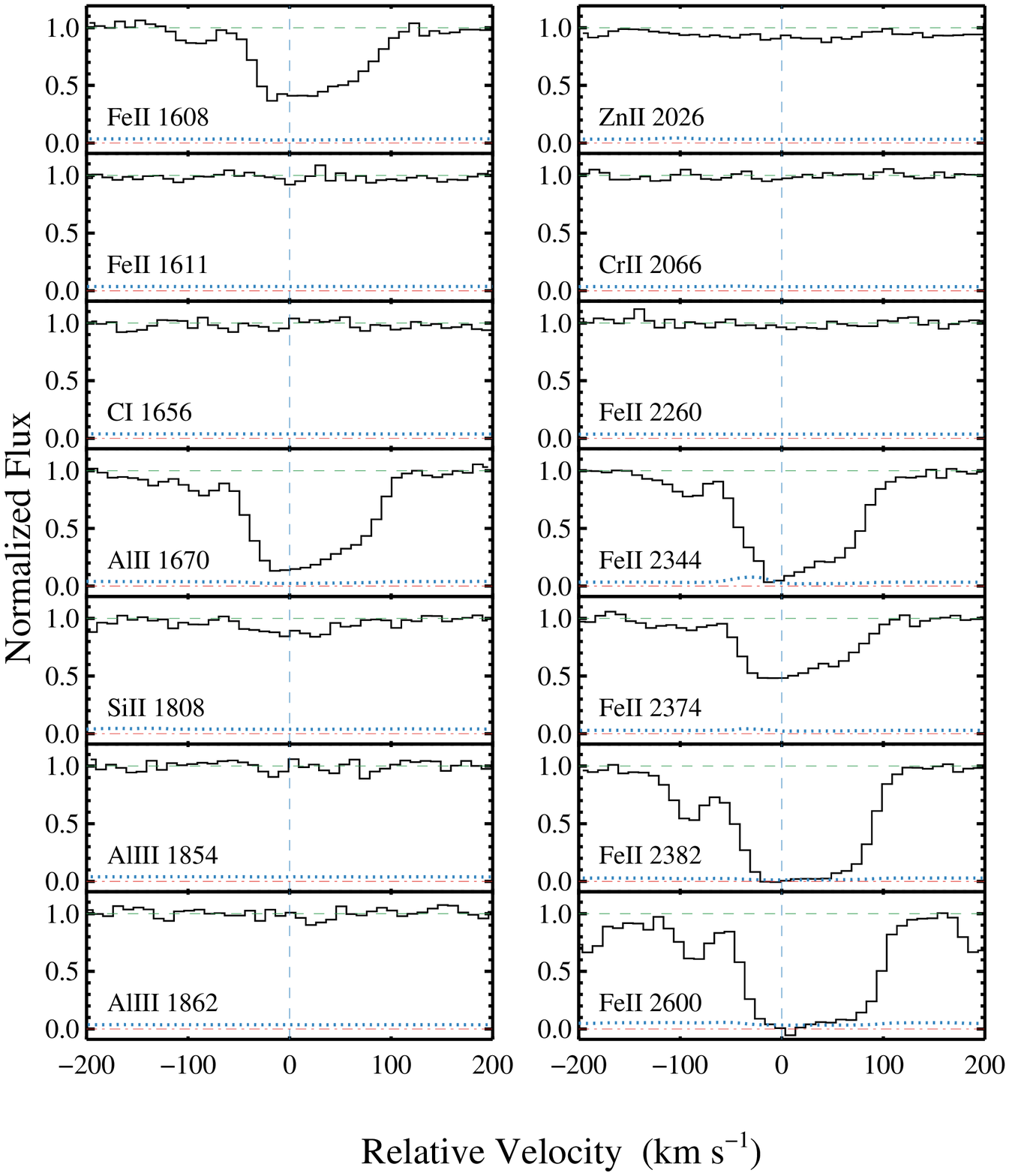}
\caption{Velocity plot of the ion transitions associated to the DLA J1151+3536 (12:G13) at $z_{\rm dla}=2.5978$.
See Figure \ref{fig:vpdla01a} for an explanation of the different line colors.}
\end{figure*}

\clearpage
\begin{figure*}
\centering
\includegraphics[scale=0.8]{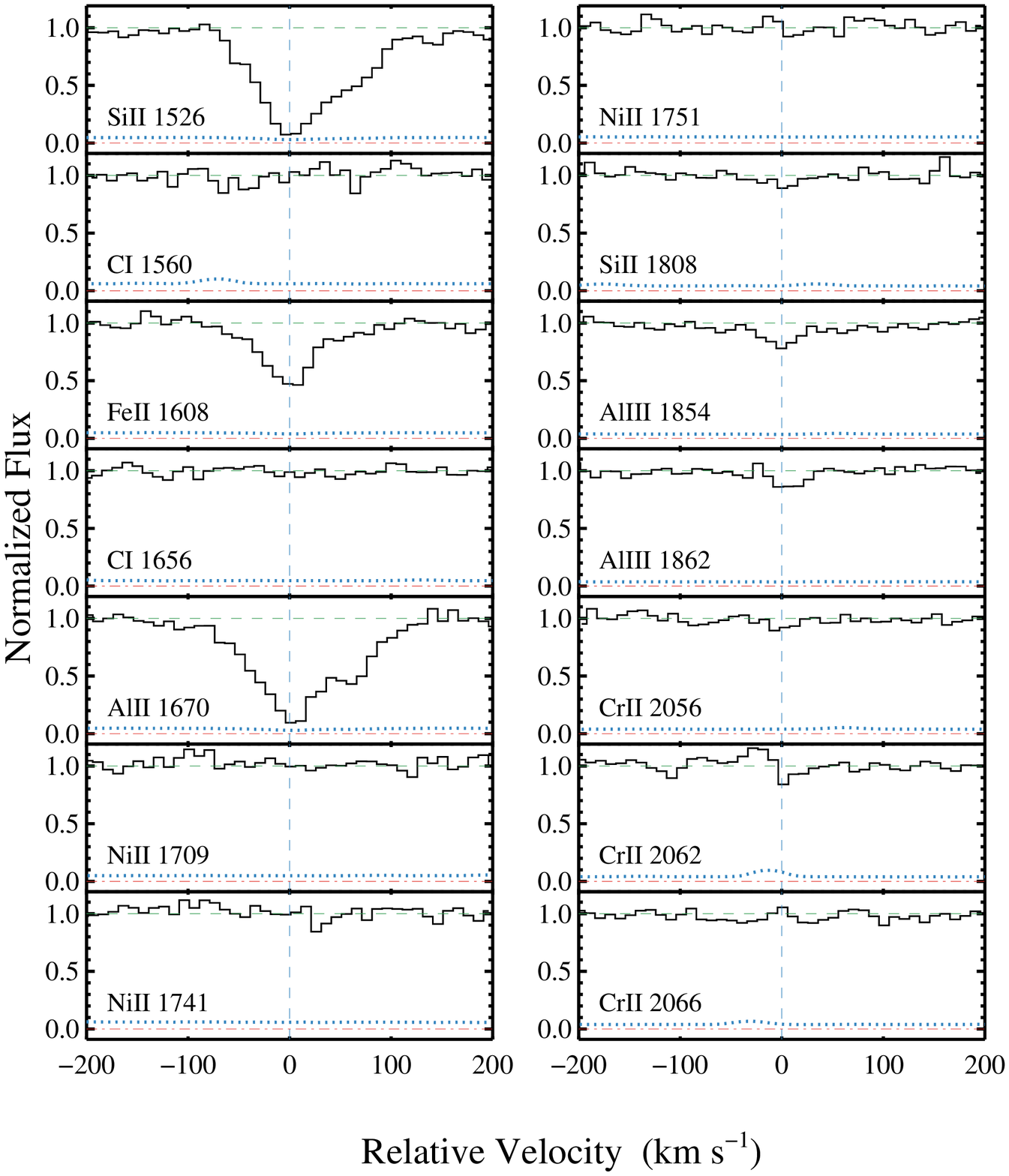}
\caption{Velocity plot of the ion transitions associated to the DLA J2123-0053 (13:H1) at  $z_{\rm dla}=2.7803$.
See Figure \ref{fig:vpdla01a} for an explanation of the different line colors.}
\end{figure*}

\clearpage
\begin{figure*}
\centering
\includegraphics[scale=0.8]{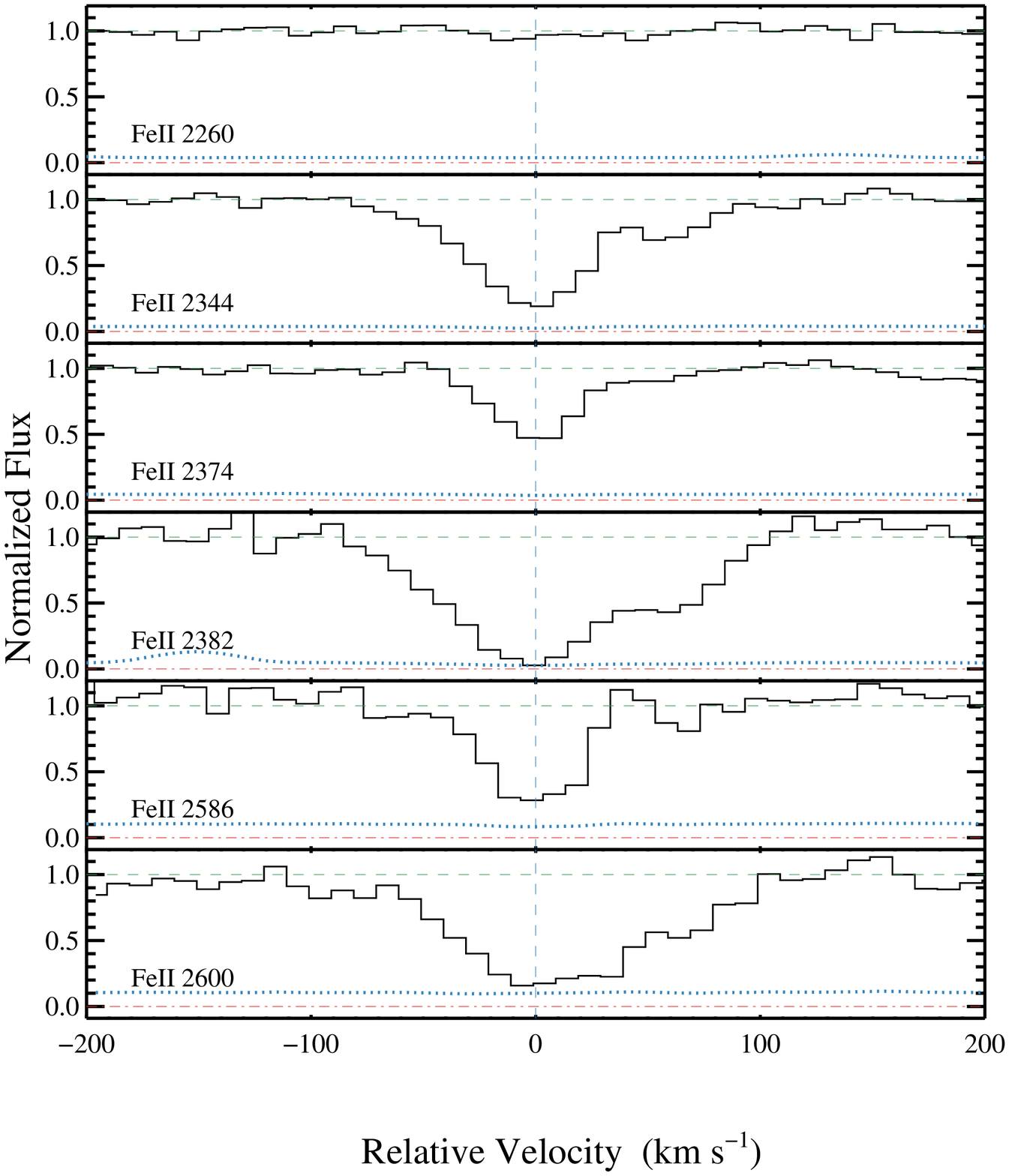}
\caption{Velocity plot of the ion transitions associated to the DLA J2123-0053 (13:H1) at  $z_{\rm dla}=2.7803$ (continued).
See Figure \ref{fig:vpdla01a} for an explanation of the different line colors.}
\end{figure*}

\clearpage
\begin{figure*}
\centering
\includegraphics[scale=0.8]{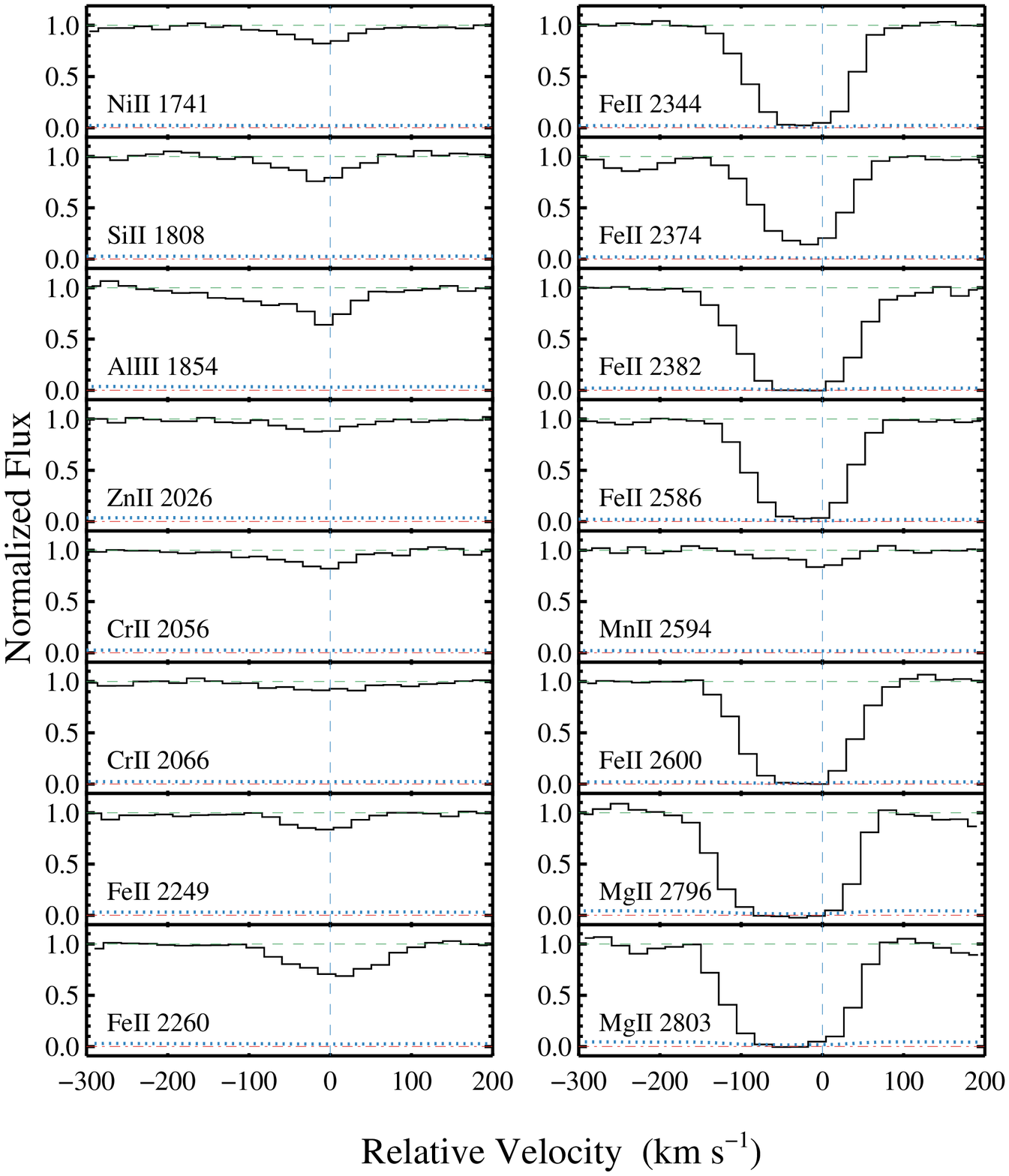}
\caption{Velocity plot of the ion transitions associated to the DLA J0407-4410 (14:H2) at   $z_{\rm dla}=1.9127$.
See Figure \ref{fig:vpdla01a} for an explanation of the different line colors.}
\end{figure*}

\clearpage
\begin{figure*}
\centering
\includegraphics[scale=0.8]{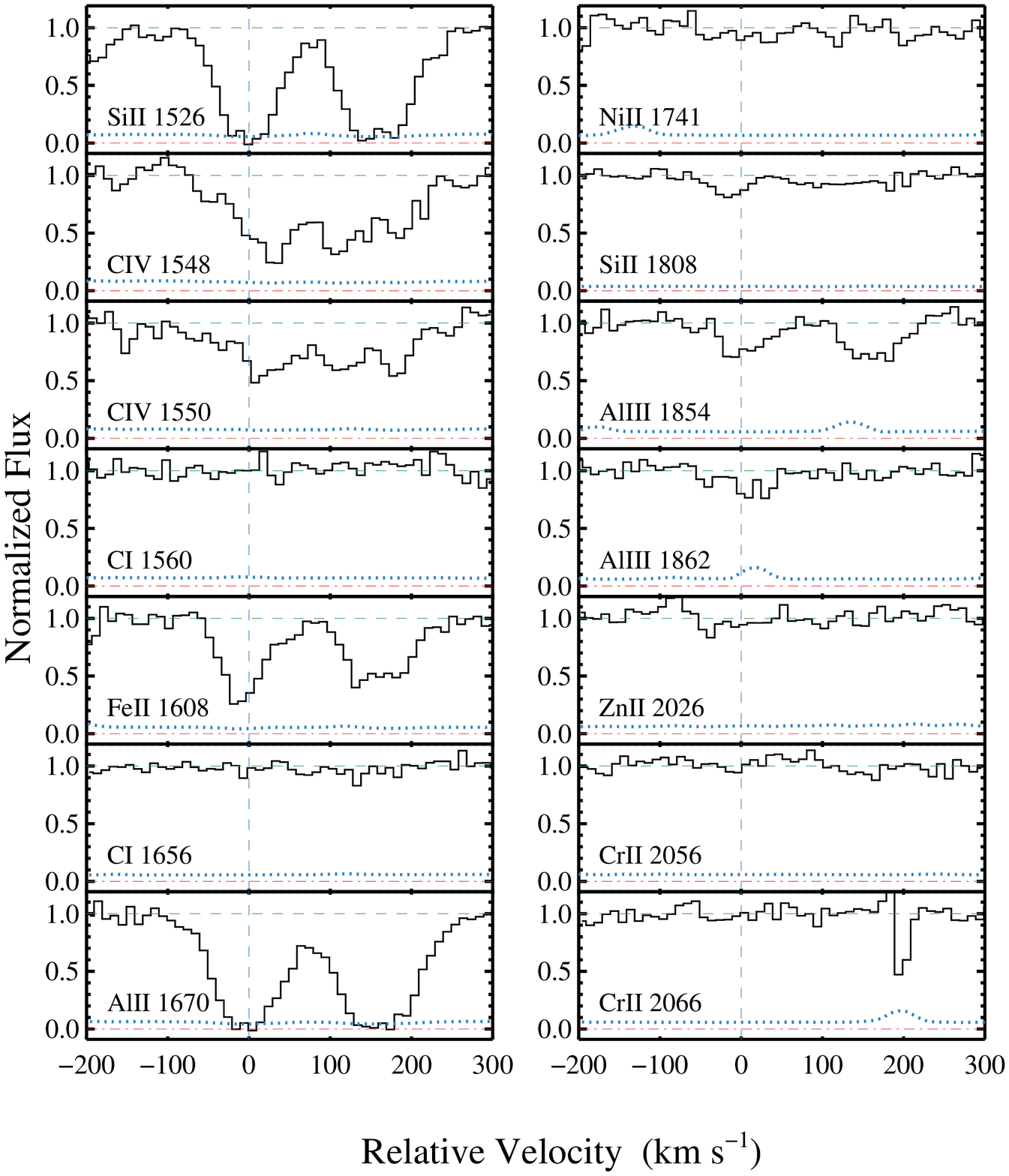}
\caption{Velocity plot of the ion transitions associated to the DLA J0255+0048 (15:H3) at   $z_{\rm dla}=3.2530$.
See Figure \ref{fig:vpdla01a} for an explanation of the different line colors.}
\end{figure*}

\clearpage
\begin{figure*}
\centering
\includegraphics[scale=0.8]{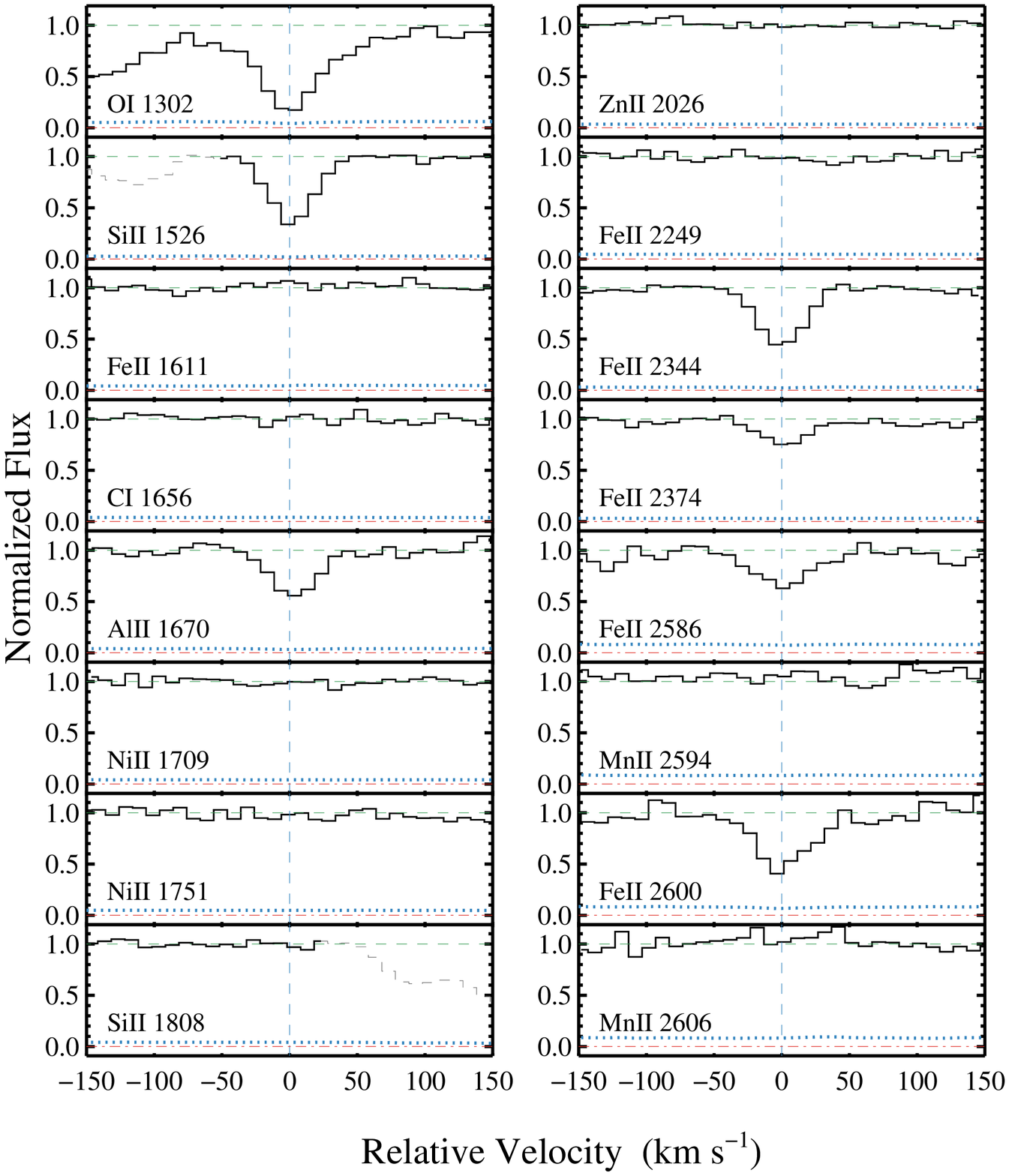}
\caption{Velocity plot of the ion transitions associated to the DLA J0816+4823 (16:H4) at  $z_{\rm dla}=2.7067$.
See Figure \ref{fig:vpdla01a} for an explanation of the different line colors.}
\end{figure*}

\clearpage
\begin{figure*}
\centering
\includegraphics[scale=0.8]{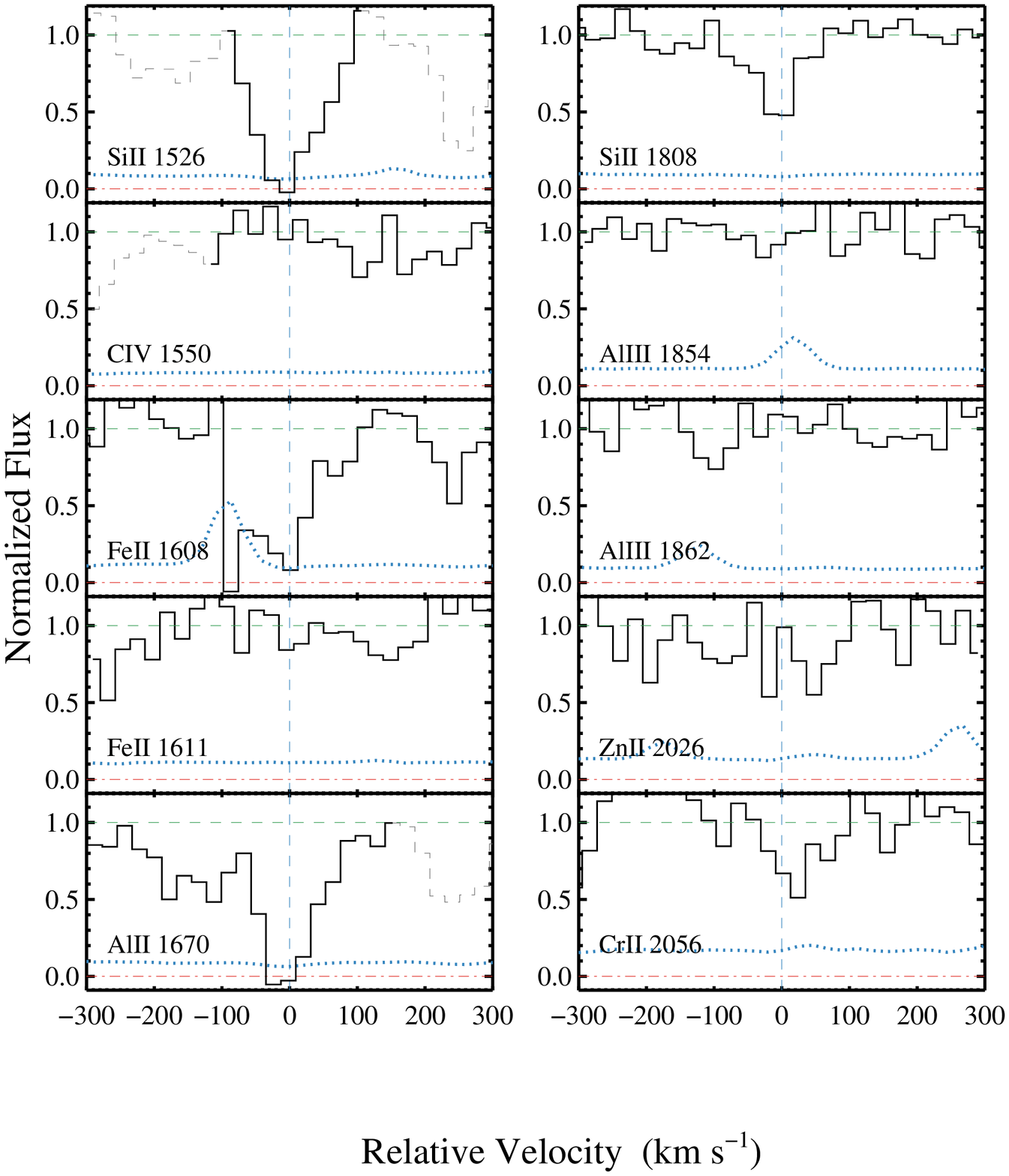}
\caption{Velocity plot of the ion transitions associated to the DLA J0908+0238 (18:H6) at  $z_{\rm dla}=2.9586$.
See Figure \ref{fig:vpdla01a} for an explanation of the different line colors.}
\end{figure*}

\clearpage
\begin{figure*}
\centering
\includegraphics[scale=0.8]{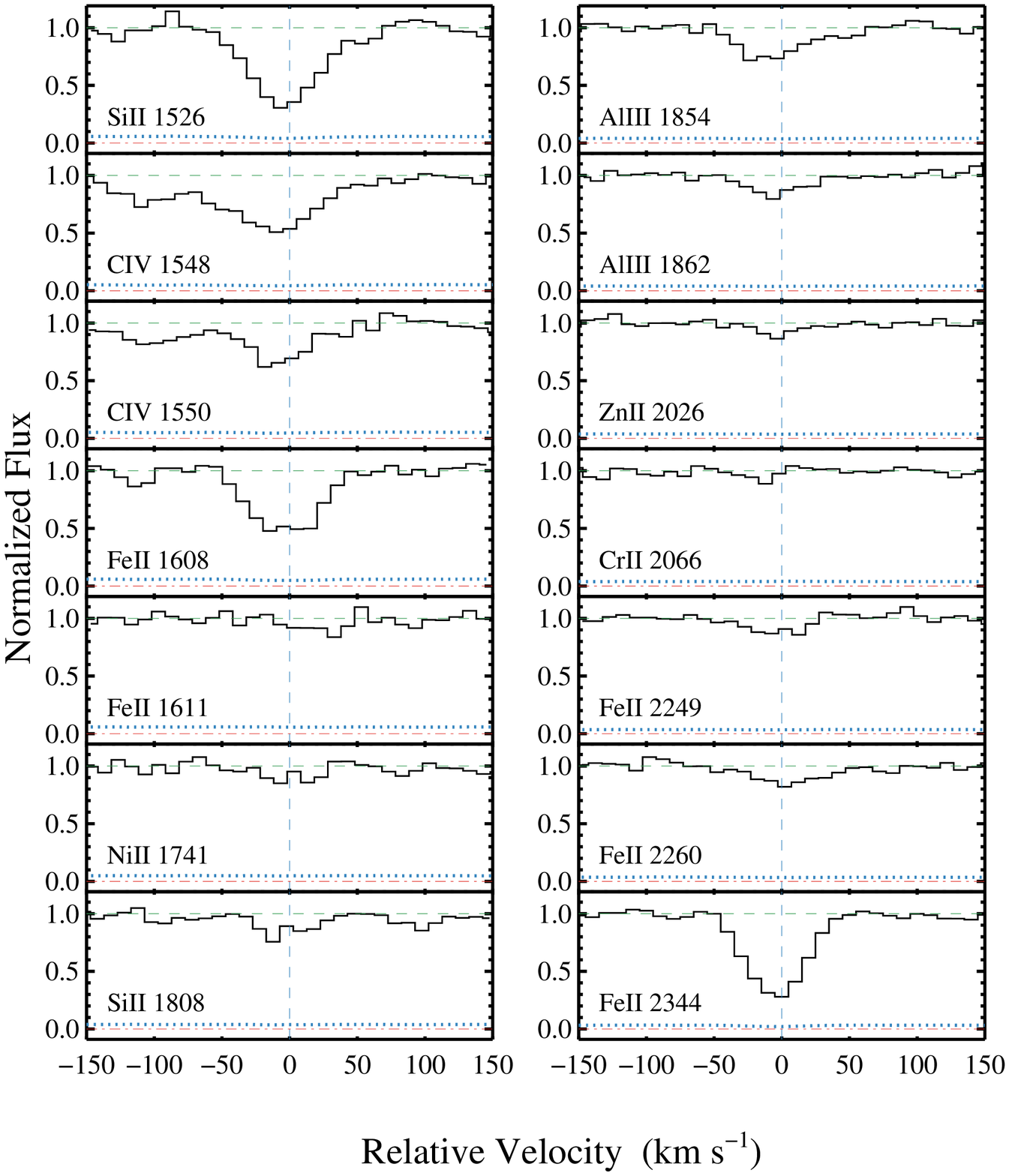}
\caption{Velocity plot of the ion transitions associated to the DLA J0844+1245 (21:H9) at   $z_{\rm dla}=1.8639$.
See Figure \ref{fig:vpdla01a} for an explanation of the different line colors.}
\end{figure*}

\clearpage
\begin{figure*}
\centering
\includegraphics[scale=0.8]{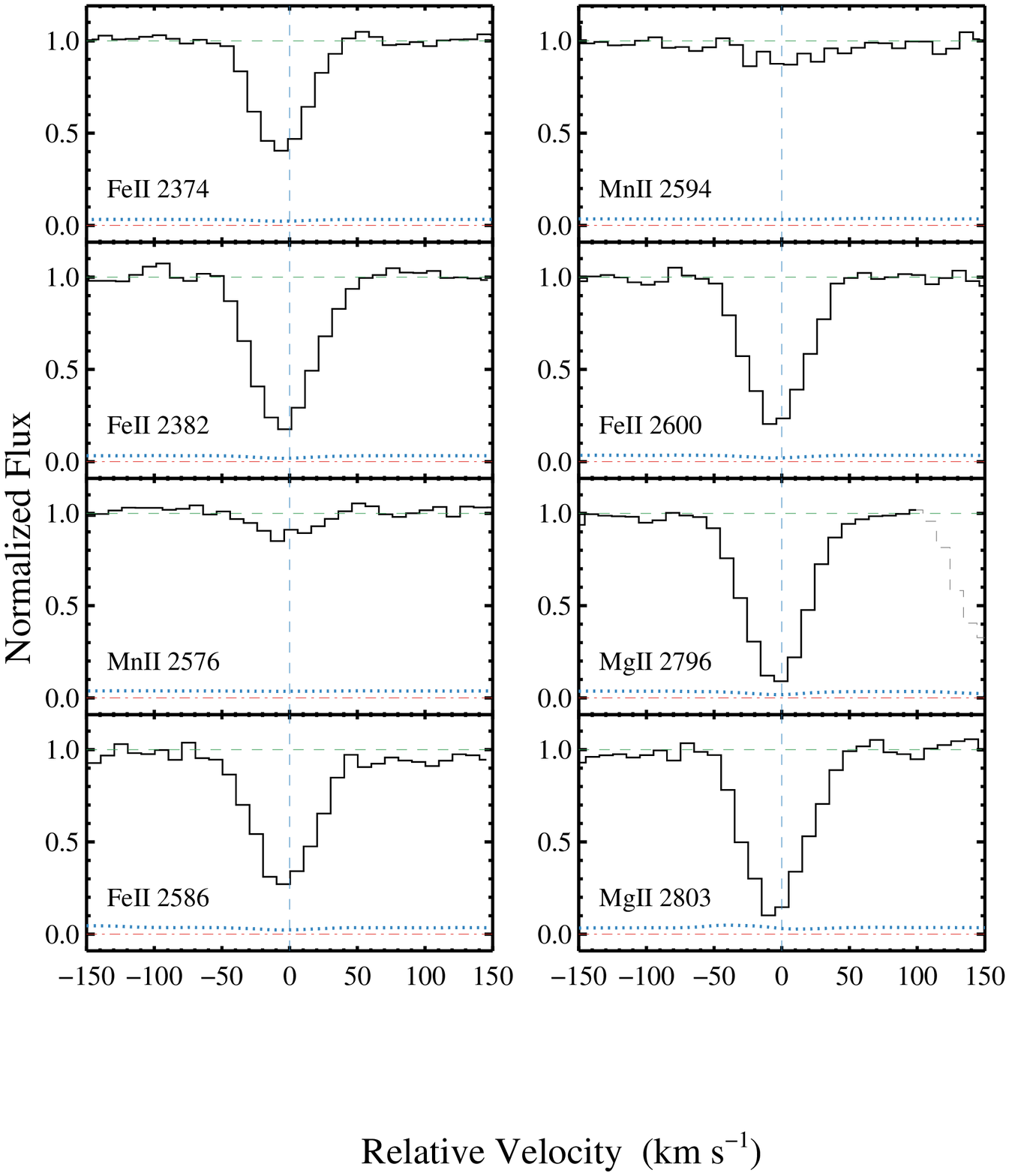}
\caption{Velocity plot of the ion transitions associated to the DLA J0844+1245 (21:H9) at   $z_{\rm dla}=1.8639$ (continued).
See Figure \ref{fig:vpdla01a} for an explanation of the different line colors.}
\end{figure*}

\clearpage
\begin{figure*}
\centering
\includegraphics[scale=0.8]{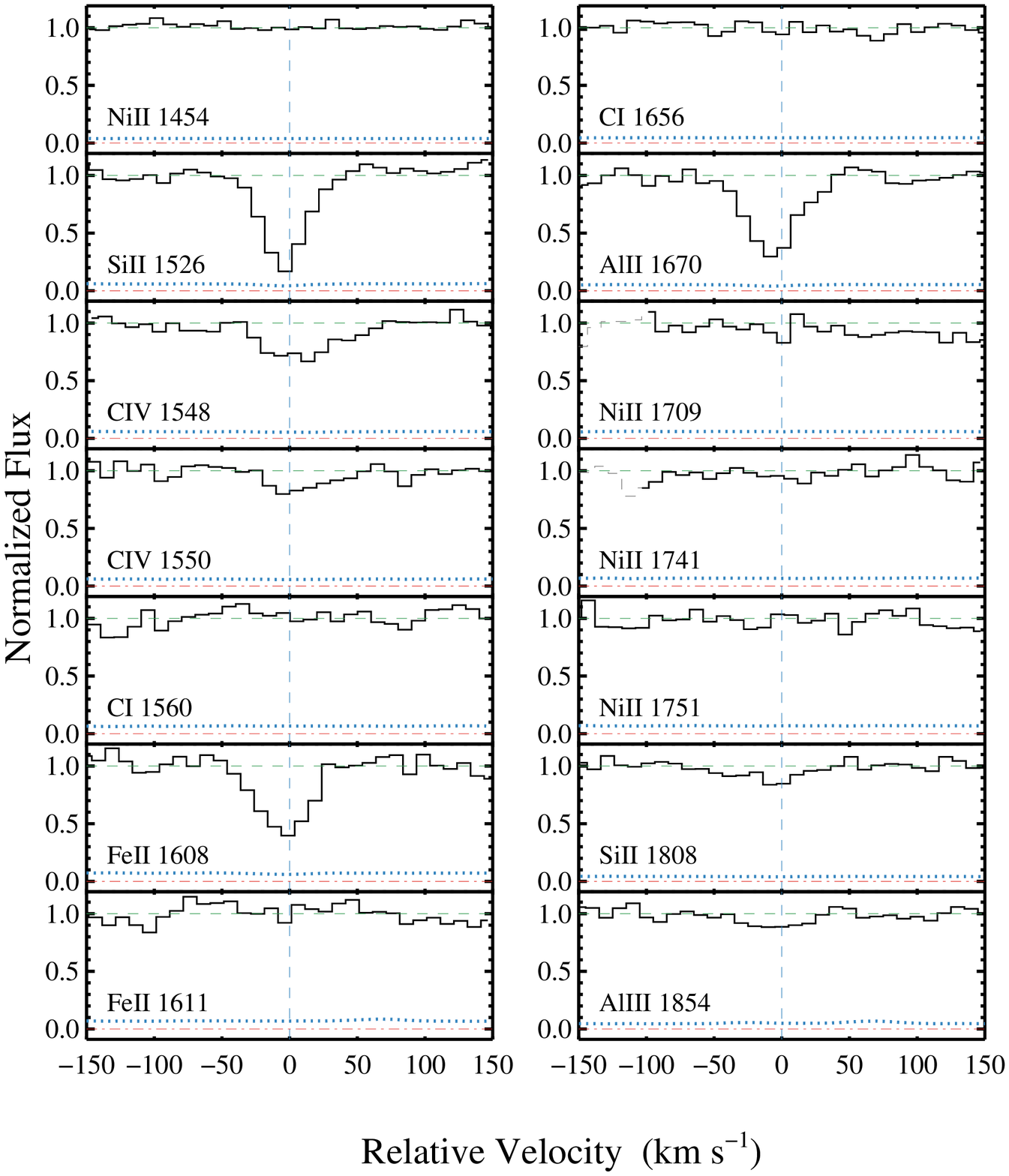}
\caption{Velocity plot of the ion transitions associated to the DLA J0751+4516 (22:H10) at   $z_{\rm dla}=2.6826$.
See Figure \ref{fig:vpdla01a} for an explanation of the different line colors.}
\end{figure*}

\clearpage
\begin{figure*}
\centering
\includegraphics[scale=0.8]{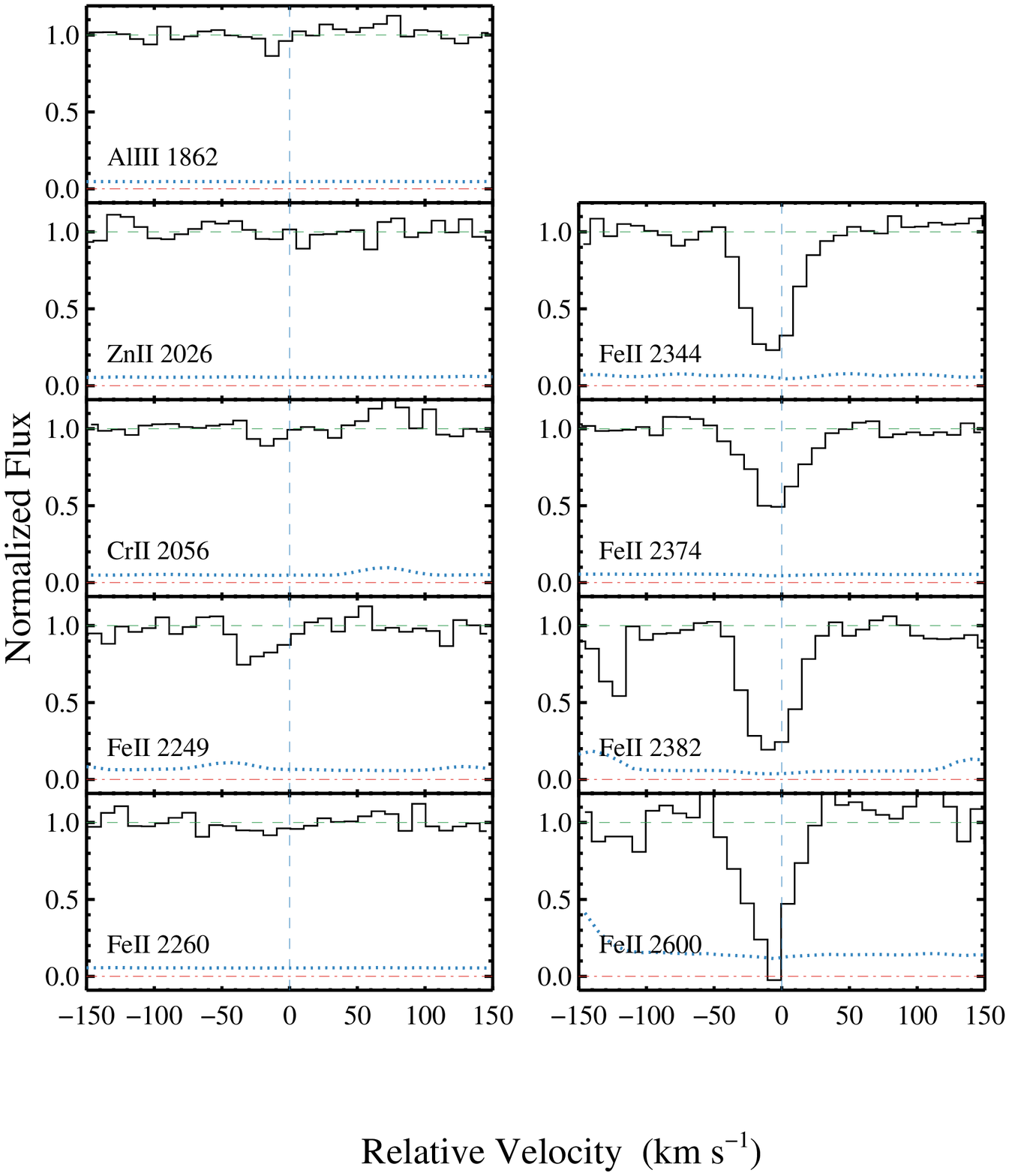}
\caption{Velocity plot of the ion transitions associated to the DLA J0751+4516 (22:H10) at   $z_{\rm dla}=2.6826$ (continued).
See Figure \ref{fig:vpdla01a} for an explanation of the different line colors.}
\end{figure*}

\clearpage
\begin{figure*}
\centering
\includegraphics[scale=0.8]{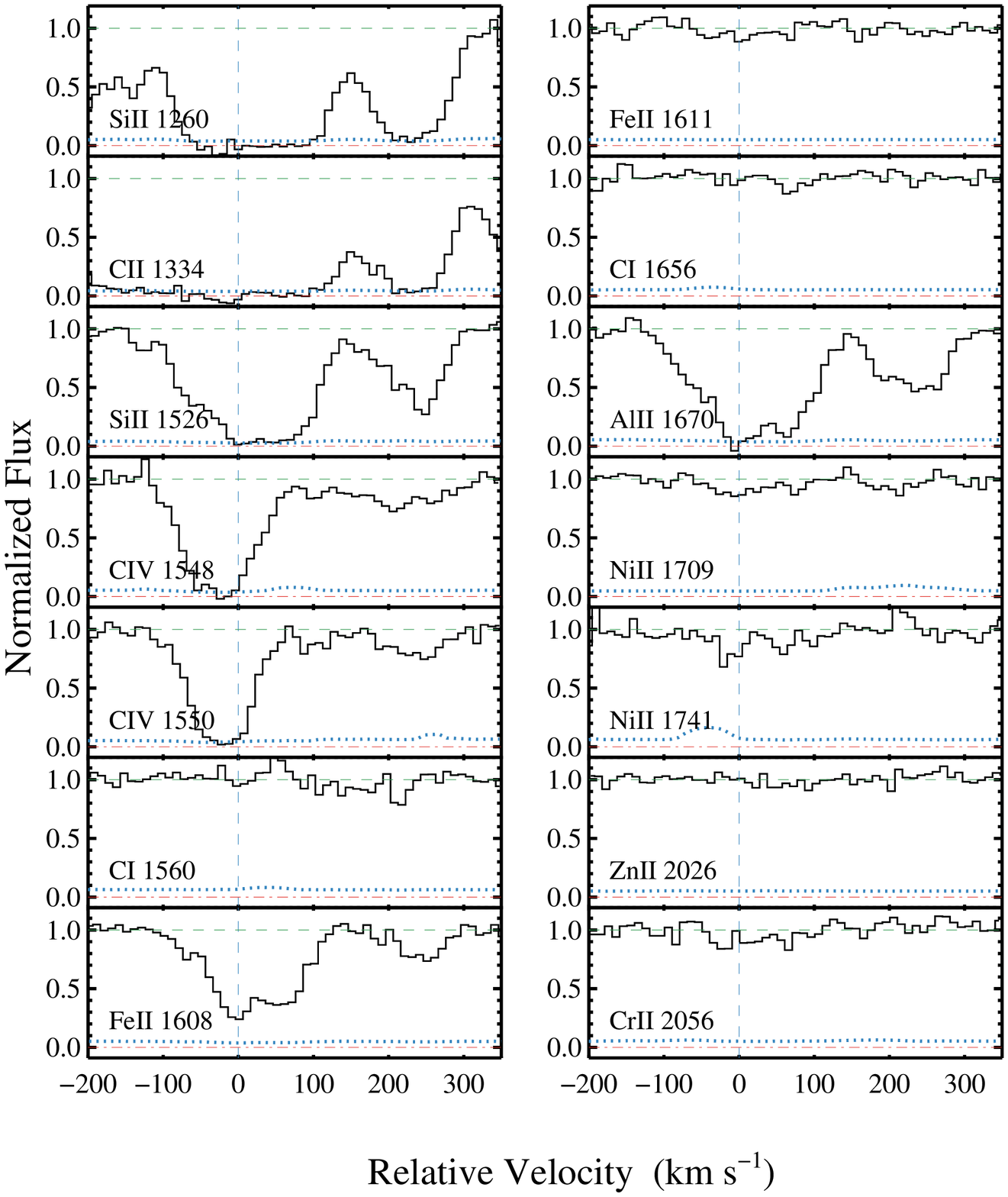}
\caption{Velocity plot of the ion transitions associated to the DLA J0818+0720 (23:H11) at   $z_{\rm dla}=3.2332$.
See Figure \ref{fig:vpdla01a} for an explanation of the different line colors.}
\end{figure*}

\clearpage
\begin{figure*}
\centering
\includegraphics[scale=0.8]{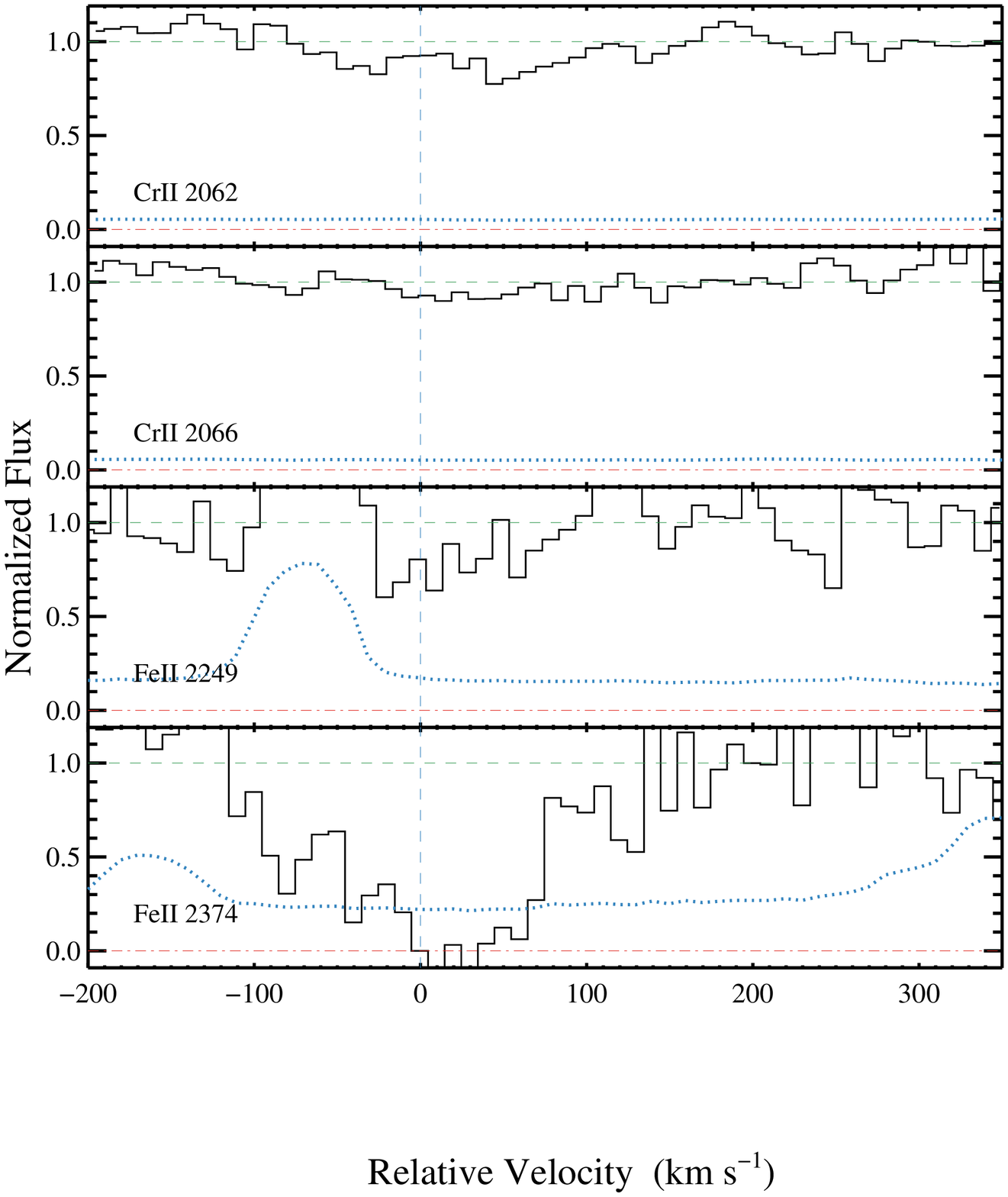}
\caption{Velocity plot of the ion transitions associated to the DLA J0818+0720 (23:H11) at   $z_{\rm dla}=3.2332$ (continued).
See Figure \ref{fig:vpdla01a} for an explanation of the different line colors.}
\end{figure*}

\clearpage
\begin{figure*}
\centering
\includegraphics[scale=0.8]{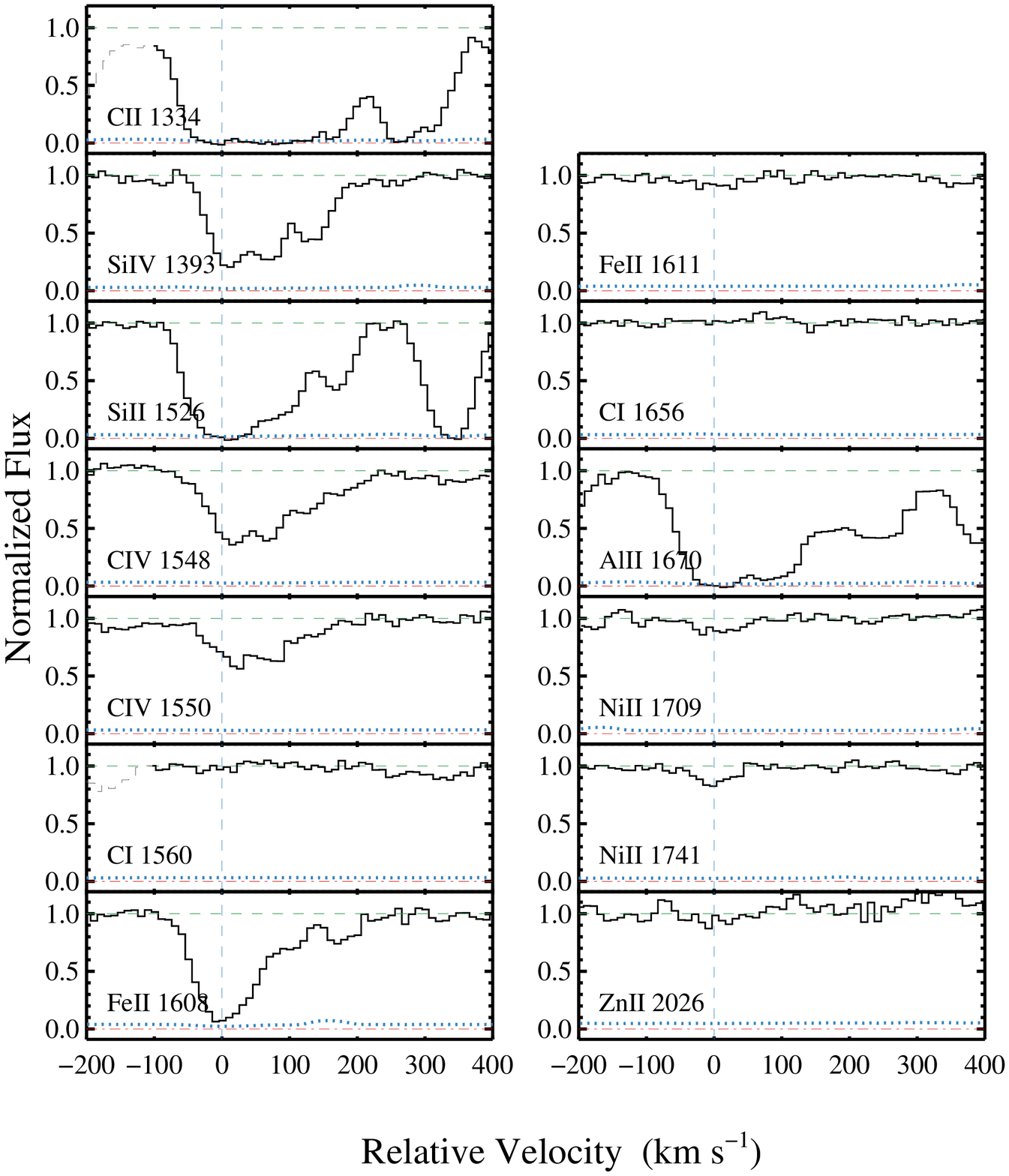}
\caption{Velocity plot of the ion transitions associated to the DLA J0818+2631 (24:H12) at   $z_{\rm dla}=3.5629$.
See Figure \ref{fig:vpdla01a} for an explanation of the different line colors.}
\end{figure*}

\clearpage
\begin{figure*}
\centering
\includegraphics[scale=0.8]{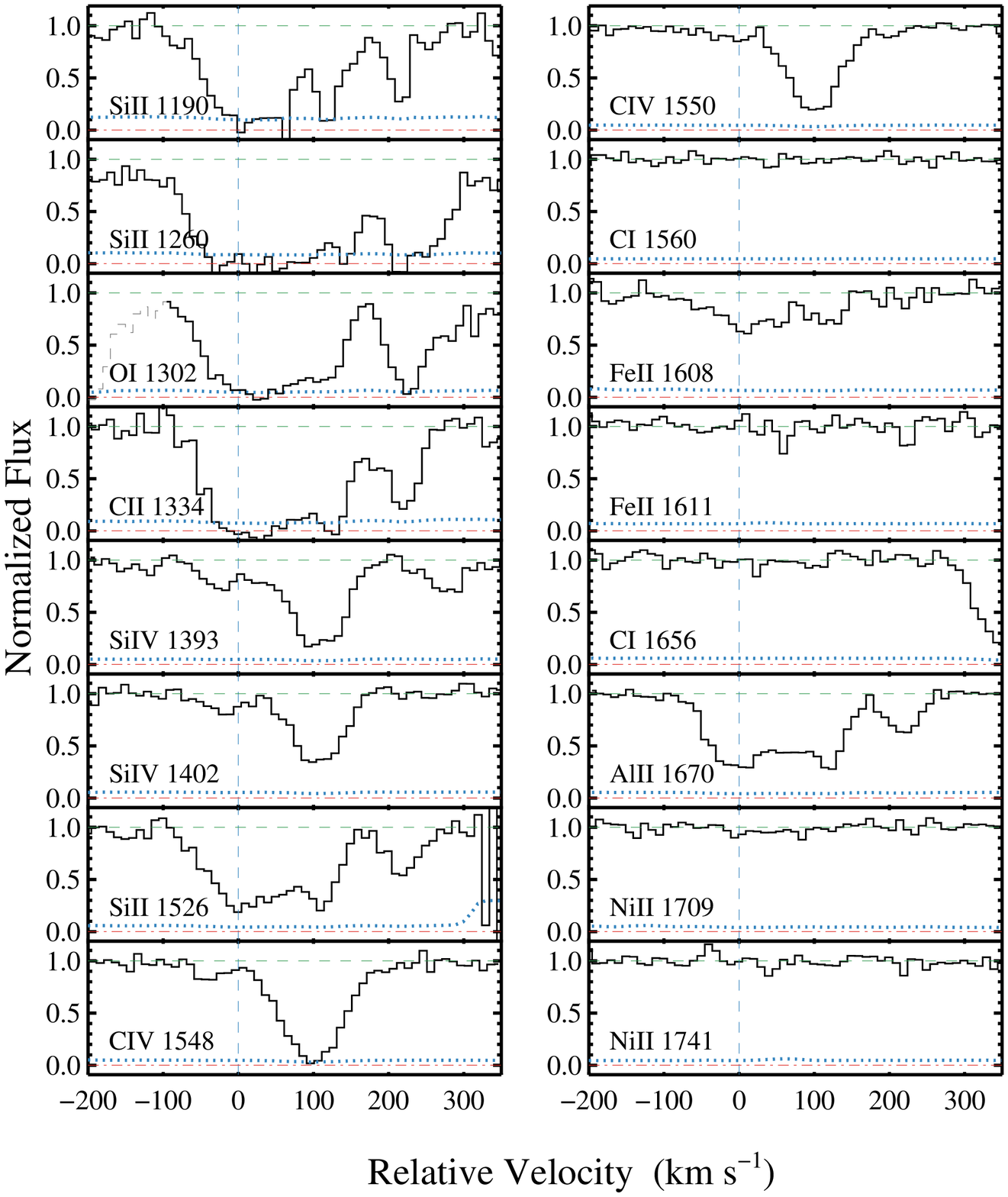}
\caption{Velocity plot of the ion transitions associated to the DLA J0811+3936 (25:H13) at  $z_{\rm dla}=2.6500$.
See Figure \ref{fig:vpdla01a} for an explanation of the different line colors.}
\end{figure*}

\clearpage

\end{document}